%% file: frion_thesis_ppgcosmo.tex
\documentclass{PPGCosmoThesis}
 
\usepackage{tensor}
\usepackage{physics}
\usepackage{subfig}
\usepackage{cite}
\DeclareUnicodeCharacter{2212}{-}

\begin{document}

% Provide the requested data below.

\title{Quantum effects in cosmology}
\author{Emmanuel Frion}
\thesisyear{2020}
\supervisor{Prof.~Nelson Pinto-Neto}
\supervisoraddress{Cosmo, Centro Brasileiro de Pesquisas Físicas (Brazil)}
\cosupervisor{Prof.~David Wands}
\cosupervisoraddress{ICG, University of Portsmouth (UK)}

\dedicatory{0} 
% The value of 0 does not print the dedicatory page. 
% Here is an example: \dedicatory{To whoever made this wonderful thesis template.}
% You may use more than one line. Just use \\ to break the lines.

\maketitle

%%% THE CHAPTERS
% Attention: check if the file names below match those in the \includeonly command.
% All the files must be saved with ".tex", but below the extension must be omitted.
\include{sections/chapter1}

\include{sections/chapter2}
\include{sections/chapter3}
\include{sections/chapter4}
\include{sections/chapter5}

\include{sections/chapter6}
\include{sections/chapter7}
\appendix
\include{sections/appendix1}

\include{sections/appendix2}
\include{sections/appendix3}
\include{sections/appendix4}

\include{sections/appendix5}

\pagebreak

%%% BIBLIOGRAPHY

\pagestyle{biblioHeader}
{
	\hypersetup{urlcolor=darkblue} 
	\bibliography{frionthesis}  % put the name of your bib file here.
}

\end{document}

%% file: sections/chapter1.tex
\chapter{Introduction}

Quantum effects play an essential role in modern cosmology. Perhaps the most striking example comes from large-scale structures, generally assumed to originate from vacuum quantum fluctuations and stretched by an expansion phase. Inflation is the leading paradigm in explaining this process. The various observational successes of inflationary models  drive the scientific community into elaborating more and more stringent tests, which can simultaneously be used to probe beyond the simple slow-roll, single field inflation. However, inflation is not a theory, since it does not arise from first principles, and going beyond inflation is a necessity. 

Among the difficulties inflation does not overcome, the easiest to notice is the presence of an initial singularity. Various alternatives and/or complementary mechanisms to inflation have been invoked in the literature. The best-known cosmological models endowed with the capacity of explaining large-scale observations while avoiding the singularity form a class called \textit{non-singular bouncing models}. The main features of these models differing from inflation are the presence of a contraction phase before expansion, and a never-vanishing scale factor. A non-singular bounce generally appears when quantum effects are part of the model, playing the role of a regulator leading to the avoidance of singularities.

This thesis is a compilation of the works realized conjointly at the Universidade Federal do Espírito Santo, Centro Brasileiro de Pesquisas Físicas and the Institute of Cosmology and Gravitation between October 2016 and September 2020, under the supervision of Nelson Pinto-Neto in Brazil and David Wands in the United Kingdom. We begin with the formulation of General Relativity using the ADM foliation of spacetime. Concretely, chapter \ref{basics GR} provides the basis of the modern cosmology picture presented in chapter \ref{modern cosmology} as well as the background necessary for understanding a Hamiltonian formulation of stochastic effects explored in chapter \ref{stochastic effects in cosmology}. Furthermore, it also serves as the basis for quantising general relativity, what will be discussed in chapter \ref{quantum cosmology} where we show that quantum effects in the de Broglie-Bohm interpretation leads to the resolution of the initial spacetime singularity. We also explore the astrophysical consequences of adding a non-minimal coupling of gravity with electromagnetism, and we show that the generation of magnetic fields is possible. Finally, the Hamiltonian formalism is also used to introduce the method of affine quantisation, a minimalist quantisation procedure, in chapter \ref{affine quantisation and cosmology}. We perform the quantum analysis of the Brans-Dicke Theory, the prototype of modified gravity theories, and we discuss the quantum equivalence of the Jordan and Einstein frames within this framework. We show that in both frames a smooth bounce is expected, and that equivalence between frames holds at the quantum level.

In this thesis, all quantities with a subscript $0$ are evaluated today. We use the unit systems $c=\hbar=1$, unless explicitly said otherwise.

%% file: sections/chapter2.tex
\chapter{Basics of General Relativity}
\label{basics GR}

The notion of energy is the basis of physics. Classical mechanics is conveniently expressed in terms of energy, represented by the Hamiltonian function, through the Hamiltonian formalism \cite{Landau:1960,Goldstein:2002}. However, the generalisation of the Hamiltonian formalism to general relativity (GR) is not trivial. We will show in this chapter how to accomplish this, starting from the Einstein-Hilbert action \cite{Einstein:1915by,Einstein:1915ca,Hilbert:1915tx}, cornerstone of GR. In its simplest form, it is expressed as
\begin{align}
\mathcal{S}_G=\frac{1}{16 \pi G} \int \textup{d}^{4}x  \sqrt{-g} \, R \;,
\label{Einstein-Hilbert action}
\end{align}
with $G$ is the Newton gravitational constant and $g$ the determinant of the spacetime metric. In the remainder of this chapter, we will use $16\pi G :=1$. $R$ is the Ricci scalar derived from
\begin{align}
R &:= g^{\mu \nu} R_{\mu \nu} \;, \\
R_{\mu \nu} &= 2 \tensor{\Gamma}{^\rho_\mu_[_\nu_,_\rho_]} + 2 \tensor{\Gamma}{^\rho_\lambda_[_\rho} \tensor{\Gamma}{^\lambda_\mu_]_\nu} \;, \\
\tensor{\Gamma}{^{\alpha}_{\mu}_{\nu}} &:= \frac{1}{2} g^{\alpha \lambda} \left(g_{\mu \lambda,\nu}+ g_{\lambda \nu,\mu}-g_{\mu  \nu, \lambda}\right)\;,
\end{align}
with square brackets meaning antisymmetric indices. In a Riemannian spacetime, $\mathcal{S}_G$ can be rewritten as an integral of first derivatives of $g_{\mu\nu}$ plus an integral of a total derivative
%\begin{align*}
%\left( \sqrt{-g} g^{\beta \nu} \right)_{;\alpha}= \left( \sqrt{-g} g^{\beta \nu} \right)_{,\alpha}  + \sqrt{-g} g^{\epsilon \nu} \tensor{\Gamma}{^{\beta}_{\alpha \epsilon}} +\sqrt{-g} g^{\beta \epsilon} \tensor{\Gamma}{^{\nu}_{\alpha \epsilon}} - \sqrt{-g} g^{\beta \nu} \tensor{\Gamma}{^{\lambda}_{\lambda \alpha}} = 0,
%\end{align*}
%
%where the last term comes from the covariant derivation of a tensor density.\\
%
%As may be seen in \eqref{1.2}, the action \eqref{1.1} depends on second derivatives of the metric. However, it can be rewritten as an integral of first derivatives of $g_{\mu\nu}$ plus an integral of a total derivative, as follows: 
%
\begin{eqnarray}
\mathcal{S}_G &=& \int d^4x \sqrt{-g} g^{\beta \nu} R_{\beta \nu} =\int d^4x \sqrt{-g} g^{\beta \nu} \left( \tensor{\Gamma}{^{\alpha}_{\beta \nu, \alpha}} - \tensor{\Gamma}{^{\alpha}_{\beta \alpha, \nu}} + \tensor{\Gamma}{^{\alpha}_{\alpha \epsilon}} \tensor{\Gamma}{^{\epsilon}_{\beta \nu}} - \tensor{\Gamma}{^{\alpha}_{\nu \epsilon}} \tensor{\Gamma}{^{\epsilon}_{\beta \alpha}} \right), \nonumber \\
&=& \int d^4x \left( \sqrt{-g} g^{\beta \nu} \tensor{\Gamma}{^{\alpha}_{\beta \nu}} -\sqrt{-g} g^{\beta \alpha} \tensor{\Gamma }{^{\lambda}_{\beta \lambda}} \right)_{,\alpha} - \int d^4x \left[ \left( \sqrt{-g}g^{\beta \nu} \right)_{,\alpha} \tensor{\Gamma}{^{\alpha}_{\beta \nu}} - \left( \sqrt{-g} g^{\beta \alpha} \right)_{,\alpha} \tensor{\Gamma}{^{\lambda}_{\beta \lambda}} \right] \nonumber \\
&\space& + \int  d^4x \sqrt{-g} g^{\beta \nu} \left( \tensor{\Gamma}{^{\alpha}_{\alpha \epsilon}} \tensor{\Gamma}{^{\epsilon}_{\beta \nu}} - \tensor{\Gamma}{^{\alpha}_{\nu \epsilon}} \tensor{\Gamma}{^{\epsilon}_{\beta \alpha}} \right) \;.
\end{eqnarray}
Let us define the following quantities :
\begin{align}
C^{\alpha} &:= \sqrt{-g}g^{\beta \nu} \tensor{\Gamma}{^{\alpha}_{\beta \nu}} - \sqrt{-g}g^{\beta \alpha} \tensor{\Gamma}{^{\lambda}_{\beta\lambda}},\\
\bar{S}  &:= \int  d^4x \sqrt{-g} g^{\beta \nu} \left( \tensor{\Gamma}{^{\alpha}_{\epsilon \beta}} \tensor{\Gamma}{^{\epsilon}_{\alpha \nu}} - \tensor{\Gamma}{^{\alpha}_{\alpha \epsilon}} \tensor{\Gamma}{^{\epsilon}_{\beta \nu}} \right) \;,
\end{align}
where, sometimes, $\bar{S}$ is called the \textit{Gamma-Gamma action}. Therefore the action could be rewritten in the form:
\begin{align}
\mathcal{S}_G = \bar{S} + \int d^{4} x \; \tensor{C}{^{\alpha}_{,\alpha}} \;.
\end{align}
Since the difference  $\mathcal{S}_G - \bar{S}$  is proportional to the integral of a total derivative, they both reproduce the vacuum Einstein field equations (EFE) by the application of the variational principle:
\begin{align}\label{1.7}
G_{\mu \nu} = R_{\mu \nu} - \frac{1}{2}g_{\mu\nu}R = 0 \quad  \rightarrow \quad R_{\mu \nu } = 0 \;.
\end{align}
Dirac gave a GR Hamiltonian formulation based on the $\bar{S}$ action and the definition of the momenta $\Pi^{\mu \nu} = \frac{\partial\cal{L}}{\partial(g_{\mu \nu,0})} $, since only first derivatives of the metric tensor appear in $\bar{S}$ \cite{Dirac:1950pj}. However, to get the final result, he had to use clever tricks and go through complicated calculations. We show in the remainder of this chapter a different Hamiltonian formulation, though related to Dirac's approach by a non-canonical transformation of phase space variables, in which the gravitational variables have a clear geometrical interpretations.

\section{ADM Variables}

In GR, space and time are treated on equal footing. However, the Hamiltonian formalism makes a clear distinction between these two notions, and considering them independently is necessary to express GR in a Hamiltonian form. This is done using the \textit{ADM formalism}, named after its authors Arnowitt, Deser and Misner, also called the \textit{3+1 foliation of spacetime}. One of the main reasons why the ADM decomposition is widely used is because the canonical variables defined in this framework have a clear physical and geometrical interpretation. This chapter is mainly based on the material found in references \cite{Arnowitt:1959ah,Gourgoulhon:2007ue,Hanson:1976cn,camacho2013geometric,Sundermeyer:1982gv,Kuchar} 

First, we need to choose a parameter describing the dynamical evolution of the system, acting as time. A physical state at a given $t$ is then a state on a three-dimensional space-like hypersurface defined by $t=$ constant. This implies that the spacetime $\mathcal{M}^4$ needs to be foliated in constant hypersurfaces, which is only possible if the topology of the $\mathcal{M}^4$ spacetime is a disjoint product
\begin{align}
{\cal{M}}^{4} = {\cal{M}}^{3}\otimes \mathbb{R} \;,
\end{align}
where $\mathcal{M}^3$ may have any topology. Therefore, this procedure cannot be applied to manifolds with a different topology from ${\cal{M}}^{3}\otimes \mathbb{R}$ (despite the fact that the manifest covariance of the theory would break), as for example, in Gödel-like spacetimes \cite{Godel:1949ga}.\footnote{ A detailed explanation about this kind of issues can be found in \cite{Schmut}.}

A well-defined constant hypersurface is expressed by the parametric equations
\begin{align}\label{1.8}
X^{\alpha} = X^{\alpha}(x^a) \;,
\end{align}
with Greek letters running from $0$ to $3$. Latin letters, running from $1$ to $3$, denote only spatial components. We can find the three tangent vectors to this hypersurface by deriving $X^{\alpha}$ along spatial directions, \textit{i.e.} we define  $X^{\alpha}_a := \tensor{X}{^\alpha _{,a}}$. Now, let us consider the unit vector  $n^{\alpha}$, normal to the hypersurface, such that 
\begin{eqnarray}\label{1.9}
g_{\alpha\beta} X^{\alpha}_{a}n^{\beta} &=& 0 \;, \\
g_{\alpha\beta} n^{\alpha}n^{\beta} &=& -1 \;. 
\end{eqnarray}
The set of constant hypersurfaces foliating spacetime is spanned by $X^\alpha = X^\alpha (x^a, t)$, with a different hypersurface for each $t$. The infinitesimal distance between two points with the same coordinate $x^{a}$ on two hypersurfaces $\Sigma (t)$ and $\Sigma (t + dt)$ separated by an infinitesimal time-lapse is represented by the deformation vector
\begin{align}
N^{\alpha} := \dot{X}^\alpha = \frac{\partial X^{\alpha}(x^a, t)}{\partial t} \;,
\end{align}
and can be decomposed on the $\{ n^{\alpha}, X^{\alpha}_{a} \}$ basis
\begin{align}\label{1.11}
N^{\alpha} = N n^{\alpha} + N^{a}X^{\alpha}_a \;.
\end{align}
The $N$ and $N^{a}$ functions are called respectively the \textit{lapse function} and the \textit{shift function}. Note that these functions appear because the time lines (to which $N^{\alpha}$ is tangent) need not be orthogonal to the hypersurface.

\begin{figure}[h!]
	\centering
	\includegraphics[scale=0.4]{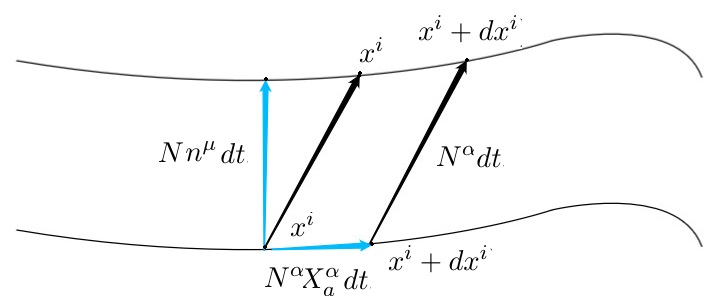}
	\caption[Sketch of two constant hypersurfaces]{Sketch of two constant hypersurfaces $t$ and $t+dt$. The vector $N^{\alpha}$ connects two points with the same coordinates $x^{i}$ on two adjacent hypersurfaces. The quantity $ds^{2}$ measures the interval between two events $(t,x^{i})$ and $(t+dt,x^{i}+dx^{i})$. $N^{\alpha}$ and $n^{\mu}$ are time-like vectors $(N_{\alpha} N^{\alpha}<0$ and $n^{\mu}n_{\mu}=-1<0)$. The time measured between two events by the observer $n^{\mu}$ is $d\tau $ and the spatial separation between them is $dl^{i}$. \label{fig1}}
\end{figure}

In order to determine the dynamics of a field in this decomposition, we have to consider its projections along the parallel and tangential directions to the hypersurfaces. Consequently, it is necessary to evaluate how the field components evolve when deformations on the hypersurfaces occur between $t$ and $t+dt$. Since $N^{\alpha}$ and $X_{a}^{\alpha}$ are tangent vectors to $t$ and $x^{a}$, respectively, we can use \eqref{1.11} to define the metric components. We then have
\begin{subequations}
	\begin{align}
	g_{ij}  &= g_{\alpha\beta}X^{\alpha}_{i}X^{\beta}_{j} := h_{ij} \;,\\
	g_{0i} &= g_{\alpha\beta}N^{\alpha}X^{\beta}_{i} := h_{ij}N^j  := N_i \;,  \\
	g_{00} &= g_{\alpha\beta}N^{\alpha}N^{\beta}= -N^2 + N^{a}N^{b}h_{ab}  = -N^2 + N^{a}N_{a} \;.  
	\end{align}
\end{subequations}
From these expressions, we can write the metric and its inverse as
\begin{subequations}\label{1.13}
	\begin{align}\label{1.13a}
	g_{\mu \nu}=\begin{pmatrix}
	-N^2 + N^{a}N_{a} &   N_i  \\
	N_j   & h_{ij}
	\end{pmatrix} \;,
	\end{align}
	\begin{align}\label{1.13b}
	g^{\mu \nu}=\frac{1}{N^2}\begin{pmatrix}
	-1 &   N^i  \\
	N^j  & h^{ij} N^2 - N^i N^j
	\end{pmatrix} \;,
	\end{align}
\end{subequations}
where $h^{ik} h_{kl}=\delta^{i}_{l}$. A geometrical interpretation of  \eqref{1.13a} using the line element invariance is sketched in figure \ref{fig1}.

The interval $ds^2$ measured by the observer $n^{\mu}$ is his proper   spatial interval, $dl^{2}=h_{ij}dl^{i}dl^{j}$, subtracted by his proper time interval, $d\tau^{2}$. Defining proper time as $\tau=N't$, with $N'$ a real function, the line element is
\begin{eqnarray}
\label{1.14}
g_{\mu \nu} dx^{\mu} dx^{\nu} := ds^2 &=& -d\tau^2 + h_{ij}dl^i dl^j \nonumber \\
&=& -N'^2 dt^2 + h_{ij}\left(N'^i dt + dx^i\right)		\left(N'^j dt + dx^j\right) \nonumber \\
&=&-\left( N'^2 - N'^iN'_i \right)dt^2  + 2N'_idx^idt + h_{ij}dx^idx^j 
\end{eqnarray}

Comparing \eqref{1.14} with the metric components  \eqref{1.13a} ($dx^{0}=dt$, in the units system where $c=1$), we see that $N' = N$ and $N'^i = N^i$. As a consequence, from \eqref{1.11}, \eqref{1.13a} and \eqref{1.14}, we see that $N dt=d\tau$. Also, we interpret $N^{i}$ as the variation between two points with the same coordinate $x^{i}$ on two constant hypersurfaces $t$ and $t+dt$. Using \eqref{1.11} to rewrite $n^{\alpha}$ in terms of $N^{\alpha} $ and $X^{\alpha}_{a}$, we obtain
\begin{align}
n^{\alpha} = \frac{1}{N}N^{\alpha} - \frac{N^{a}}{N}X^{\alpha}_{a} \;.
\end{align}
In the coordinate system  $(t,x^{i})$, where the respective curves for these coordinates are $N^{\alpha}$ and $X^{\alpha}_{a}$, the $n^{\alpha}$ vector \eqref{1.13a} takes the form
\begin{subequations}\label{1.15a}
	\begin{align}
	n^{\alpha} =& \left( \frac{1}{N} , -\frac{N^a}{N} \right) \;,\\
	n_{\alpha} =& \left( -N , 0 , 0 , 0 \right) \;. \label{1.15b}
	\end{align}
\end{subequations}

\begin{figure}[h!]
	\centering
	\includegraphics[scale=0.4]{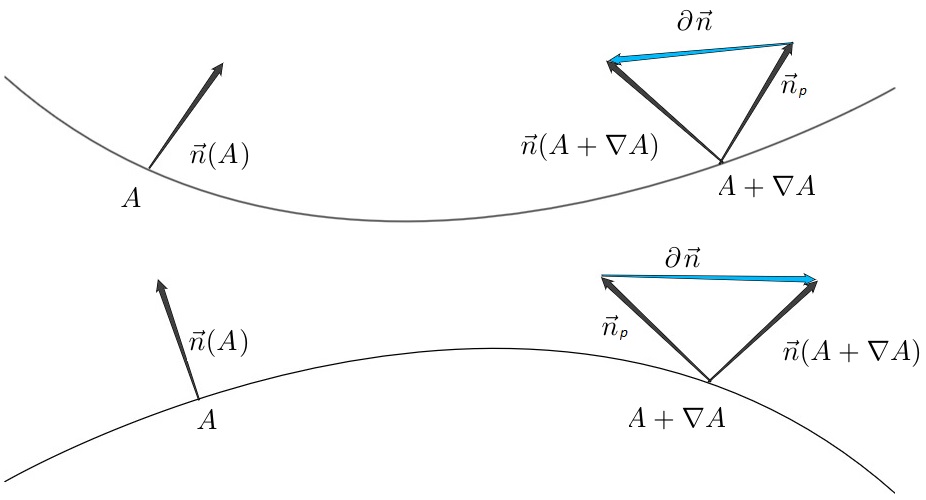}
	\caption[Hypersurfaces with different curvature]{Two hypersurfaces with the same metric $h_{ab}(\vec{x})$ but different curvatures. $\vec{n}$ is a normal vector to the hypersurface and $\delta \vec{n}$ is the difference between the vector $\vec{n}(A + \nabla A)$ and the vector $\vec{n}(A)$ parallel transported, using the connection $\Gamma^\alpha_{\mu \nu}$  from $A$ to $A + \nabla A$.}
	\label{fig2a}
\end{figure}

We now wish to express $g_{\mu \nu}$ and its inverse $g^{\mu \nu}$ as functions of the ten variables $N$, $N^{i}$ and $h_{ij}$ given by \eqref{1.13} in the Einstein-Hilbert action. Before doing so, it is necessary to characterise the curvature associated to hypersurfaces, which we recall are embedded in a four-dimensional manifold. This type of curvature is called \textit{extrinsic curvature}, or \textit{second fundamental form}. The vector normal to the hypersurface and its variation, depicted in figure \ref{fig2a}, will be our best tools to determine curvature.
Rigorously, the extrinsic curvature $K:T_p\Sigma \times T_p\Sigma \mapsto \mathbb{R}$ is a tensor mapping two vectors from the tangent plane $T_p\Sigma$ into a number, such that 
\begin{align}
K(u,v) =  -u \cdot \nabla_{v} \vec{n} \quad ; \quad \forall u,v \in T_p\Sigma \; 
\end{align}
Another, more straightforward definition is
\begin{align}\label{1.16}
K_{\mu \nu}:= -\frac{1}{2} \tensor{\perp}{^{\alpha}_{\mu}} \tensor{\perp}{^{\beta}_{\nu}} \nabla_{(\alpha} n_{\beta)} \;,
\end{align}
where
\begin{align*}
\tensor{\perp}{^{\alpha}_{\mu}} = \tensor{\delta}{^{\alpha}_{\mu}} + n^{\alpha}n_{\mu} \;,
\end{align*}
is the projector onto the three-dimensional hypersurface \footnote{The brackets stand for the symmetric combination $\nabla_{(\mu \nu)}=\left( \nabla_{\mu \nu} + \nabla_{\nu \mu} \right)$ }. A third definition consists in taking the projection of the $g_{\mu \nu}$ Lie derivative onto the hypersurface $h_{ij}(\vec{x})$ along the normal vector $n^{\alpha}$
\begin{align}\label{1.17}
K_{\mu \nu} = -\frac{1}{2} \tensor{\perp}{_{\mu}^{\alpha}} \tensor{\perp}{_{\nu}^{\beta}} \underset{n^{\lambda}}{\pounds} g_{\alpha\beta} \;,
\end{align}
since $\underset{n^{\lambda}}{\pounds} g_{\alpha\beta} = \nabla_{(\alpha} n_{\beta)}$.

In appendix \ref{section curvature components}, we show that, in our case, the extrinsic curvature takes the form
\begin{align}\label{1.22}
K_{ij}=-\frac{1}{2N}\left[\dot{h}_{ij} -N_{(i;j)}\right] \;,
\end{align}
while its inverse reads
\begin{align}
\label{1.23}
K^{ij}  = \frac{1}{2N}\left( \dot{h}^{ij} + N^{(i;j)} \right) \;.
\end{align}
Note the covariant derivatives are with respect to the hypersurface. We are now able to write the Einstein-Hilbert action  in terms of $K_{ij}, h_{ij} , N^{i}$ and $N$, all of them with well-defined geometrical meaning. In appendix \ref{section adm action}, we show explicitly that the Einstein-Hilbert action written in the ADM formalism takes the form
\begin{align} \label{2.1}
S = \int dt d^3x \, Nh^{\frac{1}{2}}\left(-K^2 + K_{ij}K^{ij} + \,^3R \right) \;,
\end{align}
after discarding surface terms, since they do not play any role in the equations of motion. Note that no time derivatives of the shift vector $N^{i}$ or of the lapse function $N$ appear\footnote{This fact is not a simple coincidence, indeed, $N$ and $N^{i}$ play the role of Lagrange multipliers for the Hamiltonian constraint and the momentum constraints.}. On the other hand, the time derivative of the spatial metric appears in the combination $K_{ij}K^{ij} - K^2$, and it can be interpreted as the kinetic part of the action, such that the unique variable responsible for  the dynamics of this theory is $h_{ij}$, while the spatial Ricci tensor plays the role of a potential term.

The equations of motion derived in appendix \ref{section equations of motion} have the form
\begin{align}\label{1.36}
-K^2 + K_{ij}K^{ij} - {^{3}R} &= 0 \;, \\
2h^{\frac{1}{2}} \left(K^{ij} - h^{ij}K  \right)_{;\, j} &= 0 \;, \label{1.38} \\
\dot{K}_{ij} - N\left( ^{3}{R_{ij}} + KK_{ij} -2\tensor{K}{_{i}^{m}} K_{mj}\right) + N_{,i;j} - N_{m;i}\tensor{K}{^{m}_{j}} - N_{m;j} \tensor{K}{^{m}_{i}} - N^{m}K_{ij;m} & = 0 \;. \label{1.41}
\end{align}
Looking at the above equations, we note that the ones related to the shift vector and the lapse function do not contain second derivatives; they are not dynamical in the proper sense and this clarifies our interpretation of them being Lagrange multipliers of constraint equations. Therefore,the two equations \eqref{1.36} and \eqref{1.38} tell us that the choice of hypersurfaces  $\Sigma (t) $ is not completely arbitrary, and each foil must have extrinsic curvature and intrinsic metric such that these constraints are satisfied. Such conditions ensure that $g_{\mu \nu}$ remains a solution of the EFE \eqref{1.7} when dynamically evolving.

Additionally, it is possible to show that if a given hypersurface $\Sigma (\bar{t})$, satisfying \eqref{1.36} and \eqref{1.38}, evolves according to \eqref{1.41}, then all the hypersurfaces $\Sigma (t), t > \bar{t}$, are still solutions of the Lagrange equations. Consequently, the set of equations \eqref{1.36}, \eqref{1.38} and \eqref{1.41} is redundant, and we actually are using more variables than required to explain the evolution of the system. Note that we can check equations \eqref{1.36}, \eqref{1.38} and \eqref{1.41} are indeed correct by comparing them to the projection of the EFE \eqref{1.7} along the orthogonal and tangential directions of constant hypersurfaces $t$, explicitly given by the conditions
\begin{align}
G_{\mu \nu}n^{\mu}n^{\nu} = 0, \qquad G_{\mu \nu}n^{\mu} \tensor{\perp}{^{\nu}_{\alpha}} = 0, \qquad G_{\mu \nu} \tensor{\perp}{^{\mu}_{\alpha}}\tensor{\perp}{^{\nu}_{\beta}}=0 \;.
\end{align}

\section{The Hamiltonian Formulation of General Relativity}

Up until now, we have developed the technology for foliating spacetime, what led us to a Lagrangian density in terms of ADM variables
\begin{align} \label{lagrangian gr}
\mathcal{L} \left(N,N^{i},h_{ij},\dot{N}, \dot{N}^{i}, \dot{g}_{ij},t\right) = N h^{\frac{1}{2}}\left( K_{ij} K^{ij} - K^{2} +\,^3R \right) \;.
\end{align}
Before going further, let us recall that in order to find an Hamiltonian density reproducing the correct field equations, we must perform a Legendre transformation with additional constraints
\begin{align}
\label{legendre transformation}
\mathcal{H}\left(q^{i},p_i,t\right) = \sum_{i} \dot{q}^{i} p_i - \mathcal{L} \left(q^{i},\dot{q}^{i},t\right) + \sum_{k=0}^{m} \lambda^{k} \phi_k \left(q^{i},p_i\right)\;,
\end{align}
for generalised velocities $q^{i}$ and momenta $p_i$ defined by
\begin{align}
p_i := \frac{\partial \mathcal{L}}{\partial \dot{q}^{i}} \;.
\end{align}
Each constraint is weakly zero, \textit{i.e.} $\phi_k \simeq 0$ for all $k$. This means that even though we have $\phi_k=0$ for each $k$, the Poisson brackets defined by
\begin{equation}
\label{2.10}
\left\lbrace A(x),B(x') \right\rbrace_{t=constant} := \int d^{3}z \left( \frac{\delta A(x)}{\delta q^{a}(z)} \frac{\delta B(x')}{\delta p_{a}(z)}-\frac{\delta A(x)}{\delta p_{a}(z)} \frac{\delta B(x')}{\delta q^{a}(z)} \right)
\end{equation}
applied to $\phi_k$ and any function depending on the canonical variables, might be different from zero. This type of constraint is named \textit{primary constraint}. Note that primary constraints appear when
\begin{align}
\textup{det} \left(\frac{\partial p_i}{\partial \dot{q}^{j}}\right) = \textup{det} \left(\frac{\partial^2 \mathcal{L}}{\partial \dot{q}^{i} \partial \dot{q}^{j}}\right) =0 \;.
\end{align}
Strictly speaking, the coefficients $\lambda^k$ in front of the primary constraints in \eqref{legendre transformation} are Lagrange multipliers. Hence, for the system to be classically consistent, each contraint must be conserved in time, so that
\begin{align}
\dot{\phi}_k \left(q^{i},p_i\right) = \left\lbrace \phi_k, H_T\right\rbrace =0 \;,
\end{align}
with $H_T$ the plain, total Hamiltonian.

In GR, the canonically conjugated momenta for the canonical variables $N$, $N^{i}$ and $h_{ij}$ are deduced from \eqref{lagrangian gr} \footnote{In this thesis, we use $P$ to denote classical variables leading to constraints, $\Pi$ for classical variables governing the dynamics, and $\pi$ for their quantised versions.}
\begin{subequations}
	\label{2.2}
	\begin{eqnarray}
	P &:= & \frac{\partial \mathcal{L}}{\partial \dot{N}}=0 \;, \label{2.2a}\\
	P_i &:= & \frac{\partial \mathcal{L}}{\partial \dot{N}^i}=0 \;, \label{2.2b} \\
	\Pi^{ij} &:=&   \frac{\partial \mathcal{L}}{\partial \dot{h}_{ij}}= \frac{\partial \mathcal{L}}{\partial K_{ab}} \frac{\partial K_{ab}}{\partial \dot{h}_{ij}}
	=-\frac{1}{2N}\frac{\partial \mathcal{L}}{\partial K_{ab}} \delta^{ij}_{ab}
	=-\frac{1}{2N}\frac{\partial \mathcal{L}}{\partial K_{ij}}\nonumber \\
	&=&-h^{\frac{1}{2}}\left(K^{ij}-h^{ij}K \right) \;.\label{2.2c}
	\end{eqnarray}
\end{subequations}
To compute equation \eqref{2.2c}, we used equations \eqref{1.22} and \eqref{1.34} as well as the fact that $\mathcal{L}$ depends on $\dot{h}_{ij}$ only through $K_{ij}$.
Note that it is impossible to get the generalized velocities $\dot{N}_{i}$, $\dot{N}^{i}$ in terms of momenta from  \eqref{2.2}. In fact, equations \eqref{2.2a} and \eqref{2.2b} are constraints of the form $\Phi_{m}(q,p)=0$, that $P$ and $P_{i}$ must satisfy. %To take care of this problem, Dirac proposed a solution \cite{Dirac1,Dirac2,Dirac3}

Applying the Legendre transform, we obtain the canonical Hamiltonian density
\begin{align}
\label{2.32}
\mathcal{H}_{C}&=P\dot{N}+P_{i}\dot{N}^{i}+\Pi^{ij}\dot{h}_{ij}-\mathcal{L} \\
&=P\dot{N}+P_{i}\dot{N}^{i}+\Pi^{ij}\dot{h}_{ij}- Nh^{\frac{1}{2}}\left( K_{ij}K^{ij}-K^{2}+\,^3R \right) \;.
\end{align}
Plugging the different momenta in \eqref{2.32} as done in appendix \ref{section hamiltonian gr}, the total Hamiltonian is
\begin{align}
H_{T}&= \int d^{3}x \mathcal{H}_{C} + \int d^{3}x \left(\lambda P+\lambda^{i}P_{i} \right) \nonumber \\
&= \int d^{3}x \left( N\mathcal{H}_{0}+N_{i} \mathcal{H}^{i}+\lambda P+\lambda^{i}P_{i} \right) \;,
\label{total hamiltonian}
\end{align}
with
\begin{subequations}
	\begin{align}
	\mathcal{H}_{0} &= G_{ijkl} \Pi^{ij} \Pi^{kl}-h^{\frac{1}{2}} \,^3R \;,  \label{h0 constraint}\\
	\mathcal{H}^{i} &= -2 \tensor{\Pi}{^{ij}_{;j}} \;. 
	\end{align}
\end{subequations}
and the \textit{DeWitt metric}
\begin{align}
G_{ijkl} = \frac{1}{2} h^{-\frac{1}{2}} \left( h_{ik}h_{jl}+h_{il}h_{jk}-h_{ij}h_{kl}\right) \;.
\end{align}
We verify explicitly in appendix \ref{section hamiltonian gr} that all constraints in GR are first-class constraints. In particular, we show that
\begin{align}
\label{gr constraints}
\dot{P} &= -\frac{\partial \mathcal{H}}{\partial N} =  \mathcal{H}_0 \simeq 0 \;, \\
\dot{P}_i &= -\frac{\partial \mathcal{H}}{\partial N^{i}} =  \mathcal{H}_i \simeq 0 \;,
\end{align}
implying that $N$ and $N^{i}$ are Lagrange multipliers. More precisely, $\mathcal{H}_0 \simeq 0$ comes from the fact that GR is invariant under time re-parametrisation, while $\mathcal{H}_i \simeq 0$ reflects invariance under diffeomorphisms. In the end, $\Pi^{ij}$ and $h_{ij}$ are the only dynamical variables of the system.

It is straightforward to show that when a scalar field is present in the action, the total Hamiltonian \eqref{total hamiltonian} is modified with the addition of the terms \cite{Langlois:1994ec}
\begin{align}
\mathcal{H}_0^{\varphi} &= \frac{1}{2} h^{-\frac{1}{2}} \Pi^2_{\varphi} + \frac{1}{2} h^{\frac{1}{2}} h^{ij} \varphi_{,i} \varphi_{,j} + h^{\frac{1}{2}} V(\varphi) \;, \\
\mathcal{H}_i^{\varphi} &= \Pi_{\varphi} \varphi_{,i} \;.
\end{align}
Then, the Hamilton equations for a scalar field read \cite{Grain:2017dqa}
\begin{align}
\dot{\varphi} &= N h^{-\frac{1}{2}} \Pi_{\varphi} + h^{ij} N_i \varphi_{,j} \;, \\
\dot{\Pi}_{\varphi} &= -N h^{\frac{1}{2}} V_{,\varphi} + \left(Nh^{\frac{1}{2}} h^{ij} \varphi_{,j}\right)_{,i} + h^{ij} N_i \Pi_{\varphi,j} + \Pi_{\varphi} h^{ij} N_{i,j} \;.
\label{scalar field adm}
\end{align}
The Hamiltonian formulation of a scalar field will be useful to introduce the stochastic formalism in chapter \ref{stochastic effects in cosmology}.

%% file: sections/chapter3.tex
\chapter{Modern Cosmology}
\label{modern cosmology}

\section{Era of High-Precision Cosmology}

In this chapter, we give an overview of our current understanding of the Universe. The present technology has made possible observations on very large scales and at very early times ($\approx 280.000$ light-years), revealing isotropy on scales of order $\approx \order{100 \textup{Mpc}}$. The main evolution of the Universe is described by the Hot Big Bang model, but the high degree of homogeneity and isotropy in the sky cannot be explained by this model. This motivated the introduction of cosmological inflation, a phase of rapid expansion in the very early Universe, that we will describe in this chapter. The material used in this chapter is based on \cite{Vennin:2014lfa, Weinberg:2008zzc, Piattella:2018hvi, Mukhanov:2005sc,Hardwick:2019zee,Dodelson:2003ft}.

\subsection{Homogeneity and Isotropy}

Cosmological data indicate there is a very high degree of homogeneity and isotropy on very large scales. While conceptually difficult to comprehend, this is convenient for cosmologists. Indeed, a complete description of Nature with mathematical tools is a very hard (if feasible at all!) task. From the material developed in the previous chapter, we find that a homogeneous and isotropic cosmological model is obtained by simply putting $N_i=0$ and $h_{ij}=a^2(t) \delta_{ij}$ in \eqref{1.14}. We obtain the metric familiar to all cosmologists:
\begin{align}
\textup{d}s^2= 	-N^2(t)\textup{d}t^2+a^2(t)\left[\frac{\textup{d}r^2}{1-kr^2}+r^2\textup{d}\theta^2+r^2 \sin^2\theta \textup{d}\phi^2\right] \;.
\label{flrw adm}
\end{align}
This is the Friedmann-Lemaître-Robertson-Walker (FLRW) metric in spherical coordinates and arbitrary time, with the signature convention $\left(-,+,+,+\right)$.  While physical results do not depend on the choice of signature, it is worth noting that the usual convention in quantum field theory (QFT) is $\left(+,-,-,-\right)$.  Several choices are possible for the lapse function. Choosing $N=1$ means we are working in cosmic time, $N=a$ is for conformal time, and with $N=1/H$ (where we defined the \textit{Hubble rate} $H=\dot{a}/a$), we are using the number of e-folds. We shall consider $N=1$ for the remainder of this chapter, thus we obtain
\begin{align}
\textup{d}s^2= -\textup{d}t^2+a^2(t)\left[\frac{\textup{d}r^2}{1-kr^2}+r^2\textup{d}\theta^2+r^2 \sin^2\theta \textup{d}\phi^2\right] \;.
\label{flrw}
\end{align}
The FLRW metric reflects clearly the symmetries of spacetime. The global geometry is contained in the parameter $k=\left\lbrace -1,0,1 \right\rbrace$, with each value describing a hyperbolic, Euclidean or elliptic universe, respectively. The dynamics is driven by the scale factor, $a(t)$, which is the quantity allowing for an expansion or a contraction of the spatial hypersurfaces. 

Modern cosmology makes use of the FLRW metric with flat spatial curvature, \textit{i.e.} $k=0$. Indeed, data from the combined datasets of Planck mission and baryonic acoustic oscillations give strong hints that the spatial curvature is almost null \cite{Akrami:2018odb}. However, some results combining Planck data with supernovae or with giant red stars find a negative value for $k$ is preferred, indicating the Universe could possibly be closed \cite{DiValentino:2020hov}.

\subsection{Redshift}

Most of the information in cosmology comes from the light we receive. From the metric \eqref{flrw}, we deduce that the radial path ($\textup{d}\theta=\textup{d}\phi=0$) traveled by a photon ($\textup{d}s=0$ since photons follow null geodesics) in a homogeneous and isotropic spacetime is given by
\begin{align}
\frac{\textup{d}t}{a(t)} =  \frac{\textup{d}r}{\sqrt{1-kr^2}} \;.
\end{align}
Thus, a photon travelling in spacetime will see its frequency changing in time, because of the time dependence of the scale factor. We say the photon will be \textit{redshifted}. The redshift is defined as the difference in wavelengths from photons of the same source received by the same observer at two different times. It is generally expressed as
\begin{align}
1+z  := \frac{\lambda_0}{\lambda_1} = \frac{a(t_0)}{a(t_1)} \;.
\label{redshift}
\end{align}
Then, the redshift is equivalently defined as the ratio between scale factors at two different times. This implies that a photon, and in fact, any particle, loses momentum as $p \propto a^{-1}$ when travelling in a FLRW universe. In turn, all massive particles slow down asymptotically, and eventually stop. This allows to define \textit{comoving coordinates}, representing a frame where the particle follows the evolution of the geometry, and possesses a proper velocity allowing it to be at rest.

\subsection{Cosmological Observations}

\begin{figure}[h!]
	\centering
	\includegraphics[scale=0.4]{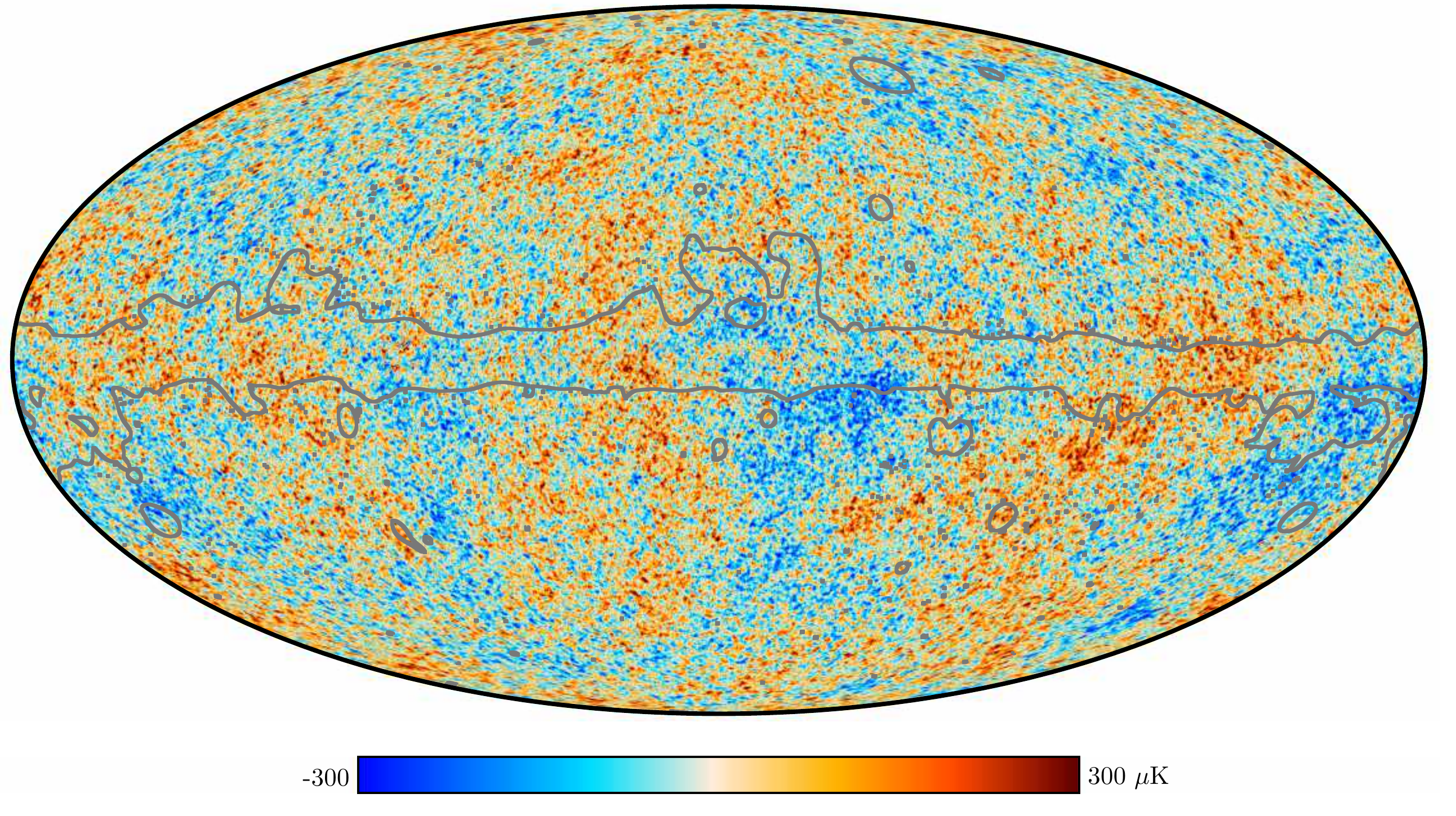}
	\caption[Planck 2018 temperature anisotropies map]{The 2018 Planck map of the temperature anisotropies of the CMB\cite{Akrami:2018odb}. Blue and red spots represent respectively colder and warmer deviations from the mean temperature. The gray outline shows the extent of the confidence mask.}
	\label{cmb sky}
\end{figure}

The Planck mission offers the most recent temperature map of the observable sky, shown in figure \ref{cmb sky}. This map is a picture of what is called the \textit{Cosmic Microwave Background} (CMB). The CMB can be described by an overall black body radiation with an average temperature of 2.7 K, with the strongest radiation in the microwave band, hence the name.
The blue and red spots outline the presence of temperature inhomogeneities. Temperature variations in the CMB sky for multipoles higher than the dipole are believed to be the product of density perturbations before recombination. On the one hand, Thomson scattering between electrons and photons induced radiative pressure. On the other hand, gravitation caused overdensities to form. These overdensities are composed of dark matter, baryons and photons, and are the seeds of large-scale structures observed today. The interplay between gravitation and pressure led overdensities to oscillate. 

\begin{figure}[h!]
	\centering
	\includegraphics[scale=0.4]{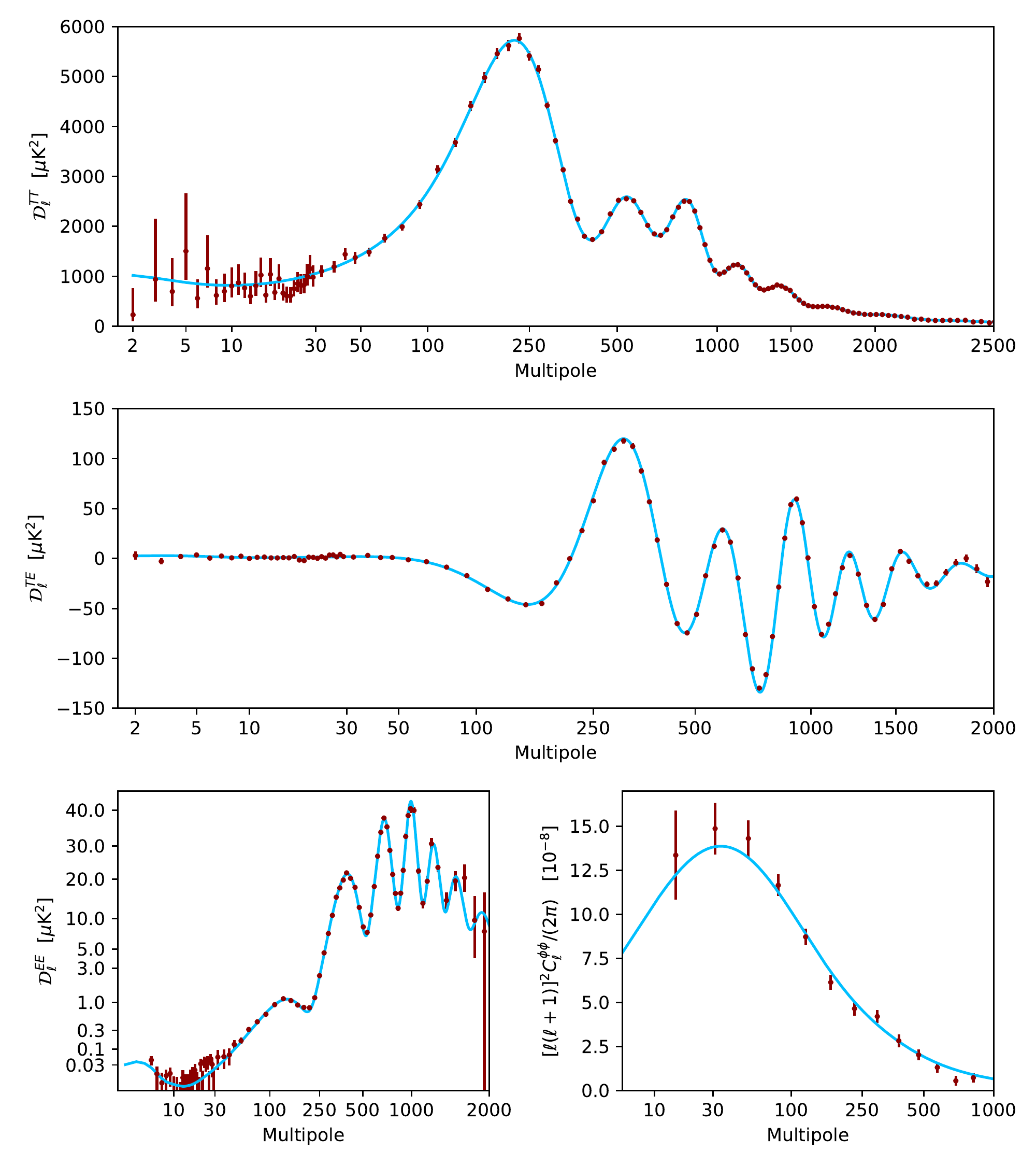}
	\caption[Planck 2018 angular power spectra]{The Planck 2018 \cite{Akrami:2018odb} angular power spectra of the CMB (TT, TE, EE). The E-modes of polarisation are determined through gravitational lensing. The lensing potential is shown in the bottom right. The blue line is a best-fit model to temperature and polarisation data.}
	\label{cmb spectra}
\end{figure}

\begin{figure}[h!]
	\centering
	\includegraphics[scale=0.4]{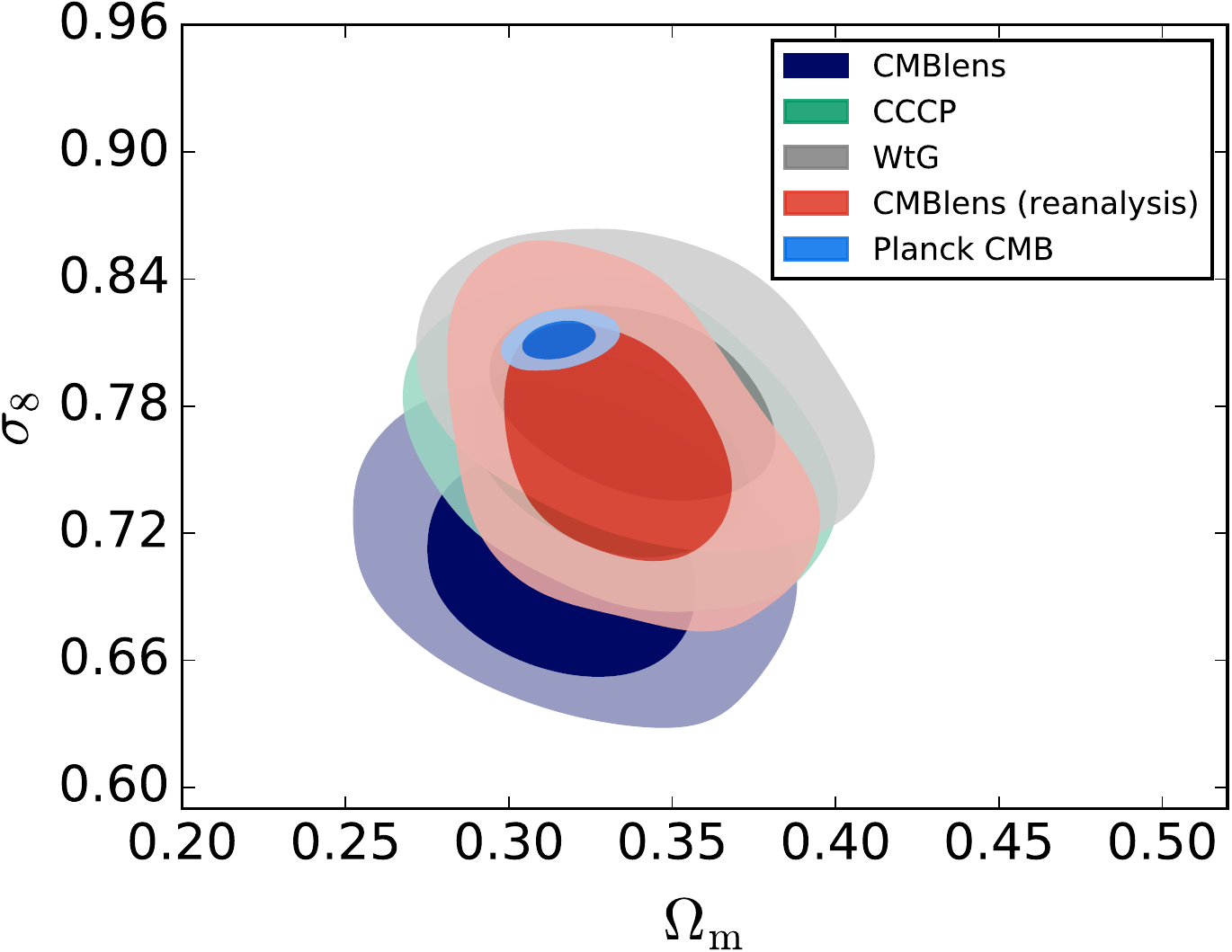}
	\caption[Probabilities in the $\left(\Omega_m,\sigma_8\right)$ plane]{Probabilities in the $\left(\Omega_m,\sigma_8\right)$ plane for different versions of the scaling relations between Compton distortion parameter and cluster mass. Here “WTG” is Weighing the Giants, “CCCP”is the Canadian Cluster Comparison Project, “CMBlens” refers to  the  CMB  lensing  method  as  analysed  by  Melin \& Bartlett (2015)  and  re-analysed  by  Zubeldia \& Challinor  (2019).  Blue contours are constraints from CMB anisotropies\cite{Akrami:2018odb}.}
	\label{matter content planck}
\end{figure}

The structure of these oscillations contains important physical information. Each peak corresponds to a resonance where photons decouple of a baryonic mode at its peak amplitude. For example, we can extract from the first peak  information on the curvature of the Universe, the second allows to determine the baryon density, and we can infer the dark matter density with the third. The first peak has the power to constrain cosmological models of the early Universe, since they typically predict a curvature power spectrum of primordial perturbations. In this sense, the power spectra provided by Planck given in figure \ref{cmb spectra} are a powerful tool to constrain inflationary and alternative models. 
Complementary observations from supernovae \cite{Riess:1998cb} suggest the Universe is currently in a phase of accelerated expansion. The origin of this expansion is still unknown, and as a consequence, this expansion was labeled as \textit{Dark Energy}. The matter content of the Universe derived from CMB anisotropies is shown together with the density fluctuation power spectrum amplitude $\sigma_8$ in figure \ref{matter content planck}. The data seems to favour a composition of baryons and dark matter of about 31\% of the total energy density, resulting in around 69\% of dark energy.

\section{Energy Content of the Universe}

We introduce in this section the classical background necessary for presenting the standard model of cosmology, which explains most of the observations described in the previous section in a consistent way. 

Let us start by deriving the Einstein equations of a FLRW spacetime from the Einstein-Hilbert action supplemented with a cosmological constant $\Lambda$
\begin{align}
\mathcal{S}_G=\frac{1}{16 \pi G} \int \textup{d}^{4}x \sqrt{-g} \left(R-2\Lambda\right) \;,
\label{Einstein-Hilbert action cosmological constant}
\end{align}
to which we also add a matter sector
\begin{align}
\mathcal{S}_m= \int \textup{d}^{4}x \sqrt{-g} \mathcal{L}_{m} \;.
\end{align}
Only those two contributions are present, thus their dynamics are entangled. The EFE read
\begin{align}
\frac{16\pi G}{\sqrt{-g}} \frac{\partial \mathcal{S}_G}{\partial g_{\mu \nu}} := G_{\mu \nu} + \Lambda g_{\mu \nu} =8\pi G T_{\mu \nu} =: -\frac{2}{\sqrt{-g}} \frac{\partial \mathcal{S}_m}{\partial g_{\mu \nu}} \;.
\label{einstein equations}
\end{align}
Plugging in the FLRW metric \eqref{flrw}, we show in appendix \ref{AppFLRWEinstein} the geometric parts give
\begin{align}
G_{00} = 3 \left(H^2 + \frac{k}{a^2}\right) \;, \quad G_{ij} = -\left(H^2+2\frac{\ddot{a}}{a}+\frac{k}{a^2}\right) g_{ij} \;.
\label{einstein equations components}
\end{align}
To describe the matter part, we consider a perfect fluid. This is convenient since perfect fluids are isotropic in their rest frame. The energy-momentum of a perfect fluid can be written
\begin{align}
T_{\mu \nu} = (P+\rho) u^{\mu} u^{\nu} +P g_{\mu \nu} \;,
\label{perfect fluid}
\end{align}
with $P$ and $\rho$ the pressure and energy density as measured in the rest frame, respectively, and $u^{\mu}$ is the four-velocity of the fluid. Note that the four-velocity can be chosen as $u^{\mu}=\left(1,0,0,0\right)$ using comoving coordinates. The conservation of energy $\nabla_{\mu} T^{\mu \nu} = 0$ then gives
\begin{align}
\dot{\rho} +3H\left(\rho+P\right) = 0 \;.
\label{continuity equation}
\end{align}
It is interesting to show this result can also be derived from the first law of thermodynamics $\textup{d}U=-P\textup{d}V$, with $U=PV$ and $V=a^3$. Finally, let us mention briefly that the behaviour of a fluid can in general be complicated. Considering a barotropic fluid, we assume pressure and energy density are simply related by
\begin{align}
P = w \rho \;,
\label{equation of state}
\end{align}
with $w$ the so-called \textit{equation of state} (EOS). It can be incredibly difficult to find the correct EOS for some fluids, \textit{e.g.} for fluids composing a neutron star. Fortunately, relevant fluids in  cosmology can be modelled with a constant EOS parameter.

In the case of a single perfect fluid, solving \eqref{continuity equation} gives us the scaling solution
\begin{align}
\rho = \rho_{init} \left(\frac{a}{a_{init}}\right)^{-3(1+w)} \;,
\end{align}
with the subscript ``$init$" meaning an initial value. In this way, we have obtained the evolution for fluids of different nature. Pressureless (or dust, or cold, or non-relativistic) matter is obviously defined by $w=0$, so the matter energy density $\rho_m$ will lose energy as $\rho_m \propto a^{-3}$, \textit{i.e.} it is inversely proportional to the volume of the spatial hypersurface. A fluid made of relativistic particles such as photons (we can also include neutrinos and other highly energetic particles) has an EOS $w=1/3$, hence the radiation energy density $\rho_{rad}$ goes as $\rho_{rad}\propto a^{-4}$. The difference with ordinary matter is that a photon loses energy not only because of expansion, but also because it is redshifting.

\begin{figure}
	\centering
	\includegraphics[scale=0.5]{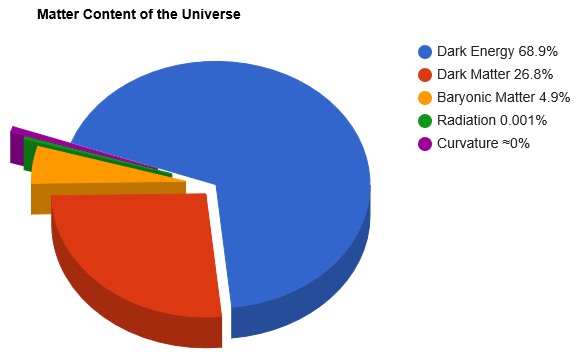}
	\caption[Current composition of the Universe]{Current composition of the Universe. Neutrinos are enclosed in ``Radiation".}
	\label{matter content universe}
\end{figure}

By plugging the Einstein tensor \eqref{einstein equations components} and the momentum-energy tensor \eqref{perfect fluid} in the EFE \eqref{einstein equations}, we find the Friedmann \cite{friedmann1922125} and the Raychaudhuri \cite{Raychaudhuri:1953yv} equations
\begin{align}
H^2 &= \frac{8 \pi G}{3} \rho - \frac{k}{a^2} + \frac{\Lambda}{3} \;, \label{friedmann equation} \\
\frac{\ddot{a}}{a} &= - \frac{4\pi G}{3} \left(\rho+3P\right) + \frac{\Lambda}{3} \;. \label{raychaudhuri equation}
\end{align}
The Friedmann equation gives the temporal evolution of the Universe in terms of the matter content, while the Raychaudhuri equation is related to the acceleration of the Universe. A second form of the Friedmann equation in terms of energy densities is given by
\begin{align}
H^2 = \frac{8 \pi G}{3} \left(\rho_m + \rho_k + \rho_{\Lambda}\right) \;;  \quad \textup{with} \;\; \rho_k := \frac{-3k}{8 \pi G a^2} \;, \quad \rho_{\Lambda} := \frac{\Lambda}{8 \pi G} \;.
\label{friedmann equation density}
\end{align}
with $\rho_k$ the density associated to the apparent spatial curvature, and $\rho_{\Lambda} $ the density associated to the cosmological constant. We see then that for curvature $w=-1/3$, and for dark energy $w=-1$. We define the \textit{critical density}
\begin{align}
\rho_c := \frac{3}{8 \pi G} H^2 \;.
\label{critical density}
\end{align}
This allows us to introduce the dimensionless quantity $\Omega := \rho / \rho_c$, called the density parameter. This quantity is particularly useful to describe the matter content of the Universe, and to compare between different cosmological models. Then, a third version of the Friedmann equation is
\begin{align}
\frac{H^2}{H_0^2} = \left(\Omega_{rad,0}+\Omega_{\nu,0}^{rel}\right) \left(\frac{a_0}{a}\right)^4 + \left(\Omega_{m,0}+\Omega_{\nu,0}^{nrel}\right) \left(\frac{a_0}{a}\right)^3 + \Omega_{k,0} \left(\frac{a_0}{a}\right)^2 + \Omega_{\Lambda,0} \;,
\label{friedmann equation density parameter}
\end{align}
where photons are contained in $\Omega_{rad,0}$, ultra-relativistic neutrinos in $\Omega_{\nu,0}^{rel}$, non-relativistic neutrinos in $\Omega_{\nu,0}^{nrel}$ and non-relativistic matter is contained in $\Omega_{m,0}$. We give in table \ref{cosmological parameters} a short list of cosmological parameters measured by the Planck collaboration \cite{Aghanim:2018eyx}, including BAO from galaxy surveys (SDSS, 2dFGRS, 6dFGRS, WiggleZ, BOSS, etc...). The joint analysis allows to relieve degeneracies from only one set of data. Most of the parameter values are consistent within different data sets, even though there is a discrepancy concerning the measurement of the Hubble rate today, the so-called $H_0$ tension. The relative importance of each matter component is depicted in figure \ref{matter content universe}.

\begin{figure}[h!]
	\centering
	\begin{tabular}{c c}
		Parameter & TT,TE,EE+lowE+lensing+BAO, 68\% limits \\
		\hline \\
		$H_{0}$ $\left[km.s^{-1} Mpc^{-1}\right]$ & $67.66 \pm 0.42$ \\
		$\Omega_{\Lambda,0}$ & $0.6889 \pm 0.0056$ \\
		$\Omega_{m,0}$ & $0.3111 \pm 0.0056$ \\
		$\Omega_{k,0}$ & $0.0007 \pm 0.0019$ \\
		$\Omega_{b,0} h^2$ & $0.02242 \pm 0.00014$ \\
		$\Omega_{c,0} h^2$ & $0.11933 \pm 0.00091$ \\
		$\Omega_{r,0}$ \footnote{Planck used WMAP temperature $T=2.7260 \pm 0.0013 K$ \cite{Fixsen:2009ug}} & $ \approx 0.000092$ \\
		$z_{eq} $ & $3387 \pm 21$ \\
		$z_{de} $ & $0.303 \pm 0.002$ \\
		Age $\left[Gyr\right]$ & $13.787 \pm 0.020$ \\
		%$n_s $ & $0.9665 \pm 0.0038$ \\
		\hline
	\end{tabular}
	\caption[Planck 2018 cosmological parameters]{Planck 2018 cosmological parameters given by a joint analysis with BAO and Planck collaboration\cite{Aghanim:2018eyx}.}
	\label{cosmological parameters}
\end{figure}

Before describing the minimal model that explains the data presented above, let us note that the dynamics of a perfect fluid can also be used to restore the time-dependence of a quantised model, as we will discuss later the problem of time when we present the Wheeler-DeWitt equation. We show explicitly how to do so using the \textit{Schutz formalism} in appendix \ref{schutz}.

\section{The Hot Big Bang Model}

\begin{figure}[h!]
	\centering
	\includegraphics[scale=0.2]{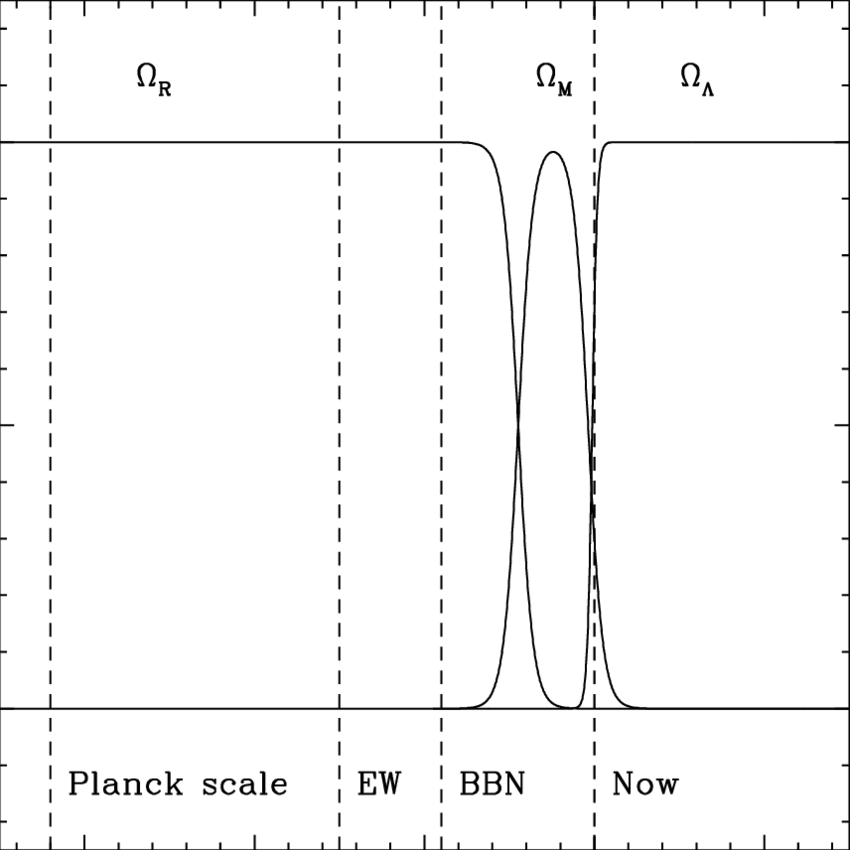}
	\caption[Evolution of density parameters]{Evolution of density parameters. Credit goes to S.~Carroll \cite{Carroll:1999iy}.}
	\label{eras}
\end{figure}

In the previous section, we laid down the cosmological foundations of a homogeneous, isotropic spacetime. We also described the current picture of the observable Universe. This picture justifies the choice of using a FLRW metric with $k=0$ to outline the evolution of the Universe from near the Planck scale to today via the Friedmann equation. The Friedmann equation is now given in terms of three non-interacting components, and each of them leads the dynamics at a different epoch, characterised by their different scaling. As we see from \eqref{friedmann equation density parameter}, matter scales with the scale factor as $a^{-3}$, and radiation has a steeper scaling than matter, going as $a^{-4}$. Going backwards in time, there is a time when matter is dominating, and going even further, radiation will eventually be dominating. We show this behaviour in figure \ref{eras}. The redshift at which these transitions happen can be found using the relation redshift/scale factor \eqref{redshift} and using the cosmological parameters given in table \ref{cosmological parameters}. Their approximate value are respectively
\begin{align}
z_{de} = \left(\frac{\Omega_{\Lambda,0}}{\Omega_{m,0}}\right)^{1/3} -1 \approx 0.3\;, \quad z_{eq} = \left(\frac{\Omega_{m,0}}{\Omega_{rad,0}}\right) -1 \approx 3387 \;.
\end{align}
We now detail the evolution of the Universe, starting from today and going back in time. 

Dark energy, associated with a cosmological constant, is the dominant component today. We observe numerous large-scale structures and galaxies in the Universe, and the oldest galaxy was found at $z=11.1$ \cite{oesch2016remarkably}. The first stars are thought to have appeared around $z=16$, marking the beginning of reionisation. The epoch between reionisation and the photon decoupling at $z \approx 1090$ (CMB) is called the Dark Ages, since only gas was present and light emitted at that time only comes from the 21-cm line of neutral hydrogen. 

When temperature rises, the photon-to-electron scattering rate is enhanced and matter dominates. The cosmic web and haloes form during this period. After $z_{eq} \approx 3387$, the radiation energy density overcomes the matter energy density, and only light elements are present. Up to around $z\approx 10^8$, or equivalently between the first ten seconds and twenty minutes after the Big Bang, the Universe is dense and hot enough to initiate nuclear reactions leading to the creation of the first elements, the most abundant ones being $^{4}He$, $D$, $^{3}He$ and $^{7}Li$ \cite{Coc:2017pxv}. This phase is called Big Bang Nucleosynthesis (BBN). Earlier than BBN, around one second after the Big Bang, the Universe is filled with leptons and neutrinos. These neutrinos are still present today, and current projects aim at the construction of a sky map similar to the CMB but with neutrinos, the Cosmic Neutrino Background \cite{Betti:2019ouf}. Around $10^{-6}$ seconds, hadrons are formed. At about $10^{-12}$ seconds, the Universe was in a state of hot quark-gluon plasma. 

The very early Universe has much more uncertainties. The standard model of particles predicts that when $T_{EW}=159.5 \pm 1.5 GeV$, the electroweak symmetry breaks \cite{DOnofrio:2015gop}. Then, at higher temperatures, all particles are effectively massless, and the weak nuclear force and electromagnetism behave in a similar way. Similarly, the symmetry emerging from combining the electromagnetic, weak and strong forces must be broken, but we have no prediction on when it should happen. It could happen before or after the accelerated expansion of space that must take place around $10^{-32}$ seconds. This expansion is called cosmological inflation in the cosmological standard model. We will describe this particular phase in section \ref{inflation}. We note that inflation could start when the symmetry of electromagnetism, weak and strong forces breaks, marking the end of the Grand Unification era. Finally, there is the Planck era for times smaller than $10^{-43}$ seconds, where all fundamental forces of Nature are believed to be unified. General Relativity predicts a singularity when time tends to zero, and physics beyond the standard model must be developed to study this era in detail. This current understanding of our Universe is called the $\Lambda$ CDM model. In the next section, we recall some issues related to the Hot Big Bang model. 

\section{The Hot Big Bang Model Problems}
\label{big bang problem}

Since the birth of the theory of relativity, it is well accepted that the speed of light be constant. It is generally believed that, in a local frame, any tachyonic particle is unphysical. Hence, information cannot propagate at a speed greater than $c$. Thus, if an event occurs at some point in spacetime, it cannot influence events happening at a distance greater than $c = \Delta x / \Delta t$. Then, every point in spacetime susceptible to be influenced by an event forms a region called the causal region. We say that two events in such region are causally connected. 

In GR, we define the \textit{Hubble Radius} as the boundary between particles moving slower and faster than light, for a given observer at a given time. It is the inverse of the Hubble rate, \textit{i.e.} $R_H = H^{-1}$, in cosmic time, or $R_H = (aH)^{-1}$ for a comoving definition. $R_H$ is of order the 4-curvature scale, therefore characterises the size of the local inertial frame \cite{Mukhanov:2005sc}. In other words, the Hubble radius today is equal to the patch size of the observable Universe. Thus, the horizon size is simply given by
\begin{align}
d_H = a(t) \int_{0}^{t} \frac{\textup{d}t^{\prime}}{a(t^{\prime})} = a \int_0^{a} \frac{\textup{d} a^{\prime}}{H(a^{\prime})a^{\prime 2}} \;.
\label{horizon distance}
\end{align}
Note that this relation is similar to the proper comoving distance defined by the distance travelled by a photon between a source and the observer, also called the angular diameter distance, given by
\begin{align}
d_A= a(t) \int_{t}^{t_0} \frac{\textup{d}t^{\prime}}{a(t^{\prime})} \;,
\label{angular diameter distance}
\end{align}
where $t$ is the time at which the source emits the photon. Putting $t=0$ gives the horizon size today.

\subsection{The Flatness Problem}

We have seen above that matter energy density, be it relativistic matter or not, will grow when the scale factor goes to zero. This implies that in a FLRW spacetime, the contributions from curvature and dark energy to the Friedmann equation \eqref{friedmann equation density} can be neglected at early times. Thus, density will be equal to the critical density \eqref{critical density} at those times. However, the total energy density today is close enough from the critical density. It is difficult to find an explanation of why the density is still close to the critical density today, after a cosmological history of billions of years. This problem, first mentioned by Dicke in 1969, \cite{dicke1970gravitation} is referred to as the \textit{flatness problem}.

A simple solution is that $k=0$ from the start, and cannot change. Another solution is given by an inflationary phase, where a rapid growth implies that $\rho/ \rho_c \rightarrow 1$ very quickly, independently of the value of $k$. In any case, the idea of a spatially flat Universe has been given strong support with the results from Planck combined with BAO \cite{Akrami:2018odb}, as recalled in table \ref{cosmological parameters}.

Quantitatively, if the scale factor grows exponentially as $a \propto e^{N}$ (with $N$ the \textit{number of e-folds}) during inflation, the contribution from the curvature density parameter $|k|/a^2 H^2$ at the end of inflation would be of order $e^{-2N}$, and consequently
\begin{align}
|\Omega_{k,0}| = \frac{|k|}{a_0^2 H_0^2} = e^{-2N} \left(\frac{a_I H_I}{a_0 H_0}\right)^2 \;.
\end{align}
Thus, there is no flatness problem if
\begin{align}
e^N > \frac{a_I H_I}{a_0 H_0} \;.
\label{lower bound inflation}
\end{align}

If we suppose inflation ends in the radiation-dominated era, we have from the Friedmann equation \eqref{friedmann equation density parameter} $H^2_I = H_0^2 \Omega_{rad,0} / a_I^2$. Plugging in this result in \eqref{lower bound inflation} gives
\begin{align}
e^N > \Omega_{rad,0}^{1/4} \sqrt{\frac{H_I}{H_0}} = \Omega_{rad,0}^{1/4} \left(\frac{\rho_I}{\rho_0}\right)^{1/4} = \frac{\rho_I^{1/4}}{0.037 \; h \; \textup{eV}} \;.
\end{align}
Choosing $\rho_I > 1$ MeV to avoid spoiling BBN, we obtain $N>17$. At the GUT scale, we obtain $N>62$, and at the Planck scale, $N>68$.	

\subsection{The Isotropy (Horizon) Problem}

The horizon represents the maximal causally connected region. The size of a causal patch was initially smaller by a ratio $a/a_0$. This means that, if we look into the past, a patch will be increasingly disconnected to other patches. In particular, the horizon at photon decoupling would be much smaller than the patch size. This is in disagreement with the overall isotropic picture we get from the CMB sky \ref{cmb sky}. This is called the \textit{horizon problem}. 

We can describe this problem more precisely comparing the proper particle-horizon size $d_H$ and the angular diameter distance $d_A$ at last scattering. The ratio $d_H/d_A$ gives the angular radius at a given scale, and evaluating it at the last scattering surface will give the size of a causal patch in the CMB. At the last scattering surface, the Universe is dominated by radiation and matter, so the horizon distance \eqref{horizon distance} and the angular diameter distance \eqref{angular diameter distance} become
\begin{align}
d_H &= \frac{2a}{H_0 \Omega_{m,0}} \left(\sqrt{\Omega_{m,0}a+\Omega_{r,0}}-\sqrt{\Omega_{r,0}}\right) \;, \\
d_A &= \frac{2a}{H_0 \Omega_{m,0}} \left(\sqrt{\Omega_{m,0}+\Omega_{r,0}}-\sqrt{\Omega_{m,0}a-\Omega_{r,0}}\right) \;.
\end{align}
Finally, using the scale factor at recombination ($a_{rec}\approx 10^{-3}$) and the cosmological parameters given in table \ref{cosmological parameters}, we obtain
\begin{align}
\frac{d_H}{d_A} \bigg{\rvert}_{rec} = 0.018 \;,
\end{align}
leading to $4 \pi / (0.018)^2 \approx 4\times 10^4$ causally disconnected regions in the sky. In other words, the size of a causal region at recombination is approximately $4\times 10^4$ smaller than the horizon patch.

The condition to solve the horizon problem is then to have $d_H > d_A$ at the time of last scattering, $t_L$. Considering once again a scale factor growing exponentially with $N$ within the approximation $z_L \ll 1$, we obtain
\begin{align}
d_H(t_L) &\approx \frac{a(t_L)}{a_I H_I} \left(e^{N}-1 \right) \;, \\
d_A(t_L) &\approx \frac{a(t_L)}{a_0H_0} \;,
\end{align}
resulting in the same condition as in \eqref{lower bound inflation}
\begin{align}
e^N > \frac{a_I H_I}{a_0 H_0} \;.
\end{align}
Therefore, an inflationary phase solves the horizon and flatness problems at the same time.

%\subsection{The Inhomogeneity Problem}
%
%While the CMB measurements show that the density and temperature are mainly uniform over scales greater than the horizon, they also provide a power spectrum of small amplitude density fluctuations which is nearly scale-invariant. These fluctuations include hot and cold spots depicted in \ref{cmb sky}. Some of these spots are of size substantially larger than the horizon size at the last scattering surface, and constitutes the \textit{inhomogeneity problem}.

\subsection{The Homogeneity Problem}

Solving the horizon problem is not sufficient to explain uniformity. Indeed, the long wavelength gradient energy in all degrees of freedom scales as $a^{-2}$, but the sky appears mostly uniform \cite{Ijjas:2018qbo}. This is called the \textit{homogeneity problem}, and is still an open problem in cosmology.

\subsection{The Monopole Problem}

Topological defects are a prediction of theories beyond the standard model of particles, such as the grand unification theory mentioned in the thermal history of the Universe. A scalar field described in GUT is very symmetric, but as temperature cools down, symmetries are spontaneously broken. This results in the vacuum acquiring expectation values corresponding to the minima of the field potential \cite{Mukhanov:2005sc}. The topology of the vacuum manifold depends on the solution of the fields classical equations of motion. These solutions, called \textit{topological defects}, can be categorised by the number of vacua they involve. The simple case of $n$ real scalar fields with $\varphi^4$ potential and vacuum expectation value $\sigma$ is described by
\begin{align}
\mathcal{L}_{\varphi} = \frac{1}{2} \partial_{\mu} \varphi \partial^{\mu} \varphi - \frac{\lambda}{4} \left(\varphi^2-\sigma^2\right)^2 \;.
\end{align} 
If $n=1$, we obtain two vacua, and the defect is called a \textit{domain wall}. With two real scalar fields or, equivalently, a complex field, we have four vacua, leading to \textit{cosmic strings}. \textit{Monopoles} arise when the vacuum has the topology of a two-dimensional sphere, \textit{i.e.} when $n=3$. These three types of defects could have, in principle, observational imprints in the sky. Domain walls are in accordance with CMB measurements if $\lambda$ and $\sigma$ are very small, but there is no theoretical ground as to why this should be the case. On the other hand, cosmic strings are not in conflict with the CMB, and could have a cosmological role in the early Universe. They are actively hunted to probe beyond the standard model of particle theories. Let us mention that $n=4$ (and higher) objects are called  \textit{textures}, which are objects discarded by CMB measurements. 

Global monopoles, \textit{i.e.} monopoles emerging from a theory with no gauge field, and local monopoles created after inflation have very little effect on the CMB, provided the temperature of reheating is lower than the GUT scale. Traces of these defects are actively being searched today. However, local monopoles created at the GUT scale are problematic. Since Grand Unified Theories incorporate electromagnetism, massive bosons are produced when the symmetry break. Then monopoles, an inevitable product of such theories, acquire a mass. In the case of local monopoles, two gauge fields are produced with mass $m_W=e\sigma$, and one remains massless. Noting $A$ the massless gauge field with covariant derivative
\begin{align}
\mathcal{D} \varphi =  \partial \varphi + e A \varphi \;,
\end{align}
the dominant contribution to the mass of the monopole is
\begin{align}
M \approx  \frac{m_W}{e^2} \;.
\end{align}
All defects are obtained from causally connected fields, implying there must be at least one defect created per horizon volume.  This allows for a rough estimate of the number of monopoles produced during symmetry breaking. Considering the GUT scale $T_{GUT}=10^{15}GeV$, the horizon scales as $t_H \approx 1/T_{GUT}^2$, then the ratio monopoles per photons is
\begin{align}
\frac{n_M}{n_{rad}} \gtrsim \frac{1}{T_{GUT}^3 t_H^3} \simeq T^3_{GUT} \;.
\end{align}
This leads to the present energy density
\begin{align}
\rho^{0}_M \approx M  n_M (t_0) \approx \frac{m_W}{e^2} T^3_{GUT} T_0^3 \approx 10^{-16} \left(\frac{m_W}{10^{15}GeV}\right) \left(\frac{T_{GUT}}{10^{15}GeV}\right)^3 \; g.cm^{-3} \;.
\end{align}
This is much higher than the critical density \eqref{critical density}, $\rho_c \approx 10^{-29} g.cm^{-3}$, and is therefore incompatible with observations. To solve the monopole problem, it is enough to have a number of e-folds greater than $10^{10}$ to dilute the monopoles. Therefore, an inflationary phase with $N>23$ is enough to get rid of the monopole problem.

\subsection{The Cosmic Singularity}

There is a geometric singularity when the scale factor goes to zero, indicating a physical divergence of temperature and density when going back in time. Therefore, the Hot Big Bang model is \textit{geodesically incomplete}. This problem is not solved with an inflationary phase.

\section{Inflationary Cosmology}
\label{inflation}

A common ansatz for the very early evolution of our Universe is cosmological inflation \cite{Starobinsky:1980te, Sato:1980yn, Guth:1980zm, Linde:1981mu,Albrecht:1982wi,Linde:1983gd,Mukhanov:1981xt,Mukhanov:1982nu,Starobinsky:1982ee,Guth:1982ec,Hawking:1982cz,Bardeen:1983qw}. This accelerated expansion in the very early universe is a well-tested paradigm since its predictions for the primordial power spectrum can be successfully compared with observations of the cosmic microwave background \cite{Hinshaw:2012aka, Akrami:2018odb}. In the presentation of the inflationary paradigm, we will use the material found in \cite{Baumann:2009ds,Piattella:2018hvi,Vennin:2014lfa,Weinberg:2008zzc}.

\subsection{Classical Inflationary Dynamics}

An inflationary phase can be obtained by considering the dynamics of a single scalar field with a very flat potential, and slowly rolling down the potential. Such a field is called the \textit{inflaton}, and its dynamics can be derived from a minimal coupling between gravity and a scalar field given by the action
\begin{align}
S = \int \sqrt{-g}\left[\frac{1}{2\kappa^{2}}R -\frac{1}{2} \partial^{\mu} \varphi \partial_{\mu} \varphi - V(\varphi)\right]d^{4}x\;.
\label{action inflation}
\end{align} 

Considering a FLRW metric, the inflaton should only depend on time at the background level. From the definition of the energy-momentum tensor \eqref{einstein equations}, we can write the energy density $\rho_{\varphi}$ and pressure $P_{\varphi}$ for the scalar field as
\begin{align}
\label{rhovarphi}
\rho_{\varphi} &= \frac{1}{2}\dot{\varphi}^{2} + V(\varphi)\;,\\
\label{Pvarphi}
P_{\varphi} &= \frac{1}{2}\dot{\varphi}^{2} - V(\varphi)\;.
\end{align}
Thus the parameter $w$ of the equation of state for $\varphi$ is given by
\begin{align}
w = \frac{P_{\varphi}}{\rho_\varphi} = \frac{\dot{\varphi}^{2} - 2V(\varphi)}{\dot{\varphi}^{2} + 2V(\varphi)}\;.\label{eq:eos1}
\end{align}

We obtain the inflaton dynamics by plugging \eqref{rhovarphi} and \eqref{Pvarphi} into the continuity equation \eqref{continuity equation}. We obtain the Klein-Gordon equation
\begin{align}
\ddot{\varphi} + 3 \frac{\dot{a}}{a} \dot{\varphi} + \frac{\partial V(\varphi)}{\partial\varphi} &= 0\;,\label{eq:KGback} 
\end{align}
subject to the Friedmann constraint coming from \eqref{friedmann equation}
\begin{align}
\left(\frac{\dot{a}}{a}\right)^{2} &= \frac{1}{3 M_{pl}^2}\left[\frac{1}{2}\dot{\varphi}^{2}+ V(\varphi)\right]\;.\label{eq:friedmann1}
%-2\frac{\ddot{a}}{a} - \left(\frac{\dot{a}}{a}\right)^{2} &= \kappa^{2}\left[\frac{1}{2}\dot{\varphi}^{2} - V(\varphi)\right]\;,\label{eq:friedmann2}\\
\end{align}
Here, we have used $G=1/8\pi M_{pl}^2$, with $M_{pl}$ the reduced Planck mass. Under this form, we can constrain the scale at which inflation takes place, namely between the GUT scale $\approx 10^{15}$ GeV and $\approx 10^{3}$ GeV \cite{Martin:2006rs}. 

The Raychaudhuri equation \eqref{raychaudhuri equation} implies that the scale factor accelerates as soon as $\rho_{\varphi}+3P_{\varphi}<0$, meaning that inflation starts when
\begin{align}
w_{\varphi} < -\frac{1}{3} \;.
\label{condition inflation eos}
\end{align}
Equivalently, we have inflation if 
\begin{align}
V>\dot{\varphi}^2 \;.
\label{slow-roll condition}
\end{align}
The equation of motion \eqref{eq:KGback} can be solved analytically  considering a potential largely dominating the dynamics, $V\gg \dot{\varphi}^2$. This is called the \textit{slow-roll condition}.  In the slow-roll regime, we see from the energy density \eqref{rhovarphi} and the pressure \eqref{Pvarphi} that $\rho_{\varphi} \simeq -P_{\varphi}$. Hence, the EOS to obtain slow-roll inflation turns out to be $w_{\varphi} \simeq -1$. In turn, the conservation equation \eqref{continuity equation} implies that $\rho_{\varphi}$ is almost constant in time, and as a consequence, $H$ is also almost constant in time. Then, for an inflationary phase in the slow-roll regime, the scale factor evolves as
\begin{align}
a(t) = a_{in} \exp\left[H\left(t-t_{in}\right)\right] \;,
\label{scale factor inflation}
\end{align}
with the subscript $in$ meaning some initial value. Then, inflation is similar to a quasi-de Sitter evolution. 

The departure from pure de Sitter solution is quantified by a set of parameters. This set can be introduced by the horizon flow functions \cite{Schwarz:2001vv,Schwarz:2004tz}, providing model-independent quantities, valid even beyond the slow-roll approximation. These functions are defined by the hierarchy
\begin{align}
\epsilon_{n+1} := \frac{\textup{d} \ln{|\epsilon_n|}}{\textup{d}N} \;, \quad \epsilon_0 := \frac{H_{in}}{H} \;, \quad N:= \ln(\frac{a}{a_{in}}) \;.
\label{slow-roll parameters}
\end{align}
The time evolution is given in terms of the number of e-folds. All $\epsilon_n$ are of the same order of magnitude. Since inflation is an accelerated phase, we get an acceleration for the scale factor under the condition
\begin{align}
\frac{\ddot{a}}{a} = \dot{H} + H^2 = H^2 \left(1-\epsilon_1\right) >0 \;.
\end{align}
From the weak energy condition $\rho_{\varphi}\geq 0$, we get that $\epsilon_1>0$. Inflation then occurs if $0<\epsilon_1<1$. The slow-roll condition is respected iff $\epsilon_n \ll 1$ for all $n\geq 1$, since inflation is a quasi-de Sitter phase, hence only small, smooth deviations from the de Sitter phase should take place. 

\subsection{Linear Perturbations}

We now describe in this section the dynamics of perturbations during an inflationary phase. One of the advantages of inflationary models is that the expansion phase of inflation stretches quantum fluctuations, providing the seeds of today large-scale structures. To describe these fluctuations, we introduce small perturbations in the inflaton that will be swept up to large scales, leaving imprints in the CMB. Since small fluctuations in the geometry can also be present, we must include them in the metric.

Let us then consider inhomogeneous scalar field perturbations, $\varphi = \varphi(t) + \delta\varphi(t,\vec{x})$, in a linearly-perturbed FLRW metric \cite{Peter:2013avv,Malik:2008im,Piattella:2018hvi}
\begin{align}
ds^2 = -(1+2A)dt^2+2a\partial_{i}B dx^{i}dt+a^2(t)\left[(1-2\psi)\delta_{ij}+2\partial_{ij}E+h_{ij}\right] dx^{i} dx^{j} \; ,
\label{perturbed flrw}
\end{align}
with $A$, $B$, $\psi$ and $E$ the scalar potentials and $h_{ij}$ the tensor perturbations. At linear order, the scalar and tensor perturbations decouple, while vector perturbations quickly vanish during inflation \cite{Mukhanov:1990me}. To be able to compare theoretical predictions to observations, we need to define the power spectrum.

\subsubsection{Definition of the Power Spectrum}

The power spectrum is a quantity giving the distribution over frequency of a signal. It is defined as the Fourier transform of the two-point correlation function. In cosmology, we are mostly interested in describing statistical fluctuations of a given quantity, hence the terms correlation functions usually refers to autocorrelation functions. %The CMB sky provides observational tests of cosmological models through  temperature fluctuations, \textit{i.e.} through correlations between hot and cold spots seen in figure \ref{cmb sky}. 

Let us consider a random variable $G$. The expectation value of $G$ at some point $\boldsymbol{x}_i$ is defined via the ensemble average \cite{Piattella:2018hvi}
\begin{align}
\langle G (\boldsymbol{x}_i) \rangle := \int_{\Omega} \textup{d}g_i \, g_i p_i(g_i) \;. 
\end{align}
This represents the probability distribution of $G$ over all possible random values $g_i\in \Omega$, with $p_i$ is their associated probability density function and where the subscript $i$ refers to the point $x_i$. If we have $g_i p_i(g_i) = g_j p_j(g_j), \forall i,j \in \mathbb{N}^{\star}$, the distribution is said to be \textit{statistically homogenenous}. Similarly, the two-point correlation function of $G$, which is simply the two-dimensional probability distribution, is given by 
\begin{align}
\langle G (\boldsymbol{x}_i) G (\boldsymbol{x}_j) \rangle := \int_{\Omega} \textup{d}g_i \textup{d}g_j \, g_i g_j p_{ij}(g_i,g_j) \;. 
\end{align}
Note the probability density $p_{ij}(g_i,g_j) \neq p_{ji}(g_i,g_j)$ in general, since the probability of $G(\boldsymbol{x}_i)$ being $g_i$ is different from $G(\boldsymbol{x}_i)$ being $g_j$.

We now perform a Fourier transformation on $G$ at some point $\boldsymbol{x}$,
\begin{align}
G(\boldsymbol{x}) = \int \frac{\textup{d}^3 \boldsymbol{k}}{\left(2\pi\right)^{3}}  \hat{G}(\boldsymbol{k}) e^{i\boldsymbol{k} \cdot \boldsymbol{x}} \;.
\end{align}
Note that if $G(\boldsymbol{x})$ is real, we can impose the \textit{reality condition} $\hat{G}(-\boldsymbol{k})=\hat{G}^{\star}(\boldsymbol{k})$. Then, we obtain the general two-point autocorrelation function in terms of the wavenumber
\begin{align}
\langle \tilde{G}(\boldsymbol{k}) \tilde{G}^{\star}(\boldsymbol{k}^{\prime}) \rangle = \int \textup{d}^{3} \boldsymbol{x} \int \textup{d}^{3}  \boldsymbol{x}^{\prime} \, \langle G(\boldsymbol{x}) G(\boldsymbol{x}^{\prime}) \rangle e^{-i \boldsymbol{k} \cdot \boldsymbol{x}} e^{i \boldsymbol{k}^{\prime} \cdot \boldsymbol{x}^{\prime}} \;.
\end{align}
We now assume statistical homogeneity, since constraints on inflation mainly come from the CMB sky, which is supposed to be a black body radiation in first approximation. In this case, we have
\begin{align}
\langle G (\boldsymbol{x}) G (\boldsymbol{x}^{\prime}) \rangle = \langle G (\boldsymbol{x}-\boldsymbol{x}^{\prime}) \rangle \;,
\end{align}
and the two-point autocorrelation function becomes
\begin{align}
\langle \tilde{G}(\boldsymbol{k}) \tilde{G}^{\star}(\boldsymbol{k}^{\prime}) \rangle = \left(2\pi\right)^{3} \delta^{(3)} \left(\boldsymbol{k}-\boldsymbol{k}^{\prime}\right) P_G(\boldsymbol{k}) \;,
\label{two-point correlation homogeneous}
\end{align}
with $P_G(\boldsymbol{k})$ the \textit{power spectrum}, defined as
\begin{align}
P_G(\boldsymbol{k}) := \int \textup{d}^3 \boldsymbol{x} \, \langle G (\boldsymbol{x}) \rangle e^{-i \boldsymbol{k} \cdot \boldsymbol{x}} \;.
\label{power spectrum}
\end{align}

The power spectrum is a prediction of theoretical models, but is not what we observe. From the CMB, we can measure the fluctuation of a given quantity between two points separated on the two-dimensional map by a distance $\boldsymbol{r}$. Concretely, we measure
\begin{align}
\langle G(\boldsymbol{x}) G(\boldsymbol{x}+\boldsymbol{r}) \rangle = \int \frac{\textup{d}^{3} \boldsymbol{k}}{\left(2\pi\right)^3} \int  \frac{\textup{d}^{3}  \boldsymbol{k}^{\prime}}{\left(2\pi\right)^3} \langle  \tilde{G}(\boldsymbol{k}) \tilde{G}^{\star}(\boldsymbol{k}^{\prime}) \rangle e^{i\boldsymbol{k} \cdot \boldsymbol{x}-i\boldsymbol{k}^{\prime} \cdot (\boldsymbol{x}+\boldsymbol{r})}  \;.
\end{align}
Performing an ensemble average, we insert \eqref{two-point correlation homogeneous} in the previous equation and obtain
\begin{align}
\langle G(\boldsymbol{x}) G(\boldsymbol{x}+\boldsymbol{r}) \rangle = \int \frac{\textup{d}^{3} \boldsymbol{k}}{\left(2\pi\right)^3} P_G(\boldsymbol{k}) e^{-i\boldsymbol{k}\cdot \boldsymbol{r}} \;.
\label{two-point correlation function real}
\end{align}
If we assume statistical isotropy, we can work out the power spectrum \eqref{power spectrum} in terms of the magnitude only, such that
\begin{align}
P_G(k) = 2\pi \int \textup{d} r \, r^2 \, \langle G (r) \rangle  \int_{-1}^{1} \textup{d}u \, e^{-i kru} \;,
\end{align}
with $u:=\cos(\boldsymbol{k}\cdot \boldsymbol{x})$. Performing the integral over $u$ leads to
\begin{align}
P_G(k) = 4\pi \int \textup{d} r \, r^2 \, \langle G (r) \rangle  \frac{\sin(kr)}{kr} \;.
\end{align}
Plugging this result in \eqref{two-point correlation function real} gives
\begin{align}
\langle G(\boldsymbol{x}) G(\boldsymbol{x}+\boldsymbol{r}) \rangle = \int_{0}^{\infty} \textup{d} k \, \frac{k^2 P_G(k)}{2\pi^2}  \frac{\sin(kr)}{kr} \;.
\label{power spectrum spherical}
\end{align}
Since $P_G(k)$ has dimension of a volume, it is convenient to defined a dimensionless power spectrum
\begin{align}
\Delta^2_G(k) := \frac{k^3 P_G(k)}{2\pi^2} \;.
\label{power spectrum dimensionless} 
\end{align}
With this definition, the two-point autocorrelation function for a homogeneous, isotropic field is
\begin{align}
\langle G(\boldsymbol{x}) G(\boldsymbol{x}+\boldsymbol{r}) \rangle = \int_{0}^{\infty} \frac{\textup{d} k}{k} \, \Delta^2_G(k)  \frac{\sin(kr)}{kr} \;.
\end{align}

\subsubsection{Scalar Power Spectrum}

The wave equation for first-order scalar field perturbations after Fourier transform is
\begin{align}
\ddot{\delta\varphi} + 3H \dot{\delta\varphi} + \frac{k^{2}}{a^{2}}\delta\varphi + m^{2}\delta\varphi = - 2 V_{,\varphi} A
+ \dot{\varphi} \left[\dot{A} + \frac{k^{2}}{a^{2}}\left(a^{2}\dot{E} - a B\right)\right]\;,\label{eq:pertKG1}
\end{align}
where $k$ is the comoving wavenumber and $V_{,\varphi}$ is the derivative of $V$ with respect to the scalar field, and $m^2 := V_{,\varphi \varphi}$. However, the perturbed metric \eqref{perturbed flrw} has extra degrees of freedom, hence we can rewrite \eqref{eq:pertKG1} in a simpler form.  To do that, we can form gauge-invariant combinations to express physical quantities avoiding the redundancy from this freedom. One particular combination is the \textit{Mukhanov-Sasaki variable} \cite{Mukhanov:1981xt,Mukhanov:1988jd,Kodama:1985bj}
\begin{align}
v := a \left(\delta \varphi + \frac{\dot{\varphi}}{H} \psi \right)\;.
\label{Mukhanov-Sasaki variable}
\end{align}
Choosing the so-called spatially-flat gauge $E=\psi=0$ \cite{Mukhanov:1985rz,Sasaki:1986hm}, the Mukhanov-Sasaki variable is nothing more than the field perturbations.

First, we notice that by perturbing the Einstein equations \eqref{einstein equations components}, we can relate scalar perturbations to matter perturbations through the constraints (in the spatially-flat gauge) \cite{Wands:2008tv}
\begin{align}
- 4\pi G\left(\dot{\varphi}\dot{\varphi} - \dot{\varphi}^{2}A + V^{\prime} \delta\varphi\right) &= 3H^{2}a + \frac{k^{2}}{a^{2}}H\left(a^{2}\dot{E} - aB\right)\;,\label{eq: energyconst}\\
4\pi G \dot{\varphi}\delta\varphi &= HA\;,\label{eq: momentconst}
\end{align}
and using the first derivative of (\ref{eq: momentconst})
\begin{align} 
\dot{A} = 4\pi G\left(\frac{\ddot{\varphi}\delta\varphi + \dot{\varphi}\dot{\delta\varphi}}{H}-\frac{\dot{\varphi}\dot{H}\delta\varphi}{H^{2}}\right)\;,
\end{align} 
we can rewrite the evolution of perturbations (\ref{eq:pertKG1}) as
\begin{align}
&\ddot{\delta\varphi} + 3H\dot{\varphi} \nonumber\\
&+\left(\frac{k^{2}}{a^{2}} + m^{2} + 12\pi G \frac{\dot{\varphi V^{\prime}}}{H} - 4\pi G\frac{\dot{\varDelta\phi}\ddot{\varphi}}{H} + 4\pi G\frac{\dot{\varphi} \dot{H}}{H^{2}} + 12\pi G \dot{\varphi}^{2} - 16\pi^{2}G^{2}\frac{\dot{\varphi}^{4}}{H^{2}}\right)\delta\varphi = 0\;.
\end{align}
Using the background Klein-Gordon equation \eqref{eq:KGback} and $\dot{H} = -4\pi G \dot{\varphi}^{2}$, the above equation becomes
\begin{align}
\ddot{\delta\varphi} + 3H\dot{\varphi} + \left(\frac{k^{2}}{a^{2}} + m^{2} - 16\pi G\frac{\dot{\varphi}\ddot{\varphi}}{H} - 24\pi G\dot{\varphi}^{2} + 8\pi G\frac{\dot{H}\dot{\varphi}^{2}}{H^{2}}\right)\delta\varphi = 0\;.
\end{align}
Then, the equation of motion for scalar perturbations (\ref{eq:pertKG1}) in the spatially-flat gauge can be written \cite{Mukhanov:1988jd,Sasaki:1986hm,Wands:2008tv}
\begin{align}
&\ddot{\delta\varphi} + 3H\dot{\varphi} + \left[\frac{k^{2}}{a^{2}} + m^{2} - \frac{8\pi G}{a^{3}}
\frac{d}{dt}
\left(\frac{a^{3}\dot{\varphi}^{2}}{H}\right) \right]\delta\varphi = 0\;.
\end{align}
Finally, we can use the Mukhanov-Sasaki variable \eqref{Mukhanov-Sasaki variable} along with a new variable $z:=a\dot{\varphi}/H$ to cast the above equation in the form of a harmonic oscillator
\begin{align}
\label{eq:perteq}
\frac{d^2 v_{\boldsymbol{k}}}{d\eta^2}+\left(k^2-\frac{1}{z}\frac{d^2  z}{d\eta^2}\right) v_{\boldsymbol{k}} = 0\; .
\end{align}

In order to find the general solution of the Mukhanov-Sasaki equation, we need to specify a set of initial conditions. Assuming that perturbations come from vacuum fluctuations, we need to quantise the theory. Let us expand $v$ and its canonical conjugate $\Pi := \partial v / \partial \eta$ \footnote{The formal definition comes from the fact we can derive the Mukhanov-Sasaki equation \eqref{eq:perteq} from the second-order action \cite{Mukhanov:1988jd}
	\begin{align}
	S^{\left(2\right)} = \int \textup{d}^{4}x \; \frac{1}{2} \left[\left(\frac{\partial v}{\partial \eta}\right)^2-\delta^{ij} \frac{\partial v}{\partial x^{i}} \frac{\partial v}{\partial x^{j}}+\frac{1}{z}\frac{d^2  z}{d\eta^2}v^2\right] \;.
	\end{align}
	Hence, we have that $\Pi = \partial \mathcal{L}^{(2)} / \partial \left(\partial v /\partial \eta \right)$, giving the definition.}  in a Fourier basis
\begin{align}
\hat{v} &= \int \frac{d^3 {\boldsymbol{k}}}{\left(2\pi\right)^{3}} \left(\hat{a}_{\boldsymbol{k}} v_{\boldsymbol{k}}(\eta) e^{-i {\boldsymbol{k}}\cdot {\boldsymbol{x}}} + \hat{a}_{\boldsymbol{k}}^{\dagger} v_{\boldsymbol{k}}^{\star}(\eta) e^{i {\boldsymbol{k}}\cdot \boldsymbol{x}}\right) \;,\\
\hat{\Pi} &= \int \frac{d^3 \boldsymbol{k}}{\left(2\pi\right)^{3}} \left(\hat{a}_{\boldsymbol{k}} \Pi_{\boldsymbol{k}}(\eta) e^{-i\boldsymbol{k} \cdot \boldsymbol{x}} + \hat{a}_{\boldsymbol{k}}^{\dagger} \Pi_{\boldsymbol{k}}^{\star}(\eta) e^{i \boldsymbol{k} \cdot \boldsymbol{x}}\right) \;.
\end{align}
The operators $\hat{a}_{\boldsymbol{k}}^{\dagger}$ and $\hat{a}_{\boldsymbol{k}}$ are called \textit{creation} and \textit{annihilation} operators, respectively. In standard quantum mechanics, we impose the canonical commutation relations
\begin{align}
\left[\hat{a}_{\boldsymbol{k}}^{\dagger},\hat{a}_{\boldsymbol{k}^{\prime}}\right]=\delta^{(3)} \left(\boldsymbol{k}-\boldsymbol{k}^{\prime}\right) \;, \quad \left[\hat{a}_{\boldsymbol{k}}^{\dagger},\hat{a}_{\boldsymbol{k}^{\prime}}^{\dagger}\right]=\left[\hat{a}_{\boldsymbol{k}},\hat{a}_{\boldsymbol{k}^{\prime}}\right]=0 \;,
\end{align}
resulting in the canonical commutation relations for the Mukhanov-Sasaki variable
\begin{align}
\left[\hat{v}_{\boldsymbol{k}}\left(x_i,\eta\right),\hat{\Pi}\left(x_i^{\prime},\eta\right)\right]= i\,\delta^{(3)} \left(x_i-x_i^{\prime}\right) \;, \quad \left[\hat{v}\left(x_i,\eta\right),\hat{v}\left(x_i^{\prime},\eta\right)\right]=\left[\hat{\Pi}\left(x_i,\eta\right),\hat{\Pi}\left(x_i^{\prime},\eta\right)\right]=0 \;.
\label{commutation relations v}
\end{align}
Note these relations must hold at equal time to enforce causality.

In the case of second-order differential equations, as this is the case with the Mukhanov-Sasaki equation \eqref{eq:perteq}, the general solution is normalised by the wronskian $W(\eta):=\left(v_{\boldsymbol{k}}^{\prime} v_{\boldsymbol{k}}^{\star}-v_{\boldsymbol{k}}^{\star \prime}v_{\boldsymbol{k}}\right)(\eta)$. At early times, we suppose vacuum fluctuations to be well inside the Hubble radius, in accordance with the causal picture we get from observations. Then, we work in a quasi-Minkowskian spacetime, allowing to neglect altogether the potential term in \eqref{eq:perteq}. The solution can now be decomposed into positive and negative frequency parts
\begin{align}
v_{\boldsymbol{k}}(\eta) = A_{\boldsymbol{k}} e^{ik\eta}+ B_{\boldsymbol{k}} e^{-ik\eta} \;, \quad k \gg R_H \;,
\end{align}
with $A_{\boldsymbol{k}}$ and $B_{\boldsymbol{k}}$ integration constants depending only on $\boldsymbol{k}$. The Wronskian condition, coupled with the commutation relations \eqref{commutation relations v}, implies that $W=i$. Then,  normalising $v_{\boldsymbol{k}}$ at early times to the zero-point fluctuations of a free field in flat spacetime \cite{Chernikov:1968zm,Schomblond:1976xc,Bunch:1978yq} gives $i= W = 2ik \left(|A_{\boldsymbol{k}}|^2+|B_{\boldsymbol{k}}|^2\right)$. Furthermore, in the present situation where spacetime is Minkowski-like, the equivalence principle allows to define vacuum as an unique quantum state invariant under Poincaré transformations \cite{Streater:1989vi}. This implies $B_{\boldsymbol{k}} = 0$, obtaining $|A_{\boldsymbol{k}}|=1/\sqrt{2k}$. The invariance by rotation entails we can drop the three-dimensional $\boldsymbol{k}$ for its magnitude $k$, and we eventually obtain
\begin{align}
\lim\limits_{k\rightarrow \infty} v_k = \frac{1}{\sqrt{2k}} e^{ik\eta}\;.
\label{Bunch-Davies vacuum}
\end{align}
This state is most often referred to as the \textit{Bunch-Davies vacuum}. The power spectrum predicted by a single-field inflationary model is, using \eqref{power spectrum dimensionless}, 
\begin{align}
\Delta^2 _v (k) = \frac{k^3}{2\pi^2} |v_k|^2 \;,
\end{align} 
and all that remains is calculating $|v_k|^2$. Before doing so, we point out that $\Delta^2 _v (k)$ can be related to the power spectrum of curvature perturbations $\Delta^2 _{\zeta} (k)$, which is a conserved quantity on large scales \cite{Lyth:1984gv,Martin:1997zd}. Therefore, $\Delta^2 _{\zeta} (k)$ is unsensitive to the different processes occurring before recombination. It can be related to the power spectrum of the Mukhanov-Sasaki variable as
\begin{align}
\Delta^2 _{\zeta} (k) = \frac{1}{2a^2 M_{pl}^2 \epsilon_1} \Delta^2 _v (k) =  \frac{k^3}{4\pi^2 a^2 M_{pl}^2 \epsilon_1} |v_k|^2 \;.
\label{power spectrum curvature}
\end{align}

In order to find the solution $v_k$ using the slow-roll approximation, we need to express the Mukhanov-Sasaki equation \eqref{eq:perteq} in terms of the epsilon parameters \eqref{slow-roll parameters}. The variable $z$ in terms of the parameter $\epsilon_1=-\dot{H}/H^2$ becomes $z=a\sqrt{2M_{pl}^2 \epsilon_1}$, and the Mukhanov-Sasaki equation transforms as
\begin{align}
\label{eq:perteq slow-roll}
\frac{d^2 v_{k}}{d\eta^2}+\left(k^2-\frac{1}{\sqrt{a\epsilon_1}}\frac{d^2  \sqrt{a\epsilon_1}}{d\eta^2}\right) v_{k} = 0\;.
\end{align}
To go further, we also need to express the scale factor in the same way. We introduce a pivot scale $k_{\star}$ at which we evaluate the power spectrum. Other starred quantities in what follows are evaluated at the same scale. The change in the scale factor can be written
\begin{align}
\ln(\frac{a}{a_{\star}}) = \int_{\eta_{\star}}^{\eta} \textup{d} \ln(a) = \int_{\eta_{\star}}^{\eta} \mathcal{H} \textup{d} \eta^{\prime} = \int_{\eta_{\star}}^{\eta} \mathcal{H} \eta^{\prime} \textup{d} \ln(\eta^{\prime}) \;,
\label{scale factor epsilon}
\end{align}
where we defined the conformal Hubble rate $\mathcal{H} := aH$. Remembering the relation between conformal and cosmic time
\begin{align}
\eta = \int \frac{\textup{d}t}{a} = \int \frac{\textup{d}a}{a \mathcal{H}} \;,
\end{align}
we can show that at first-order in slow-roll
%we can integrate the last term by part to obtain \cite{Vennin:2014lfa}
%\begin{align}
%	\int \frac{\textup{d}a}{a \mathcal{H}} = -\frac{1}{\mathcal{H}} + \int \frac{\textup{d}a}{a \mathcal{H}} \epsilon_1 \;.
%\end{align}
%Integrating once again by part the last term
%\begin{align}
%	\int \frac{\textup{d}a}{a \mathcal{H}} \epsilon_1 = -\frac{\epsilon_1}{\mathcal{H}} + \int \textup{d}a \left[\frac{\epsilon_1}{a \mathcal{H}}\left(\epsilon_1+\epsilon_2\right)\right] \;,
%\end{align}
%we obtain the conformal time
%\begin{align}
%	\eta = - \frac{1+\epsilon_1}{\mathcal{H}} + \int \textup{d}a \left[\frac{\epsilon_1}{a \mathcal{H}}\left(\epsilon_1+\epsilon_2 \right)\right] \;.
%\end{align}
%At first-order in slow-roll, we can neglect the integral term. We plug this result in \eqref{scale factor epsilon}, obtaining
%\begin{align}
%	\ln(\frac{a}{a_{\star}}) \simeq \left(1+\epsilon_{1\star} \right) \left(\frac{\eta_{\star}}{\eta}\right) \;.
%\end{align}
%Finally,
\begin{align}
a \simeq - \frac{1}{H_{\star}\eta} \left[1 + \epsilon_{1\star}-\epsilon_{1\star} \ln(\frac{\eta}{\eta_{\star}})\right] \;.
\end{align}
Performing a Taylor expansion of $\epsilon_1$ around $\eta_{\star}$
\begin{align}
\epsilon_1 \simeq \epsilon_{1\star} + \epsilon_{1\star} \epsilon_{2\star} \ln(\frac{a}{a_{\star}}) \simeq \epsilon_{1\star} \left[1-\epsilon_{2\star}\ln(\frac{\eta}{\eta_{\star}})\right] \;,
\end{align}
the Mukhanov-Sasaki equation becomes
\begin{align}
\frac{d^2 v_{k}}{d\eta^2}-\left[k^2-\frac{2}{\eta^2}\left(1+\frac{3}{2}\epsilon_{1\star}+\frac{3}{4}\epsilon_{2\star}\right) \right] v_{k} = 0\;.
\end{align}

To see this equation has an analytic solution, let us rewrite it as a Bessel equation
\begin{align}
\label{Mukhanov-Sasaki equation nu}
\frac{d^2 v}{d\eta^2}+\left(k^2-\frac{\nu^2-1/4}{\eta^2}\right) v = 0 \;,
\end{align}
with Bessel index
\begin{align}
\nu := \frac{3}{2} \sqrt{1+\frac{4}{3} \epsilon_{1\star}+\frac{2}{3} \epsilon_{2\star}} \;.
\end{align}
The generic solution of such a Bessel equation can be given as a combination of the Hankel functions $H_{|\nu|}^{(1)}$ and $H_{|\nu|}^{(2)}$ \cite{NIST:DLMF} such that
\begin{align}
v_k = \sqrt{|k\eta|} \left[V_+H_{|\nu|}^{(1)}(|k\eta|)+V_-H_{|\nu|}^{(2)}(|k\eta|) \right] \;,
\label{Bessels}
\end{align}
We normalise the solution with the Bunch-Davies vacuum  \eqref{Bunch-Davies vacuum} on small scales (early times), $v_k\to e^{-ik\eta}/\sqrt{2k}$ for $\eta \rightarrow -\infty$, so we set $V_{+} = 0$ and $V_{-} = \sqrt{\pi /4k}$. The solution (\ref{Bessels}) then provides us with the corresponding solution on large scales (late times) for $k\eta \rightarrow 0$
\begin{align}
\label{eq:delta_phi}
v_{k} = i\sqrt{\frac{1}{4\pi k}}\frac{\Gamma(\lvert\nu\rvert)2^{\lvert\nu\rvert}}{\lvert k\eta\rvert^{\lvert\nu\rvert - 1/2}}\;,
\end{align}
where $\Gamma(\lvert\nu\rvert)$ is the Gamma function. Expanding this expression, we finally obtain the curvature power spectrum \eqref{power spectrum curvature} in slow-roll \cite{Stewart:1993bc}
\begin{align}
\Delta^2 _{\zeta} (k) = \frac{H_{\star}^2}{8\pi^2 \epsilon_{1\star} M_{pl}^2} \left[1-2\left(C+1\right)\epsilon_{1\star}-C\epsilon_{2\star}-\left(2\epsilon_{1\star}+\epsilon_{2\star}\right) \ln(\frac{k}{k_{\star}})\right] \;,
\label{power spectrum scalar}
\end{align}
with $C:=\ln 2 + \gamma_E -2$, $\gamma_E$ being the Euler-Mascheroni constant.

In summary, the inflationary scenario amounts to considering the evolution of a quantum field in a quasi-de Sitter slow-roll expansion. The quantum field starts in a Bunch-Davies vacuum which is subsequently replaced by a distribution function and injected into the power spectrum to describe the field with a classical phase space. The validity of this procedure has been widely questioned \cite{Guth:1985ya,Grishchuk:1990bj,Albrecht:1992kf,Martin:2015qta,Martin:2018zbe,Martin:2018lin}. In \cite{Ashtekar:2020mdv}, three notions of classicality are discussed: ``fading" of non-commutativity, quantum squeezing and the replacement of the quantum field by a distribution function. If the context of standard inflation, these notions are indistinguishable, but classicality can manifest itself in only one or two ways when going beyond standard inflation \footnote{Note that this a different issue from the quantum-to-classical transition; nothing is said about the collapse of the wave function, for instance.}. 

\subsubsection{Tensorial Power Spectrum}

Since tensor perturbations are transverse ($\partial^{i} h_{ij}=0$) and trace-free ($\delta^{ij}h_{ij}=0$), as well as  gauge-invariant, finding their power spectrum is particularly easy. We can expand tensor perturbations into eigenmodes of the spatial Laplacian, $\nabla^2 e_{ij} = -(k^2/a^2) e_{ij}^{\left(+,\times\right)}$, with $+,\times$ two polarisation states. We obtain \cite{Wands:2008tv}
\begin{align}
h_{ij} = h(t) e_{ij}^{\left(+,\times\right)}(x) \;,
\end{align}
with $h$ the gravitational wave amplitude. The wave equation is then simply given by
\begin{align}
\ddot{h} + 3H \dot{h} + \frac{k^2}{a^2}h = 0 \;.
\end{align}
Following a similar procedure as in the previous section, the power spectrum for tensor modes leaving the horizon is
\begin{align}
\Delta^2 _{h} (k) = \frac{2H_{\star}^2}{\pi^2  M_{pl}^2} \left[1-2\left(C+1\right)\epsilon_{1\star}-2\epsilon_{1\star} \ln(\frac{k}{k_{\star}})\right] \;.
\label{power spectrum tensor}
\end{align}

\subsection{Inflation and Observations}

The power spectra \eqref{power spectrum scalar} and \eqref{power spectrum tensor} derived previously have a logarithmic dependence on the scale. Therefore, any observed deviation from a scale-invariant spectrum should be weak, and this deviation can be used to constrain inflationary models. We can write the scalar spectrum for a generic model as a power-law
\begin{align}
\Delta^2_{\zeta} := A_s \left(\frac{k}{k_{\star}}\right)^{n_s-1} \;,
\end{align}
with $A_s$ the amplitude of scalar perturbations, from which we derive the \textit{scalar spectral index}
\begin{align}
n_s = 1 + \frac{\textup{d}\ln \Delta^2_{\zeta}(k)}{\textup{d}\ln k} \;.
\label{spectral index scalar}
\end{align}
Analogously, we can define a tensor spectrum
\begin{align}
\Delta^2_{h} := A_T \left(\frac{k}{k_{\star}}\right)^{n_T} \;,
\end{align}
with \textit{tensor spectral index}
\begin{align}
n_T = \frac{\textup{d}\ln \Delta^2_{h}(k)}{\textup{d}\ln k} \;.
\label{spectral index tensor}
\end{align}
Finally, the ratio between the tensor and scalar power spectra is given by
\begin{align}
r := \frac{\Delta_h^{2}(k_{\star})}{\Delta_{\zeta}^{2}(k_{\star})} \;.
\label{ratio tensor scalar}
\end{align}
We can relate primordial perturbations to the dynamics of the Hubble parameter during inflation (or equivalently, to the inflaton potential), therefore constraining these quantities directly. The values of different parameters given by Planck 2018 are displayed in table \ref{inflation parameters}. Interestingly, the running of the scalar spectral index $\textup{d} n_s / \textup{d} \ln k$ is consistent with a vanishing value. While the running of the running is not so constrained, adding polarisation data helps constraining it further \cite{Akrami:2018odb}. The spectral indices for a single field in slow-roll are given by $n_s \simeq 1-2\epsilon_{1\star}-\epsilon_{2\star}$ and $n_T \simeq -2\epsilon_{1\star}$, and the production of gravitational waves is $r \simeq 16\epsilon_{1\star}$.

\subsection{A Hint for Modified Gravity?}
\label{hint modified gravity}

\begin{figure}[h!]
	\centering
	\includegraphics[scale=0.6]{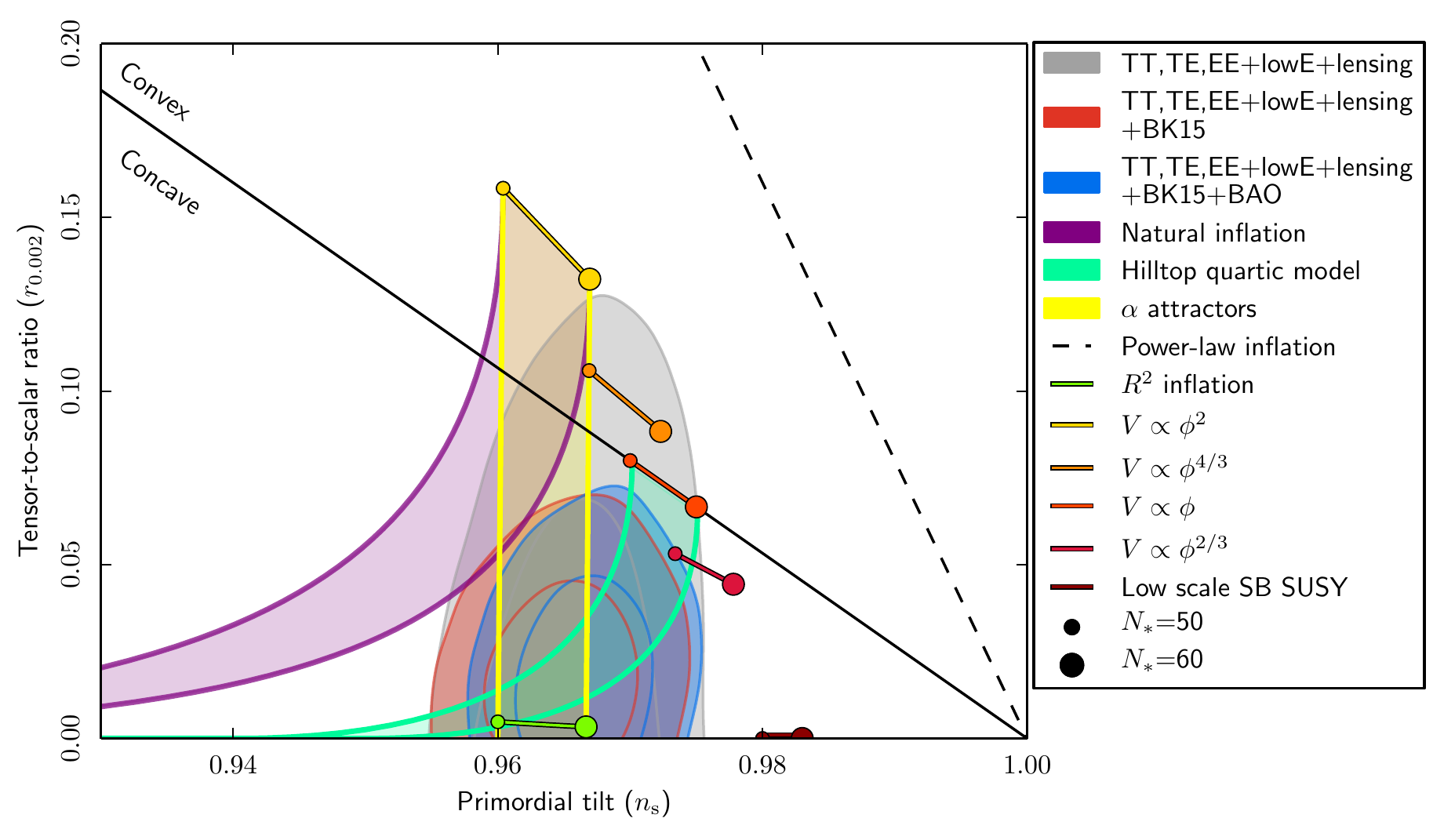}
	\caption[Planck 2018 constraints on inflationary models]{Marginalized joint 68\% and 95\% CL regions for $n_s$ and $r$ at $k = 0.002$ Mpc$^{−1}$ from 2018 Planck data alone and in combination with BK15 or BK15+BAO data, compared to the theoretical predictions of selected inflationary models \cite{Akrami:2018odb}. Note that the marginalized joint $68\%$ and $95\%$ CL regions assume $\textup{d}n_s/\textup{d}\ln{k}=0$.}
	\label{constraints inflation}
\end{figure}

Predictions of various inflationary models are summarised in figure \ref{constraints inflation}. Interestingly, the best-fit for single-field models is the Starobinsky model, an $R^2$-modification of the Einstein-Hilbert action \eqref{Einstein-Hilbert action} with potential \cite{Starobinsky:1980te}
\begin{align}
V(\varphi) = \frac{3}{4} M_{pl}^2 M^2 \left(1-e^{-\sqrt{\frac{2}{3}}\frac{\varphi}{M_{pl}}}\right)^2 \;,
\label{starobinsky potential}
\end{align}
where $M$ is a mass scale to be fixed by observations. The Starobinsky model predicts that the parameters $n_s$ and $r$ in terms of numbers of e-folds are \cite{Hwang:2001pu}
\begin{align}
n_s = 1-\frac{2}{N} \;, \quad r = \frac{12}{N^2} \;,
\end{align}
and provides the best fit to Planck data for $50<N<60$.

The modification of the Einstein-Hilbert action with a higher-order curvature term opens a window on theoretical modifications of general relativity. This class of models pertains to the class of \textit{modified gravity} models. There is a straightforward generalisation of the Einstein-Hilbert action replacing the Ricci scalar by an arbitrary function depending on the Ricci scalar. This class of theories is named \textit{$f(R)$ modified gravity} \cite{Koyama:2015vza}, and constitutes a very active field of research.  However, it is worth noting there is no fundamental explanation of why this function should be truncated to keep only the $R^2$ contribution in order to fit the data. Intuitively, if the squared term becomes important, higher-order terms should also become important. This issue aside, we also note that an interesting feature of $f(R)$ theories is that they can provide an accelerated expansion phase, and can be accommodated to explain inflation as well as the present accelerated expansion attributed to a cosmological constant.

\begin{figure}[h!]
	\centering
	\begin{tabular}{c c}
		Parameter & TT,TE,EE+lowE+lensing, 68\% limits \\
		\hline\\
		$\ln(10^{10}A_s) $ & $3.044 \pm 0.014 $ \\ 
		$n_s $ & $0.9649 \pm 0.0042$ \\
		$\textup{d} n_s / \textup{d} \ln k $ & $0.013 \pm 0.012$ \\
		$\textup{d}^2 n_s / \textup{d} \ln k^2 $ & $0.022 \pm 0.012$  \\
		$r$ & $<0.16$  \\
		$\epsilon_1 $ & $<0.0063$ (95\% CL) \\
		$\epsilon_2 $ & $0.030^{+0.007}_{-0.005}$ \\
		\hline
	\end{tabular}
	\caption[Planck 2018 inflationary parameters]{Inflationary parameters given by Planck 2018 + lensing \cite{Akrami:2018odb}. Note that $r<0.16$ is dependent on whether we consider additional parameters (scale dependence, curvature, neutrinos, etc...). We give the highest upper limit.}
	\label{inflation parameters}
\end{figure}

\section{Conceptual Challenges Within the Inflationary Paradigm}

Despite the observational tests passed by inflation, the mechanisms generating an inflationary phase are not well-posed, and inflation faces some theoretical challenges that need to be solved. We now discuss some of them.

\subsection{The Flat Potential Problem}

We saw that we need the slow-roll condition \eqref{slow-roll condition} to initiate an inflationary phase. For inflation to solve the flatness and horizon problems, we need at least 60 e-foldings. Then, the potential needs to drive the inflaton dynamics during this period, and the form of the potential in most inflationary models is constrained to be very flat. This flatness is not predicted by any standard particle, and even scalar fields beyond the Standard Model struggle to reproduce such flatness. Furthermore, only a Higgs-like scalar particle has been detected up until now, and Higgs inflation requires a non-minimal coupling with gravity \cite{Bezrukov:2007ep}. Therefore, the minimal model of inflation presented here cannot be a definitive answer to what generated primordial fluctuations. However, it is worth noting some models do not present this fine-tuned potential. For example, inflation with a Coleman-Weinberg potential as used in \cite{Albrecht:1982wi} is free from this problem.

\subsection{Quantum-to-Classical Transition Problem}
\label{quantum to classical problem}

We have seen that inflation stretches quantum fluctuations to large scales. While the predictions on the power spectrum are in agreement with observations, there is a conceptual issue with the transition from microscopic to macroscopic scales. The initial quantum vacuum state for perturbations is supposed to be invariant by translation and rotation. However, we observe inhomogeneities in our Universe, contradicting the translational invariance. We could propose a way-out by supposing the initial quantum state can be decomposed into a sum of pure states, with some or all of each non-invariant by translation. Then, the phenomenon of decoherence would disentangle those states, selecting one particular state giving the inhomogeneous Universe we observe. But we would not be able to give a reasonable explanation of how would this selection occur \cite{Pinto-Neto:2013npa,Pinto-Neto:2013toa}.

The collapse of the wave function is usually invoked to measure observables. Quantum effects in cosmology cannot be explained by the standard interpretation, because a classical apparatus, external to the measured system, is required. Since observations are made inside the physical system, standard inflation with canonical quantum fluctuations is conceptually ill-defined. We discuss this more in detail in section \ref{the born rule and measure problem}.

\subsection{Trans-Planckian Modes Problem}

There is another issue with the duration of inflation, associated with the modes of primordial fluctuations. All the modes observed today are inside the Hubble radius, but a long period of inflation implies that the initial physical wavelengths were smaller than the Planck scale. The problem is sketched in the left panel of figure \ref{trans-planckian}. This is in conflict with the semi-classical description of matter used to derive linear perturbations, in turn implying that a perturbative approach is valid at arbitrarily high-energy scales \cite{Brandenberger:2011gk}. It was argued that if fluctuations are non-adiabatic, the effects of new physics would leave imprints in the primordial power spectrum leading to constraints on inflationary models \cite{Brandenberger:2004kx,Tanaka:2000jw,Starobinsky:2001kn}. One could impose a naive cut-off in the UV regime, but this procedure typically leads to a loss of unitarity in the evolution of the  modes \cite{Weiss:1985vw,Jacobson:1999zk}. 

Recently, a stronger proposal banishing all UV modes to ever cross the horizon was made, known as the Trans-Planckian Censorship Conjecture \cite{Bedroya:2019snp,Bedroya:2019tba}. The consequences for inflation are dramatic, as a consistent inflationary model must occur at an energy scale of $10^9$ GeV, and an upper bound on the slow-roll parameter  $\epsilon_1 < 10^{-31}$ is set.

On a side note, this condition is thought to be similar to the de Sitter conjecture evoked in the Swampland \cite{Ooguri:2006in}, stating that any consistent effective theory with UV completion must comply with a set of conjectures. However, there is still doubt on whether the censorship really is similar to the de Sitter conjecture \cite{Saito:2019tkc}, and if the censorship as initially stated can be refined to accept some inflationary models \cite{Cai:2019dzj}. The actual debate concerns whether inflation, displaying the Trans-Planckian problem, is indeed part of the Swampland or not \cite{Berera:2020dvn}.

\subsection{Fine-Tunings}

Two others fine-tuning problems concern the shape of inflationary potentials and the tilt \cite{Ijjas:2015hcc}. The amplitude of primordial density fluctuations is of order $\Delta_{\zeta}^2 \approx 10^{-10}$, and is used to constrain the landscape of viable potentials through the relation
\begin{align}
\Delta_{\zeta}^2 = \frac{V^3}{12 \pi^2 M_{pl}^6 V^{\prime 2}} \;.
\end{align}
Therefore, the parameters of each inflationary scenario must be fine-tuned to comply with observations. For the case of Starobinsky inflation, where the potential is given in \eqref{starobinsky potential} and since the number of e-folds can be estimated as
\begin{align}
N = \frac{3}{4} e^{\sqrt{\frac{2}{3}}\frac{\varphi}{M_{pl}}} \;,
\end{align}
we obtain
\begin{align}
V &= \frac{3}{4} M_{pl}^2 M^2 \left(1-\frac{3}{4N}\right)^2 \;,\\
V^{\prime} &= \frac{9}{8} \sqrt{\frac{2}{3}} \frac{M_{pl}M^2}{N} \left(1-\frac{3}{4N}\right) \;,
\end{align}
and eventually we find
\begin{align}
\Delta_{\zeta,\textup{Star}}^2 = \frac{M^2}{24\pi^2 M_{pl}^2} N^2 \left(1-\frac{3}{4N}\right)^4 \approx 10^{-10} \;.
\end{align}
We see clearly that the mass scale must be fine-tuned. For $50<N<60$, we find $M\approx 10^{14.5}$ GeV. Since inflation was primarily introduced to avoid fine-tuning of the Hot Big Bang model, this fine-tuning is especially troubling.  %Considering the simple potential $V(\varphi)=\lambda \varphi^4$, where $\lambda$ is thought to take values around unity, this implies a tuning of 15 orders of magnitude to explain the amplitude.  Since inflation was primarily introduced to avoid fine-tuning of the Hot Big Bang model, this fine-tuning is especially troubling. 

The tilt predicted by a model depends on the equation of state $w$ used to describe the inflaton. For a generic inflationary scenario, it can be written as \cite{Mukhanov:2005sc}
\begin{align}
n_s-1 \simeq -3\left(1+w\right) -\frac{1}{H} \frac{\textup{d}}{\textup{d}t} \ln(1+w)  -\frac{1}{H} \frac{\textup{d}}{\textup{d}t} \ln c_s  \;,
\end{align}
where $c_s$ is the speed of sound, defined as
\begin{align}
c_s^2 := \frac{\partial P}{\partial \rho} \;.
\end{align}
The terms on the right are evaluated at the time of horizon crossing. A fine-tuned tilt is therefore necessary to accommodate the observations. Since inflationary models are constrained by $n_s$ and the tensor-to-scalar ratio $r$, it is important to know whether inflation produces a natural range of tensor modes. Forthcoming satellite missions such as LiteBIRD aim at measuring CMB B-modes with enough sensitivity to reach $r \simeq 10^{-3}$ \cite{Sugai:2020pjw}, which is the effective lower limit for the detection of primordial gravitational waves. However, it is thought to be unlikely to define criteria giving generic models robust against small changes with a definite gravitational wave prediction \cite{Bird:2008cp}. This makes the inflationary paradigm somewhat unnatural.

\subsection{Ambiguity of the Bunch-Davies Vacuum}

When we imposed initial conditions for the quantum perturbations, we defined creation and annhilation operators, hence we defined a vacuum state. We expressed the general solution for the perturbations with a normalisation leading to the Bunch-Davies vacuum \eqref{Bunch-Davies vacuum}, which states that all modes with positive frequency asymptotically match a Minkowski-like vacuum for arbitrarily small wavelengths. This was possible because Minkowski spacetime isometries, the Poincaré group, is a consequence of the equivalence principle. However, there is no timelike Killing vector in a FLRW spacetime, and vacuum is not uniquely defined in the asymptotic past limit \cite{Chung:2003wn,Scardua:2018omf}. If we could take the limit $\eta \rightarrow -\infty$ in \eqref{Bunch-Davies vacuum}, the ambiguity would vanish. In the physical situation where inflation begins at a given time $\eta_{init}$, we can only know the coefficients $A_{\boldsymbol{k}}$ and $B_{\boldsymbol{k}}$ from the Hankel expansion as early as $\eta_{init}$, but not before. This reflects in an ambiguity in the super-horizon inflaton power spectrum \eqref{power spectrum scalar} reading
\begin{align}
\bigg\rvert \frac{\delta \Delta^2_v(k)}{\Delta^2_v(k)} \bigg\rvert \simeq 2 | \textup{Re} B_k| \simeq \order{e^{-(N_{vac}-N_k)}} \;,
\end{align} 
where $N_{vac}$ and $N_k$ denote the number of e-folds before the end of inflation when spacetime is considered a vacuum, and when modes leave the horizon in a de Sitter phase, respectively. Note that this ambiguity is independent of the Trans-Planckian physics we mentioned; adding a cut-off in momentum space does not aliviate the ambiguity. 

We should note that another well-known vacuum defined by the absence of particles can be used, called the \textit{adiabatic vacuum} \cite{Parker:1974qw}. The number operator $n_k$ counts how many particles with mode $k$ is present in a given state. It is defined under three assumptions; $n_k$ must be hermitian, if expansion stops then $n_k$ matches the Minkowski number operator, and the state defined as vacuum should produced a negligible number of particles during cosmic evolution. Therefore, an adiabatic vacuum is well-defined if there is no particle production at two different times. In the adiabatic formalism, we define boundary conditions as an asymptotic expansion for each mode $v_k$, to all orders in the asymptotic parameters. In a de Sitter spacetime, the Bunch-Davies vacuum and the adiabatic vacuum match. Since inflation begins at a finite time, the two vacua definition have different asymptotic behaviour. Therefore, one could think the adiabatic vacuum can define unambiguously a vacuum at finite initial time, and this would be the case if the asymptotic expansion converges. Unfortunately, this is not the case since the infinite order adiabatic vacuum boundary conditions do not distinguish between a vacuum and a redefinition of the vacuum. Note that for a massless field in de Sitter, the uncertainty is erased by
\begin{align}
\bigg\rvert \frac{\delta \Delta^2_{v}(k)}{\Delta^2_{v}(k)} \bigg\rvert  \simeq \order{\exp\left(-e^{N_{vac}-N_k}\right)} \;.
\end{align} 
In the case of slow-roll inflation, the ambiguity in the power spectrum of curvature perturbation is given by
\begin{align}
\bigg\rvert \frac{\delta \Delta^2_{\zeta}(k)}{\Delta^2_{\zeta}(k)} \bigg\rvert \simeq 2 | \textup{Re} B_k| \simeq \order{p\exp\left[-2\sqrt{2}e^{(N_{vac}-N_k)/2}-\left(N_{vac}-N_k\right)/2\right]} \;,
\end{align}
where $p:=3\left(2\epsilon_1+\frac{\epsilon_2}{2}\right)$. In the case that $p \rightarrow 0$, the best estimate is given by the de Sitter case. In summary, there is a theoretical uncertainty in the inflationary power spectrum.

\subsection{Infrared Divergences Problem}

In de Sitter, infrared divergences naturally appear for massless and some massive scalar fields \cite{Miao:2010vs}. Therefore, taking the pure de Sitter limit of the free field \eqref{eq:delta_phi}, \textit{i.e.} expanding the Hankel functions to first order
\begin{align}
\varphi_k(\eta) = \frac{H}{\sqrt{2k}} \left(\eta-\frac{i}{k}\right) e^{-ik\eta} \;,
\label{scalar field expanded}
\end{align}
and plugging the result in the correlation function \eqref{power spectrum spherical} assuming a Bunch-Davies vacuum \eqref{Bunch-Davies vacuum}, we obtain
\begin{align}
\bra{0} \varphi(x,\eta) \varphi^{\star}(x^{\prime},\eta) \ket{0} \xrightarrow[k\eta \ll 1, k\eta^{\prime} \ll 1]{} \frac{H^2}{4\pi^2} \left(\ln k - \frac{k^2 |x-x^{\prime}|^2}{12}\right) \;.
\label{infrared divergence}
\end{align}
Therefore, a logarithmic infrared divergence appears in the correlation function.

%\subsubsection{The Cosmological Constant Problem}
%
%Observations of the local Universe show we are currently in a phase of re-acceleration attributed to the presence of a cosmological constant in Einstein equations.

\section{Solving Hot Big Bang and Some Inflation Problems with a Bounce}

Despite all the success of inflation in explaining large-scale observations, there are models seeking to give an alternative explanation of the initial conditions for the Big Bang and even the initial singularity. The most common such theories are called bouncing models, which consist of an existing universe before the Big Bang, often with a very long contracting phase, see \textit{e.g.} \cite{Novello:2008ra, Cai:2014bea, Battefeld:2014uga,Lilley:2015ksa, Brandenberger:2016vhg, Peter:2008qz,Galkina:2019pir}, and possibly an inflationary phase after the bounce, see \textit{e.g.} \cite{Biswas:2012bp, Piao:2003zm, Xia:2014tda, Liu:2013kea, Qiu:2015nha, Frion:2018oij}. An early proposal in which a pre-Big Bang phase could set the initial conditions for a subsequent post-Big Bang phase was made by Gasperini and Veneziano \cite{Gasperini:1992em}. Since then, several models have been presented, see e.g. \cite{Khoury:2001wf,Wands:2008tv}. Another interesting aspect of a bouncing universe lies in a possible explanation of the origin of supermassive black holes observed at very high redshift \cite{Banados:2017unc}. Such black holes are too young to accrete sufficient matter to explain its present mass in the $\Lambda$CDM model, but this problem is alleviated if they originate from before the Big Bang.

Adding a contracting phase and a bounce changes the causal structure. In the far past, the causal patch size was much smaller than the horizon, so the entire observable Universe was causally connected in the past. Since the horizon size can grow arbitrarily large in the past, it will eventually surpass the patch size. Hence, there is no horizon problem in bouncing cosmologies. Also, bouncing models are geodesically complete by construction, hence there is no initial singularity.

Constructing a classical bounce in GR using a FLRW metric poses a theoretical challenge, as it requires the addition of exotic matter violating the null energy condition \cite{Battefeld:2014uga}. A way-out is easy to find, as one can generate a bounce in anisotropic spacetimes \cite{Gonzalez:1994ny,Dechant:2008pb}, in modified gravities \cite{Oikonomou:2015qha}, in metric-affine theories \cite{Barragan:2009sq} or any combination of these possibilities, to name a few. A quantum bounce is even easier to obtain, and quantising GR can also lead to a bounce when using the de Broglie-Bohm interpretation of quantum mechanics \cite{Pinto-Neto:2013toa}, or using affine quantisation \cite{Bergeron:2013ika}, for example. A quantum bounce is a common feature of Loop Quantum Cosmology \cite{Wilson-Ewing:2017vju,Olmo:2008nf}, where triads are used instead of a metric.

Another interesting point is that the quantum-to-classical transition is not needed in classical non-singular bouncing models, by construction \cite{Ijjas:2018qbo}. In quantum bouncing model, this is no longer true within the Copenhagen interpretation. However, as we explained before, describing the Universe with standard Quantum Mechanics does not make sense. For instance, a quantum bounce within the de Broglie-Bohm theory is conceptually well-defined.

In a bouncing model, initial perturbations giving rise to large-scale structures can arise from quantum fluctuations in the far past \cite{Pinho:2006ym,Peter:2006hx,Vitenti:2012cx,Falciano:2013uaa,Bacalhau:2017hja}. The curvature scale tends to infinity in the asymptotic past. As a consequence, vacuum initial conditions for cosmological perturbations can be imposed in the dust-dominated contracting phase. Since the contraction starts at a slow-paced rate, choosing an adiabatic vacuum is quite natural since there will be almost no particle production initially \cite{Vitenti:2020???}. 

\begin{figure}[h!]
	\centering
	\includegraphics[scale=0.3]{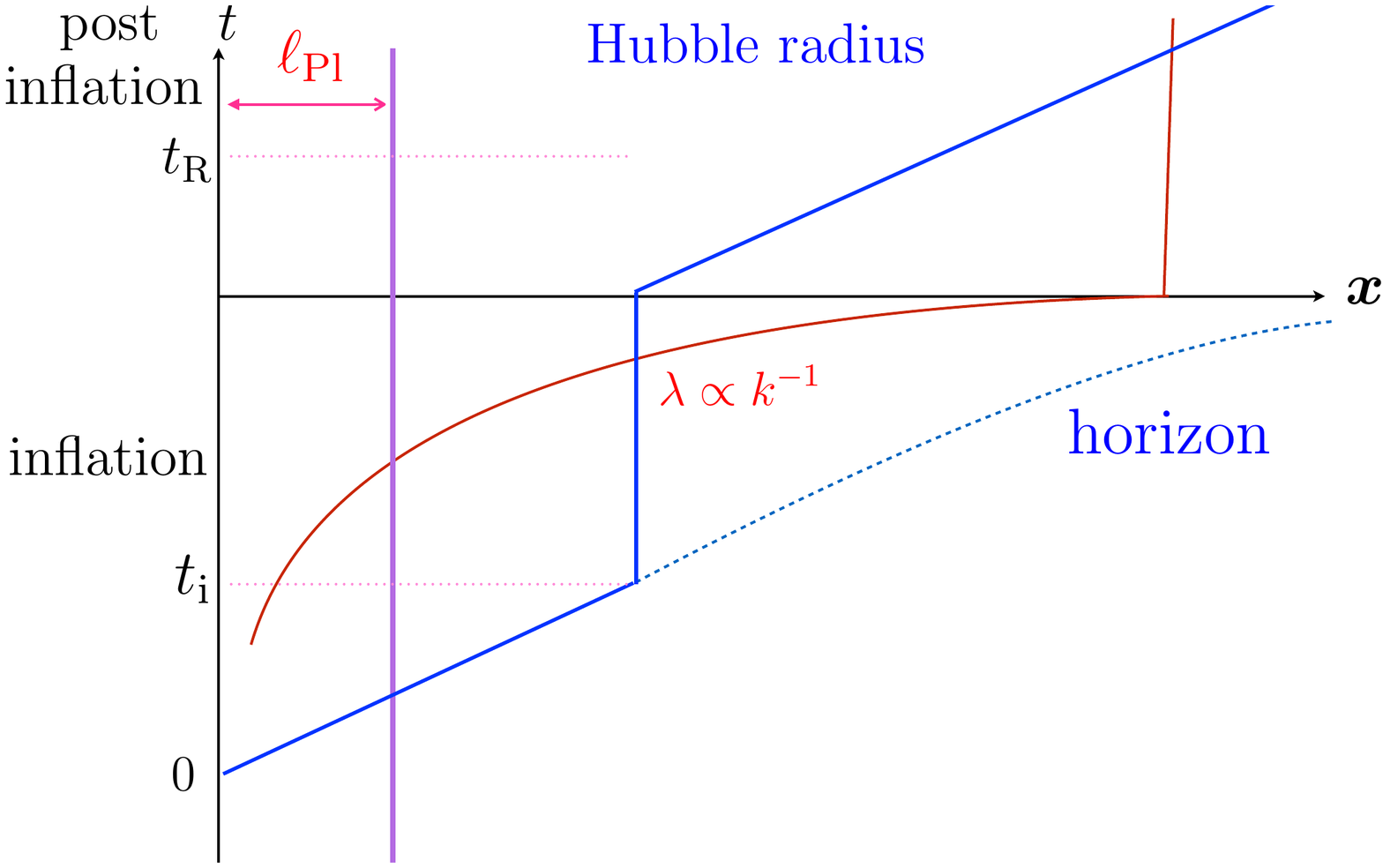}
	\hspace{1cm}
	\includegraphics[scale=0.3]{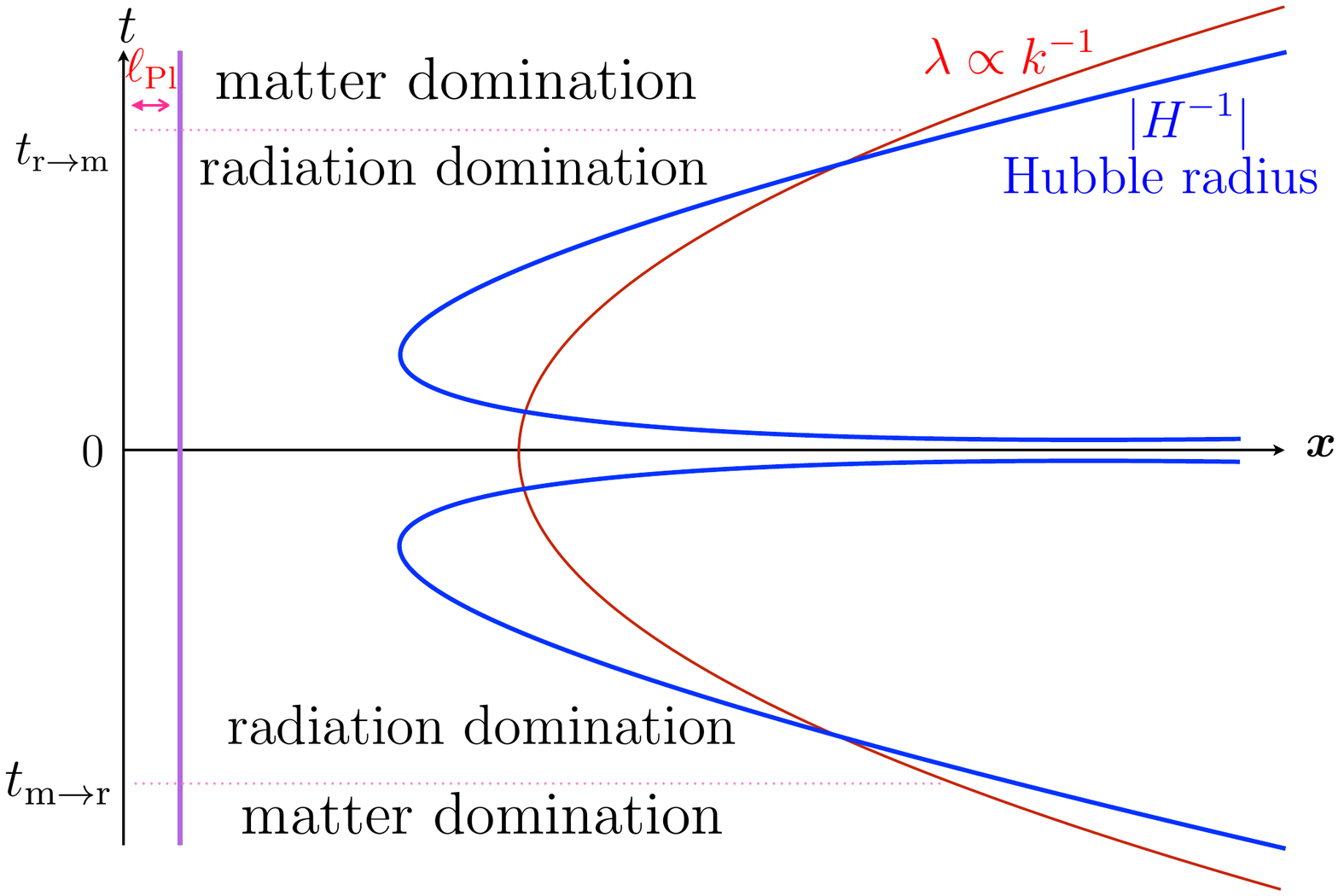}
	\caption[Trans-Planckian problem in inflation and matter bounce]{Trans-Planckian modes $k$ with corresponding wavelengths $\lambda$ in inflationary (left) and matter bounce (right) cosmologies \cite{Finelli:2001sr}. Relevant cosmological scales observed today originate from below the Planck length in inflation, while this cannot be the case with a bounce. Credit goes to R.~Brandenberger and P.~Peter \cite{Brandenberger:2016vhg}.}
	\label{trans-planckian}
\end{figure}

The Trans-Planckian problem is also naturally avoided in bouncing cosmologies. If the energy scale of a bounce is of the same order as the energy scale of inflationary models, the wavelengths corresponding to the anisotropies observed in the CMB have a size of about 1 mm \cite{Brandenberger:2016vhg}. This is sketched in the right panel of figure \ref{trans-planckian}.

In a contraction phase, local monopoles are created in abundance if temperature is high enough, typically at the GUT scale. As long as the bounce does not occur at a much higher energy scale, all monopoles will be diluted in the same fashion as in inflationary models.

\subsection{Shear Instabilities}

In order to solve the remaining Hot Big Bang problems, a bouncing model must also be able to provide sufficiently small gradients of energy, and in fact of any form of vector perturbations after inflation, as they are at most very tiny in the CMB sky. This can be a problem in bouncing models, since the energy density associated with anisotropic stress grows as $a^{-6}$ in a homogeneous cosmology, and in turn causes strong instabilities in the contracting phase \cite{Belinsky:1970ew}, spoiling the homogeneity and isotropy of the background. This behaviour is known as the \textit{BKL instability}, or mixmaster dynamics. In particular, explaining the origin of large-scale structures from quantum fluctuations can be challenging as they generate anisotropic stress \cite{Miranda:2019ara,Grain:2020wro}. In an initially homogeneous background filled with perfect fluid, the vacuum stress can grow strong enough to backreact and modify the classical dynamics.

Another source of instabilities comes from the growth of perturbations. Even if scalar and tensor perturbations can remain linear throughout the bounce \cite{Vitenti:2011yc,Pinto-Neto:2013zya}, it is often believed that vector perturbations, scaling as $a^{-2}$ in the presence of a perfect fluid, will grow uncontrollably towards the bounce, spoiling the background dynamics. However, it was shown that vector perturbations do not diverge in non-singular bounces \cite{Bari:2019yvk}, even in the more general case of imperfect fluids \cite{Pinto-Neto:2020xmb}.

\subsection{Distinguishing Bouncing and Inflationary Models}

The detection of primordial gravitational waves (PGW) would serve as a basis for distinguishing inflationary models, that require $r\neq 0$, from some bouncing models such as the ekpyrotic model and cyclic scenarios with an extremely tiny gravitational waves production \cite{Boyle:2003km}. However, it should be noted that if PGW are observed, standard inflation would happen at an energy scale a few times $10^{16}$ GeV \cite{Lyth:1996im}, hence the potential would prove incompatible with the standard model of particles. In this case, the inflaton would present Trans-Planckian fluctuations. While the behaviour of a scalar at such energy scales is unknown, and maybe undesirable as explained earlier, only the energy scale of inflation itself matters when observing the sky. Hopefully, future surveys will be able to distinguish between the different scenarios.

Many bouncing models predict large non-gaussian contributions stemming from interaction terms ensuring the bounce mechanism. Curvature perturbations can be expanded in a linear and non-linear parts as
\begin{align}
\zeta = \zeta_{gauss} (\boldsymbol{x}) + \frac{3}{5} f_{NL} \zeta^2_{gauss} (\boldsymbol{x}) + \order{\zeta_{gauss}^3} \;,
\end{align}
where $\zeta_{gauss}$ is linear and gaussian. The $f_{NL}$ term is called the \textit{non linearity parameter}. The factor $3/5$ comes from the definition of non-gaussianities in terms of the gauge-invariant Newtonian potential $\Phi$. During matter domination, we have $3\zeta =5\Phi$ \cite{Maldacena:2002vr}. Depending on the configuration, several limits have been given by the Planck collaboration \cite{Akrami:2019izv} within the 68\% confidence level as
\begin{align}
f_{NL}^{\textup{local}} &= -0.9 \pm 5.1 \\
f_{NL}^{\textup{equil}} &= -26 \pm 47 \\
f_{NL}^{\textup{ortho}} &= -38 \pm 24 \;.
\end{align}
The local term contains primordial fluctuations usually produced during inflation from an isocurvature field, but also from the ekpyrotic scenario \cite{Buchbinder:2007at,Koyama:2007if,Lehners:2010fy}. The equilateral type results from modifications of the kinetic Lagrangian of a general single-field inflation model, or from higher-order derivative operators, while the orthogonal type, orthogonal to the local and equilateral types, mainly comes from the latter possibility. Note that a fourth type exist, the folded type, mixing local and orthogonal terms. It appears in some models with vacuum different from the Bunch-Davies vacuum \cite{Chen:2006nt}. Non-gaussianities for various inflationary models can be found in \cite{Wang:2013zva}. As an example, the single-field slow-roll inflation gives $f_{NL}^{\textup{local}} < 1$ \cite{Maldacena:2002vr}, while the matter bounce scenario in \cite{Cai:2009fn} gives $f_{NL}^{\textup{local}} = -35/8$.

%% file: sections/chapter4.tex
\chapter{Stochastic Effects in Cosmology}
\label{stochastic effects in cosmology}

\section{Stochastic Formalism}

When we discussed the importance of inflationary perturbations, we have supposed that the inflaton is split into a homogeneous and inhomogeneous parts. Inhomogeneities, supposedly small, were treated as quantum perturbations and with no influence on the background. It is natural to wonder if quantum fluctuations can affect background trajectories. The stochastic formalism models the effect of quantum vacuum fluctuations by introducing a cut-off scale (the so-called coarse-graining scale) splitting the fluctuations into two parts: quantum vacuum modes (below the coarse-graining scale) and the long wavelength field. The long wavelength field behaves classically, and its dynamics can be modified by quantum fluctuations under the form of a stochastic noise. The stochastic formalism was introduced in cosmology by Starobinsky \cite{Starobinsky:1986fx} to describe the effects of random vacuum fluctuations on inflationary dynamics \cite{Nambu:1987ef,Nambu:1988je,Nambu:1989uf,Kandrup:1988sc,Nakao:1988yi,Mollerach:1990zf,Linde:1993xx,Starobinsky:1994bd,Glavan:2017jye,Hardwick:2019uex,Saitou:2019jez,Tokuda:2018eqs,Hardwick:2018sck,Ando:2018qdb,Rigopoulos:2016oko,Nambu:2009zb,Tye:2008ef,Kunze:2006tu,Geshnizjani:2004tf,Matarrese:2003ye,Bellini:2001jm,Bellini:2000er,Bellini:1999zb,Calzetta:1999zr,Casini:1998wr,Ramsey:1997fz,Boyanovsky:1996ab,GarciaBellido:1994vz,Mijic:1994vv,Habib:1992ci,Mizutani:1991xh,Graziani:1988yd,Pattison:2017mbe,Pattison:2018bct}. More recently, this formalism was used in \cite{Grain:2017dqa} to show that slow-roll inflation is a stochastic attractor. The stochastic formalism also comes with the appealing feature that it reproduces results from QFT in curved spacetime at the perturbative level and beyond \cite{Starobinsky:1994bd,Finelli:2008zg,Finelli:2010sh,Garbrecht:2013coa}.  All modes with frequency below the cut-off scale are treated as quantum operators, while super-horizon modes are classical. In some sense, we can say that we are applying out-of-equilibrium QFT for open systems to an inflationary context.

We first illustrate how the stochastic formalism applies in cosmology by showing the divergence found in \eqref{infrared divergence} can be lifted introducing a time-dependent comoving cut-off scale (the coarse-graining scale)
\begin{align}
\label{cgscale}
k_{\sigma} = \sigma a H \;,
\end{align}
below which small (sub-Hubble) wavelengths are integrated out. We thus derive an effective theory for the long wavelength part, with $\sigma$ the ratio between the Hubble radius and the cut-off wavelength. Note that we require the coarse-graining scale to be larger than the Hubble scale, hence $\sigma <1$, to neglect gradient terms \cite{Pattison:2019hef}. The reason is that no causal process can prepare the initial state in a coherent Bunch-Davies vacuum over an infinite spatial section \cite{Prokopec:2007ak}. 

We will characterise the noise emerging from sub-Hubble quantum fluctuations evolving into the super-Hubble regime following the stochastic formalism described in \cite{Grain:2017dqa}. 
Classically, the time derivative of the field, $\dot{\varphi}$, can be related to its conjugate momentum, $\pi_{\varphi}\equiv \partial \mathcal{L}/\partial \dot{\varphi}$. Using the ADM formulation of a scalar field \eqref{scalar field adm} in a homogeneous and isotropic spacetime, the scalar field reads
\begin{align}
\dot{\varphi} = \frac{N}{a^{3}}\pi_{\varphi}\;,\label{eq:phipi}
\end{align}
and the evolution for $\pi_{\varphi}$ is given by
\begin{align}
\dot{\pi}_{\varphi} = - N a^{3} V_{,\varphi} +a\delta^{ij}\partial_{i}\partial_{j}\varphi \;.
\label{eq:pidot}
\end{align}
In the stochastic approach, one splits the scalar field and its momentum into a long-wavelength part and small-wavelength part as
\begin{align}
\varphi = \overline{\varphi} + \varphi_{Q}\,, \qquad \pi = \overline{\pi} + \pi_{Q}\;,
\end{align}
where the subscript ``$Q$" describes the small-wavelength quantum part. Explicitly, we define the coarse-grained field and momentum in terms of the window function $W$, hence integrating out the UV modes, as
\begin{align}
\varphi_Q \left(\boldsymbol{x},\eta \right) = \int \frac{\textup{d}^3 k}{\left(2\pi\right)^3} W \left(\frac{k}{k_{\sigma}}\right) \left(a_{\boldsymbol{k}} \varphi_{\boldsymbol{k}} (\eta) e^{-i \boldsymbol{k}\cdot \boldsymbol{x}} + a_{\boldsymbol{k}}^{\dagger} \varphi_{\boldsymbol{k}}^{\star} (\eta) e^{i \boldsymbol{k}\cdot \boldsymbol{x}}\right) \;, \\
\pi_Q \left(\boldsymbol{x},\eta \right) = \int \frac{\textup{d}^3 k}{\left(2\pi\right)^3} W \left(\frac{k}{k_{\sigma}}\right) \left(a_{\boldsymbol{k}} \pi_{\boldsymbol{k}} (\eta) e^{-i \boldsymbol{k}\cdot \boldsymbol{x}} + a_{\boldsymbol{k}}^{\dagger} \pi_{\boldsymbol{k}}^{\star} (\eta) e^{i \boldsymbol{k}\cdot \boldsymbol{x}}\right) \;,
\label{small-wavelength quantum part}
\end{align}
where $W\simeq 0$ for $k\ll k_{\sigma}$ and  $W\simeq 1$ for $k\gg k_{\sigma}$. Plugging the quantum parts in \eqref{eq:phipi} and \eqref{eq:pidot}, we can write the Langevin equations for the stochastic quantities $\varphi$ and $\pi$
\begin{align}
\dot{\overline{\varphi}} &= \frac{N}{a^3} \overline{\pi} + \xi_{\varphi} (\boldsymbol{x},\eta) \;, \\
\dot{\overline{\pi}} &= -Na^3 V_{,\varphi}\left(\overline{\varphi}\right) + \xi_{\pi} (\boldsymbol{x},\eta) \;,
\label{Langevin equation}
\end{align}
where the quantum noises $\xi_{\varphi}$ and $\xi_{\pi}$ are given by \cite{Nakao:1988yi,Habib:1992ci,Rigopoulos:2005xx,Tolley:2008na}
\begin{align}
\xi_{\varphi}(\boldsymbol{x},\eta) &= -\int \frac{\textup{d}^3 k}{\left(2\pi\right)^3} \dot{W} \left(\frac{k}{k_{\sigma}}\right) \left(a_{\boldsymbol{k}} \varphi_{\boldsymbol{k}} (\eta) e^{-i \boldsymbol{k}\cdot \boldsymbol{x}} + a_{\boldsymbol{k}}^{\dagger} \varphi_{\boldsymbol{k}}^{\star} (\eta) e^{i \boldsymbol{k}\cdot \boldsymbol{x}}\right) \;, \\
\xi_{\pi}(\boldsymbol{x},\eta) &= - \int \frac{\textup{d}^3 k}{\left(2\pi\right)^3} \dot{W} \left(\frac{k}{k_{\sigma}}\right) \left(a_{\boldsymbol{k}} \pi_{\boldsymbol{k}} (\eta) e^{-i \boldsymbol{k}\cdot \boldsymbol{x}} + a_{\boldsymbol{k}}^{\dagger} \pi_{\boldsymbol{k}}^{\star} (\eta) e^{i \boldsymbol{k}\cdot \boldsymbol{x}}\right) \;.
\label{stochastic noises}
\end{align}
Note that this result can be also derived by integrating out the small-wavelength degrees of freedom, analogously as in renormalisation schemes in QFT. Also, the fact that $\varphi$ and $\pi$ are still operators becomes irrelevant.

It becomes clear the infrared divergence identified in \eqref{infrared divergence} disappears. Indeed, inserting the long-wavelength counterpart of \eqref{small-wavelength quantum part} by switching $W(k/k_{\sigma}) \leftrightarrow W(k_{\sigma}/k)$ into the solution \eqref{scalar field expanded}, we would now find \eqref{infrared divergence} is finite. The cancellation of the divergence is not a coincidence. In fact, the leading IR divergence vanishes at all orders of perturbation. However, we shall see another problem emerges when including an interaction term in the Klein-Gordon background equation \eqref{eq:KGback}. In conformal time, it is expressed as
\begin{align}
\frac{\partial^2 \varphi}{\partial \eta^2} + 2 \mathcal{H} \frac{\partial \varphi}{\partial \eta} - \nabla^2 \varphi + a^2 \frac{\partial V}{\partial \varphi} = 0 \;.
\end{align}
Noting the free field as $\varphi^{(0)}$, the solution to this inhomogeneous Klein-Gordon equation can be found using the Green retarded propagator
\begin{align}
\left( \frac{\partial^2 \varphi}{\partial \eta^2} + 2 \mathcal{H} \frac{\partial \varphi}{\partial \eta} - \nabla^2 \varphi + a^2 \frac{\partial V\left(\varphi^{(0)}\right)}{\partial \varphi^{(0)}} \right) G\left(x_i,\eta;x_i^{\prime},\eta^{\prime}\right) = \frac{1}{a^3} \delta^3 \left(x_i-x_i^{\prime}\right) \;,
\end{align}
where the Green function is 
\begin{align}
G\left(x_i,\eta;x_i^{\prime},\eta^{\prime}\right) = \frac{i}{a^2}  \Theta \left(\eta - \eta^{\prime}\right) \left[\varphi^{(0)}\left(x_i,\eta\right),\varphi^{(0)}\left(x_i^{\prime},\eta^{\prime} \right)\right] \;.
\end{align}
Integrating all interactions of the field, we construct the Yang-Feldman equation \cite{Yang:1950vi}
\begin{align}
\varphi(x_i,\eta) = \varphi^{(0)}(x_i,\eta) - \int \textup{d} \eta^{\prime} \int \textup{d}^3 x^{\prime}  \sqrt{-g} G\left(x_i,\eta;x_i^{\prime},\eta^{\prime}\right) a^2(\eta^{\prime}) \frac{\partial V\left(\varphi(x_i^{\prime},\eta^{\prime}) \right)}{\partial \varphi^{(0)}} + \order{G^2} \;.	
\end{align}
Note that the vertex $\sqrt{-g}G a^2 \approx \textup{d} \ln a$ adds another infrared divergence, breaking down the perturbative expansion after some time \cite{Burgess:2010dd}. This secular effect can emerge, for example, when studying perturbations of non-conformal fields \cite{Tsamis:2005hd}.

An interesting aspect of the stochastic formalism is the quantum-to-classical transition naturally occurring when considering super-horizon modes. Let us consider the noise power spectrum  
\begin{align}
\Xi_{f,g}(\boldsymbol{x},\eta;\boldsymbol{x}^{\prime},\eta^{\prime}) := \left\langle 0 | \xi_f (\boldsymbol{x},\eta) \xi_g (\boldsymbol{x}^{\prime},\eta^{\prime}) |0 \right\rangle\;,
\label{noise ps}
\end{align}
with $f$, $g$ and $\xi_f$, $\xi_g$ being shorthand notation for the field or its momentum and their respective noises. Recalling the generic power spectrum \eqref{power spectrum spherical} and inserting the noises \eqref{stochastic noises} into \eqref{noise ps} leads to
\begin{align}
\Xi_{f,g} = \int_{0}^{\infty} \frac{\textup{d}k \,k^2}{2\pi^2} \dot{W} \left(\frac{k}{k_{\sigma}(\eta)}\right) \dot{W} \left(\frac{k}{k_{\sigma}(\eta^{\prime})}\right) f_k(\eta) g_k^{\star}(\eta^{\prime}) \frac{\sin(k|\boldsymbol{x}^{\prime}-\boldsymbol{x}|)}{k|\boldsymbol{x}^{\prime}-\boldsymbol{x}|} \;.
\end{align}
For simplicity, we choose the window function to be a Heaviside function, whose derivative is
\begin{align}
\dot{\Theta} \left(\frac{k}{k_{\sigma}(\eta)} -1\right)= \delta \left(\frac{k}{k_{\sigma}(\eta)} -1\right)= k_{\sigma}(\eta) \delta \left(k-k_{\sigma}(\eta)\right) \;.
\end{align}
Therefore, the product of window functions means the noise correlation is evaluated at equal time, and the noises are white. In turn, $\varphi$ is a Markovian process. This would not be the case had we chosen a smooth function over a step function, and would lead to non-Gaussianities in the final power spectrum \cite{Liguori:2004fa}. It follows that
\begin{align}
\Xi_{f,g} = \frac{1}{6\pi^2} \frac{\textup{d}k_{\sigma}^3(\tau)}{\textup{d}\tau} \bigg\rvert_{\tau=\eta} f_{k=k_{\sigma}(\eta)} g^{\star}_{k=k_{\sigma}(\eta)} \frac{\sin(k_{\sigma}(\eta)|\boldsymbol{x}^{\prime}-\boldsymbol{x}|)}{k_{\sigma}(\eta)|\boldsymbol{x}^{\prime}-\boldsymbol{x}|} \delta(\eta - \eta^{\prime}) \;.
\end{align}
As we suppose the noises are autocorrelated, we have $\boldsymbol{x}^{\prime}=\boldsymbol{x}$, and we can rewrite this result in terms of the dimensionless power spectrum \eqref{power spectrum dimensionless}
\begin{align}
\Xi_{f,g} = \frac{\textup{d}\ln(k_{\sigma}(\eta))}{\textup{d} \eta} \Delta_{f,g}^2 \left(k_{\sigma}(\eta);\eta\right) \;, \quad \textup{with} \quad \Delta_{f,g}^2 \left(k_{\sigma}(\eta);\eta\right) = \frac{k_{\sigma}^3(\eta)}{2\pi^2} f_{k=k_{\sigma}(\eta)} g^{\star}_{k=k_{\sigma}(\eta)} \;.
\label{ps white noise}
\end{align}

We notice that hermiticity of the noise correlator is evident, hence $\Xi_{f,g}=\Xi_{g,f}^{\star}$. The only non-trivial antisymmetric combination in this case is given
\begin{align}
\Xi_{\varphi, \pi}(\eta)-\Xi_{\pi, \varphi}(\eta) &=\frac{1}{6\pi^2} \frac{\textup{d}k_{\sigma}^3(\tau)}{\textup{d}\tau} \left[\varphi_{k=k_{\sigma}(\eta)} \pi^{\star}_{k=k_{\sigma}(\eta)} - \pi_{k=k_{\sigma}(\eta)} \varphi^{\star}_{k=k_{\sigma}(\eta)}\right] \nonumber \\
&= -\frac{i}{6\pi^2} \frac{\textup{d}k_{\sigma}^3(\tau)}{\textup{d}\tau} \;;
\end{align}
where canonical quantisation was used in the second line. This term contains all the information about the quantum nature of the fluctuations, and disappears on large scales. This brings the interesting feature that, when dealing with stochastic processes, we lose all information about commutators and the outcome is classical. The probability distribution of finding the coarse-grained field in a state $\varphi$ at time $\eta$, the \textit{Fokker-Planck distribution}, can be obtained using the Itô interpretation \footnote{The description of stochastic processes depends on the chosen discretisation scheme. In stochastic calculus, the solution of a differential equation is a function evaluated at each step $i$ as $\left(1-\alpha\right)N_i+\alpha N_{i+1}$, where $N_i$ and $N_{i+1}$ are time variables, and $0<\alpha<1$ a parameter. The Itô scheme corresponds to the choice $\alpha=0$, and possesses the advantage that we can express variables at time $N_{i+1}$ in terms of known values at $N_i$. However, covariance can be lost in this scheme, though it can be retrieved by parallel transporting vielbeins from the Stratonovich scheme, equivalent to $\alpha=1/2$ \cite{Pinol:2020cdp}.}. We now wish to write the distribution associated to the Langevin equations \eqref{Langevin equation}, denoted $P(\overline{\varphi},\overline{\pi},\eta)$. The Fokker-Planck evolution equation yields \cite{risken1996fokker}
\begin{align}
\frac{\partial P(\overline{\varphi},\overline{\pi},\eta)}{\partial \eta} = &-\frac{\partial}{\partial \overline{\varphi}} \left[\frac{N}{a^3} \overline{\pi} P(\overline{\varphi},\overline{\pi},\eta)\right] + \frac{\partial}{\partial \overline{\pi}} \left[Na^3 V_{,\varphi}\left(\overline{\varphi}\right) P(\overline{\varphi},\overline{\pi},\eta)\right] \nonumber\\
&+ \left[\frac{\partial^2}{\partial \overline{\varphi}^2} \Xi_{\varphi, \varphi} + \frac{\partial^2}{\partial \overline{\varphi} \partial \overline{\pi}} \Xi_{\varphi, \pi} +  \frac{\partial^2}{\partial \overline{\pi} \partial \overline{\varphi}} \Xi_{\pi, \varphi} + \frac{\partial^2}{\partial \overline{\pi}^2} \Xi_{\pi, \pi} \right] P(\overline{\varphi},\overline{\pi},\eta)  \;.
\end{align} 
The first line is nothing more than the deterministic evolution, and each term is called a drift term. The second line collects all terms sourced by the noise, and compose the diffusion term. Since none of the noise correlators depend on the deterministic variables, we rewrite the Fokker-Planck equation as
\begin{align}
\frac{\partial P(\overline{\varphi},\overline{\pi},\eta)}{\partial \eta} = &-\frac{N}{a^3} \overline{\pi}\frac{\partial}{\partial \overline{\varphi}}  P(\overline{\varphi},\overline{\pi},\eta) + Na^3 V_{,\varphi}\left(\overline{\varphi}\right) \frac{\partial}{\partial \overline{\pi}}  P(\overline{\varphi},\overline{\pi},\eta) \nonumber \\
&+ \left[\Xi_{\varphi, \varphi} \frac{\partial^2}{\partial \overline{\varphi}^2}  + \Xi_{\varphi, \pi} \frac{\partial^2}{\partial \overline{\varphi} \partial \overline{\pi}}  +  \Xi_{\pi, \varphi} \frac{\partial^2}{\partial \overline{\pi} \partial \overline{\varphi}} +  \Xi_{\pi, \pi}\frac{\partial^2}{\partial \overline{\pi}^2} \right] P(\overline{\varphi},\overline{\pi},\eta)  \;.
\label{Fokker-Planck equation}
\end{align}
Let us put the noises in a matrix form. Be the deterministic vector and the matrix noise
\begin{align}
\boldsymbol{\Phi} = \begin{pmatrix}
\overline{\varphi} \\
\overline{\pi}
\end{pmatrix} \quad \textup{and} \quad 
\boldsymbol{\Xi} = \begin{pmatrix}
\Xi_{\varphi, \varphi} & \Xi_{\varphi, \pi} \\
\Xi_{\varphi, \pi} & \Xi_{\pi, \pi}
\end{pmatrix} \;.
\end{align}
The Hessian $\boldsymbol{H}$ of the probability density function is formed from the matrix entries
\begin{align}
H_{ij} = \frac{\partial^2 P(\boldsymbol{\Phi},\eta)}{\partial \Phi_{i} \partial \Phi_{j}} \;.
\end{align}
Thus, the diffusion term in \eqref{Fokker-Planck equation} can be written as the trace $\textup{Tr}\left[\boldsymbol{H}\boldsymbol{\Phi}\right]$. Recalling the matrix noise is hermitian, we can decompose the noise matrix on the basis $\left(I_2,J_x,J_y,J_z\right)$, where $I_2$ is the two-dimensional identity matrix and the other three are Pauli matrices. Note that all four matrices but $J_y$ are symmetric. This gives
\begin{align}
\boldsymbol{\Xi} = \frac{1}{2} \left(\Xi_{\varphi, \varphi} + \Xi_{\pi, \pi}\right) I_2 + \frac{1}{2} \left(\Xi_{\varphi, \pi} + \Xi_{\pi, \varphi}\right) J_x + \frac{i}{2} \left(\Xi_{\varphi, \pi} - \Xi_{\pi, \varphi}\right) J_y + \frac{1}{2} \left(\Xi_{\varphi, \varphi} - \Xi_{\pi, \pi}\right) J_z \;.
\end{align}
Since the Hessian is symmetric under indices permutations while $J_y$ is antisymmetric, the third term, built out of the commutator of quantum fluctuations, vanishes. Hence, quantum diffusion is described by a classical distribution through correlated noises.

%First, we note
%\begin{align}
%	\boldsymbol{\Phi} = \begin{pmatrix}
%			\overline{\varphi} \\
%			\overline{\pi}
%		\end{pmatrix} \;, \quad 
%		\boldsymbol{\xi} = \begin{pmatrix}
%		\xi_{\varphi} \\
%		\xi_{\pi}
%		\end{pmatrix} \;,
%\end{align}
%leading to the vectorial Langevin equation
%\begin{align}
%	\dot{\boldsymbol{\Phi}} = \boldsymbol{A} \boldsymbol{\Phi} + \boldsymbol{\xi} \;, \quad \textup{with} \quad \boldsymbol{A} = \begin{pmatrix}
%		0 & N/a^3 \\
%		-Na^3 V_{,\varphi} \left(\overline{\varphi}\right) & 0 
%	\end{pmatrix} \;.
%\end{align}

\section{Stochastic Collapse with Exponential Potential}

We now apply the stochastic formalism to a phase of contraction, in which we consider a scalar field with an exponential potential. The remaining of this chapter serves as a presentation and discussion of the results we obtained in \cite{Miranda:2019ara}. This simple choice provides a model for an accelerated expansion and inflation in the early universe \cite{Lucchin:1984yf, Kitada:1992uh, Wands:1993zm} or dark energy at the present epoch \cite{Wetterich:1994bg, Amendola:1999qq}. But it can also drive a collapsing universe and this configuration is particularly interesting since due to their scale-invariant form, exponential potentials are simple to study analytically. Such a configuration in the case of bouncing models with the bounce due to quantum effects was studied in \cite{Bacalhau:2017hja}. In order to explore this simplicity, we will focus our attention on models with a scalar field $\varphi$ and scalar potential
\begin{align}
V=V_0 \exp{- \kappa \lambda \varphi } \; ,
\end{align}
where $\kappa = \sqrt{8\pi G}$ and $\lambda$ is the slope of the potential. We can identify three scalar field collapse scenarios based on the form of the potential. In terms of energy density $\rho$ and pressure $P$ of the scalar field, we have 
\begin{itemize}
	\item Non-stiff/Matter/Pressureless collapse ($P < \rho$ with $V > 0$);
	\item Pre-Big Bang collapse ($P = \rho $ with $V = 0$);
	\item Ekpyrotic collapse ($P \gg \rho$ with $V < 0$);	
\end{itemize}

Before studying how collapse models behave in the presence of quantum fluctuations, it is interesting to show that the classical dynamics of a scalar field cosmology with an exponential potential can be reduced to an one-dimensional problem. We discuss the phase space portrait for this theory as well as the stability of the fixed points which represent power-law expansion or collapse. We will use the single-field inflation action \eqref{action inflation} in a flat FLRW spacetime.

\begin{figure}[h!]
	\centering
	\includegraphics[scale=0.2]{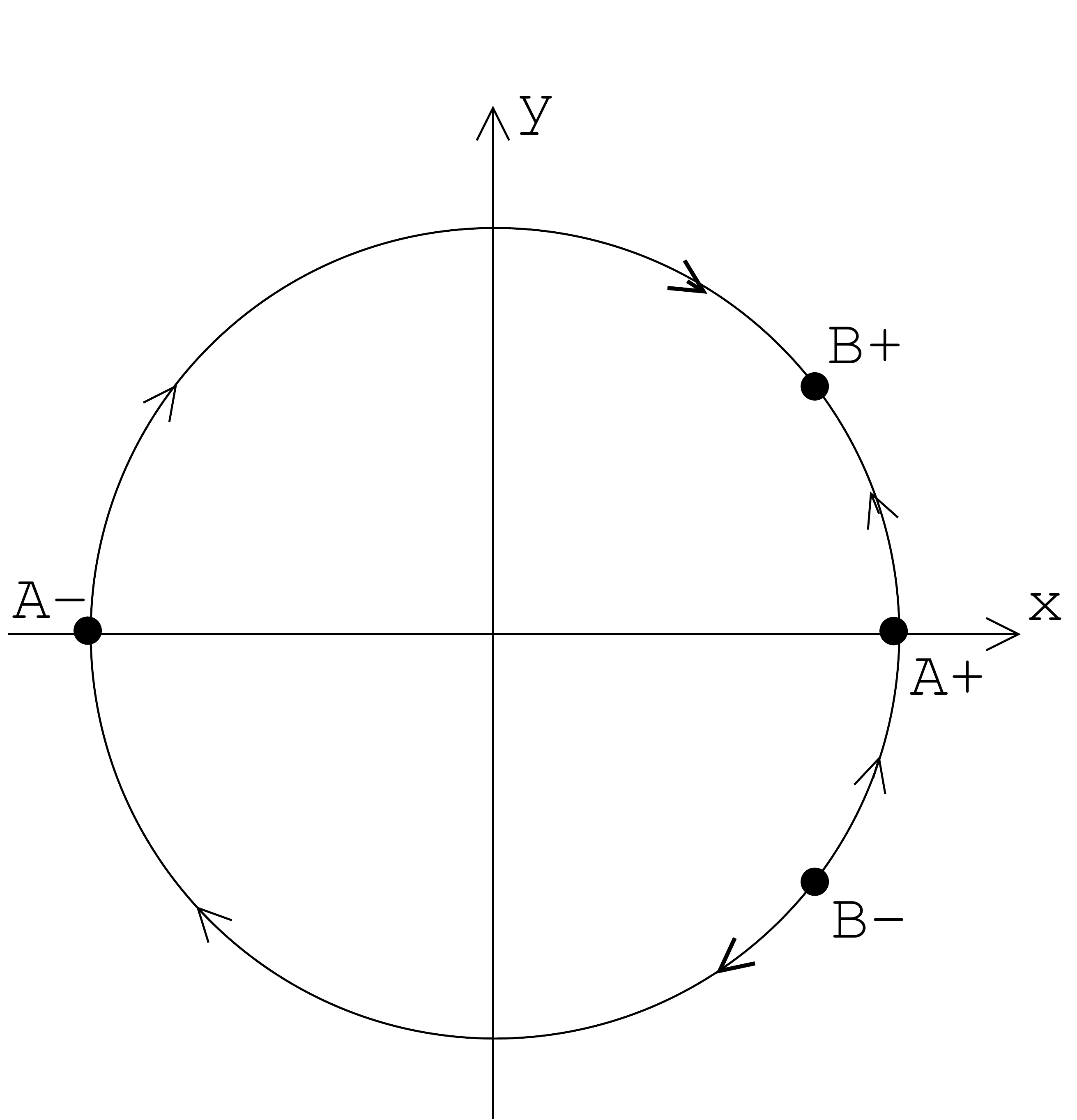}
	\hspace{1cm}
	\includegraphics[scale=0.2]{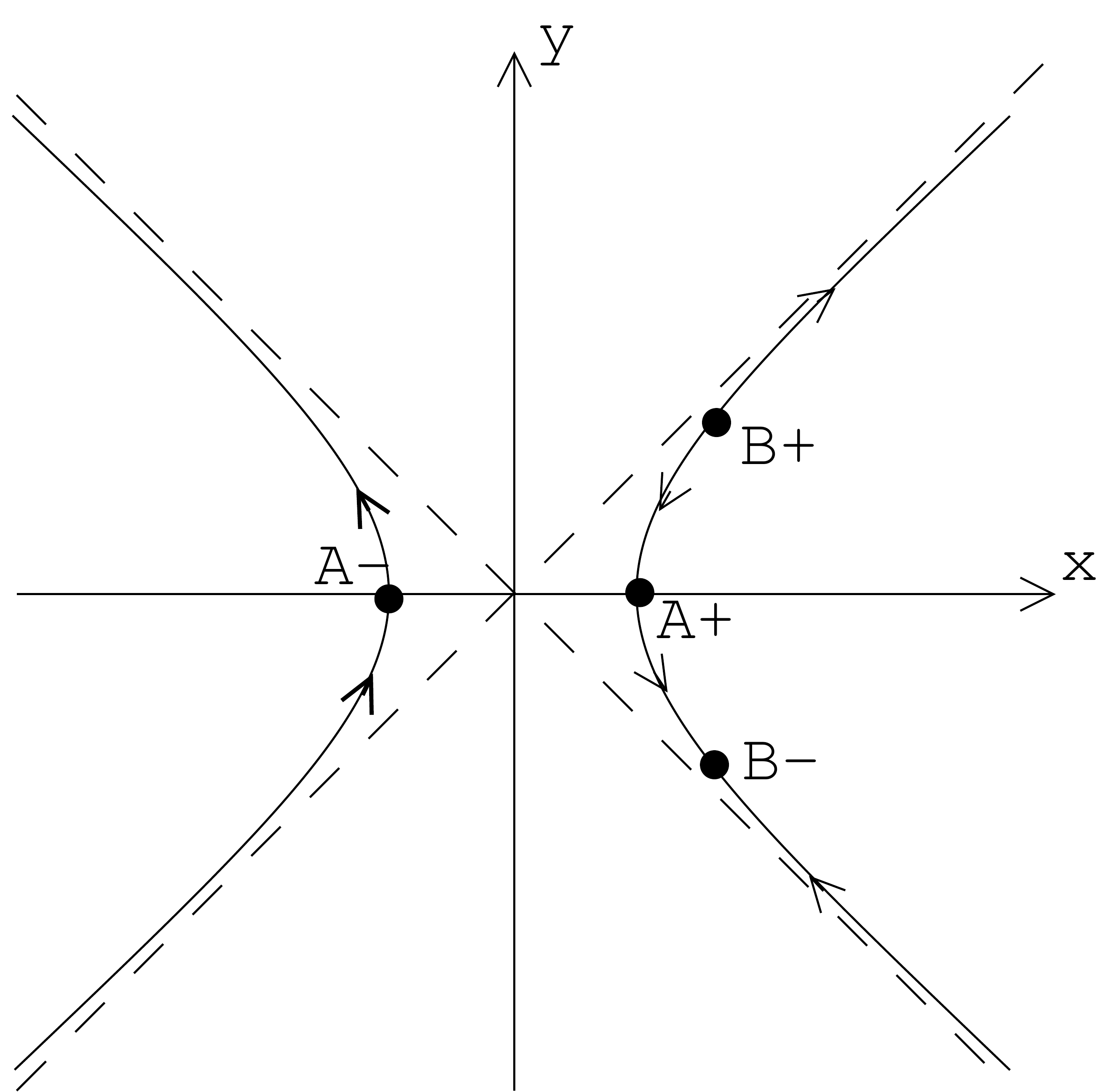}
	\caption[Dynamics of a scalar field with flat ($\lambda^2<6$) positive exponential potential ($x^2+y^2=1$) and steep ($\lambda^2>6$) negative ($x^2-y^2=1$) exponential potential.]{Dynamics of a scalar field with flat ($\lambda^2<6$) positive exponential potential ($x^2+y^2=1$) and steep ($\lambda^2>6$) negative (($x^2-y^2=1$)) exponential potential. The upper half-plane is for $H>0$, \textit{i.e.} an expanding Universe, and the lower half-plane for $H<0$. The circle and hyperbole represent the Friedmann constraint \eqref{eq:FriedmannConstraint}. $A\pm$ are saddle points representing the kinetic-dominated region. $B-$ and $B+$ represent the same potential-kinetic solution in two different phases. During contraction, $B$ is a repeller, and turns out to be a late-time attractor during expansion. Note the sign of $x$ flips when $H$ crosses 0. Credit goes to I.~Heard and D.~Wands \cite{Heard:2002dr}.}
	\label{heard1}
\end{figure}

In order to perform a qualitative analysis of the system described by the Klein-Gordon equation (\ref{eq:KGback}) and the Friedmann constraint (\ref{eq:friedmann1}), new variables can be introduced as
\begin{align}
\label{eq:dimensionless variables}
x = \frac{\kappa \dot{\varphi}}{\sqrt{6} H} \;, \quad y = \frac{\kappa \sqrt{\pm V}}{\sqrt{3} H},
\end{align}
where we use $\pm$ for positive and negative scalar potentials, $\pm V>0$. With these variables, we can write the Friedmann constraint~(\ref{eq:friedmann1}) as
\begin{align}
x^2 \pm y^2 = 1\;,\label{eq:FriedmannConstraint}
\end{align}
and the equation of state~(\ref{eq:eos1}) becomes
\begin{align}
w = \frac{x^{2}\mp y^{2}}{x^{2}\pm y^{2}}\;.\label{eq:eos2}
\end{align}
Then, we are able to rewrite the equation of motion (\ref{eq:KGback}) in terms of the autonomous system
\begin{align}
x^{\prime} &= -3x(1-x^{2})\pm \lambda\sqrt{3/2}y^{2}\;,\label{eq:xback}\\
y^{\prime} &= xy(3x-\lambda\sqrt{3/2})\;,
\end{align} 
where a prime is a derivative with respect to the logarithm of the scalar factor and $N = \ln a$ counts the number of e-folds to the end of inflation or the collapse phase. We identify critical points of the system with fixed points where $x^{\prime}=0$ and $y^{\prime}=0$. The dynamics is pictorially represented in figure \ref{heard1}

There are two kinetic-dominated solutions 
\begin{align}
x_{a} = -1\ {\textup or}\ +1\;, \quad y_{a} = 0\;,\label{eq:critp1}
\end{align}
with equation of state~(\ref{eq:eos2}) $w_a=1$. These fixed points therefore correspond to solutions where $a\propto t^{1/3}$ in an expanding universe for $t>0$, or $a\propto (-t)^{1/3}$ for $t<0$ in a contracting universe.

There is also a potential-kinetic-scaling solution for $\pm(6-\lambda^{2})>0$ (\textit{i.e.}, a sufficiently flat positive potential, $\lambda^2<6$ for $V>0$, or a sufficiently steep negative potential, $\lambda^2>6$ for $V<0$) which is given by
\begin{align}
x_{b} = \frac{\lambda}{\sqrt{6}}\;, \quad y_{b}=\sqrt{\frac{\pm(6-\lambda^{2})}{6}}\;;\label{eq:critp2}
\end{align}

This scaling solution corresponds to a solution with constant equation of state~(\ref{eq:eos2})
\begin{align}
%w &= 2x^{2} - 1\label{eq:xomega}\\
w_{b} &= \frac{\lambda^{2}}{3}-1\;,
\end{align}
and thus a power-law solution for the scale factor
\begin{align}
a(t) \propto |t|^{p}\;, \quad \varphi(t) = \sqrt{\frac{4}{3\kappa^{2}(1+w_b)}}\ln |t| + C\;,\label{eq:phisol}
\end{align}
where $C$ is an arbitrary constant of integration and
\begin{align}
\label{pval}
p = 
%2/3(1+w_b)=
\frac{2}{\lambda^2} \,.
\end{align}

First-order perturbations around this critical point yield the linearised equation~\cite{Heard:2002dr}
\begin{align}
\label{classical x}
x^{\prime} = \frac{(\lambda^{2}-6)}{2} (x-x_{b})\;.
\end{align}

We see that in an expanding universe, $H>0$, the scaling solution (\ref{eq:critp2}) is stable for $\lambda^2<6$, corresponding to $p>1/3$ from (\ref{pval}). Thus the scaling solution is stable whenever it exists for a positive potential in an expanding universe, but it is never stable for a negative potential in an expanding universe. 
Conversely, in a collapsing universe, since $N$ decreases with cosmic time, $H<0$, the scaling solution is stable for $\lambda^2>6$, corresponding to $p<1/3$. Thus the scaling solution is stable whenever it exists for a negative potential in a collapsing universe, but it is never stable for a positive potential for $H<0$.

In summary:
\begin{itemize}
	\item Expanding universe ($N\to+\infty$): \begin{itemize}
		\item[$\diamond$] The scaling solution exists and is stable for a positive, flat potential $p>1/3$ (including inflation, $p>1$).
		\item[$\diamond$] The scaling solution exists but is unstable for a negative, steep potential $p<1/3$.
	\end{itemize}
	\item Contracting universe ($N\to-\infty$): \begin{itemize}
		\item[$\diamond$] The scaling solution exists and is stable for a negative steep potential $p<1/3$ (including ekpyrosis, $p\ll 1$).
		\item[$\diamond$] The scaling solution exists but is unstable for a positive flat potential $p>1/3$ (including matter collapse, $p\simeq2/3$).
	\end{itemize}
\end{itemize}

\section{Linear Perturbations}
\label{sec:3}

We now study linear perturbations about the background solutions and in particular the solutions of the perturbed field in a collapsing scenario. Note that for a power-law cosmology, $z''/z=a''/a\propto (aH)^2$.
%$(d^2z/d\eta^2)/z=[(2p-1)/p](aH)^2$. 
Hence the solutions for small (sub-Hubble) and  large (super-Hubble) scales of the Mukhanov-Sasaki equation \eqref{eq:perteq} are respectively
\begin{eqnarray}
\label{vacuumsoln}
\delta\varphi &\simeq& \frac{e^{-ikt/a}}{a\sqrt{2k}} \quad \textrm{for}\ k^{2}/a^2 \gg H^2
%\quad (\textrm{sub-Hubble scales})
\,,\\
\delta\varphi &\simeq& \frac{C\dot{\varphi}}{H} + \frac{D\dot{\varphi}}{H} \int\frac{H^{2}}{a^{3}\dot{\varphi}^{2}}dt \quad \textrm{for}\ k^{2}/a^2\ll H^2
%\quad (\textrm{super-Hubble scales})
\,,
\end{eqnarray}
where we have chosen the quantum vacuum normalisation for the under-damped oscillations on sub-Hubble scales (\ref{vacuumsoln}).	

A characteristic feature of an inflating spacetime is that the comoving Hubble length decreases in an accelerating expansion with $\dot{a}>0$ and $\ddot{a}>0$. The same is true for the comoving Hubble length, $|H|^{-1}/a=1/|\dot{a}|$ in a decelerating, collapsing universe with $\dot{a}<0$ and $\ddot{a}<0$.
%	
%	During both inflationary expansion and decelerating collapse models, the comoving horizon size $\lvert k/aH\rvert \approx \lvert k/\dot{a}\rvert$ shrinks because $\lvert\dot{a}\rvert$ is increasing. 
%	
As a result quantum vacuum fluctuations, on sub-Hubble scales at early times (\ref{vacuumsoln}), lead to well-defined predictions for the power spectrum of perturbations on super-Hubble scales for potential-kinetic-scaling solutions with $\lambda^2<2$ (and hence $p>1$) in an expanding cosmology, or with $\lambda^2>2$ (and hence $p<1$) in a collapsing cosmology.

The characteristics of the inflation and collapse models for different values of $p$ are summarised in table \ref{comparison}.
\begin{table}[h]
	\centering
	\begin{tabular}{c c}
		\large Power-law inflation \quad & \quad \large Decelerated collapse \vspace{0.1cm} \\ 
		$H>0$ & $H<0$ \vspace{0.1cm}\\
		$\dot{a}>0$, $\ddot{a}>0$ & $\dot{a}<0$, $\ddot{a}<0$ \vspace{0.1cm}\\ 
		$p>1$ & $0<p<1$
		\\
	\end{tabular}
	\caption[Comparison between $H$, $\dot{a}$, $\ddot{a}$ and $p$ for power-law inflation and collapse.]{Comparing the quantities $H$, $\dot{a}$, $\ddot{a}$ and $p$ for power-law inflation and collapse. Although $\dot{a}$ is negative in the collapse case, its magnitude $\lvert\dot{a}\rvert$ is increasing. $p < 0$ is not allowed since this requires $\rho_\varphi+P_\varphi<0$, which is inconsistent with (\ref{rhovarphi}) and (\ref{Pvarphi}) for a canonical scalar field. }
	\label{comparison}
\end{table}	

\subsection{Scalar Field Perturbations in a Power-Law Collapse}\label{sec:ztochasticollapse}

We now investigate the effects of small scale quantum fluctuations on the stochastic evolution of the coarse-grained field above the Hubble scale
Let us consider a collapsing universe with the scale factor being a power-law scaling solution with $a \propto (-t)^p$ and $t<0$,
\begin{align}
a \propto (-t)^p\; \quad \textrm{where}\quad p = \frac{2}{3\left(1+w\right)}\;.\label{eq:powerlawsol}
\end{align}
We can re-express the scale factor in terms of conformal time as
\begin{align}
\label{eq:scale factor}
a(\eta) \propto (-\eta)^{p/(1-p)}\;.
\end{align}
%and using the relation for the Hubble rate in conformal time $\mathcal{H} = a H$, we find that $a$ can be re-expressed exactly as
Using the relation for the Hubble rate in conformal time, $H=a^{\prime}/a^2$, we find that $a$ can also be expressed as
\begin{align}
a= \left( \frac{p}{1-p} \right) \frac{1}{H\eta} 
% = \left(\nu - \frac{1}{2}\right)\frac{1}{H\eta}
\;,
\end{align}
where $\eta<0$.

Since $\dot{\varphi}/H$ is constant in this case we have $z\propto a$, which allows us to rewrite (\ref{eq:perteq}) as a Bessel equation
\begin{align}
\label{harmonic oscillator}
\frac{d^2 v}{d\eta^2}+\left(k^2-\frac{\nu^2-1/4}{\eta^2}\right) v = 0 \;,
\end{align}
where
\begin{align}
\label{def:nu}
\nu = \frac{3}{2}+\frac{1}{p-1} = -\frac{3}{2}\left[\frac{1+(3p-2)}{1-(3p-2)}\right]\;.
\end{align}
Note that for power-law collapse with $p<1$ we have $\nu<3/2$\footnote{In Appendix \ref{AppMapping} we explicitly show the mapping between the quantities $p$, $\nu$, $\lambda^{2}$ and $w$.}.

We recall from \eqref{Bessels} the general solution for a given $k$ can be expressed as a linear combination of Hankel functions
\begin{align}
v_k = \sqrt{|k \eta|} \left[V_{+}H_{\lvert\nu\rvert}^{(1)}(|k\eta|) + V_{-}H_{\lvert\nu\rvert}^{(2)}(|k\eta|)\right]\;.
\end{align}
As discussed in \cite{Wands:2008tv}, this solution generates a scale invariant spectrum, $|\delta\varphi_k^2|\propto k^{-3}$, not only for slow-roll inflation ($P/ \rho \rightarrow -1$ and $\nu = 3/2$) but also for a pressureless collapse ($P/ \rho \rightarrow 0$ and $\nu = -3/2$).

Combining the solution for $v_k$ (\ref{eq:delta_phi}) and the scale factor (\ref{eq:scale factor}), we see that in this large-scale limit the field perturbations are constant for $\nu >0$, since
\begin{align}
\delta\varphi_k \propto \frac{1}{a} (-\eta)^{\frac{1}{2}-|\nu |} \propto (-\eta)^{\nu - |\nu |} \;.
\end{align}
Conversely, for $\nu<0$ we see that the scalar field perturbations can grow rapidly on super-Hubble scales and diverge as $\eta\to0$.

%[DELETE THE FOLLOWING?:]
%At leading order in $-k\eta$, we can express the field and its conjugate momentum as
%\begin{align}
%\delta\varphi &= \frac{i}{\sqrt{4\pi}}\left(\frac{2}{2\nu -1} \right) \frac{2^{|\nu|}\Gamma(\lvert\nu\rvert)}{k^{|\nu|}} \frac{H}{(-\eta)^{|\nu|-3/2}} \bigg\rvert_{a = \left(\nu -\frac{1}{2}\right) \frac{1}{(-\eta) H}}\;,\label{eq:deltaphisol}\\
%\delta\pi_{\varphi} &= \frac{i}{\sqrt{4\pi}}\frac{2^{|\nu|}\Gamma(\lvert\nu\rvert)}{k^{\lvert\nu\rvert}}\left(\nu - \frac{1}{2}\right)^{4}\left[\left(\frac{2}{2\nu - 1}\right)\left(\lvert\nu\rvert - \nu\right) -  \frac{\kappa^2 \dot{\varphi}^{2}}{2H^{2}}\right]\frac{1}{H(-\eta)^{\lvert\nu\rvert+3/2}} \bigg\rvert_{a = \left(\nu -\frac{1}{2}\right) \frac{1}{(-\eta) H}} \;.\label{eq:pisol}
%\end{align}

\subsection{Perturbations in Phase Space Variables}

%The objective of this section is to relate the quantum noises to the evolution equation for the equation of state (EOS). The EOS can be decomposed into an adiabatic and a non-adiabatic parts as $w\rightarrow w+\delta w$. The first term contains all perturbations leaving the EOS constant (\textit{i.e.} fixed background), and the second represents deviations from a constant EOS. We show here we can  relate perturbations of the EOS to perturbations of the variables $x$ and $y$ at the critical point.

%\textcolor{blue}{We show schematically how this reflects in Fig.\eqref{pot}}.

Introducing first order perturbations of the dimensionless phase space variables (\ref{eq:dimensionless variables}), we obtain
\begin{align}
%x+\delta x &=  \frac{\kappa }{\sqrt{6} } \frac{1}{H+\delta H} \frac{1}{1+A} \frac{d}{dt} (\varphi + \delta \varphi) \nonumber \\
%&=  \frac{\kappa }{\sqrt{6} } \frac{1}{H} \left(1-\frac{\delta H}{H}\right) (1-A) (\dot{\varphi}+\dot{\delta\varphi})  \nonumber \\
%x+\delta x &=  \frac{\kappa }{\sqrt{6} } \frac{1}{H} \left(\dot{\varphi}+\dot{\delta\varphi}-A\dot{\varphi}-\frac{\dot{\varphi}}{H}\delta H\right)\;,
%\end{align}
%So, considering only the perturbed part, we have
%\begin{align}
%\label{eq:delta x total}
%\delta x = \frac{\kappa }{\sqrt{6} } \frac{1}{H} \left(\dot{\delta\varphi}-A\dot{\varphi}-\frac{\dot{\varphi}}{H}\delta H\right)\;.
%\end{align}
%Analogously, we get the perturbation for $y$
%\begin{align}
%\label{eq:perturbed y}
%\delta y = \frac{\kappa }{\sqrt{3}} \frac{\sqrt{V}}{H} \left(\frac{V_{,\varphi}}{2V}\delta\varphi-\frac{\delta H}{H}\right)\;.
%\end{align}
\label{eq:delta x total}
\delta x &= \frac{\kappa }{\sqrt{6} } \frac{1}{H} \left(\dot{\delta\varphi}-A\dot{\varphi}-\frac{\dot{\varphi}}{H}\delta H\right)\;,\\
\label{eq:perturbed y}
\delta y &= \frac{\kappa }{\sqrt{3}} \frac{\sqrt{V}}{H} \left(\frac{V_{,\varphi}}{2V}\delta\varphi-\frac{\delta H}{H}\right)\;,
\end{align}
where we are also including the metric perturbations $t\rightarrow(1+A)t$ and $H\rightarrow H+\delta H$ as described in \cite{Pattison:2019hef}. By perturbing the Friedmann equation, we obtain 
\begin{align}
\delta H =\frac{\kappa^{2}}{6H}(V_{,\varphi}\delta\varphi +\dot{\varphi}\dot{\delta\varphi}-\dot{\varphi}^{2}A)\;,
\end{align}
and since we are working in the spatially-flat gauge we can use the momentum constraint \cite{Mukhanov:1990me,Bassett:2005xm,Wands:2008tv} to write the perturbed lapse function in terms of the scalar field perturbation as
\begin{align}
A = \frac{\kappa^{2}\dot{\varphi}\delta\varphi}{2H}\;.
\end{align}
%In order to make a more intuitive study of the perturbed EOS, we  write (\ref{eq:delta x total}) 

At the critical point $x=x_{b}$ given by (\ref{eq:critp2}), the large-scale limit for the scalar field perturbations (\ref{eq:delta_phi}) then gives\footnote{The kinetic dominated solution analysis is briefly discussed in Appendix \ref{AppKinetic}, since it provides $\delta x = 0$ without regard to the solution $\delta\varphi$.}
\begin{align}
\delta x_k = \frac{i\kappa }{\sqrt{24\pi}}\left(1-\frac{\lambda^{2}}{6}\right)\left(\frac{2}{2\nu - 1}\right)^{2}\frac{\Gamma(\lvert\nu\rvert)2^{\lvert\nu\rvert}}{k^{\lvert\nu\rvert}}\left(\lvert\nu\rvert - \nu\right)\, H\, (-\eta)^{-\lvert\nu\rvert + 3/2}\;.
\end{align}
%
%To find the stochastic version of the first-order perturbations equation given by (\ref{classical x}), given in terms of number of e-folds. Let us then
We can re-express $\delta x$ with respect to the Hubble rate at a given time
\begin{align}
\label{pivot}
H &= 
%H_{\star} \left(\frac{-t_{\star}}{-t}\right)= H_{\star} \left(\frac{a_{\star}}{a}\right)^{\frac{1}{p}}=
H_{\star} \exp{\left[\left(\frac{\nu-3/2}{\nu-1/2}\right) \left(N_{\star}-N\right)\right]} \; .
\end{align}
where $N_{\star}$ is evaluated at some initial time
%with all quantities with a star evaluated at some initial time. The first equality comes from $p=Ht$. The second equality comes from (\ref{eq:powerlawsol}). In the third one we used the definition of the number of e-folds $N=\ln{(a)}$. Also, 
and the conformal time can be expressed using (\ref{eq:scale factor}) as
\begin{align}
(-\eta)=(-\eta_{\star}) \exp{\left[\frac{1}{\nu-1/2}\left(N_{\star}-N\right)\right]} \;,
\end{align}
which gives
\begin{align}
\label{deltaxN}
\delta x_k = \frac{i\kappa }{\sqrt{24\pi}}\left(1-\frac{\lambda^{2}}{6}\right)\left(\frac{2}{2\nu - 1}\right)^{2+|\nu|}\frac{\Gamma(\lvert\nu\rvert)2^{\lvert\nu\rvert}}{\sigma^{\lvert\nu\rvert}}\left(\lvert\nu\rvert - \nu\right) H_{\star} (-\eta)^{3/2}_{\star} \exp{\left[\frac{\nu}{\nu-1/2} \left(N_{\star}-N\right)\right]}\;.
\end{align}
We see that for $\nu>0$, which includes power-law inflation ($\nu = 3/2$) and ekpyrotic collapse ($\nu = 1/2$), the scalar field perturbations at the leading-order on large scales leave the phase space variable unchanged, $\delta x=0$.

Perturbing the Friedmann constraint \eqref{eq:FriedmannConstraint} requires $\delta y = -(x/y) \delta x$
%\begin{align}
%\label{eq:perturbed constraint}
%\delta y = -\frac{x}{y} \delta x \;.
%\end{align}
and, as a consequence, we can write the perturbation of the equation of state \eqref{eq:eos2} as
%\begin{align}
%\delta w = 4 \left[xy^2\delta x-x^2 y\left(-\frac{x}{y}\right)\delta x\right] = 4\left(xy^2+x^3\right) \delta x = 4x\delta x\;.
%\end{align}
\begin{align}
\delta w = 4x\delta x\;.
\end{align}
Hence for $\nu>0$ the scalar field perturbations at leading-order on large scales correspond to adiabatic perturbations which leave the equation of state unperturbed, $\delta w=0$. More generally, adiabatic perturbations on large scales correspond to local perturbations forwards or backwards in time along the background trajectory \cite{Wands:2000dp} which correspond to a fixed point in phase space. % separate universe paper
In appendix \ref{appnextorder} we consider the next-to-leading order scalar field perturbations on large-scales which give rise to a finite perturbation $\delta x\neq0$ on finite scales and finite time, $k\eta\neq0$ for $\nu>0$.

Conversely, for $\nu<0$ initial quantum field perturbations on sub-Hubble scales give rise to non-adiabatic perturbations on super-Hubble scales at late times, $\delta w\neq0$, which correspond to perturbations away from the fixed point in the dimensionless phase space.
We will now consider the effect of these quantum vacuum fluctuations evolving into the super-Hubble regime and giving rise to a stochastic diffusion in the phase space.

\section{Stochastic Noise from Quantum Fluctuations}
\label{linear noises}

We are now in measure to describe stochastic effects in our collapse scenario. To do so, we will study the deviations about the classical solution in phase space, as well as the maximum lifetime of collapsing scenarios in the presence of stochastic fluctuations. The stochastic noise, associated with the small-wavelength quantum fluctuations modes crossing the coarse-graining scale into the long-wavelength field at each time step, $d\tau$, is described by the two-point correlation matrix \eqref{noise ps}. In the following we will choose the time variable, $\tau$, to be the number of e-folds, $N$. For white noise, we can rewrite these entries in terms of the power spectrum \eqref{ps white noise} as
\begin{align}
\Xi_{f,g}(N) &= \frac{1}{6\pi^{2}}\frac{dk^{3}_{\sigma}(N)}{dN}f_{k}\left(N\right)g_{k}^{*}\left(N\right) \label{eq:Xifg} \;.
\end{align}

\subsection{Quantum Diffusion in Phase Space}	

%\subsection{Variance in phase space}

Near the critical point $x=x_b$ given by (\ref{eq:critp2}), the stochastic version of (\ref{classical x}) is
\begin{align}
\frac{d(\bar{x}-x_b)}{dN} &= m (\bar{x}-x_b) +\hat{\xi}_{x},
\end{align}
where the eigenvalue $m=(\lambda^2-6)/2$, whose solution is given by considering an It\^o process \cite{oksendal2013stochastic}
%\begin{align}
%d X_N &= m X_N dN + \hat{\xi}_{x} dN.
%\end{align}
%Multiplying the previous equation by $e^{-m N}$and using the relation
%\begin{align}
%d\left(e^{-m N} X_N\right) &= -m e^{-m N} X_N dN + e^{-m N} dX_N
%\end{align}
%we find
%\begin{align}
%d\left(e^{-m N} X_N\right) &= e^{-mN } \hat{\xi}_{x} dN \,
%\end{align}
%and eventually,
%\begin{align}
%X_N &= e^{m\left(N-N_{\star}\right)} X_{N_{\star}} + \int_{N_{\star}}^{N} e^{m\left(N-S\right)} \hat{\xi}_{x} dS,
%\end{align}
%Putting back the initial variables, we get
\begin{align}
\label{stochastic solution x}
\bar{x}(N)-x_c &= e^{m\left(N-N_{\star}\right)} \left(\bar{x}(N_{\star})-x_c\right) + \int_{N_{\star}}^{N} e^{m\left(N-S\right)} \hat{\xi}_{x} dS.
\end{align}

We define the variance associated with the coarse-grained field $\bar{x}$ as
\begin{align}
\sigma_x^2 := \left\langle \left(\bar{x}(N)-x_c\right)^2 \right\rangle \;;
\end{align}
whose evolution equation is given by
\begin{align}
\label{variancedifeq}
\frac{d\sigma_x^2}{dN} = 2 m \sigma_x^2+2 \left\langle \hat{\xi}_x \left(\bar{x}-x_c\right) \right\rangle .
\end{align}
%We solve this equation with the same methodology as before, with the exception of the introduction of a factor $e^{-2mN}$ this time. 
The solution can be split into a classical part and a quantum part, given by\footnote{We show the explicit calculation in Appendix \ref{App2}.}
\begin{align}
\label{variance}
\sigma_x^2 (N) &= \sigma_{x,cl}^2 (N)+\sigma_{x,qu}^2 (N)\nonumber\\
&=\sigma_x^2(N_{\star}) e^{2m\left(N-N_{\star}\right)} + \int_{N_{\star}}^{N}  dS \, e^{2m(N-S)} \Xi_{x,x}(S)  \;,
\end{align}
where the two-point correlation matrix $\Xi_{x,x}(S)$ is defined in Eq.~(\ref{eq:Xifg}), using the notation of \cite{Grain:2017dqa}. The classical part, $\sigma_{x,cl}^2 (N)$, is given by the variance at some initial time times an exponential function of the number of e-folds. The quantum part, $\sigma_{x,qu}^2 (N)$, is the accumulated noise between the initial time and a later time. 
%Then, the variance can be written either using \eqref{eq:Xixx} (or \eqref{xix} near the critical point). 

We find the two-point correlation function for the perturbations in the dimensionless phase space variable, $x$, by applying \eqref{eq:Xifg} to $\delta x$ given in \eqref{deltaxN} as
\begin{align}
\label{eq:Xixx}
\Xi_{x,x}(N) &= \frac{1}{2\pi^{2}}\sigma^{3} \left(\nu -\frac{1}{2}\right)^2  \frac{1}{(-\eta_{\star})^3} \exp{\left[\frac{-3}{\nu-1/2} \left(N_{\star}-N\right)\right]} |\delta x|^2 \nonumber \\
%&= g(\nu,\lambda,\sigma) \kappa^{2} H^{2}_{\star} \exp{\left[-\frac{3-2\nu}{\nu-1/2} \left(N_{\star}-N\right)\right]}
%\nonumber \\
&= g(\nu,\sigma) \kappa^{2} H^{2}(N)
\;,
\end{align}
with
\begin{align}
g(\nu,\sigma) := \frac{\Gamma^{2}(\lvert\nu\rvert)\nu^{2}2^{2\lvert\nu\rvert +4}}{(12\pi)^{3}\sigma^{2\lvert\nu\rvert - 3}} \left(\frac{2}{2\nu - 1}\right)^{2\lvert\nu\rvert + 4}(\lvert\nu\rvert -\nu)^{2}\;.
\end{align}

%	Now, using the definition given by (\ref{eq:Xifg}) with $f=\delta x$ and $g=\left(\delta x\right)^{\star}$, we are able to compute the simplified correlation matrix of the noise in $x$ as
%	\begin{align}
%	\Xi_{x, x} = \frac{\kappa^{2}\Gamma^{2}(\lvert\nu\rvert)2^{2\lvert\nu\rvert}}{(12\pi)^{3}}\frac{(6-\lambda^{2})^{2}}{\sigma^{2\lvert\nu\rvert -3}}\left(\frac{2}{2\nu -1}\right)^{2\lvert\nu\rvert +1}\left(\lvert\nu\rvert - \nu\right)^{2} \frac{H^{2}(\eta)}{(-\eta)}\;.\label{eq:Xixx}
%	\end{align}

%(\ref{eq:Xixx}) is our main result of this section. 
As previously noted, for $\nu>0$ the classical trajectory for $x$ remains preserved by the leading order perturbations in the scalar field since $\delta x = 0$ on large scales ($k\eta\to0$) \footnote{In Appendix \ref{appnextorder}, we take into account the next-to-leading order field contribution to compute $\Xi_{x,x}(N)$ and show that even in this configuration quantum diffusion should not take the system away from the fixed point. We show in particular that this is the case for quasi-de Sitter inflation.}, but the same is not true for $\nu$ negative since in this case $\delta x \neq 0$. We derive in Appendix \ref{appendixc} an alternative way to find this result using perturbations of the field and momentum. 

% Finally, note also that in the ekpyrotic case we have $\nu=1/2$ and then $aH=0$, which leads to $k_{\sigma}=0$.

%\subsubsection{Solutions of the quantum noise}
Inserting our result for the correlation function (\ref{eq:Xixx}) in the quantum part of (\ref{variance}), we find
%\begin{align}
%	%\sigma_x^2 (N) &= \int_{N_{\star}}^{N}  dS \, e^{2m(N-S)} \frac{a^{12}}{|\bar{\pi}^2+2a^6V|^{3}} \left(4|V|^2 \Xi_{\pi, \pi}+|\bar{\pi}V^{\prime}|^2\Xi_{\delta\varphi, \delta\varphi} -4VV^{\prime}Re{[\bar{\pi}]} \Xi_{\delta\varphi, \pi}\right) \nonumber \\
%    \sigma_x^2 (N)	&= \int_{N_{\star}}^{N}  dS \, e^{2m(N-S)} \frac{a^{12} V^2}{(\bar{\pi}_{\varphi}^2+2a^6V)^{3}} \left(4 \Xi_{\pi, \pi}+\bar{\pi}_{\varphi}^2 \lambda^2 \kappa^2 \Xi_{\varphi, \varphi} + 4\lambda \kappa \bar{\pi}_{\varphi} \Xi_{\varphi, \pi}\right) \;. \nonumber \\
%\end{align}
%Near the fixed point, this expression is greatly simplified and gives
\begin{align}
\label{variance final}
%\sigma_x^2 (N) &= \int_{N_{\star}}^{N}  dS \, e^{2m(N-S)} \left(\frac{6-\lambda^2}{12a^6}\right)^3 \frac{6\kappa^2}{3H^2(6-\lambda^2)} a^{12} \left(4 \Xi_{\pi, \pi}+a^6\lambda^4 H^2 \Xi_{\delta\varphi, \delta\varphi} + 4\lambda^2 Ha^3 \Xi_{\delta\varphi, \pi}\right) \nonumber \\
%&= \int_{N_{\star}}^{N}  dS \,  \frac{2\kappa^2 \left(6-\lambda^2\right)^2}{12^3} \frac{e^{2m(N-S)}}{H^2a^6} \left(4 \Xi_{\pi, \pi}+a^6\lambda^4 H^2 \Xi_{\delta\varphi, \delta\varphi} + 4\lambda^2 Ha^3 \Xi_{\delta\varphi, \pi}\right) \nonumber \\
\sigma_{x,qu}^2 (N)	&= g(\nu,\sigma) \kappa^{2} H^{2}_{\star} \exp{\left[\frac{3-2\nu}{\nu-1/2} \left(-N_{\star}\right) \right]} e^{2mN}\int_{N_{\star}}^{N}  dS \, e^{-2mS}  \exp{\left[\frac{3-2\nu}{\nu-1/2} S \right]} \; .
\end{align}
%		with
%		\begin{align}
%		g(\nu,\lambda,\sigma) &:= \frac{\kappa^{2}\Gamma^{2}(\lvert\nu\rvert)2^{2\lvert\nu\rvert}}{(12\pi)^{3}}\left(\frac{2}{2\nu -1}\right)^{2\lvert\nu\rvert +1} \frac{(6-\lambda^{2})^{2}\left(\lvert\nu\rvert - \nu\right)^{2}}{\sigma^{2\lvert\nu\rvert -3}}\;.
%		\end{align}
Re-expressing the eigenvalue $m$ in terms of the index, $m=-2\nu/(\nu-1/2)$, the solution of (\ref{variance final}) is then
%\begin{align}
%\label{variance star}
%\sigma_{x,qu}^2 (N) = & \tilde{g}(\nu,\lambda,\sigma) \kappa^{2} H^{2}_{\star}\exp{\left[\frac{3-2\nu}{\nu-1/2} \left(-N_{\star}\right) \right]}  \exp{\left(\frac{-4\nu}{\nu-1/2}N\right)} \nonumber \\  & \times \left[\exp{\left(\frac{3+2\nu}{\nu-1/2}N\right)}-\exp{\left(\frac{3+2\nu}{\nu-1/2}N_{\star}\right)}\right] \;,
%\end{align}
\begin{align}
\label{variance hn}
\sigma_{x,qu}^2 (N)
%\tilde{g}(\nu,\lambda,\sigma) \kappa^{2} H^{2}_{\star} \left[\exp{\left(-\frac{3-2\nu}{\nu-1/2}\left(N_{\star}-N\right)\right)}-\exp{\left(\frac{4\nu}{\nu-1/2}\left(N_{\star}-N\right)\right)}\right] \nonumber \\
&= \tilde{g}(\nu,\sigma) \kappa^{2} H^{2}(N) \left\lbrace 1-\exp{\left[\frac{3+2\nu}{\nu-1/2}\left(N_{\star}-N\right)\right]}\right\rbrace \;,
\end{align}
where
\begin{align}
\tilde{g}(\nu,\sigma) &= \left(\frac{\nu-1/2}{3+2\nu}\right) g(\nu,\sigma)\nonumber\\
&=\frac{\Gamma^{2}(\lvert\nu\rvert)2^{2\lvert\nu\rvert +4}}{(12\pi)^{3}\sigma^{2\lvert\nu\rvert - 3}} \left(\frac{\nu^{2}}{3+2\nu}\right)\left(\frac{2}{2\nu - 1}\right)^{2\lvert\nu\rvert + 3}(\lvert\nu\rvert -\nu)^{2}\;.
\end{align}
Equation~(\ref{variance final}) is given in terms of the Hubble scale at a fixed time, $H_\star$, while we have used \eqref{pivot} to give the variance (\ref{variance hn}) in terms of the time-dependent Hubble scale, $H(N)$.

We can compare the growth rate of the classical and quantum perturbations by comparing the time dependence from the two parts in \eqref{variance}. We note first that the time dependence of the classical term goes as
\begin{align}
\sigma_{x,cl} \propto \exp\left[\frac{4\nu }{\nu-1/2} \left(N_{\star}-N\right)\right]\;.
\end{align}
From \eqref{variance hn}, we see the time dependence of the quantum term behaves as
\begin{align}
\sigma_{x,qu} \propto \exp\left[\frac{4\nu }{\nu-1/2} \left(N_{\star}-N\right)\right] \left\lbrace\exp\left[-\frac{3+2\nu }{\nu-1/2} \left(N_{\star}-N\right)\right] -1\right\rbrace \;.
\end{align}
Remember $N_{\star}-N$ grows with time ($N$ decreases) in an expanding universe. Thus, the quantum variance decays with time if we have
\begin{align}
\frac{3+2\nu}{\nu-1/2} >0 \;.\label{conditionfornu}
\end{align}
This is the case if either $\nu>1/2$ or $\nu<-3/2$. However, a positive $\nu$ will cancel the leading order quantum diffusion, so we will consider only the second case, $\nu<-3/2$, in the following analysis. Thus, the classical perturbations grow faster than the quantum noise if $\nu<-3/2$, and the quantum noise grows faster if $\nu>-3/2$.

Also, the condition \eqref{conditionfornu} provides a shift in the spectrum, when compared to the case $\nu = -3/2$, since the scalar spectral index can be written as
\cite{Zeldovich:1983cr,Wands:1998yp,Finelli:2001sr,Peter:2006hx,Peter:2008qz,Guimaraes:2019sqf}
\begin{align}
n_{\textup s} &= 1+ \frac{12 w}{1+3w} =  1+\frac{4(2\nu+3)}{3}\;,
\end{align}
and it is clear to see that when $\nu = -3/2-\epsilon$, where $\epsilon$ is a small positive parameter, $w < 0$ and the spectrum becomes red, \textit{i.e.}, $n_{\textup s} < 1$. 

%The general solution for the variance due to quantum fluctuations is given for $\nu\neq-3/2$ by \eqref{variance hn}. However, t
To understand the behaviour around $\nu \approx -3/2$ we will consider $\nu=-3/2-\epsilon$ which for $|\epsilon(N_\star-N)|\ll1$ leads to
\begin{align}
\label{variance star pressure}
\sigma_{x,qu}^2 (N) 
%&= \frac{3}{2^7 \pi } \frac{1}{\sigma^{2\epsilon}} \frac{H^{2}_{\star}}{M_{pl}^2}   \left(N_{\star}-N\right) e^{3\left(N_{\star}-N\right)} \nonumber \\
&= \frac{3}{128 \pi} \frac{1}{\sigma^{2\epsilon}} \frac{H^{2}(N)}{M_{pl}^2}    \left(N_{\star}-N\right) \;,
\end{align}
where we have used $\kappa^{2} =8 \pi/M_{pl}^2$ with $M_{pl}$  the Planck mass, and we recall that $\sigma$ is the coarse-graining scale. The diffusion thus has the form of a random walk with $N_\star-N$ steps of equal, but growing, length $\propto |H(N)|$.

%The second equality comes from \eqref{pivot} for values close to $\nu=-3/2$. 
We see that \eqref{variance star pressure} depends weakly on the coarse-graining scale for 
%$\epsilon\ll1$,
$\nu \approx -3/2$, 
and becomes independent of $\sigma$
in the limit $\nu=-3/2$, where $\epsilon \rightarrow 0$.
%we have
%\begin{align}
%\label{variance star pressure 2}
%\sigma_{x,qu}^2 (N) &= \frac{3}{2^7 \pi}  \frac{H^{2}(N)}{M_{pl}^2}    \left(N_{\star}-N\right) \;.
%\end{align}
%Hence, the resulting expression for the variance does not depend on the coarse-graining scale, 
This is not surprising since we know that quantum fluctuations in a pressureless collapse give rise to a scale-invariant spectrum of perturbations \cite{Wands:1998yp,Finelli:2001sr,Allen:2004vz}.

%Note that by switching to, the variance diverges as $(-t)^{-12/\lambda^2}$.	

\subsection{Maximum Lifetime of the Collapsing Phase}

%It is interesting to know
We can now examine when the variance becomes large, \textit{i.e.}, when $\sigma_{x,qu}\approx1$, so that the quantum diffusion due to the stochastic noise results in a significant deviation from the critical point.

\subsubsection{Radiation-Dominated Collapse}

Consider first the case of a potential-kinetic-scaling collapse with $\lambda=2$, giving rise to an equation of state $w=1/3$, analogous to a radiation-dominated cosmology, and index $\nu=-1/2$. The variance \eqref{variance hn} in this case becomes
\begin{align}
\sigma_{x,qu}^2(N) &= \frac{\sigma^2}{54 \pi} \frac{H_{\star}^2}{M_{pl}^2} \left\{\exp{\left[4\left(N_{\star}-N\right)\right]}-\exp{\left[2\left(N_{\star}-N\right)\right]}\right\} \nonumber \\
&\approx \frac{\sigma^2}{54 \pi} \frac{H^2(N)}{M_{pl}^2} \;,
\end{align}
where to get the second line, we have neglected the second exponential term since the first one will grow much quicker. For $\sigma^2_{x,qu}(N_{\textup end})=1$, we get the straightforward result 
\begin{align}
|H_{\textup end}| \approx \frac{13}{\sigma} M_{pl} \; .
\end{align}

We conclude that a radiation-dominated collapsing phase cannot escape the fixed point due to quantum diffusion until it approaches the Planck scale. Indeed, since we require $\sigma <1$, we see that a deviation from the classical fixed point $x=x_b$ due to quantum diffusion would require the Hubble scale to become greater than the Planck scale. In practice as soon as the Hubble scale approaches the Planck scale our semi-classical analysis breaks down.

\subsubsection{Pressureless Collapse}

\begin{figure}[h!]
	\centering
	\includegraphics[scale=0.3]{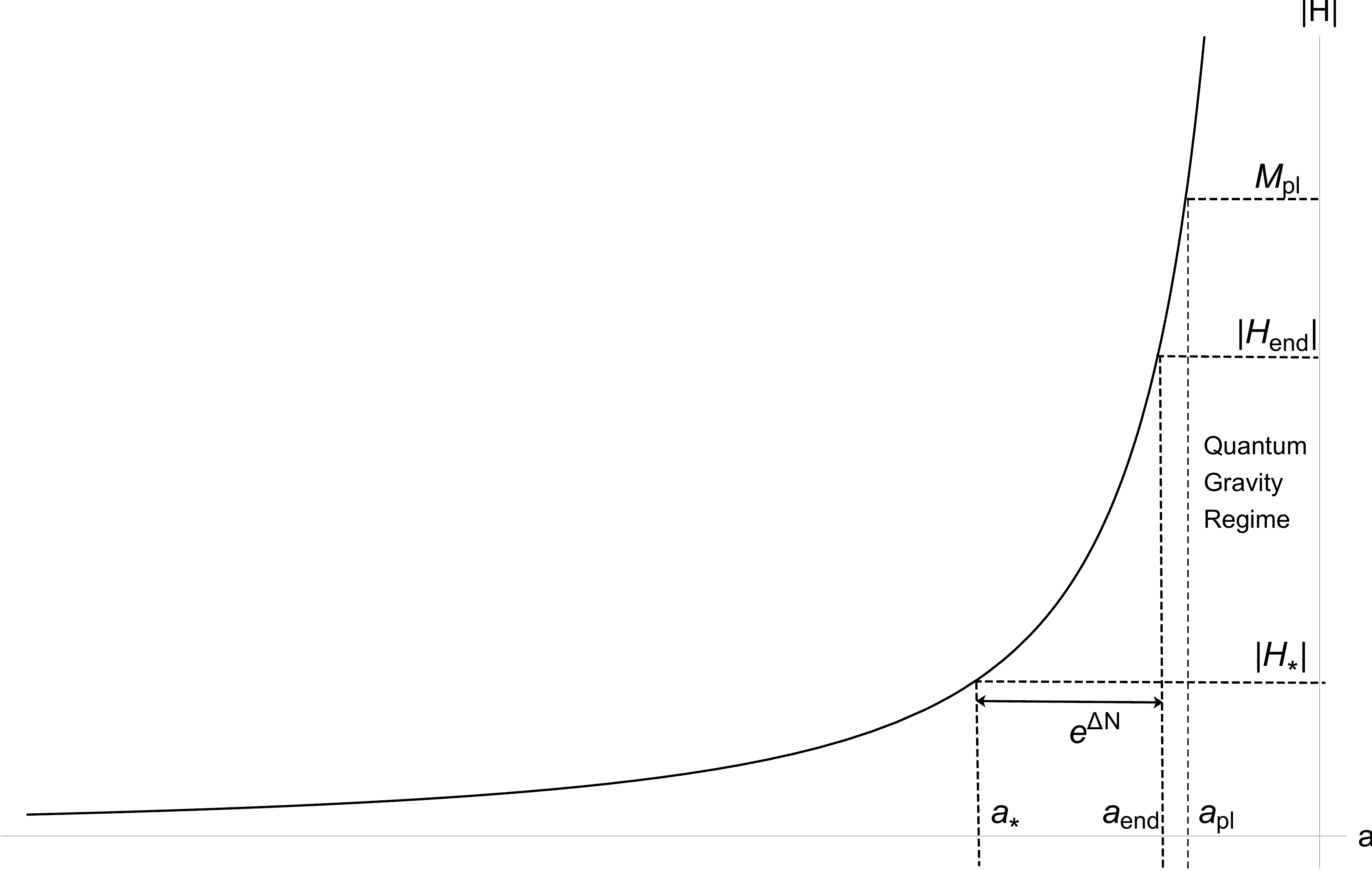}
	\caption[Evolution of the Hubble rate in a pressureless collapse]{Evolution of the Hubble rate, $\lvert H\rvert$, in a pressureless collapse. For quantum diffusion to lead to a deviation from the classical fixed point before the Hubble rate reaches the Planck scale, $|H_{\textup end}|<M_{\textup Pl}$, requires a very low initial energy scale, $|H_*|\ll |H_{\textup end}|$.
	}
	\label{scale}
\end{figure}

For the case of a pressureless collapse, $\nu = -3/2$, we find $\sigma_{x,qu}(N_{\textup end})=1$ when
\begin{align}
\label{lifetime}
|H_{\textup end}| = \sqrt{\frac{128 \pi}{3 (N_{\star}-N_{\textup end})}} M_{pl} 
% \approx \sqrt{\frac{134}{N_{\star}-N}} M_{pl}\; .
\end{align}
Thus, for pressureless collapse case, quantum diffusion gives a time, $t_{\textup end}$, at which stochastic trajectories leave the classical fixed point before we reach the Planck scale, $|H_{\textup end}|<M_{pl}$, if the number of e-folds during the collapse is greater than 134. 
We show a simple example in Fig.\eqref{scale}. 

In terms of the initial Hubble rate, using \eqref{pivot} for the Hubble rate $H(N)$ for $\nu=-3/2$, we have
\begin{align}
(N_{\star}-N_{\textup end}) \exp{\left[3(N_{\star}-N_{\textup end})\right]} =\frac{128 \pi}{3} \frac{M_{pl}^2}{ H_{\star}^2} \;,
\end{align}
from which we get an approximate number of e-folds during the collapse phase
\begin{align}
N_{\star}-N_{\textup end} \approx \frac{2}{3} \ln{\left(\sqrt{\frac{128 \pi}{3}}  \frac{M_{pl}}{ |H_{\star}|}\right) } \;.
\end{align}
Conversely we can obtain an expression for the Hubble rate at the end of the pressureless collapse starting from an initial Hubble rate $H_{\star}$ given by
\begin{align}
|H_{\textup end}| \approx \sqrt{\frac{64 \pi}{\ln{\left(\sqrt{\frac{128 \pi}{3}}  \frac{M_{pl}}{ |H_{\star}|}\right) }}} M_{pl} \;.
\end{align}

Knowing how the comoving Hubble length behaves in terms of time during pressureless collapse, we can estimate a lower limit on the number of e-folds required during pressureless collapse to solve the horizon and smoothness problems of the hot big bang:
\begin{align}
\frac{k_{\textup end}}{k_\star} = \frac{a_{\textup end}H_{\textup end}}{a_\star H_\star} = \left( \frac{t_\star}{t_{\textup end}} \right)^{1/3} = e^{(N_\star-N_{\textup end})/2} > e^{70}\;,
\end{align}
where we are considering $70$ as a ratio between the Hubble length over Planck scale compared to horizon size today. This is a similar number for inflation to solve the flatness and  horizon  problems  of  Big  Bang  cosmology. Then, we would need 
\begin{align}
N_\star - N_{\textup end} > 140\;,
\end{align}	
which is remarkably close to the estimate $N_\star-N_{\textup end}>134$ that follows from requiring $|H_{\textup end}|<M_{pl}$ (\ref{lifetime}).

%(and say why this is a good choice)
%		From \eqref{variance star} with $\nu=-3/2$, we get
%		\begin{align}
%		e^{-\frac{7}{2}N} - e^{-2N_{\star}-\frac{3}{2}N}&= - \frac{2^7 \pi}{3} \frac{M_{pl}^{2}}{H^{2}_{\star}} (-\eta_{\star}) e^{-\frac{7}{2}S_{\star}} \;.
%		\end{align}
%		Considering that $N_{\star} \gg N$, we get the time at which the perturbative analysis cannot be trusted anymore
%		\begin{align}
%		N &= -\frac{2}{7} \ln{\left[e^{-2N_{\star}} - \frac{2^7 \pi}{3} \frac{M_{pl}^{2}}{H^{2}_{\star}} (-\eta_{\star}) e^{-\frac{7}{2}S_{\star}} \right]}\;.
%		\end{align}

\section{Conclusions on Stochastic Effects in Cosmology}
\label{conclusion}

The origin of the primordial density perturbations which lead to the large-scale structures we observe in the Universe is still a topic of debate. While a lot of attention has been given to quantum fluctuations in an inflationary expansion phase, another interesting possibility is that of vacuum fluctuations in a collapse phase preceding the present cosmological expansion. Perturbations of a self-interacting scalar field can be a useful starting point to study collapsing phases since in the case of linear perturbations, we can have a duality between the expanding and contracting solutions, as shown in \cite{Wands:1998yp}.

We have presented in this chapter an analysis of the stability of a power-law solution for a scalar field with an exponential potential, both classically and in the presence of quantum perturbations which can give rise to a stochastic noise on super-Hubble scales, either during an accelerated expansion or a decelerated collapse. We considered three possible cosmological scenarios: power-law inflation, a pressureless collapse 
%(which can be thought of as the dual of inflation since the scale factor velocity and acceleration share the same characteristics with a negative sign) 
and an ekpyrotic collapse. It is well known that inflationary and ekpyrotic models are classically stable under perturbations, while the matter collapse is unstable. We have shown in this work that in inflationary and ekpyrotic models quantum perturbations at leading-order on large (super-Hubble) scales are adiabatic, and do not drive these models away from their fixed point solution in phase space. This means that the equation of state is unperturbed, $\delta w=0$, and one does not modify the classical behaviour of such models by adding quantum noise. In fact, all models with a positive index $\nu$ (the index of the Bessel function governing the evolution of the Mukhanov-Sasaki variable, see Eq.~(\ref{def:nu})) have a vanishing quantum contribution to the equation of state at leading order on super-Hubble scales. As a consequence, only models with a negative index, $\nu<0$, including the pressureless collapse case, diffuse away from the classical fixed point due to quantum noise. The perturbation of the equation of state
%, or equivalently of the kinetic variable $x$ we used throughout this paper, 
is summarised in Table \eqref{summary}.

\begin{table}[h]
	\centering
	\begin{tabular}{c|c}
		\large Inflation or ekpyrotic collapse  \quad & \quad \large Pressureless collapse  \\ 
		\hline
		$\nu > 0 \rightarrow$ adiabatic noise & $\nu < 0 \rightarrow$ non-adiabatic noise \\ 
		$\delta w = 0$  & $\delta w\neq 0 $	\\
	\end{tabular}
	\caption[Comparison of the behaviour of $\delta w$ in de Sitter inflation, ekpyrotic collapse and pressureless collapse]{Table comparing the behaviour of $\delta w$ for three cases in the super-Hubble limit $\sigma \rightarrow 0$: de Sitter inflation ($\nu = 3/2$), ekpyrotic collapse ($\nu = 1/2$) and pressureless collapse ($\nu = -3/2$).}
	\label{summary}
\end{table}

We then considered the maximum lifetime of the classical fixed point in the presence of quantum noise in a collapse phase with $\nu<0$. In the general case, we find that the collapse models are stable against quantum diffusion in the semi-classical theory since the variance with respect to the classical fixed point on super-Hubble scales is proportional to the Hubble rate and remains small while the Hubble rate remains below the Planck scale. 

However, for the particular case $\nu=-3/2$, corresponding to a pressureless collapse, the perturbations on super-Hubble scales are known to be scale-invariant; in this case we found the quantum diffusion leads to a random walk away from the fixed point. If we start the classical collapse from very low energy scales arbitrarily close to the fixed point solution, then the semi-classical collapse phase (with the Hubble scale below the Planck scale) can last for many e-folds before diffusion drives the evolution away from the fixed point solution.

Our analysis is limited to first-order perturbations about the classical fixed points, however one might expect that fluctuations about the pressureless collapse eventually end up in a kinetic-energy-dominated collapse phase with $w=+1$ since this is the stable fixed point solution in this case. Perturbations about the kinetic-dominated collapse due to quantum fluctuations are adiabatic in the super-Hubble limit and so this fixed point solution remains stable against quantum noise up to the Planck scale.

%% file: sections/chapter5.tex
\chapter{Quantum Cosmology}
\label{quantum cosmology}

%Carla's thesis \cite{Carla_tese}.
%
The previous chapter was devoted to the evolution of quantum perturbations and their backreaction on a classical background. It is natural to ask what would be the consequences of going further, applying a quantisation scheme to the whole Universe. From the covariance principle, the theory of general relativity must be invariant under diffeomorphisms. As a consequence, we cannot write generalised velocities in terms of momenta in GR. To canonically quantise gravity, it is necessary to first express the Hamiltonian constraints of the system \footnote{Note that this is not a generic feature of quantisation schemes. For instance, Loop Quantum Gravity or String Theory do not require Hamilton equations.}, then quantise those constraints. In this chapter, we first give a review on the interpretations of quantum mechanics, then we canonically quantise GR in the de Broglie-Bohm approach arguing the Copenhagen interpretation is not suited for cosmology, and we finish with an illustration of this quantum cosmology with astrophysical implications describing a process of magnetogenesis.

\section{The Born Rule and the Measurement Problem}
\label{the born rule and measure problem}

The very notion of reality at microscopic scales has never reached an agreed upon interpretation. How we can measure a physical observable at macroscopic scales has always been unambiguous. The reason is that the apparatus used for measure does not interact enough with the system to have a sensible change in the results. This is no longer true with quantum measures. We give the basic ingredients of quantum theory, and a description of different interpretations. We first lay down the notions used to define a quantum measurement. The material used in this section can be found in \cite{McMaster:1954,Kiefer:2002,Pinto-Neto:2013toa,Marchildon:2019}.

Let $\ket{\Psi}$ be the state of the quantum system. An observable $A$ is associated to a self-adjoint operator. The possible outcomes of measuring $A$ are the eigenvalues $a_i$ associated to eigenvectors $\ket{a_i}$.  We can write the state as a linear combination of the eigenvectors as
\begin{align}
\ket{\Psi} = \sum_{i} c_i \ket{a_i} \;,
\end{align}
with $c_i$ c-numbers. Then, the projector onto the Hilbert space is an arbitrary combination given by
\begin{align}
\ket{\Psi}\bra{\Psi} = \sum_{i,j} c_i c_j^{\star} \ket{a_i} \bra{a_j^{\star}} := \sum_{i,j} \rho_{ij} \ket{a_i} \bra{a_j^{\star}} \;.
\label{quantum state squared}
\end{align}
The matrix $\rho$ defined by the entries $\rho_{ij}$ is called the \textit{density matrix}, and gives a statistical description of the quantum system. It contains information not only on pure states, but also on mixed states. A first interesting property of this matrix is that for $i=j$ (pure states), we recover the Born rule, which states that if the system is in the state $\Psi$, the probability associated  with the outcome $a_i$ is $|c_i|^2$. Now, if we consider $\ket{\Psi}$ to be normalised, it comes from \eqref{quantum state squared} that
\begin{align}
\int \ket{\Psi}\bra{\Psi} \textup{d}^3 x = \sum_{i,j} \rho_{ij} \ket{a_i} \bra{a_j^{\star}} = \sum_{i} \rho_{ii} = 1 \;,
\end{align}
and 
\begin{align}
\textup{Tr}\rho = 1\;.
\end{align}
The trace operator \textup{Tr} is a map from the space of positive, linear operators acting on a Hilbert space into the real line $\mathbb{R}$. Similarly, a measurement of some variable $F$, which can be viewed as a projector acting upon a Hilbert space $\mathcal{H}$,  yields a statistical average, or probability, given by
\begin{align}
P(F) := \int \bra{\Psi}F \ket{\Psi} \textup{d}^3 x  = \sum_{i,j} F_{ij} \rho_{ji} \;, 
\end{align}
with
\begin{align}
F_{ij} := \int \bra{a_i^{\star}} F \ket{a_j} \textup{d}^3 x  \;.
\end{align}
Then,
\begin{align}
P(F) = \sum_{i} (F\rho)_{ii} = \textup{Tr} (F\rho) = \textup{Tr} (\rho F) \;,
\label{probability density matrix}
\end{align}
where the last equality comes from the fact that all matrices are Hermitian. This was first proved by Gleason in 1957 \cite{gleason1957measures}.  

Let us consider a quantum system $\mathcal{S}$ and a quantum measurement apparatus $\mathcal{M}$. The total system is then an isolated quantum system given by the direct sum $\mathcal{S}\oplus \mathcal{M}$. If we note $\ket{\alpha_0}$ the initial state of the apparatus and $\ket{\alpha_i}$ a final state, then a faithful measurement of the system is represented by a unitary evolution $U(\mathcal{S},\mathcal{M})$, symbolically represented as
\begin{align}
\ket{a_i} \otimes \ket{\alpha_0} \xrightarrow{U(\mathcal{S},\mathcal{M})} \ket{a_i}  \otimes  \ket{\alpha_i} \;.
\end{align}
Thus, if the system is in an arbitrary state $\ket{\Psi}$, its evolution will be given by the superposition
\begin{align}
\sum_{i} c_i \ket{a_i}  \otimes  \ket{\alpha_0} \xrightarrow{U(\mathcal{S},\mathcal{M})} \sum_{i} c_i  \ket{a_i}  \otimes  \ket{\alpha_i} \;.
\label{unitary evolution}
\end{align}
This superposition of states is the core of quantum theory. We note from \eqref{unitary evolution} that all possible outcomes are perfectly correlated with the observable initial values. However, this also shows that \textit{all} outcomes will be displayed, what is clearly in contradiction with a macroscopic measurement, which is unique. This is the well-known \textit{measurement problem}. The interactions of the quantum system and the measurement apparatus with the environment unavoidably leads to the appearance of classical behaviour. This emergence of classical properties, called \textit{decoherence}, is usually assumed as a good candidate from the quantum-to-classical transition. This is a dynamical effect through which the large number of degrees of freedom from the environment and the system state interact. Concretely, decoherence is the loss of phase correlations between wave functions of the environment, coming from interactions between the environment and the system, and the environment itself.

Even though the mathematical foundations have been laid down by Von Neumann \cite{VonNeumann:1927,VonNeumann:2018}, many different quantum theories exist, each with its own interpretation of reality. The superposition principle \eqref{unitary evolution} allows for \textit{entanglement} between two particles and, with the experimental violation of Bell inequalities \cite{Clauser:1969ny,Aspect:1981zz}, theories with ``hidden variables" determining the outcome deterministically are not viable unless these variables are non-local (like the de Broglie-Bohm theory). We will present two interpretations in the next sections: the Copenhagen interpretation and the de Broglie-Bohm interpretation.

\subsection{The Copenhagen Interpretation}

In the Copenhagen interpretation, the Born rule is admitted as well as the duality between wave and particle. However, the superposition of states for the apparatus is not taken into account, and a classical description of the measure must be used. Thus the apparatus is always real, while observables become real only when a measure is performed. According to this interpretation, before the measurement is made, particles do not possess properties like position or momentum. In other words, the Copenhagen interpretation assumes two distinct notions of reality to different physical systems: a classical system is always real, while a quantum system is only real upon measurement. On the other hand, the reduction of the wave packet has no dynamical significance, and the transition of a quantum system into a classical one is not explained. This is in contradiction with theoretical and experimental developments. For instance, neutrino oscillations \cite{Pontecorvo:1957cp,GonzalezGarcia:2007ib}, reversible transitions from a superfluid phase to a Mott insulator \cite{Greiner:2002} or quantum cryptography \cite{Gisin:2002zz} cannot be explained without a dynamical wave function.

The superposition obtained by unitary evolution \eqref{unitary evolution} is not universal \cite{VonNeumann:2018}. Von Neumann postulated that upon measurement, the superposition vanishes and the probability for the system to be in the state $\ket{a_i} \ket{\alpha_i}$ is given by $|c_i|^2$, retrieving the Born rule. This behaviour is called the \textit{collapse of the wave function}. In opposition to the initial Copenhagen Interpretation, the wave function is now dynamical, and the collapse can occur at any moment. The modern Copenhagen interpretation includes the collapse as a fundamental component of quantum mechanics. The behaviour of a quantum system is undetermined, and it is the Born rule that gives the probability to be in one state or another.

\subsection{The de Broglie-Bohm Interpretation}

The incompleteness of the wave function to describe physical processes led to the first realistic model by Bohm \cite{Bohm:1951xw,Bohm:1951xx}, in which the wave function is supplemented with classical particles and fields. This model is non-local, and all variables have definite position and momentum. The de Broglie-Bohm (dBB) theory can be formulated starting from the formulation of classical mechanics based on the Hamilton-Jacobi equation \cite{John:2014sea}
\begin{align}
\frac{\partial S}{\partial t} + H \left(q_i,\frac{\partial S}{\partial q_i},t\right) = 0\;,
\label{hamilton-jacobi equation}
\end{align}
where $S(q_i, \alpha_i, t)$ is the Hamilton principal function, $H$ is the Hamiltonian, $q_i$ are the configuration space variables, $\alpha_i$ are constants of integration and $t$ denotes time.

Let us write the Hamiltonian of a single non-relativistic particle in the coordinate representation 
\begin{align}
H = \frac{P^2}{2m} + V(x) \;,
\end{align}
with $P$ the particle momentum, $m$ its mass, and $V(x)$ its potential energy. Then, the Schr\"{o}dinger equation for this Hamiltonian reads
\begin{align}
i\hbar \frac{\textup{d}\Psi(x,t)}{\textup{d}t} = \left[-\frac{\hbar}{2m}\nabla^2+V(x)\right] \Psi(x,t) \;.
\label{Schrödinger equation}
\end{align}
Therefore, the wave function acquires an autonomous dynamics, and obeys the Schrödinger equation without collapse. Writing the wave function in the polar form $\Psi := R e^{iS/\hbar}$ leads to the set of equations
\begin{align}
\frac{\partial S}{\partial t} +\left[\frac{1}{2m}\left(\nabla S\right)^2+V\right] = \frac{\hbar^2}{2m} \frac{\nabla^2 R}{R} \;, \\
\frac{\partial R^2}{\partial t} + \nabla \left(R^2 \frac{\nabla S}{m}\right) = 0\;.
\label{continuity equation probability density}
\end{align}
The first equation resembles a Hamilton-Jacobi equation, but this is a purely quantum equation, with $V$ a quantum potential. The second equation is the continuity equation obeying the Born rule
\begin{align}
P \equiv \Psi^{\star} \Psi = R^2 \;.
\label{probability density dbb}
\end{align}
The idea of recasting the Schrödinger equation in a Hamilton-Jacobi-like equation originates from Louis de Broglie \cite{deBroglie:1925cca}, introducing the idea of a guiding wave as trajectory. 

These two equations bear important consequences. The first is that quantum particles follow trajectories $x(t)$, each one independent of observations. Upon measurement, the particle is confined into a particular wave packet. Since each packet is spatially separated, they cannot interfere on the state of the particle. Thus, there is an apparent ``collapse" of the wave function \footnote{Everett built a quantum theory based on the same notion of reality as Bohm, but assumed that all possible outcomes exist simultaneously \cite{Everett:1957hd}.}. A second consequence is that particles are not separated from the quantum field $\Psi$, and each one satisfies the Schrödinger equation \eqref{Schrödinger equation}. Assuming trajectories led de Broglie to postulate the \textit{guidance equation}
\begin{align}
\label{guidance equation}
m_i \dot{\boldsymbol{x}}_i = \nabla_i S \;.
\end{align}
Solving \eqref{Schrödinger equation} will then give us $x(t)$ along with an integration constant, the initial position, which is unknown. This initial condition is often referred to as the hidden variable of the dBB theory. Finally, given an initial probability density $P(t_0,x_0)$, the continuity equation \eqref{continuity equation probability density} ensures we recover all statistical predictions of standard quantum mechanics.

The Born rule is not a basic law in the dBB theory, since the link between the probability density and wave function is additional to the basic principles of the wave function. Instead, the Born rule is a dynamical feature called \textit{quantum equilibrium}. An initially out-of-equilibrium system, where $P(t_0,x_0) \neq R^2(t_0,x_0)$,  rapidly reaches equilibrium through the evolution of \eqref{Schrödinger equation}, and stays in this state. This is an analog of the H-theorem in classical statistical mechanics, which shows that a system with a large number of degrees of freedom reaches thermal equilibrium, but applied to the pilot-wave theory \cite{Valentini:1990zq,Valentini:1991fia}. As a consequence, all statistical predictions from standard quantum mechanics are recovered in the dBB theory. Furthermore, the possibility of an out-of-equilibrium initial state not relaxing to $P(t,x) \neq R^2(t,x)$ would result in a different prediction from the Copenhagen interpretation, hence leading to eventual distinguishability criteria between dBB theory and other interpretations. This has been investigated in pre-inflationary and inflationary context \cite{Valentini:2006yj,Valentini:2008dq,Colin:2013rwa,Colin:2014pna,Colin:2015tla} as well as in information flows from black holes \cite{Kandhadai:2019ztn}.

In 1952, Bohm also suggested to take the derivative of the guidance equation \eqref{guidance equation} to obtain the first Newton's law of motion
\begin{align}
m \ddot{\boldsymbol{x}}_i = -\nabla_i \left(V+Q\right) \;,
\end{align}
where the particle is submitted to an additional potential $Q$ of quantum origin, namely
\begin{align}
Q := - \sum_{i} \frac{\hbar^2}{2m_i}\frac{\nabla_i^2 R}{R}\;.
\end{align}
This potential characterises the non-locality inherent to the dBB theory. Indeed, in the limit $R \rightarrow 0$, the potential can be very large while the wave function is small. The Bell test experiments  state that a quantum theory must be whether non-local or non-ontological. Since dBB is ontological, it must be non-local, which is indeed the case. As an example, we could cite the two-slit experiment, where the wave function gives informations about the two slits to the photon. In consequence, the photon will not travel in a straight line any more and, where the destructive interference fringes lie, no photon can travel in these fringes. This reflects the presence of an infinite quantum potential, forbidding the presence of photons inside these regions. 

We close this section with a couple of remarks. First, the classical limit in the dBB theory is simply obtained when the quantum potential is negligible compared to the classical kinetic and potential energies. Therefore, the classical limit is straightforward to obtain. However, we stress that even though the quantum potential gives an intuitive explanation of Bohmian mechanics, it is not needed to derive quantum mechanical results.

\section{Quantising General Relativity}
\label{quantising GR}

The canonical quantisation of a physical system without second-class constraints can be summarised as follows
\begin{enumerate}
	\item Promote the canonical variables $p$ and $q$ to quantum operators $\hat{p}$ and $\hat{q}$;
	\item These quantum operators, and functions explicitly depending on them, must satisfy the commutation relations obtained from the classical formalism through the substitution
	\begin{align}
	\left\{ A(p,q), B(p,q) \right\} \to - \frac{i}{\hbar} \left[ \hat{A}(\hat{p},\hat{q}), \hat{B}(\hat{p}, \hat{q}) \right] \;,
	\label{4.1}
	\end{align}
	with 
	\begin{align}
	\left[ \hat{A}, \hat{B} \right] = \hat{A}\hat{B} - \hat{B}\hat{A}\nonumber \;.
	\end{align}
	\item Once a particular representation satisfying \eqref{4.1} has been chosen, the dynamical evolution of the wave function $\psi$ is given by the Schr\"{o}dinger's equation
	\begin{align} \label{4.2}
	i \hbar \frac{\partial \psi}{\partial t} = \hat{H}_c (\hat{q}, \hat{p}) \psi \;,
	\end{align}
	where $\hat{H}_c = \hat{H}_T - \lambda_m \hat{\phi}_m$.
	\item The first-class constraints of the theory, $\phi_m (q, p) \approx 0$, are imposed as restrictive equations to the possible wave functions of the system
	\begin{align} \label{4.3}
	\phi_m (\hat{q}, \hat{p})\psi = 0 \;.
	\end{align}
	We cannot consider first-class constraints as operators identities or, consequently, \eqref{4.1} would be violated.
\end{enumerate}

Clearly, the quantum states  $\ket{\psi}$ of the  system (with wave representations given by $\braket{x}{\psi} = \psi(x)$ or $\braket{p}{ \psi} = \psi(p)$), refer to a Hilbert space with scalar product $\braket{\psi}{\psi'}$ coming from the probabilistic interpretation of the theory, and also from the differential equation \eqref{4.2}. In GR, we only have
\begin{align} \label{4.14}
\left\{ h_{ij}(x), \Pi^{kl}(x') \right\} = \delta^{kl}_{ij} \delta^3 (x-x') \Rightarrow \left[ \hat{h}_{ij}(x), \hat{\pi}^{kl}(x') \right] = i \hbar \delta^{kl}_{ij} \delta^3 (x-x') \;,
\end{align}
with every other commutator being zero. A possible choice of representation satisfying equations \eqref{4.14} is
\begin{align}
\hat{h}_{ij}(x) = h_{ij}(x) \;, \quad \text{and} \quad \hat{\pi}^{ij}(x) = -i\hbar \frac{\delta}{\delta h_{ij}(x)} \;,
\end{align}
with $\psi (h_{ij}(x),t)$. We can show that $\psi$ is independent of $N$ and $N^i$ since
\begin{align}
\left[ N(x), P (x') \right] = \delta^3 (x-x') \;, && \left[ N^i(x), P_j (x') \right] = \delta^i_j \delta^3 (x-x') \;,
\end{align}
and, considering the representation
\begin{align}
P(x) = -i \hbar \frac{\delta}{\delta N(x)} \;, && P_i (x) = -i\hbar \frac{\delta}{\delta N^i(x)} \;,
\end{align}
along with the use of \eqref{4.3} and the canonically conjugated momenta \eqref{2.2}, we obtain
\begin{subequations} \label{4.16}
	\begin{align}
	\frac{\delta \psi}{\delta N}(N,N^i,h_{kl}) = 0 \;,\label{4.16a} \\
	\frac{\delta \psi}{\delta N^i}(N,N^i,h_{kl}) = 0 \label{4.16b} \;.
	\end{align}
\end{subequations}

The Hamiltonian $H_c$ of GR is null, hence equation \eqref{4.2} applied to GR  indicates that $\psi$ does not depend on the parameter $t$. This is the so-called \textit{problem of time}. This was expected since GR is invariant through time re-parametrisation. Only the constraints \eqref{gr constraints} are left, and they give
\begin{align}
\hat{\mathcal{H}}_0 \psi = 0 =	\left( \hat{G}_{ijkl} \hat{\pi}^{ij} \hat{\pi}^{kl} - \hat{h}^{\frac{1}{2}} \,^{3}\hat{R} \right)\psi \;, &\implies \; G_{ijkl}\frac{\delta^{2}\psi}{\delta h_{ij}\delta h_{kl}} + h^{\frac{1}{2}} \,^3\hat{R} \psi = 0 \;, \label{4.17} \\
\hat{\mathcal{H}}^i \psi = 0 = \left( \tensor{\hat{\pi}}{^{ij}_{,j}} + \tensor{^3\hat{\Gamma}}{^i_{ab}}\hat{\pi}^{ab} \right)\psi \;, &\implies \; \left(\frac{\delta \psi}{\delta h_{ij}}\right)_{;j} = 0 \label{4.18} \;.
\end{align}

%Equations \eqref{4.17} and \eqref{4.18} both display ordering problem. The proposed order for $\mathcal{H}_i \psi =0$ in \eqref{4.18} seems the most reasonable, because in this way the physical meaning of that equation is obviously related to the classical constraint $\mathcal{H}_i \approx 0$, examined in the last chapter. Then, let an infinitesimal spatial coordinate transformation
%\begin{align}
%\bar{h}_{ij} = h_{ij} + \xi_{i;j} + \xi_{j;i} = h_{ij} + \delta h_{ij} \;,
%\end{align}
%be applied to the wave function $\psi (\bar{h}_{ij})$ as
%\begin{align}
%\nonumber\psi (\bar{h}_{ij}) &= \psi (h_{ij} + \delta h_{ij}),\\
%\nonumber&= \psi (h_{ij}) + \int d^3x \frac{\delta \psi}{\delta h_{ij}} \delta h_{ij} \\
%\nonumber&= \psi(h_{ij}) + \int d^3x \frac{\delta \psi}{\delta h_{ij}} 2 \xi_{i;j} \\
%\nonumber&= \psi(h_{ij}) +\int d^3x \; 2 \left(\frac{\delta \psi}{\delta h_{ij}}\right)_{;j} \\
%&= \psi(h_{ij}) \;. \label{4.20}
%\end{align}
%From \eqref{4.20}, it can be seen that the functions $\psi(h_{ij}(x))$ satisfying \eqref{4.18} do not change in relation to the spatial metric.
% that are pure spatial coordinate change (the gauge transformation in that theory), that is, metric wave functions that differ only by a spatial coordinate transformation have the same value.
The ordering of equation \eqref{4.17} is a debate in the literature \cite{Hawking:1985bk,Christodoulakis:1986pm,Vela,Komar:1979vd}, and its solution is not well established. Even worse, its physical interpretation is still unclear. %As we discussed earlier, 
On the other hand, the constraint $\mathcal{H}_0 \approx 0$, when applied to the wave function through the use of \eqref{4.3}, can be cast in a form remembering a Klein-Gordon equation.%e given by \eqref{4.6}, which gives the dynamical evolution of the wawefunction $\psi (h_{ij}(x))$ dynamics. 

%Let us now consider equation \eqref{4.17}, which is known as the WdW equation. This equation will be the one governing the wave function $\psi (h_{ij})$ dynamics. One of the variables $h_{ij}(x)$ is the physical time whereby the wave function evolves (as $q^0$ between \textcolor{blue}{among?} the variables $q^\mu$ in \eqref{4.6}). However, finding such variable in \eqref{4.17} is not an easy task as it was in \eqref{4.6}, where the conjugated \textcolor{red}{*canonical*} momenta to the time $t$ appears linearly and isolated. This is the famous question of time in quantum gravity. One way to solve this problem would be to find a transformation of the canonical variables that allows to write \eqref{4.17} in a Schr\"{o}dinger's equation-like \textcolor{blue}{like my like?} \textcolor{red}{*YES!!*} form \cite{Kuchar,Vela,Teitel}. Some Schr\"{o}dinger-like equations have been found already, however they are un-treatable, on the practical point of view due, basically, to the complicated structure of \eqref{4.17}, mainly because of the $\,^3{R}(h)$ term. \textcolor{blue}{R(h)?} \textcolor{red}{*Recursos Humanos*}
%
%Some proposals have been presented in the literature \cite{Casta,Rovelli,Kucha,Unruh,Unruh2}, but none of them is definitive.

A way to consider equation \eqref{4.17} would be to understand the quantity $G_{ijkl}$ as being the metric in the space of metrics $h_{ij}(x)$, known as \textit{superspace} (with no relation to super-symmetry). In that case, what would be the signature of $G_{ijkl}$? It is important to notice that it is symmetric under interchanges of $(i,j)$ with $(j,i)$, $(k,l)$ with $(l,k)$ and also $(i,j)$ with $(l,k)$. Hence, $G_{ijkl}$ may be written as a bi-dimensional $6 \times 6$ symmetric matrix $M_{AB}$ through the  indices mapping
\begin{align}
(1,1) \to (1) \;, && (1,2) \to \frac{1}{\sqrt{2}}(4) \;,\\
(2,2) \to (2) \;, && (2,3) \to \frac{1}{\sqrt{2}}(5) \;,\\
(3,3) \to (3) \;, && (3,1) \to \frac{1}{\sqrt{2}}(6) \;.
\end{align} 
The factors $\frac{1}{\sqrt{2}}$ appear for the mapping $\delta^{ij}_{kl} \to \delta^A_B$ to be possible. In the particular case that $h_{ij} = \delta_{ij}$, the matrix $M_{AB}$ associated to $G_{ijkl}$ is
\begin{align}\nonumber
M_{AB}=\left(
\begin{array}{c c c c c c}
1/2 & -1/2 &-1/2 & 0 & 0 & 0 \\
-1/2 & 1/2 & -1/2 & 0 & 0 & 0 \\
-1/2 & -1/2 & 1/2 & 0 & 0 & 0 \\
0 & 0 & 0 & 1 & 0 & 0 \\
0 & 0 & 0 & 0 & 1 & 0 \\
0 & 0 & 0 & 0 & 0 & 1 \\
\end{array}
\right) \;,
\end{align}
Thus, finding the eigenvalues of $M_{AB}$, consists in finding the eigenvalues of the first block of $M_{AB}$, called  $N$. The eigenvalues equation is 
\begin{align}
det(N-\lambda I)=0 \Rightarrow \lambda^3-\frac{3}{2}\lambda^2+\frac{1}{2}=0 \Rightarrow (\lambda-1)^2(\lambda+\frac{1}{2})=0 \;.
\end{align}
Therefore, the eigenvalues of $N$ are (-1/2,1,1) and the eigenvalues of $M_{AB}$, and consequently of $G_{ijkl}$, are (-1/2,1,1,1,1,1). It can be concluded from this analysis that equation \eqref{4.17} has a Klein-Gordon-like structure with potential $R(h_{ij})$ on the superspace. It can be shown that there exists an unique time-like coordinate on this superspace associated to $\textup{det}(h_{ij})$ \cite{DeWitt:1967yk,Vela}. However, considering the Wheeler-De Witt equation as being a Klein-Gordon-like equation endangers the probabilistic interpretation of $\psi(h_{ij})$ because, as it is known, a conserved current of probability cannot be built from the Klein-Gordon equation with the zeroth component positively defined. One solution could be trying to extract a square root of the Wheeler-De Witt equation, in the same way as the Dirac equation is the square root of Klein-Gordon equation. Nevertheless, this method clashes with the non-positively definition of the potential $R(h_{ij})$. 

A second solution would be to quantise the wave function $\psi(h_{ij})$ itself, which would be a third quantisation of gravity. It is argued that in the semi-classical approximation, such probabilistic interpretation can be restored \cite{Halliwell:1990uy}. All these problems are still without solution.

\section{The Minisuperspace Approximation of General Relativity}
\label{minisuperspace approximation}

We have derived in the previous section the quantisation of general relativity through the Hamiltonian formalism and, in particular, we have shown that the dynamics is obtained through the constraint \eqref{h0 constraint}. However, this constraint depends on the superspace metric, and a general solution for the quantum system \eqref{4.17} is very unlikely to be found, since the problem involves a set of coupled partial differential equation for each spacetime point. Then, a partial solution is to reduce the number of degrees of freedom by freezing out those of gravity and matter. This configuration with reduced degrees of freedom for homogeneous cosmologies, which can be understood as a projection of the whole superspace containing only long-wavelength modes of the size of the Universe \cite{Kerbrat:1992zy}, is called the \textit{minisuperspace approximation}. Even though we do not consider the full superspace, working in the minisuperspace approximation can give some insight on what a full theory of quantum gravity should is, and is therefore still relevant for the study of quantum gravity \cite{Halliwell:1990uy,Ryan}. Also, the high degree of homogeneity in the observable Universe, as described in the first chapter, might be a hint that this approximation is quite reasonable in quantum cosmology.

We first recall that a homogeneous and isotropic metric is given by the FLRW metric \eqref{flrw adm} 
\begin{align}
\textup{d}s^2= 	-N^2(t)\textup{d}t^2+a^2(t)\left[\frac{\textup{d}r^2}{1-kr^2}+r^2\textup{d}\theta^2+r^2 \sin^2\theta \textup{d}\phi^2\right] \;.
\end{align}
The only dynamical variable in this model is the scale factor $a(t)$. After discarding total derivatives, the gravitational lagrangian can be reformulated as
\begin{align}
\mathcal{L}_G = \frac{6}{N} a \dot{a}^2 \;.
\end{align}
To palliate the problem of time arising in the quantisation of the Hamiltonian, we add a degree of freedom through a perfect fluid implemented with the Schutz formalism described in appendix \ref{schutz}. Equivalently, we could introduce a scalar field called $K$-essence \cite{Almeida:2017xhp}. The advantage of using the Schutz formalism is that the Wheeler-de Witt equation can be cast as a Schrödinger-like equation. With the notation used in the appendix, the action reads \cite{Lapchinsky:1977vb}
\begin{align}
S = \int \textup{d}t \left(\Pi_a \dot{a} + \Pi_{\epsilon} \dot{\epsilon} + P_{\theta} \dot{{\theta}}+\Pi_S \dot{S} - N \mathcal{H}^0\right) \;, 
\end{align}
with
\begin{align}
N\mathcal{H}^0 = N \left( - \frac{\Pi_a^2}{24a} - 6 k a+ \frac{\Pi_{\epsilon}^{1+w}}{(16\pi)^w} \frac{e^S}{a^{3w}} \right) \;,
\label{hamiltonian gr}
\end{align}
We shall now set $k=0$, thus specialising to flat hypersurfaces. % since we neglected various boundary terms, implying we are working with closed spatial sections. Boundary terms are energy terms in some asymptotic limit, and we would work in this case with open spatial sections. Depending on the asymptotic limit, different values for the same terms can be obtained. Consequently, the ADM formalism is not adapted to deal with open spatial sections.  

Following the technology developed at length in the previous sections, the momenta for the scale factor and $\epsilon$ take the form
\begin{align}
\Pi_a  = -\frac{12}{N} a \dot{a} \;, \quad \Pi_{\epsilon} = -N \rho_{0} U^{0} a^3	 \;.
\label{momenta scale factor perfect fluid}
\end{align}
and the constraints
\begin{align}
P_{\theta}=0 \;, \quad \Pi_S = \theta \Pi_{\epsilon} \;.
\end{align}
We can express the matter section in a simple way by performing the canonical transformations
\begin{align}
T = -(16\pi)^w\frac{\Pi_S}{\Pi_{\epsilon}^{1+w}} e^{-S} \;, &\quad \Pi_T = \frac{\Pi_{\epsilon}^{1+w}}{(16\pi)^w} e^S \;, \\
\epsilon^{\prime} = \epsilon - (1+w) \frac{\Pi_S}{\Pi_{\epsilon}} \;, & \quad \Pi_{\epsilon^{\prime}} = \Pi_{\epsilon} \;,
\end{align}
turning the matter super-Hamiltonian into
\begin{align}
H_M = \frac{N}{a^{3w}} \Pi_T \;.
\label{matter superhamiltonian}
\end{align}
Finally, summing the gravitational and matter sectors leads to the total super-Hamiltonian
\begin{align}
H_T = N\mathcal{H}_T = N	\left(-\frac{\Pi_a^2}{24a}  +\frac{\Pi_T}{a^{3w}}\right) = 0 \;.
\label{constrain total hamiltonian}
\end{align}
In particular, the Poisson bracket of $T$ is
\begin{align}
\left\lbrace T, \mathcal{H}^0 \right\rbrace_{\mathcal{H}^0=0} = a^{-3w} > 0 \;,
\end{align}
what is reassuring since there is no dependence on the canonical momentum. We say that $T$ is a \textit{global time} \cite{Hajicek:1986ky}, leading to the desired featured of unitary evolution.

The perfect fluid appears linearly in the Hamiltonian, and allows us to cast the Wheeler-de Witt equation $\hat{\mathcal{H}} \psi =0$ into a Schrödinger-like equation \cite{AcaciodeBarros:1997gy,Alvarenga:1998wx,Alvarenga:2001nm}. Before quantising the constraint \eqref{constrain total hamiltonian}, we choose the ordering
\begin{align}
\frac{N}{a^{3w}} \left(-\frac{1}{24} a^{\frac{{3w-1}}{2}} \Pi_a a^{\frac{{3w-1}}{2}} \Pi_a + \Pi_T\right) = 0 \;.
\end{align}
This ordering is arbitrary, but physically interesting since after the gauge choice $N=a^{3w}$, quantising the constraint \eqref{constrain total hamiltonian} reads
\begin{align}
\frac{1}{24} \left[a^{\frac{{3w-1}}{2}} \frac{\partial}{\partial a} \left( a^{\frac{{3w-1}}{2}} \frac{\partial}{\partial a}\right) \right] \Psi  (a,T) = i \frac{\partial \Psi}{\partial T} (a,T) \;,
\label{wdw minisuperspace}
\end{align}
leaving covariance upon field redefinitions manifest. Noticing the same expression appears twice in the left-hand side, it is natural to define a new variable
\begin{align}
\chi = -\frac{2}{3w-3} a^{\frac{{3w-3}}{2}} \;, \quad \textup{such that} \quad a^{\frac{{3w-1}}{2}} \frac{\partial}{\partial a} = \frac{\partial }{\partial \chi} \;,
\end{align}
allowing us to recast the constraint in the simpler form
\begin{align}
i \frac{\partial \Psi}{\partial T} (\chi,T) = \frac{1}{24} \frac{\partial^2 \Psi }{\partial \chi^2} (\chi,T) \;.
\label{wave function schrodinger}
\end{align}
In the end, we obtain a time-reversed Schrödinger equation for a one-dimensional free particle, with mass equal to twelve. Note that the time parameter has been arbitrarily chosen. In our case, time is given by a perfect fluid, but another fluid would be likewise acceptable. However, in a realistic scenario with multiple fluids, we could wonder if only one fluid is going to drive the evolution of the Universe wave function, if each fluid dominates a given epoch, or if there is going to be an interplay between the fluids \cite{PintoNeto:2005gx,Monerat:2019zil}.

Solving equation \eqref{wave function schrodinger} requires choosing initial conditions for the wave function. For flat hypersurfaces, we show in appendix \ref{section initial conditions wave function} that solutions with unitary evolution must satisfy the condition
\begin{align}
\left(\Psi^{\star} \frac{\partial \Psi}{\partial \chi} - \Psi \frac{\partial \Psi^{\star}}{\partial \chi} \right) \bigg\rvert_{\chi=0} = 0 \;.
\end{align}
Then, choosing the normalised wave function
\begin{align}
\Psi_{init} (\chi) = \left(\frac{8}{T_b \pi}\right)^{\frac{1}{4}} \exp(-\frac{\chi^2}{T_b}) \;, 
\label{initial wave function}
\end{align}
as initial condition, with $T_b$ an arbitrary constant which fixes the
bounce timescale, we can apply the propagator procedure defined in \cite{Alvarenga:1998wx,Alvarenga:2001nm} to find the normalised wave function of the Universe at all times \cite{Pinto-Neto:2013toa}
\begin{align}
\Psi (a,T) = &\left(\frac{8 T_b}{\pi \left(T^2+T_b ^2\right)}\right)^{\frac{1}{4}} \exp(\frac{-4{T_b}a^{3(1-w)}}{9\left(T^2+T_b ^2\right)\left(1-w\right)^2}) \nonumber \\
& \times \exp \left\lbrace -i \left[\frac{4Ta^{3(1-w)}}{9\left(T^2+T_b ^2\right)\left(1-w\right)^2} + \frac{1}{2} \atan(\frac{T_b}{T}) - \frac{\pi}{4} \right]\right\rbrace \;.
\label{wave function universe}
\end{align}
Now that the wave function has been obtained, we would like to apply the quantum cosmology derived previously to a situation of physical interest, namely a model of creation of primordial magnetic fields.

\section{Primordial Magnetogenesis in Bouncing Cosmology}
\label{primordial magnetogenesis}

The existence of magnetic fields in a variety of scales in the Universe (see for instance \cite{beck2012magnetic,Beck:2013bxa,Durrer:2013pga}) calls the question of their origin. In particular, there are several observations consistent with weak $\sim 10^{-16}$ Gauss fields in the intergalactic medium, coherent on Mpc scales: the 21-cm hydrogen line \cite{Minoda:2018gxj}, the anisotropy of ultra-high energy cosmic rays \cite{Bray:2018ipq}, CMB distortions \cite{Ade:2015cva,Chluba:2019kpb}, B-mode polarisation measurements \cite{Zucca:2016iur,Pogosian:2018vfr}, magnetic reheating \cite{Saga:2017wwr}, Big Bang Nucleosynthesis (BBN) \cite{Kawasaki:2012va}, and $\gamma$-rays \cite{Barai:2018msb}, among others. Since such fields remained largely undisturbed during the cosmological evolution (as opposed to those in the presence of structure), they offer a window to  their origin, which is generally assumed to be primordial.

Primordial seed fields (which may be amplified later by the dynamo mechanism \cite{Brandenburg:2004jv}) are generated before structure formation, for instance out of the expansion of the universe, either during inflation \cite{Martin:2007ue,Ratra:1991bn,Davis:1995mv,Berera:1998hv,Kandus:1999st,Bassett:1999ta,Battaner:2000kf,Davis:2000zp,Tornkvist:2000js,Davis:2005ih,Anber:2006xt,Jimenez:2010hu,Das:2010ywa,BeltranJimenez:2011vn,Bonvin:2011dt,Elizalde:2012kz,Bonvin:2013tba,Caprini:2014mja,Choudhury:2015jaa,Sharma:2017eps,Caprini:2017vnn,Sharma:2018kgs,Kamarpour:2018ckk, Savchenko:2018pdr,Sobol:2018djj,Subramanian:2019jyd,Patel:2019isj,Sobol:2019xls,Fujita:2019pmi,Shakeri:2019mnt,Kobayashi:2019uqs,Shtanov:2019civ,Sharma:2019jtb}, or in cosmological models with a bounce \cite{Battefeld:2004cd,Salim:2006nw,Membiela:2013cea,Sriramkumar:2015yza,Chowdhury:2016aet,Qian:2016lbf,Koley:2016jdw,Chen:2017cjx,Leite:2018bbo,Chowdhury:2018blx,Barrie:2020kpt}.\footnote{Cosmological magnetic fields may also be produced during phase transitions, see for instance \cite{Grasso:2000wj}, or through the generation of vortical currents \cite{Carrilho:2019qlb}.} However, since minimally-coupled electromagnetism is conformally invariant, the expansion cannot affect its vacuum state. Hence such invariance must be broken in order to generate seed magnetic fields.

Conformal invariance can be broken in several ways: through the addition of a mass term \cite{Enqvist:2004yy}, by  coupling the electromagnetic (EM) field to a massless charged scalar field \cite{Emami:2009vd} or the axion \cite{Adshead:2016iae}, and by a non-minimal coupling with gravity. The last option has been widely studied in the case of inflationary models (see  \cite{Turner:1987bw,	Bamba:2006ga,	Campanelli:2008qp,	Kunze:2009bs, Kunze:2012rq, Savchenko:2018pdr}, among others). However, inflationary magnetogenesis is not free of problems. Among these, we can mention an exponential sensitivity of the amplitude of the generated magnetic field with the parameters of the inflationary model \cite{Subramanian:2015lua}, the strong coupling problem \cite{Demozzi:2009fu}, and the limits in the magnetic field strength coming from the gravitational backreaction of the electric fields that are produced simultaneously with the magnetic fields \cite{Green:2015fss}. Hence, instead of an inflationary model, a nonsingular cosmological model in conjunction with a coupling of the type $RF_{\mu\nu}F^{\mu\nu}$ is used here to study the production of seed magnetic fields. Nonsingular models are likely to ease both the problem of exponential sensitivity of the result and the strong coupling problem, since they expand slower than inflationary models. Moreover, we shall see below that backreaction is not an issue for the model chosen here.

Magnetogenesis in nonsingular cosmological models can be divided into two classes, depending on whether the coupling of the EM field with the scalar field is fixed on theoretical grounds (see for instance \cite{Salim:2004pv, Salim:2006nw}), or chosen in a convenient way in terms of the expansion factor (see \cite{Membiela:2013cea,Qian:2016lbf,Chowdhury:2016aet} ). The coupling between the Ricci scalar and the EM field is theoretically motivated by the vacuum polarisation described by quantum electrodynamics (QED)  in a curved background \cite{Drummond:1979pp}, and introduces a mass scale to be fixed by observations. 

The remaining of this chapter presents the results we obtained in \cite{Frion:2020bxc}.

\subsection{Background Cosmology}

We now apply the wave function \eqref{wave function universe} to magnetic fields in the dBB theory. First note the probability density is given by
\begin{align}
\rho (a,T) = a^{\frac{1-3w}{2}} |\Psi_{init} (a,T)|^2 \;.
\end{align}
Adapting the continuity equation \eqref{continuity equation probability density} to our current notations, we find
\begin{align}
\frac{\partial \rho}{\partial T} +  \frac{\partial}{\partial a} \left( -\frac{a^{1-3w}}{12} \rho \frac{\partial S}{\partial a}\right) = 0 \;.
\label{probability density wave function universe}
\end{align}
A direct consequence, remembering the guidance equation \eqref{guidance equation}, is 
\begin{align}
\frac{\textup{d} a}{\textup{d} T} = - \frac{a^{3w-1}}{12} \frac{\partial S}{\partial a} \;.
\label{guidance equation scale factor}
\end{align}
This is in accordance with the classical relation
\begin{align}
\dot{a} = \left\lbrace a, \mathcal{H}_T \right\rbrace = \frac{\partial \mathcal{H}_T }{\partial \Pi_a} = - \frac{a^{3w-1}}{12} \Pi_a \;,
\end{align}
with $\Pi_a = \partial S / \partial a$. Inserting the phase of the wave function \eqref{wave function universe} into \eqref{guidance equation scale factor} gives us the behaviour of the scale factor
\begin{align}
a (T) = a_0 \left[1 + \left(\frac{T}{T_0}\right)^2\right]^{\frac{1}{3\left(1-w\right)}} \;.
\label{scale factor}
\end{align}
Remembering that $T$ is related to cosmic time through $N dT = dt$, the gauge choice made in \eqref{wdw minisuperspace} implies  \cite{Peter:2008qz}
\begin{align}
dt=a^{3w}dT \;.
\end{align}
Note that the initial wave function \eqref{initial wave function} was chosen for simplicity. Coupled to the scale factor \eqref{scale factor}, the wave function of the Universe describes a symmetrical cosmological bounce. However, modifying the initial wave function or considering combinations of initial wave functions lead to physically different solutions, describing most likely asymmetric bounces \cite{Delgado:2020htr}.

Finally, we note that $S$ satisfies a modified Hamilton-Jacobi \eqref{hamilton-jacobi equation}
\begin{align}
\frac{\partial S}{\partial T} - \frac{a^{3w-1}}{24} \left( \frac{\partial S}{\partial a} \right)^2+ Q = 0 \;,
\end{align}
with quantum potential
\begin{align}
Q :=  \frac{a^{\frac{3w-1}{2}}}{24R}  \frac{\partial}{\partial a} \left( a^{\frac{{3w-1}}{2}} \frac{\partial R}{\partial a}\right) \;. 
\end{align}

From now on, we shall set $w=0$, leading to a scale-invariant spectrum for the curvature perturbations, and allowing us to set $t=T$. It will also be useful to express the scale factor as $a(t) := a_0 Y(t)$, with
\begin{align}
Y(t)= \frac{1}{x_b} \left( 1+\frac{t^2}{t_b^2}
\right)^{1/3} \;,
\label{yt}
\end{align}
where we have defined $x := a_0/a$ and $t_b := 2\ell_b$, 
with $\ell_b$ the curvature scale at the bounce ($\ell_b := 1/\sqrt{\vert R(0)\vert}$ where $R$ is the four-dimensional Ricci scalar) satisfying $10^3\:t_{\textup{Planck}}<t_b<10^{40}\:t_{\textup{Planck}}$.\footnote{The lower bound is set 
	by imposing the validity of the Wheeler-DeWitt equation, \textit{i.e.},
	by restricting 
	the curvature
	to values such that possible discreteness of the spacetime geometry is negligible, while quantum effects are still relevant \cite{Peter:2006hx}. Since $t_\text{Planck}\simeq 10^{-44}s$ and recalling that BBN happened around $10^{4}s$, the upper bound simply reflects the latest time at which the bounce can occur.}

The only matter content needed to describe the bounce is dust. Therefore, we consider dark matter as the only component present, and we now define parameters that are directly related to observations. Let us first write down the Friedmann equation
\begin{align}
H^2 = \frac{8\pi G}{3} \frac{\rho_m}{a^3} \;,
\label{friedrho}
\end{align}
with $\rho_m$ the dark matter density energy. The ratio between the Friedmann equation \eqref{friedrho} at some time $t$ and the same equation evaluated today leads to
\begin{align}
H^2 = H^{2}_{0} \Omega_m x^3 \;,
\label{fried}
\end{align}
with $\Omega_m$ the dimensionless dark matter density today. Note that at $x=1$ we have $H^2 = H_0^2\Omega_m$, this means that in the contraction phase, at the same scale as today $a = a_0$, the Hubble factor is $-H_0\sqrt{\Omega_m}$ due to the lack of other matter components. Then, from the expansion of $a(t)$ for large values of $t$, it follows that 
\begin{align}
H^2\approx  \frac{4}{9t_b^2}\left( \frac{x}{x_b}\right)^3 \;.
\label{larget}
\end{align}
Now, using $H_0=70 \; \textup{km\;s}^{-1}\textup{\;Mpc}^{-1}$ and the lower bound on $t_b$, it is straightforward to derive an upper limit on $x_b$ by equating Eqs.~\eqref{fried} and \eqref{larget},
\begin{align}
\Omega_m = \frac 4 9 \frac{1}{t_b^2 x_b^3 H_0^2} \; \implies x_b < \frac{10^{38}}{\Omega_m^{1/3}} \; .
\label{limit xb}
\end{align}

For later convenience, we define
$R_{H_0} := H_0^{-1}$,
$t_s := {t}/{R_{H_0}}$, and 
$
\alpha := {R_{H_0}}/{t_b}$,
and rewrite $Y(t)$
as
\begin{align}
Y(t_s) = \frac{1}{x_b} \left(1+\alpha^2t_s^2\right)^{1/3} \;,
\end{align}
with
\begin{align}
\alpha = \frac 3 2 \sqrt{\Omega_mx_b^3} \;.
\label{alpha}
\end{align}
We will see in the next section how to relate the previous quantities to the electromagnetic power spectrum, and what constraints can be derived on the parameters of the model.

\subsection{The Electromagnetic Sector}
\label{emsector}

Dark matter alone is not sufficient to produce electromagnetic field. As recalled earlier, Maxwell electromagnetism is conformally invariant. A simple way to break this invariance and produce magnetic fields is through a non-minimal coupling between dark matter and electromagnetism as
\begin{align}
\label{lagr}    
{\cal L}= -f
F_{\mu\nu}F^{\mu\nu} \;,
\end{align}
where
\begin{align}
f \equiv \frac{1}{4}+\frac{R}{m_{\star}^2} \;,
\label{coupling}
\end{align}
and $m_{\star}$ is a mass scale to be determined by observations.

Straightforwardly, the equations of motion for the electromagnetic field that follow from Eq.~\eqref{lagr} are
\begin{align}
\label{eqmov}
\partial_\mu(\sqrt{-g}\:f\:F^{\mu\nu})=0 \;,
\end{align}
where the field $F_{\mu\nu}$ is expressed in terms of the gauge potential $A_\mu$ as $F_{\mu\nu}=\partial_\mu A_\nu-\partial_\nu A_\mu$.

We now quantise the electromagnetic field in the Coulomb gauge with respect to the cosmic time foliation, where $A_0=0$ and $\partial_i
A^i=0$. The spatial part of the vector potential is the operator
\begin{align}
\label{decomp}
\hat{A}_i(t,\mathbf{x})=\sum_{\sigma=1,2}\int \frac{d^3k}{(2\pi)^{3/2}}\left[\epsilon_{i,\sigma}(\mathbf{k})\hat{a}_{\mathbf{k},\sigma} A_{k,\sigma} (t)e^{i\mathbf{k}\cdot\mathbf{x}}+H.C.\right] \;,  
\end{align}
where $\epsilon_{i,\sigma}(\mathbf{k})$ are the photons true degrees of freedom. They are orthonormal and transverse
vectors, constant on each hypersurface (their Lie derivative is zero with respect to the spatial foliation vector field). We see the first term contains the quantum operator $\hat{a}_{\mathbf{k},\sigma}$, called the annihilation operator. The second term in \eqref{decomp}, $H.C.$, stands for the Hermitian conjugate of the first term, and therefore contains the creation operator $\hat{a}^\dagger_{\mathbf{k},\sigma}$.  They satisfy the commutation relations
\begin{align}
[\hat{a}_{\mathbf{k},\sigma}, \hat{a}^\dagger_{\mathbf{k'},\sigma'}] &=
\delta_{\sigma\sigma'}\delta(\mathbf{k} - \mathbf{k'}) \;,\\
[\hat{a}_{\mathbf{k},\sigma}, \hat{a}_{\mathbf{k'},\sigma'}]&=0 \;,\\
[\hat{a}^\dagger_{\mathbf{k},\sigma},
\hat{a}^\dagger_{\mathbf{k'},\sigma'}]&=0 \;.
\end{align}
The time-dependent coefficients
$A_{k,\sigma}(t)$ and their associated momenta $\Pi_{k,\sigma} \equiv
4afA^\prime_{k,\sigma}(t)$ satisfy the wronskian condition
\begin{align}\label{vacuumnorma}
A_{k,\sigma}(t)\Pi^*_{k,\sigma}(t)
-A^*_{k,\sigma}(t)\Pi_{k,\sigma}(t) = i,
\end{align}	
for each wavenumber $k$ and helicity $\sigma$.

We emphasise that the quantisation of the gauge-fixed electromagnetic field in the absence of charges is equivalent to that of two free real scalar fields. Consequently, the choice of vacuum for each polarisation $\sigma$ corresponds to the choice of vacuum of each scalar degree of freedom. However, using the fact that we are dealing with an
isotropic background, there is no reason to make different choices of vacuum
for different polarisations. For this reason, we choose a single time-dependent
coefficient to describe both polarisations, \textit{i.e.}, $A_{k,1} = A_{k,2} \equiv{A_{k}} $. 
Therefore, the same vacuum is chosen for both polarisations. Now, inserting this decomposition in
the EOM \eqref{eqmov}, we get the equation governing the evolution of the modes
$A_k(t)$
\begin{align}
\label{pot}
\ddot{A}_k+\left(\frac{\dot a}{a}+\frac{\dot f}{f}
\right)\dot{A}_k +\frac{k^2}{a^2}A_k=0 \;.
\end{align}

This equation is most easily analysed numerically by defining the quantities
\begin{align}
k_s := kR_H,\;\;
\;\;A_{sk}(t_s) := \frac{A_k(t_s)}{\sqrt{x_bR_{H_0}}} \;,
\end{align}
where $R_H=R_{H_0}/a_0$ is the comoving Hubble radius today, 
the differential Eq.~\eqref{pot} can be written as
\begin{align}
A_{sk}^{\prime\prime}+\left(\frac{ Y^\prime}{Y}+\frac{f^\prime}{f}
\right){A}_{sk}^\prime +\frac{k_s^2}{Y^2}A_{sk}=0 \;,
\label{diffeq}
\end{align}
where a prime denotes the derivative with respect to $t_s$. The coupling \eqref{coupling} then takes the form 
\begin{align}
f=\frac 1 4 \left[ 1+C^2\frac{\alpha^2 t_s^2+3}{(\alpha^2t_s^2+1)^2}
\right] \;; \;\;\;\;\; \textup{with}\;\; C^2 := \frac{4}{3} \frac{\ell_*^2}{t_b^2} \;,
\;\;\;\;\;\ell_* := \frac{1}{m_*} \;.
\label{cc}
\end{align}
An upper limit on $C$ can be straightforwardly derived from Eq.~\eqref{cc}. Since any contribution to the usual Maxwell's equations at BBN must be negligible, we impose the second term in Eq.~\eqref{cc} to be smaller than $10^{-2}$ at BBN. Together with the fact that $\alpha^2 t_s^2 \gg 1$ at this time, we get
\begin{align}
C < 10^{-19} x_b^{3/2} \;.
\label{limit c}
\end{align}

Now that the evolution equation has been defined, we relate the gauge modes to the energy densities of the electric and magnetic fields, respectively given by
\begin{align}
\rho_E &=\frac{f}{8\pi}g^{ij}A_i^\prime A_j^\prime\;, \\
\rho_B &=\frac{f}{16\pi}g^{ij}g^{lm}(\partial_j A_m-\partial_m A_j) (\partial_i A_l-\partial_l A_i) \;,
\end{align}
where $g^{ij}=\delta^{ij}/a^2$ are the spatial components of the inverse metric. To find the spectral energy densities, we first insert expansion \eqref{decomp} into $\rho_E$ and $\rho_B$. The resulting operators $\hat{\rho}_E$ and $\hat{\rho}_B$ upon quantisation are
\begin{align}
\hat{\rho}_B &= \frac{f}{2\pi^2 R_{H_0}^4 Y^{4}} \int \dd{\ln{k}} \;  \vert A_{sk}\vert^2 k^5 \;, \label{magnetic spectral density} \\
\hat{\rho}_E &= \frac{f}{2\pi^2 R_{H_0}^4 Y^{2}} \int \dd{\ln{k}} \; \vert A^{\prime}_{sk}\vert^2 k^3 \;. \label{electric spectral density}
\end{align}
We now evaluate the expectation value of the two densities in vacuum, defined by 
$\hat{a}_{\mathbf{k},\sigma} \ket{0}=0$, and define the spectra as
\begin{align}
{{\cal P}_{i}} \equiv \frac{\textup{d}\bra{0} \hat{\rho}_i \ket{0}}{\dd{\ln{k}}} \;,  \quad i=E,B \;.
\end{align}
This yields the magnetic and electric spectra, respectively
\begin{align}
{\cal P_{B}} &\equiv B^2_\lambda = \frac{f}{2\pi^2 R_{H_0}^4}\frac{\vert A_{sk}\vert^2}{Y^4}k^5 \;, 
\label{magpow} \\ 
{\cal P_{E}} &\equiv E_{\lambda}^2=\frac{f}{2\pi^2 R_{H_0}^4}\frac{\vert A_{sk}^\prime\vert^2}{Y^2}k^3 = \frac{1}{2\pi^2 R_{H_0}^4}\frac{\vert \Pi_{sk}\vert^2}{fY^4}k^3 \;.
\label{elecpow}
\end{align}
In the last line, we also expressed ${\cal P_{E}}$ in terms of the momentum canonically conjugate to the gauge field $\Pi_{sk} = Yf A_{sk}^\prime$ (see Appendix ~\ref{sec:adiab}), which is nothing but the electric field mode itself.

Finally, we can express the magnetic and electric fields $B_{\lambda}$ and $E_{\lambda}$ in Gauss, using $H_0^2=1.15\times 10^{-64}$ G
\begin{align}
\label{est}
B_\lambda &=\sqrt{\frac{f}{2\pi^2}} 
\frac{\vert A_{sk}\vert }{Y^2}k^{5/2}\:1.15\times 10^{-64} {\textup{G}} \;, \\
E_\lambda &=\sqrt{\frac{1}{2 \pi^2 f}} 
\frac{\vert  \Pi_{sk}\vert }{Y^2}k^{3/2}\:1.15\times 10^{-64} {\textup{G}}.
\end{align}

\subsection{Adiabatic Vacuum Initial Conditions}
\label{sec:adiab}

In order to find the form of the gauge field and its momentum in vacuum, we first need to impose initial conditions for the EM field. To this end, we follow the adiabatic vacuum prescription implemented in Ref.~\cite{Vitenti:2020???}. Even though we are dealing with vector degrees of freedom, since the time-dependent coefficient $A_k(t)$ satisfies the normalisation condition~\eqref{vacuumnorma}, it follows that $A_k(t)$ has a behaviour similar to the one of the coefficient one would obtain when quantising a single free scalar field. Let us then consider the Hamiltonian
\begin{align}
\label{hamiltonian}
\mathcal{H} = \frac{\Pi_k^2}{2m} + \frac{m\nu^2 A_k^2}{2} \;,
\end{align}
where $m$ and $\nu$ can be functions of time. The Hamilton equations of motion
\begin{align}
\label{eq-hamiltonian}
A^\prime_k=\frac{\Pi_k}{m}; \quad \Pi^\prime_k = -m\nu^2 A_k 
\end{align}
lead to Eq.~\eqref{diffeq} if one identifies $m = Yf$ and $\nu = k / Y$.

A convenient choice is to express $A_{k}$ and $\Pi_k$ as the components of a particular eigenvector of the complex structure matrix (see Ref.~\cite{Vitenti:2020???} for the mathematical and physical reasons to implement this choice),
\begin{align}
\label{v:components}
A_k &\equiv \frac{1}{2}\exp{(-\gamma_k/2)} \left[\exp{(\chi_k/2)}-\mathsf{i}\exp{(-\chi_k/2)}\right] \;,\\ 
\Pi_k &\equiv -\frac{1}{2}\exp{(\gamma_k/2)} \left[\exp{(\chi_k/2)}+\mathsf{i}\exp{(-\chi_k/2)} \right)] \;.
\end{align}
The variables $\chi_k$ and $\gamma_k$ are real time-dependent functions, and can be used to represent the aforementioned matrix as
\begin{align}
\label{M:components}
M_a{}^b = \left( \begin{array}{cc} \sinh{\chi_k} & \cosh{\chi_k} \exp({-\gamma_k}) \\ -\cosh{\chi_k} \exp({\gamma_k}) & -\sinh{\chi_k} \end{array} \right).
\end{align}
Latin indices refer to the phase space vector components
defined by $v_a \equiv \left(A_k,\Pi_k\right)$, which are raised and lowered using the symplectic matrix as defined in Ref.~\cite{Vitenti:2020???}. The phase space vectors $v_a$  satisfying the normalisation condition~\eqref{vacuumnorma} (modulo a global time-dependent phase) have an one-to-one correspondence with matrices of the form shown in Eq.~\eqref{M:components}, and consequently with a pair $(\chi_k,\;\gamma_k)$. For this reason, we will denote interchangeably  $\left(A_k,\Pi_k\right)$ and $\left(\chi_k,\;\gamma_k\right)$ with the same symbol $v_a$.

The Hamilton Eqs.~\eqref{eq-hamiltonian} induce the dynamics of the matrix $M_a{}^b$, which reads
\begin{align}
\label{set:equations}
\chi_k^\prime&= -2\nu\sinh(\gamma_k-\xi) \;,\\
\gamma_k^\prime &= +2\nu\cosh(\gamma_k-\xi)\tanh(\chi_k) \;,
\end{align}
where $\xi \equiv \ln (m\nu)$. The complex structure matrix satisfies
\begin{align}
\label{M2:identity}
M_a{}^c M_c{}^b = -\delta_a{}^b \;,
\end{align}
and, the comparison of two different vacuum definitions, given respectively by $v_a$ and $u_a$, yields the Bogoliubov coefficients
\begin{align}
\label{beta-M}
\vert \beta_{v,u}\vert^2 = -\frac{1}{4} \textup{Tr} \left[\boldsymbol{I} + \boldsymbol{M}(v) \boldsymbol{M}(u)\right] \;,
\end{align}
with $\textup{Tr}$ the trace operator, $\boldsymbol{I}$ the identity matrix and $\boldsymbol{M}(v)$ ($\boldsymbol{M}(u)$) is the matrix  associated with the components $M_a{}^b(v)$ ($M_a{}^b(u)$) defined by the vector components $v_a$ ($u_a$). In this framework, a vacuum choice translates into a choice of functions $v^V_a \equiv \left(\chi_k^V(t),\;\gamma_k^{V}(t)\right)$ defined locally (with a finite number of time derivatives of the background variables), which do not necessarily satisfy the equations of motion~\eqref{set:equations} but give an approximation close enough to a solution. Moreover, the vacuum must be fixed by choosing a time $t_0$ where the variables satisfy 
$$v_a(t_0) = v^V_a(t_0), \quad\implies\quad \left(\chi_k(t_0), \gamma_k(t_0)\right) = \left(\chi^V_k(t_0), \gamma^V_k(t_0)\right).$$ In other words, if $v_a^V$ is stable in the sense that $$\Delta v_a \equiv \left(\delta\chi_k,\;\delta\gamma_k \right) = \left(\chi_k(t) - \chi^V_k(t), \gamma_k(t) - \gamma^V_k(t)\right)$$ remains small for a finite time interval, then  particle creation will also be small in this interval. This characterises the so-called adiabatic vacuum. Hence, we find the adiabatic vacuum by finding the critical points of the system \eqref{set:equations}. When $\xi$ is constant in time, the critical points of the system \eqref{set:equations} are obvious: $\chi^V_k = 0$ and $\gamma^V_k = \xi$, a choice satisfying the condition of being locally defined in terms of the background. Then, substituting into Eq.~\eqref{v:components}, and using it as initial conditons for the system~\eqref{eq-hamiltonian}, yields the following solution
\begin{align}
\label{WKB}
A_k &= \frac{e^{-\mathsf{i}\pi/4}}{\sqrt{2m\nu}}
\exp\left[-\mathsf{i}\int_{t_0}^t\nu\textup{d}t\right] \;,\\
\Pi_k &= -\mathsf{i} e^{-\mathsf{i}\pi/4}\sqrt{\frac{m\nu}{2}}
\exp\left[-\mathsf{i}\int_{t_0}^t\nu\textup{d}t\right] \;.
\end{align}
In this case, the vacuum is perfectly stable, there is never particle production because $\chi_k(t) = \chi^V_k(t) = 0$ and $\gamma_k(t) = \gamma^V_k(t) = \xi$ for any time $t$, and consequently $\vert\beta_{v,v^V}\vert^2=0$, see Eq.~\eqref{beta-M}. We have a perfect adiabatic vacuum, which coincides with the WKB solution.

In the case where $\xi$ changes in time, there is one well-known situation where adiabatic vacua can be defined: when the mode frequencies dominate the dynamics. Let us define $$F_n \equiv \left(\frac{1}{2\nu}\frac{\dd}{\dd t}\right)^n \xi \;,$$ where $F_0 = \xi$, the function $F_1$ is the ratio between the time derivative of $\xi$ and $\theta \equiv \int2\nu\dd t$, $F_2$ the ratio between the time derivative of $F_1$ and $\theta$ and so forth.  Then, in the case $1 \gg F_1 \gg\dots\gg F_n>\dots$, which means that $\xi$ slowly varies in cosmic time when compared with the variation of $\theta$, one can still find approximate critical points (\textit{i.e.} adiabatic vacua), which can be reached through successive approximations, as explained in Ref.~\cite{Vitenti:2020???}. Up to second order, the approximate critical points read
\begin{align}
\label{variables2}
\chi^V_k&= F_1 \;,\nonumber \\
\gamma^V_k&= F_0 -F_2\;.
\end{align}
If they are inserted in Eq.~\eqref{v:components}, they lead to the usual WKB expansion (modulo a time-dependent phase). As discussed in~\cite{Vitenti:2020???}, around these functions, the variables $\Delta v_a$ satisfy a forced harmonic oscillator equation of motion with force of order $\mathcal{O}(F_3)$.

In our case, we have $m\nu=kf$. In the far past of the contracting phase one gets, for $f$ given in Eq.~\eqref{cc},
\begin{align}
\label{xi1}
\left\vert\frac{\dd\xi}{\dd\theta}\right\vert \approx \frac{C^2}{x_b^3k\vert t\vert^{7/3}} \ll 1 \;,
\end{align}
which implies that $$\vert t\vert \gg \vert t_a\vert \equiv\left(\frac{C^2}{x_b^3k}\right)^{3/7}.$$ As the physically relevant parameter space we consider satisfies $C^2/x_b^3 \ll 1$, then $\vert t_a\vert  \ll 1$, and this condition is easily satisfied.

However, the other adiabaticity conditions impose a more stringent constraint on $\vert t\vert$. Indeed, 

\begin{align}
\label{xi2}
\left\vert\frac{\dd^2\xi}{\dd\theta^2}\right\vert \ll \left\vert\frac{\dd\xi}{\dd\theta}\right\vert \Rightarrow \vert t\vert  \gg \vert t_c\vert \equiv
\left(\frac{7}{3 k}\right)^{3}\approx k^{-3} \;.
\end{align}
One can easily verify that all other conditions yield, apart numerical factors of order $1$,\footnote{It starts around $1$ and grows slowly with $n$, this is a natural feature of an asymptotic expansion. In other words, for a fixed time and mode $k$ there is a maximum order $n$ from which the series starts to be a bad approximation.} the same condition \eqref{xi2}. Hence, the adiabaticity condition reads

\begin{align}
\label{adiabatic}
\vert t\vert  \gg \vert t_c\vert  \approx k^{-3}.
\end{align}
This means that modes with the size of the Hubble radius today leave (enter) the adiabatic regime in the contracting (expanding) phase for times of the order the Hubble time today, independently of the parameters $x_b$ and $C$. Smaller wavelengths leave (enter) the adiabatic regime later (earlier) than the present Hubble time, following the rule $k^{-3}$.

To summarise, one can impose adiabatic vacuum initial conditions for the electromagnetic field in the contracting phase of the present bouncing model when $\vert t\vert  \gg \vert t_c\vert  \approx k^{-3}$. In this regime, the modes read, at leading order,
\begin{align}
\label{initial-vacuum}
A_k &= \frac{e^{-\mathsf{i}\pi/4}}{\sqrt{2kf}}\exp\left(-\mathsf{i} k\eta\right)+\dots \;,\\
\Pi_k &= -\mathsf{i} e^{-\mathsf{i}\pi/4}\sqrt{\frac{kf}{2}}
\exp\left(-\mathsf{i} k\eta\right)+\dots \;,
\end{align}
where $\eta$ is the conformal time $dt = Yd\eta$.

Since $f\approx 1/4$ for $\vert t\vert  \gg \vert t_c\vert $ , it follows that 
\begin{align}
\label{initial-modulus}
\vert A_k\vert  &= \sqrt{\frac{2}{k}} +\dots \;, \\
\vert \Pi_k\vert  &= \sqrt{\frac{k}{8}} +\dots. \;,
\end{align}
and both the field and its canonical momentum are constant in this regime.

\subsection{Analytical Results}
\label{Aresults}

Now that initial conditions for the electromagnetism field have been obtained, we can start the analytic study of the time behaviour and spectra of $A_{k}$ satisfying the gauge field evolution equation \eqref{diffeq} (from now on the index $s$ on the time variable and wavenumber will be omitted), and its canonical momentum $\Pi_k$, in the different stages of the cosmic evolution. In the sequel, this analysis will be compared with the numerical results.

As shown in section \ref{sec:adiab}, the adiabatic vacuum is a consistent choice for the EM field initial conditions. The modes in vacuum are 
\begin{align}
\vert A_k\vert  &= \sqrt{\frac{2}{k}} +\dots \;, \\
\vert \Pi_k\vert  &= \sqrt{\frac{k}{8}} +\dots. \;,
\end{align}
and both the field and its canonical momentum are constant in this regime.
Now that the initial conditions for the EM field have been defined, we can move on to the analysis of the evolution of the electric and magnetic modes from the far past up to the present day. 

Three important characteristic times related to the evolution of the modes are worthy of note
. The first is the time limit of the adiabatic regime, $\vert t_c\vert $, defined in Eq.~\eqref{xi2}. The second one is the time where quantum effects leading to the bounce take place, \textit{i.e.} $\vert t_b\vert  = 1/\alpha$. Consequently,  the bounce phase takes place for $t$ such that  $-1/\alpha < t < 1/\alpha$. The third one is the characteristic time when the evolution of $f$  becomes important. Examining Eq.~\eqref{cc}, one gets the time $\vert t_f\vert  = C/\alpha$, up to $\vert t_b\vert $, which means that the evolution of $f$ is important when $-C/\alpha < t < -1/\alpha$, and  $1/\alpha < t < C/\alpha$. The domain of physically allowed parameters imposes that
\begin{align}
\label{times}
\vert t_c\vert  \gg \vert t_f\vert  \gg \vert t_b\vert  \;.
\end{align}

For $\vert t\vert <\vert t_c\vert $, the solution leaves the frequency-dominated
region. In this case, one can perform the usual expansion in $\nu^2$ derived
from the Hamilton Eqs.~\eqref{eq-hamiltonian} through iterative substitutions:
\begin{align}
\label{iterations1}
\Pi_k(t) &= -\int^t m(t_1)\nu^2(t_1) A_k(t_1)\dd t_1 + A_2(k) = mA^\prime_k(t)\Rightarrow\\
A_k(t) &= -\int^t \frac{\dd t_2}{m(t_2)}\int^{t_2}m(t_1)\nu^2(t_1) A_k(t_1) \dd t_1 + A_2(k)\int^t \frac{\dd t_1}{m(t_1)} + A_1(k)\Rightarrow\\
A_k(t) &= A_1(k)\left(1-\int^t \frac{\dd t_2}{m(t_2)}\int^{t_2}m(t_1)\nu^2(t_1) \dd t_1\right) + \\ 
& A_2(k)\left(\int^t \frac{\dd t_1}{m(t_1)} - \int^t \frac{\dd t_2}{m(t_2)}\int^{t_2}m(t_1)\nu^2(t_1)\dd t_1\int^{t_1} \frac{\dd t_3}{m(t_3)}\right) +\dots \; ,
\end{align} 
where $A_1(k)$ and $A_2(k)$ are constants in time depending only on $k$, leading to the momentum expression
\begin{align}
\label{iterations2}
\Pi_k(t) = -A_1(k)\int^{t}m(t_1)\nu^2(t_1) \dd t_1 + A_2(k)\left(1 - \int^t m(t_1)\nu^2(t_1)\dd t_1\int^{t_1} \frac{\dd t_2}{m(t_2)}\right)+\dots \;.
\end{align}
We can now evaluate the time evolution and spectra in the different phases of the cosmic evolution. 

\subsubsection{The Contracting Phase and the Bounce}

In the case of $A_k(t)$, all time-dependent terms are decaying in the contracting era up to the end of the bounce. As a consequence, $A_k(t)=A_1(k)$ is constant during all this phase. By continuity with the adiabatic phase, we conclude that
\begin{align}
\label{A1}
A_1(k)\propto k^{-1/2} \;.
\end{align}

The time-dependent terms of the momentum $\Pi_k(t)$ are also decaying, except for the one multiplying $A_1(k)$, which grows as $t^{-5/3}$ for $-C/\alpha < t < -1/\alpha$, since $f\propto 1/t^2$ in this region. Then, for $t< -C/\alpha$, $\Pi_k(t)=A_2(k)$ which, by continuity with the adiabatic phase, implies that 
\begin{align}
\label{A2}
A_2(k)\propto k^{1/2}.
\end{align}
In the period  $-C/\alpha < t < -1/\alpha$, the term multiplying $A_1(k)$ eventually surpasses the constant mode at a time $t_\pi$, and $\Pi_k(t)$ grows.

At the bounce itself $Y$ and $f$ are almost constant, therefore the modes will not evolve during this phase.

\subsubsection{The Expanding Phase}

In the expanding phase, the most important growing function related to $A_k(t)$ is the first one multiplying $A_2(k)$, which grows as fast as $t^{7/3}$ starting from some time $t_A$ in the interval $1/\alpha < t < C/\alpha$, and as $t^{1/3}$ for $C/\alpha < t < t_c$. 

In the case of $\Pi_k(t)$, as the integral multiplying $A_1(k)$ strongly decreases as $t^{-5/3}$ when $1/\alpha < t < C/\alpha$, the value of $\Pi_k(t)$ saturates in the value it gets by the end of the bounce, $t\approx 1/\alpha$. Also, $\Pi_k(t)$ acquires a $k^2$ dependence through the $\nu^2$ term. Combined with the $k$ dependence of $A_1(k)$, we obtain $\Pi_k(t) \propto k^{3/2}$.

After $t_c$, both $A_k(t)$ and $\Pi_k(t)$ begin to oscillate.

\subsubsection{Summary}

For the $A$-field, the spectra and time dependence in the different cosmic evolution phases is:
\begin{align}
\label{historyA}
-\infty < t < t_A \; &: \vert A_k(t)\vert  \propto k^{-1/2} \;, \\
t_A < t < C/\alpha \; &: \vert A_k(t)\vert  \propto k^{1/2}t^{7/3} \;, \\
C/\alpha < t < k^{-3} \; &: \vert A_k(t)\vert  \propto k^{1/2}t^{1/3}  \;, \\
t > k^{-3} \; &: \vert A_k(t)\vert  \propto k^{1/2} \; \times {\textrm{(oscillatory factors)}},
\end{align}
where $t_A \in (1/\alpha,C/\alpha)$.

For the $\Pi$-field, we have:
\begin{align}
\label{historyP}
-\infty < t < t_\pi \; &: \vert \Pi_k(t)\vert \propto k^{1/2} \;,\\
t_\pi < t < -1/\alpha \; &: \vert \Pi_k(t)\vert  \propto k^{3/2} t^{-5/3} \;,\\-1/\alpha< t < k^{-3} \; &: \vert \Pi_k(t)\vert  \propto k^{3/2} \;,\\
t > k^{-3} \; &: \vert \Pi_k(t)\vert  \propto k^{3/2} \; \times {\textrm{(oscillatory factors)}},
\end{align}
where $t_\pi \in (-C/\alpha,-1/\alpha)$.

Note that both the final spectrum of ${\cal P_{B}}$ and ${\cal P_{E}}$ (given in Eqs. \eqref{magpow} and \eqref{elecpow}) go as $k^6$.

\subsection{Numerical Results}
\label{results}

\begin{figure}[h!]
	\centering
	\includegraphics[scale=0.41]{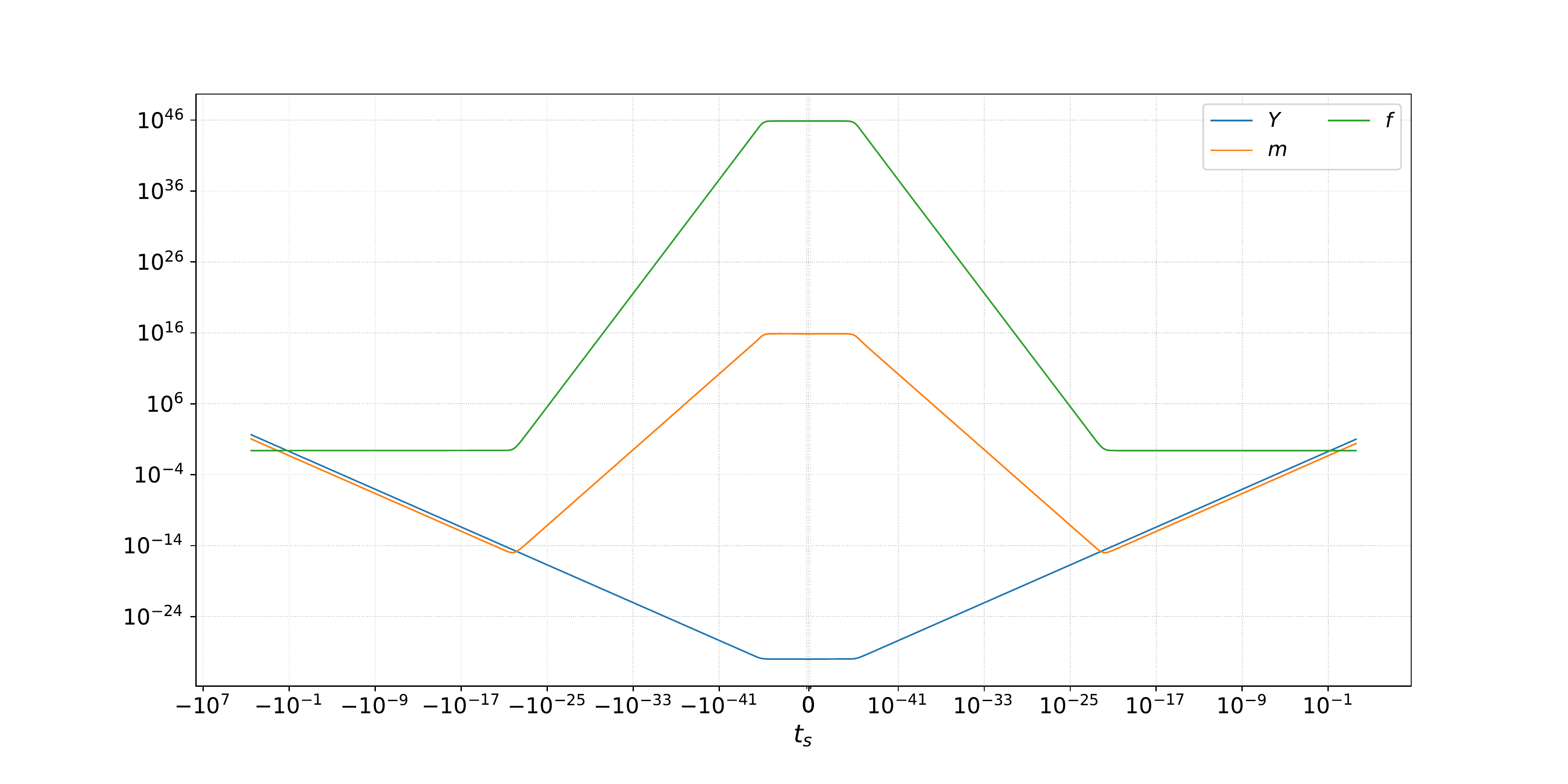}
	\caption[Evolution of the magnetic field model parameters]{Evolution of the coupling $f$, the scale factor $Y$ normalised today, and the mass $m=af$ with time. We have used $C=10^{23}$ and $x_b=10^{30}$.}
	\label{ymf}
\end{figure}

\begin{figure}[h!]
	\includegraphics[scale=0.41]{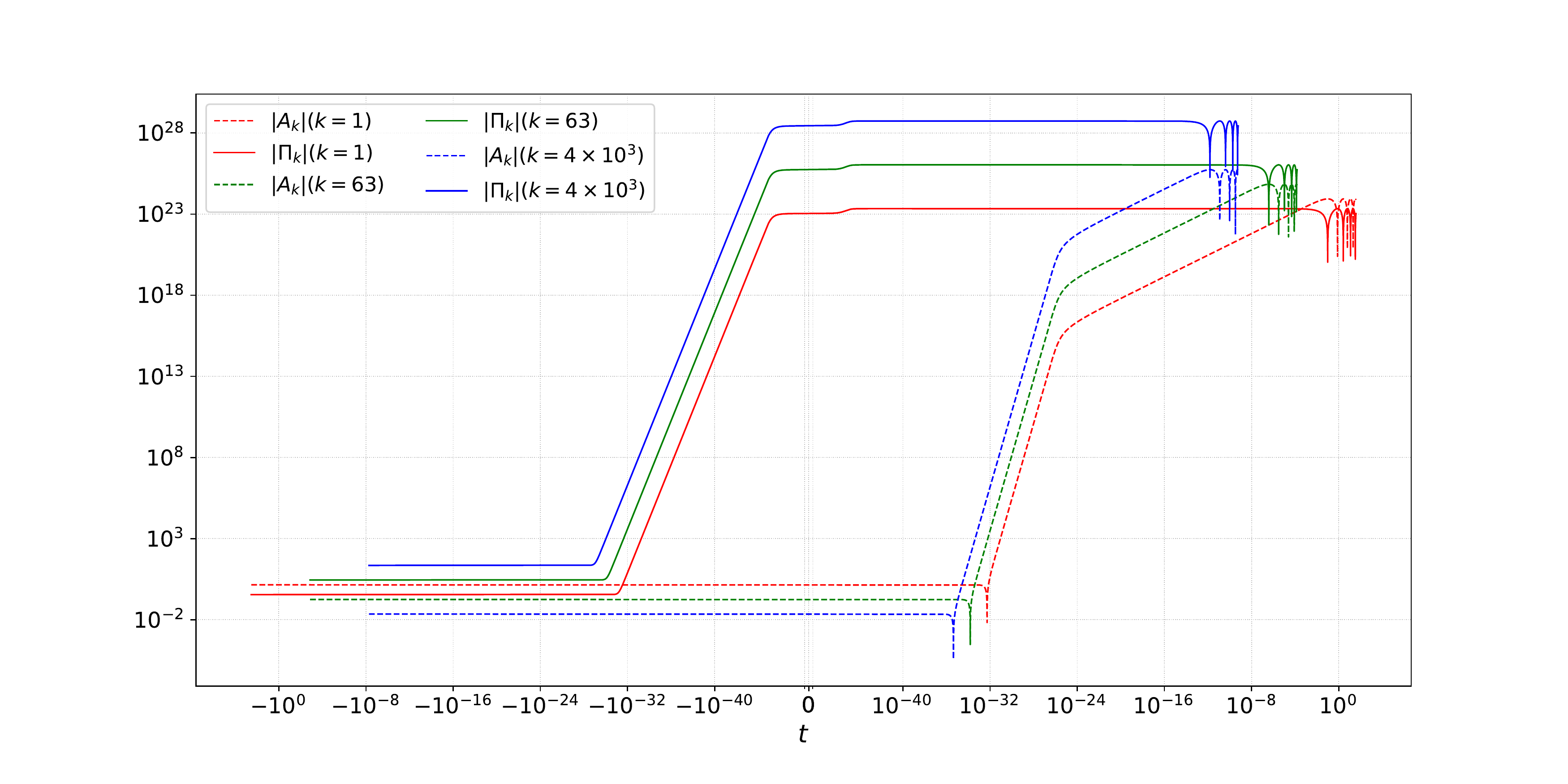}
	\includegraphics[scale=0.41]{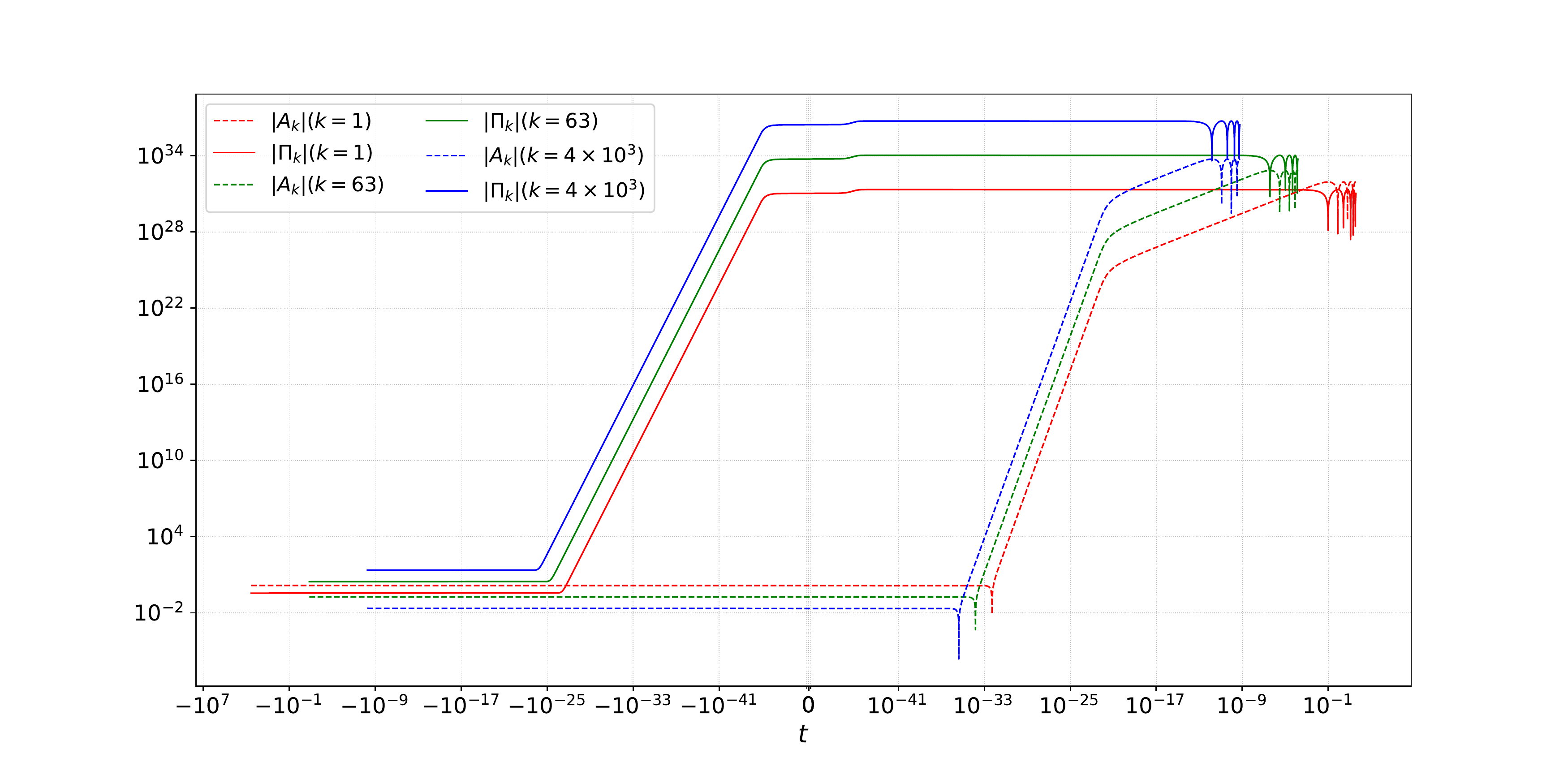}
	\caption[Evolution of the magnetic modes and momenta with variation of the coupling strength.]{Evolution of the absolute values of the magnetic modes ($A_{k}$) and their momentum ($\Pi_k $) through the bounce in a dust background for $C=10^{19}$ and $x_b=10^{30}$ (top), and for $C=10^{23}$ and $x_b=10^{30}$ (bottom). The same colour for the gauge field and its momentum evolution is chosen for a given $k_s$. We see that larger values of $C$ lead to a higher final amplitude.}
	\label{mode evolution cc}
\end{figure}

\begin{figure}[h!]
	\centering
	\includegraphics[scale=0.41]{../figures/MagDustModeEvol_cc23_xb30.pdf}
	\includegraphics[scale=0.41]{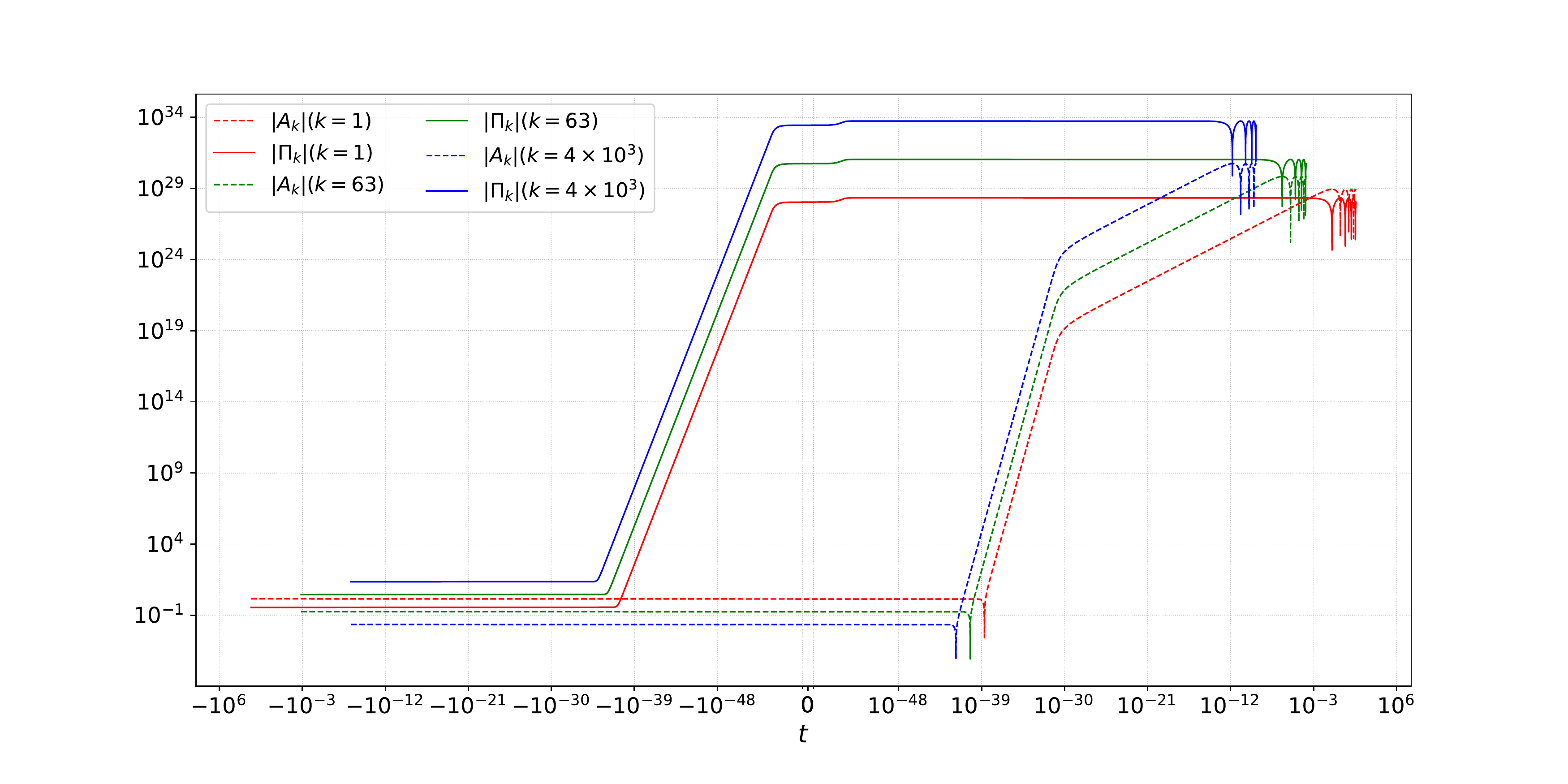}
	\caption[Evolution of the magnetic modes and momenta with variation of the bounce scale factor.]{Same as Fig.~\ref{mode evolution cc} for $C=10^{23}$ and $x_b=10^{30}$ (top), and for $C=10^{23}$ and $x_b=10^{36}$ (bottom). We see that larger values of  $x_b$ lead to a quicker evolution of the modes.}
	\label{mode evolution xb}
\end{figure}

\begin{figure}[h!]
	\centering
	\includegraphics[scale=0.4]{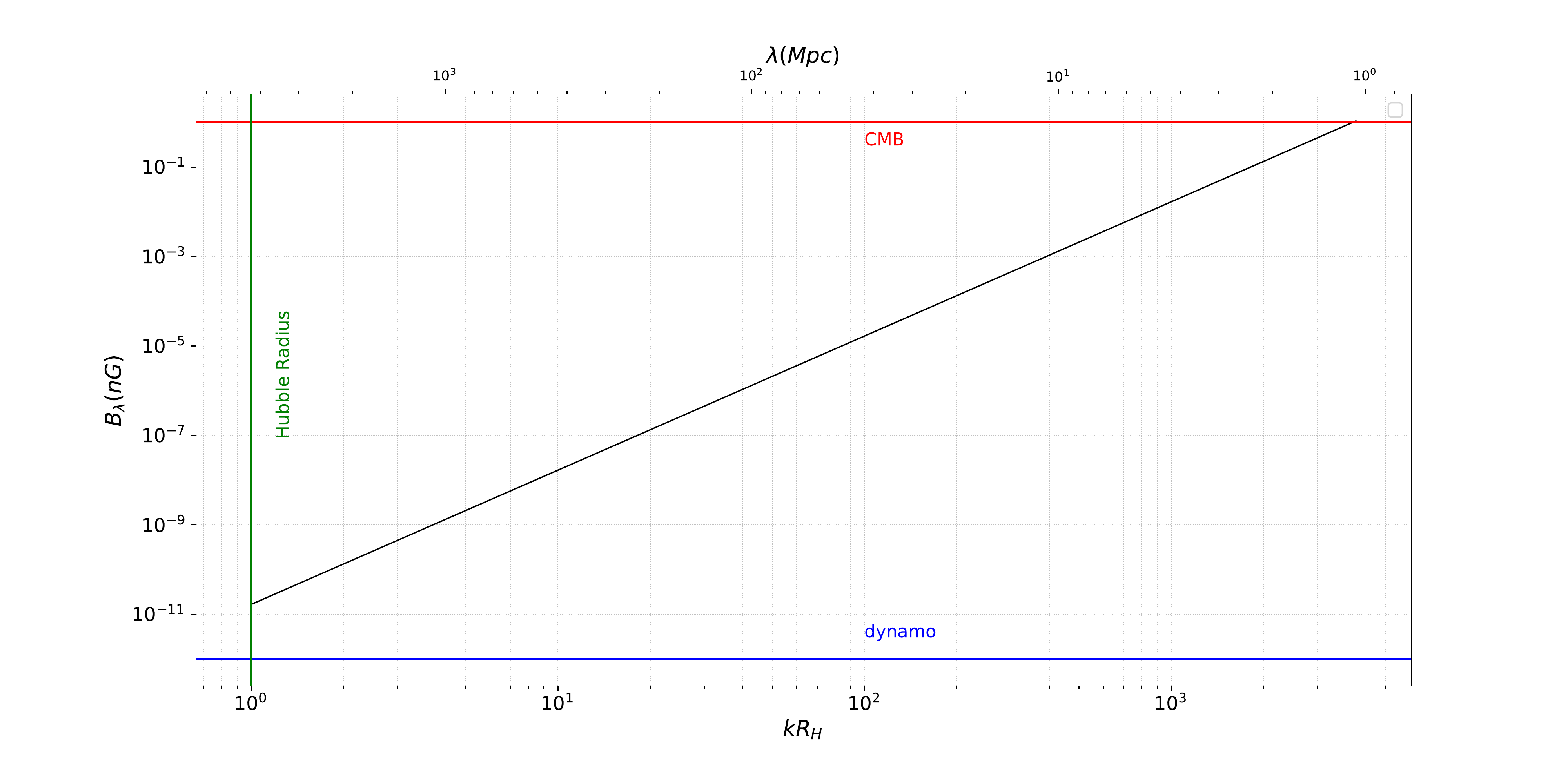}
	\includegraphics[scale=0.4]{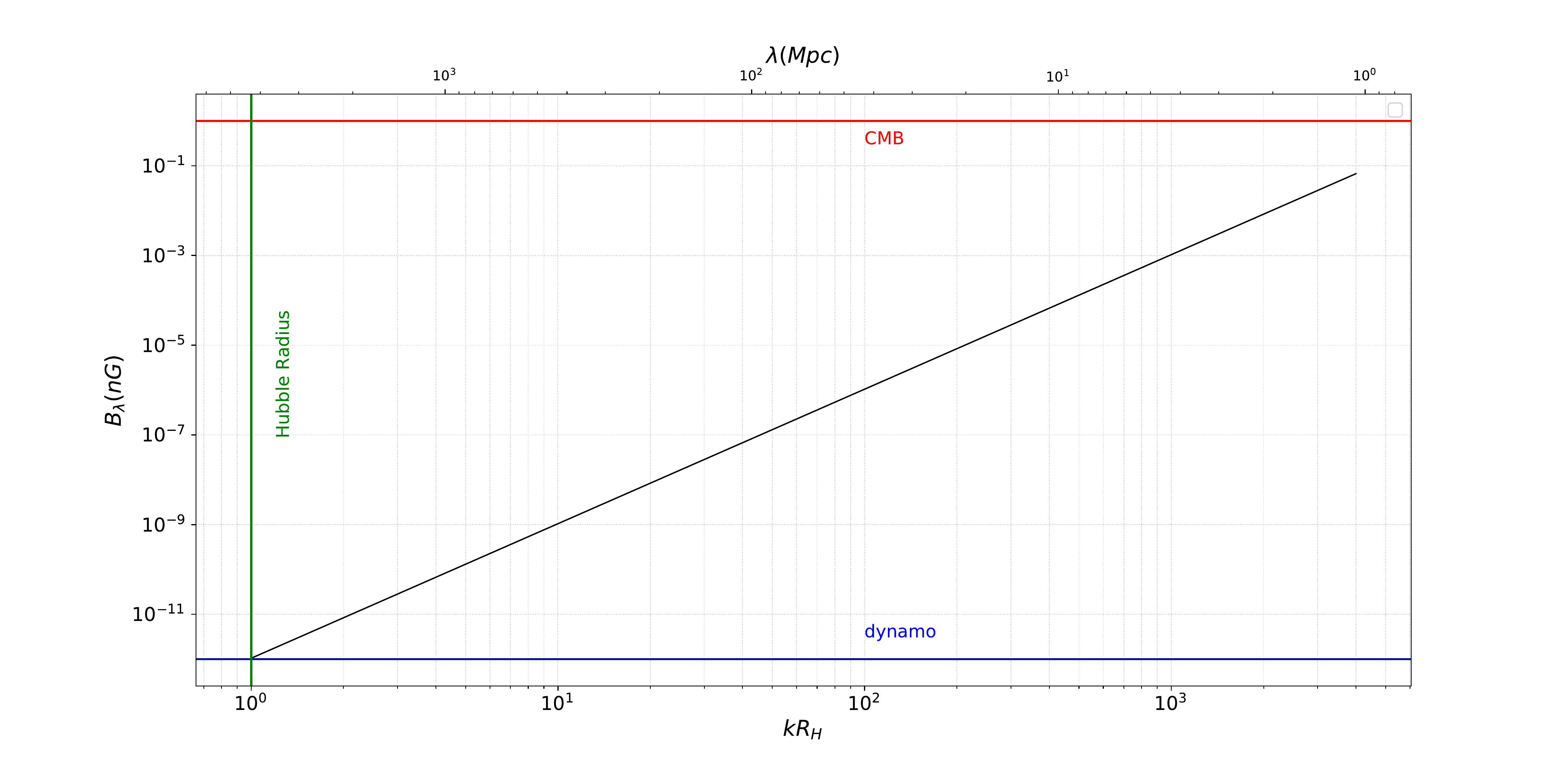}
	\caption[Magnetic field amplitude today on scales $\geq 1 Mpc$ with variation of the coupling strength.]{Magnetic field amplitude for $C=2.6 \times 10^{26}$ and $x_b=10^{38}$ (top), and $C=6.5 \times 10^{25}$ and $x_b=10^{38}$ (bottom). For these values, the seed field is sufficient to trigger the dynamo mecanism on large scales. The amplitude today is larger on all scales for larger values of $C$.}
	\label{magfield}
\end{figure}

\begin{figure}[h!]
	\centering
	\includegraphics[scale=0.4]{../figures/MagDustB_cc6-5e25_xb38.pdf}
	\includegraphics[scale=0.4]{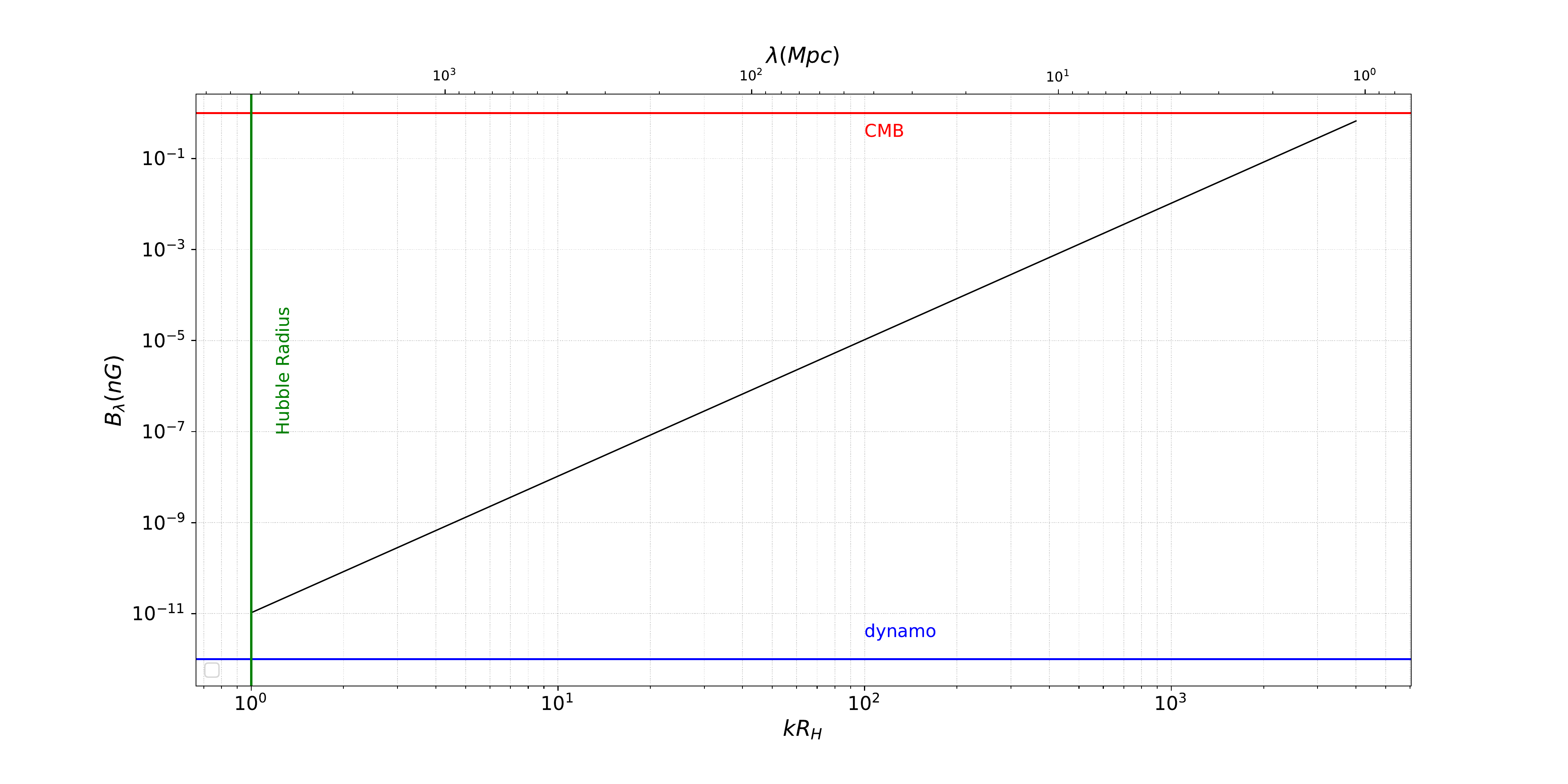}
	\caption[Magnetic field amplitude today on scales $\geq 1 Mpc$ with variation of the bounce scale factor.]{Magnetic field amplitude for $C=6.5 \times 10^{25}$ and $x_b=10^{38}$ (top) and for $C=6.5 \times 10^{25}$ and $x_b=10^{36}$ (bottom). The amplitude today is bigger at all scales when $x_b$ is smaller.}
	\label{magfield2}
\end{figure}

After these analytical considerations, let us now turn to the numerical calculations, which confirm the behaviours presented in this section, and 
allow the calculation of the field amplitude. We start this section by showing in Fig.~\ref{ymf} the time behaviour of the
coupling $f$ given in Eq.~\eqref{coupling}, the scale factor $Y=a/a_0$ from
Eq.~\eqref{yt}, and the mass $m=Yf$. From the definition of $|t_f|$ and $|t_b|$ in the previous section, and choosing $C=10^{23}$ and $x_b=10^{30}$, we obtain respectively $|t_f|\simeq 10^{-22}$ and $|t_b|\simeq 10^{-45}$. This is consistent with the behaviour shown in the figure. %
The numerical evolution of the gauge field $A_{k}$ and its momentum $\Pi_k$ is shown next. In Fig.~\ref{mode evolution cc}, the influence of the parameter $C$ on the evolution of the modes is shown explicitly for $C=10^{19}$ and $C=10^{23}$ with $x_b=10^{30}$, while the influence of $x_b$ is shown in Fig.~\ref{mode evolution xb} for $x_b=10^{30}$ and $C=10^{36}$ with $C=10^{23}$.\footnote{We choose the values of $C$ and $x_b$  to be well inside the allowed parameter space at 1 Mpc, as can be seen in Fig.~\ref{parameter space 1Mpc}. We will use the same set of values throughout this section, except for Figs~\ref{magfield} and \ref{magfield2}.} Note that in these figures, as well as in the following ones, we performed the computation for $1<k<4000$, since $k=4000$ implies a physical wavelength of about 1 Mpc (remember that $k$ is in units of Hubble radius). One can verify in these figures all time and $k$ dependence described in
Sec.~\ref{Aresults}, summarised in Eqs.~\eqref{historyA} and \eqref{historyP}.

\begin{figure}[h!]
	\centering
	\includegraphics[scale=0.41]{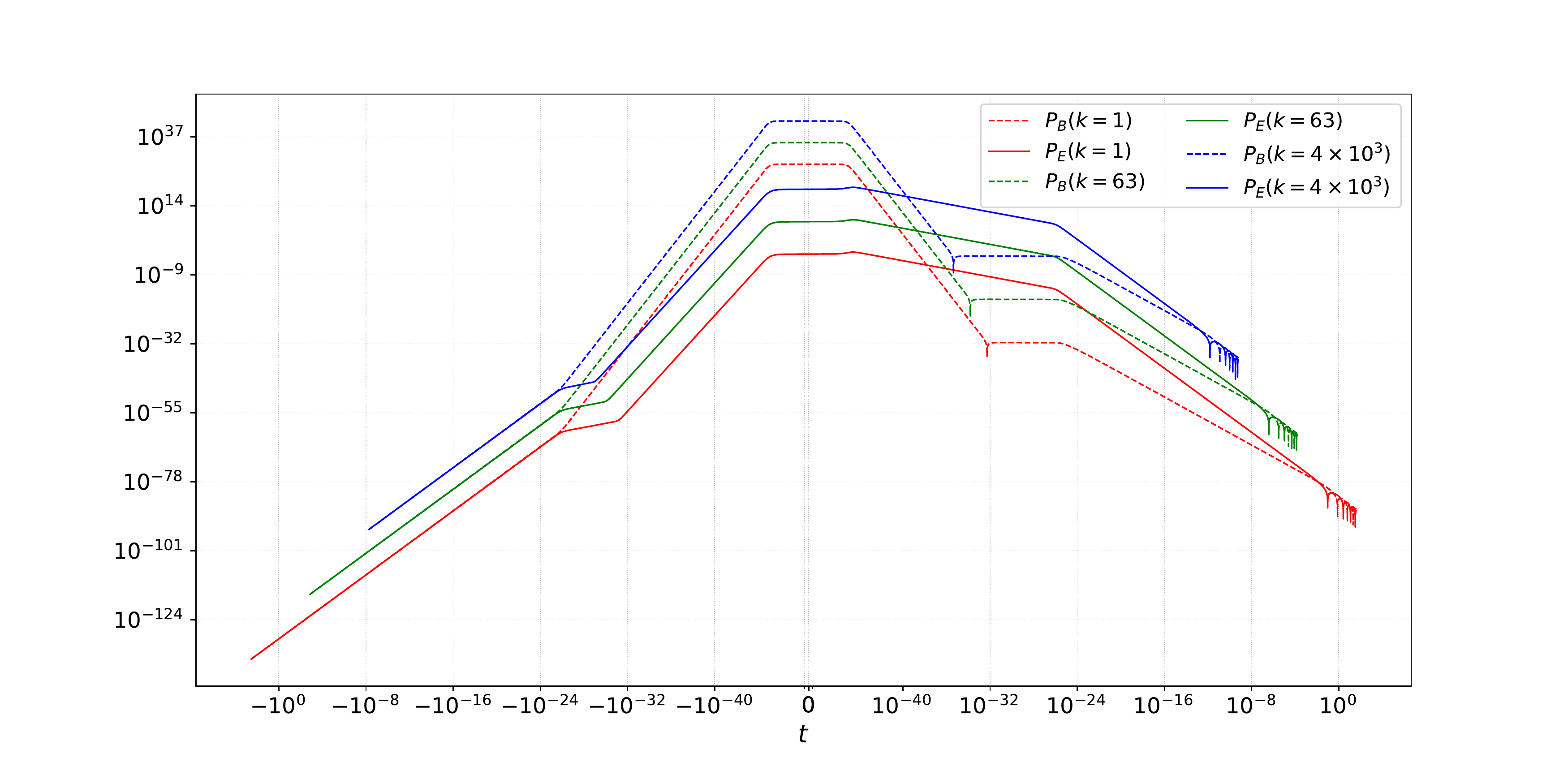}
	\includegraphics[scale=0.41]{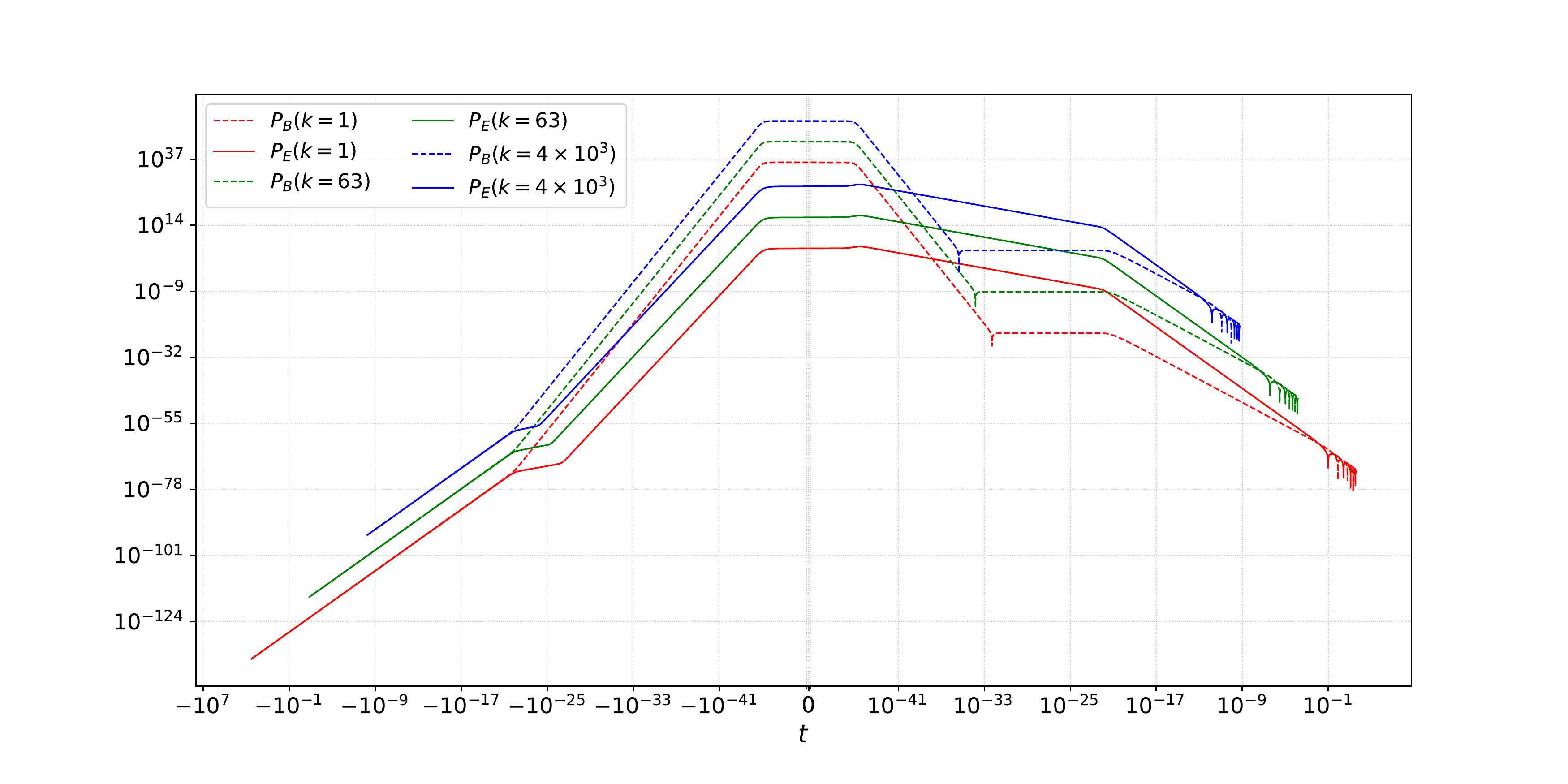}
	\caption[Evolution of the magnetic and electric power spectra with variation of the coupling strength.]{Evolution of the magnetic (dashed lines) and electric (continuous line) power spectra for $C=10^{19}$ and $x_b=10^{30}$ (top), and for $C=10^{23}$ and $x_b=10^{30}$ (bottom). We see that with larger $C$'s, the decrease of the electric contribution at late times happens later, and the total electromagnetic power spectrum is more important.}
	\label{spectra evolution cc}
\end{figure}

\begin{figure}[h!]
	\centering
	\includegraphics[scale=0.41]{../figures/MagDustPBPE_cc23_xb30.pdf}
	\includegraphics[scale=0.41]{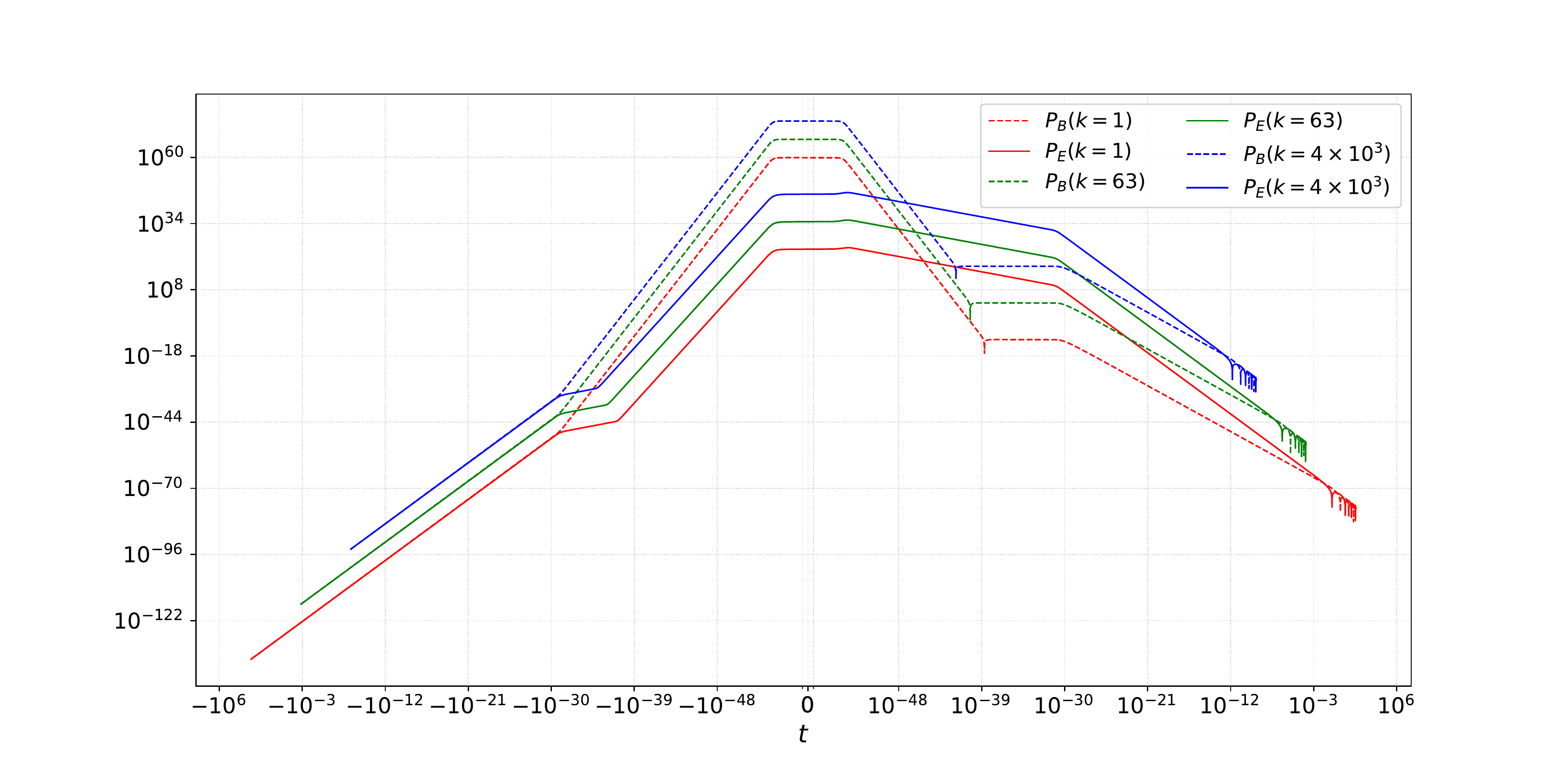}
	\caption[Evolution of the magnetic and electric power spectra with variation of the bounce scale factor.]{Same as Fig.~\ref{spectra evolution cc} for $C=10^{23}$ and $x_b=10^{30}$ (top), and for $C=10^{23}$ and $x_b=10^{36}$ (bottom). Higher values of $x_b$ imply an overall stronger total electromagnetic power spectrum, but with a stronger decrease rate at late times.}
	\label{spectra evolution xb}
\end{figure}

Now that the evolution of the modes has been described, we can use the shape of the spectra that follows from the results in Figs.~\ref{mode evolution cc} and \ref{mode evolution xb}, and Eqs~\eqref{magpow} and \eqref{elecpow}, the last one expressed in terms of the momentum, to fathom the time evolution of the magnetic and electric power spectra shown in Figs~\ref{spectra evolution cc} and \ref{spectra evolution xb}. At the beginning of the  evolution, modes are not excited. Only vacuum fluctuations are present, with the usual $k^4$ spectrum, increasing as $Y^{-4}$ due to contraction. When the coupling $f$ becomes relevant, the magnetic field power spectrum begins to increase faster, since $f$ is a growing function in the contracting phase, while the electric field power spectrum presents a slower increment, up to the time when $\Pi_k$ also begins to increase. After the bounce the situation is reversed, because $f$ is a decaying function of time in the expanding phase: the electric power spectrum decreases much slower than the magnetic one. Using Eq.~\eqref{historyP}, one can see that the decay is mild, going as $t^{-2/3}$, when $1/\alpha < t < C/\alpha$, opening a window in time where the electric spectrum has a significantly higher contribution than the magnetic one.

\begin{figure}[h!]
	\centering
	\includegraphics[scale=0.37]{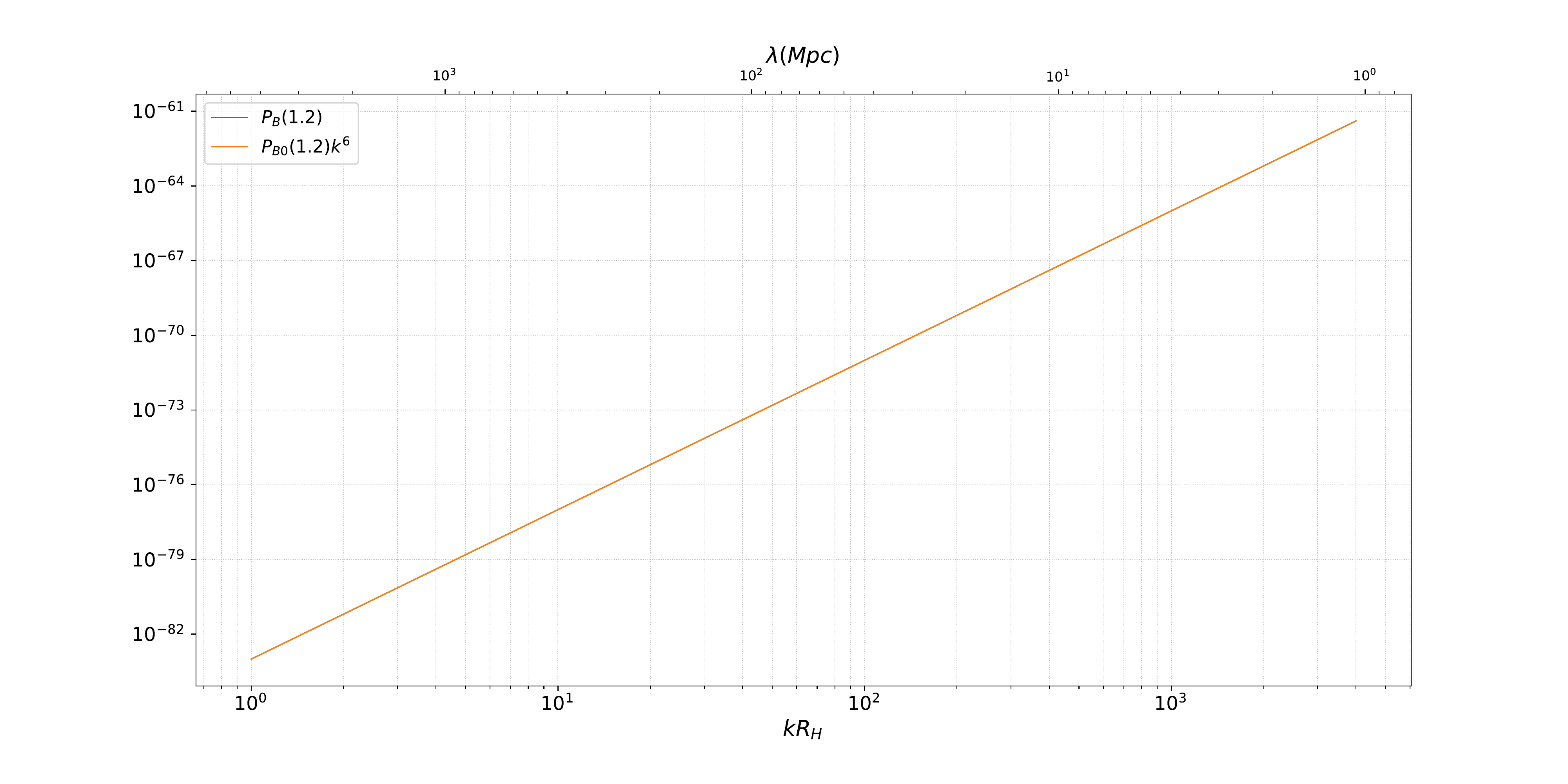}
	\includegraphics[scale=0.37]{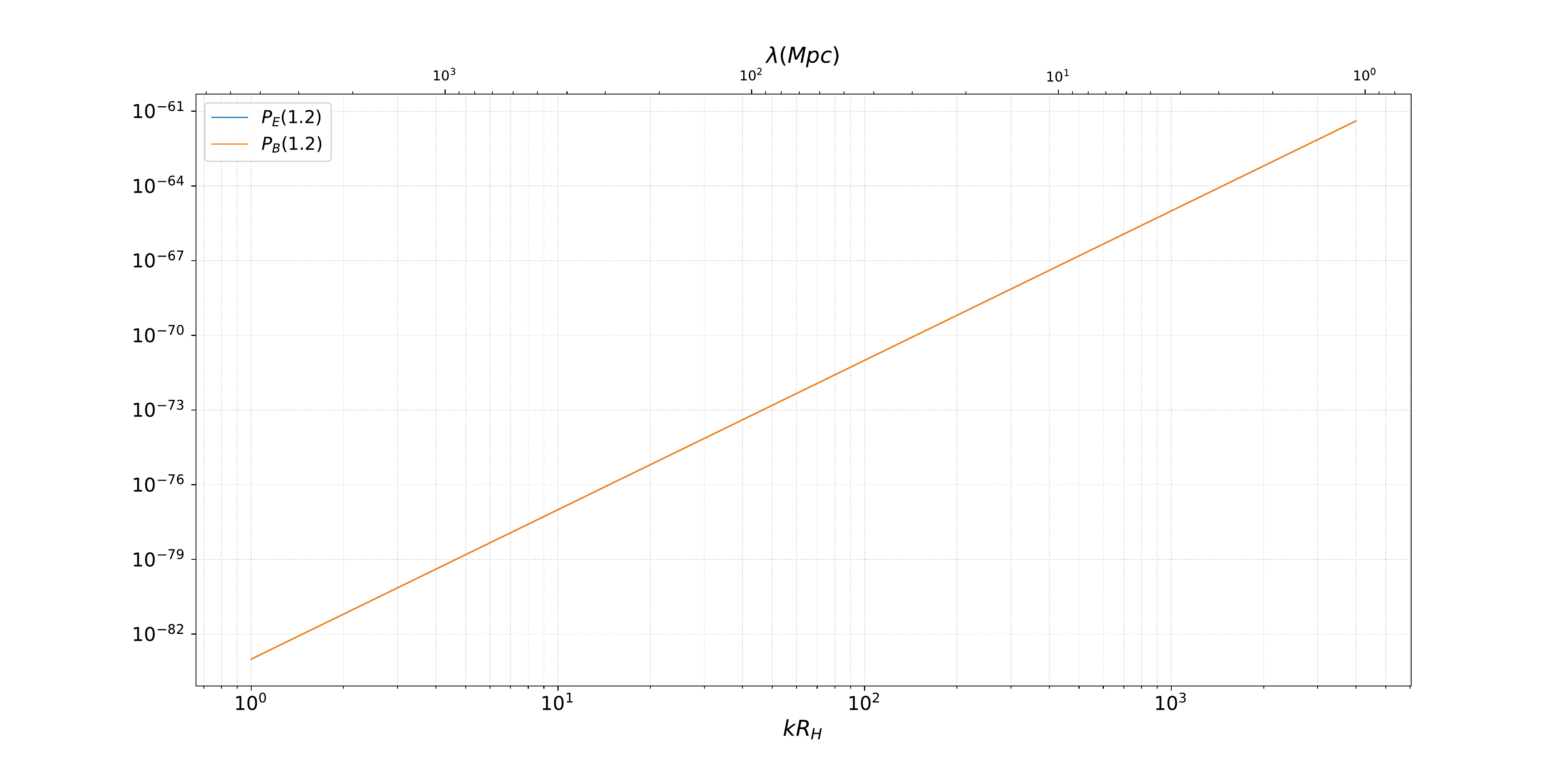}
	\caption[Magnetic power spectrum today]{Behaviour of the magnetic power spectrum today from \eqref{magpow} (blue) for $C=10^{23}$ and $x_b=10^{30}$. It is perfectly compatible with a power-law (top figure) with spectral index $n_B=6$ (orange). Note that $P_{B0}\equiv P_B(kR_H=1)$. We also show that the electric power spectrum behaves in the same fashion (bottom).}
	\label{powerspectrum}
\end{figure}

Another interesting aspect of the magnetic and electric power
spectra is their dependence in terms of $k$, shown 
in Fig.~\ref{powerspectrum}. As predicted in Sec.~\ref{Aresults}, we obtain the spectral index $n_B=6$. This
is typical of non-helicoidal and causally generated magnetic fields, as noted by
Caprini and Durrer~\cite{Caprini:2001nb, Durrer:2003ja}. %

From the power spectrum, we are able to get the amplitude of the magnetic field
\eqref{est} as a function of the scale, which is shown in Figs.~\ref{magfield}
and \ref{magfield2}.  Fig.~\ref{magfield} shows that a larger $x_b$, or
equivalently a lower scale factor at the bounce ($a_b$),  results in a lower
amplitude of the field. Thus a deeper bounce tends to generate weaker magnetic
fields. This is because electric and magnetic fields are generated when $f$
effectively changes in time, which happens for $-C/\alpha < t < C/\alpha$
(except for the short period of the bounce). Since $\alpha \propto
x_b^{3/2}$, a larger $x_b$ implies a shorter period in which the non-minimal
coupling is effective. For the same reason, a larger value of $C$ leads to a
larger amplitude of the magnetic field.

In the next section, we discuss how observations and theoretical limits can be used to constrain the parameters of our models.

\subsection{Discussion}
\label{discussion}

We now wish to confront the results of the previous section with observational
and theoretical limits found in the literature. Limits coming from several
physical processes can be invoked, as recalled in the introduction. However, it
is worth noting that many of them focus on specific models with considerable
uncertainties, or use specific priors leading to confusion on the possible upper
and lower bounds.\footnote{For instance, see
	\cite{Zucca:2016iur,Pogosian:2018vfr} for a discussion about the suppressed
	apparent limit  on the magnetic spectral index $n_B$, when assuming a different
	prior from Planck 2015 \cite{Ade:2015cva}} Since there is no unanimously accepted
limit on the spectral index, we will focus on the bounds derived considering
$n_B$ as a free parameter. Thus, we shall consider an upper bound around
$B_{\lambda}<10^{-9}G$,\footnote{See~\cite{Bray:2018ipq} and
	\cite{Safarzadeh:2019kyq} for recent limits using ultra-high-energy cosmic rays
	anisotropy and ultra-faint dwarf galaxies, respectively. See also
	\cite{Broderick:2018nqf} for a stronger upper limit of $B_{\lambda}<10^{-15}G$,
	putting detections of intergalactic magnetic fields with $\gamma$-ray under
	pressure.} and a first lower bound of around $B_{\lambda}>10^{-17}G$.\footnote{This limit comes from the non-detection of secondary GeV $\gamma$-rays
	around TeV blazars. However, there is still an ongoing debate on whether this
	lower limit should be trusted. See for example
	\cite{Broderick:2011av,Subramanian:2019jyd}.} The second lower limit we
consider concerns the minimum seed field in galaxies that would be amplified via
dynamo mechanism \cite{subramanian1994}, namely $B_{\lambda}>10^{-21}G$.

\begin{figure}[h!]
	\centering
	\includegraphics[scale=0.6]{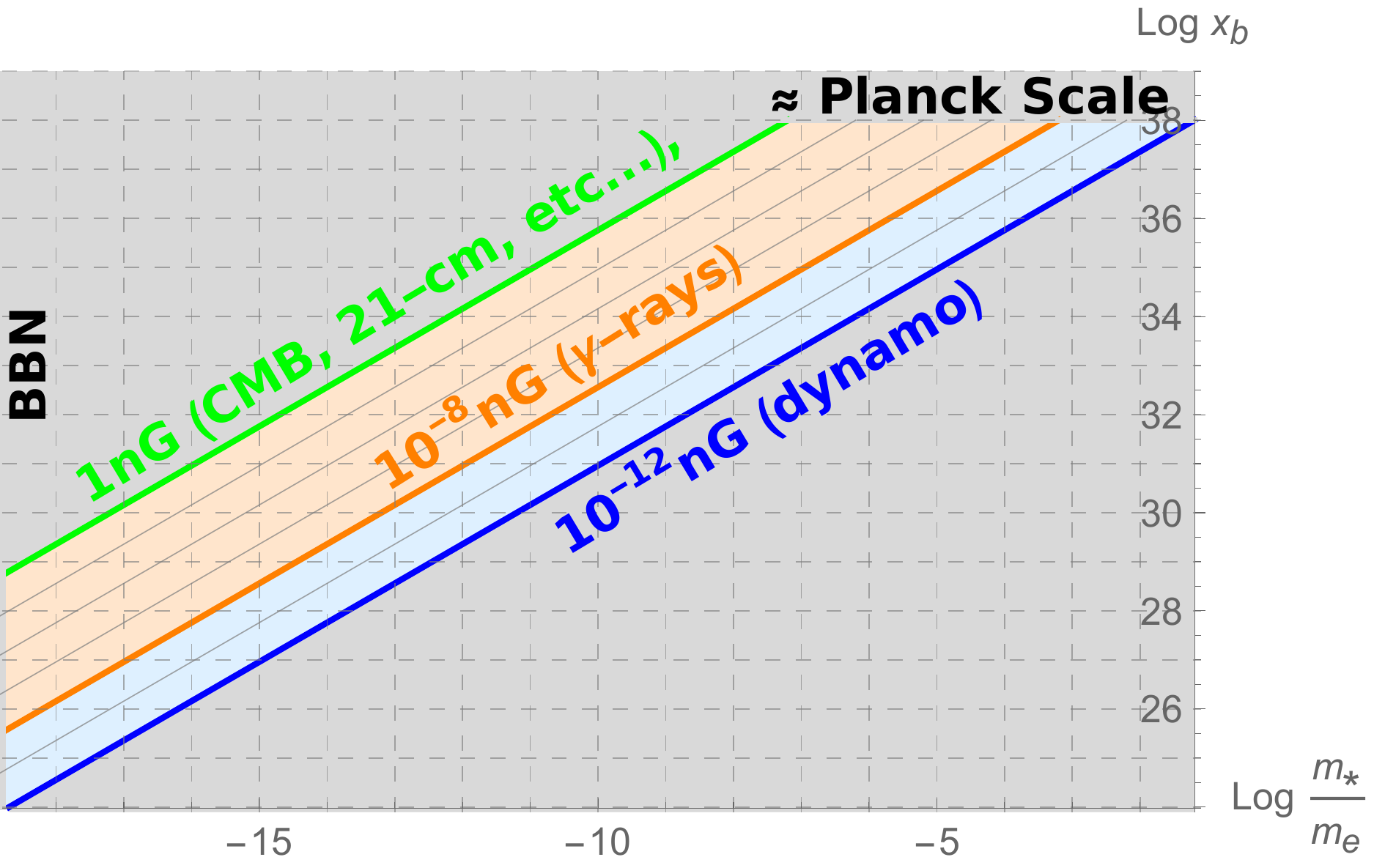}
	\caption[Parameter space with magnetic field amplitudes consistent with current limits at 1 Mpc.]{Parameter space with magnetic field amplitudes consistent with current limits at 1Mpc. The blue region represents the allowed values to initiate the dynamo effect, with the blue line a theoretical lower limit \cite{subramanian1994,Martin:2007ue}. The orange region represents allowed values by observations on large scales in voids, with the orange line a lower limit derived by blazars observations \cite{Taylor:2011bn} and the green line an upper limit derived using Ultra-High-Energy Cosmic Rays, Ultra-Faint Dwarf galaxies, 21-cm hydrogen lines, etc. \cite{Bray:2018ipq,Safarzadeh:2019kyq,Minoda:2018gxj}. Note the orange and blue regions are overlapping. The grey shaded region represents excluded values of the magnetic field. Each oblique grey line gives an amplitude for the magnetic field a hundred times higher than the lower line.}
	\label{parameter space 1Mpc}
\end{figure}

These theoretical and observational limits are used in figure~\ref{parameter space 1Mpc} to constrain the region in parameter space for which consistent values of magnetic seed fields are obtained at 1 Mpc. The upper value $x_b\lesssim10^{38}$ comes from Eq.~\eqref{limit xb} reflecting the earliest possible time for the bounce to occur. It is denoted ``Planck Scale'' in the graph. There is another limit set to preserve nucleosynthesis denoted ``BBN''. This can be derived by plugging Eq.~\eqref{limit c} into Eq.~\eqref{mass} presented below, giving $m_{\star}=10^{-19}m_e$.

In order to infer the allowed mass scales for the minimal coupling, one can use, for instance, the relation between $C$ and $m_{\star}$ coming from Eq.~\eqref{cc} to show
\begin{align}
\label{mass}
\frac{m_{\star}}{m_e} = \frac{\alpha}{10^{38} C},
\end{align}
where $m_e$ is the electron mass. The maximum mass allowed in this model is then $0.1 m_e$. Therefore, the value of the electron mass for $m_{\star}$ is not allowed by our model, a feature shared with power-law inflationary models
\cite{Turner:1987bw, Campanelli:2008qp}.

\subsection{Backreaction}
\label{backreaction}

The growth of magnetic and electric fields in primordial magnetogenesis models can become an issue. If the electromagnetic contribution is greatly enhanced to the point of being comparable with the background energy density, the background dynamics can be modified and anisotropies can appear \cite{Kanno:2009ei}.

We define the matter and radiation energy densities, respectively, as
\begin{align}
\rho_m \equiv \frac{\Omega_m}{Y^3}\;, \quad \rho_r \equiv \int \dd{\ln{k}}\; \left(\mathcal{P}_{E,0}+\mathcal{P}_{B,0}\right) \left(\frac{Y_0}{Y}\right)^{4} \;.
\end{align}
As pointed out in previous works on magnetogenesis in bouncing models, see
\textit{e.g.} \cite{Sriramkumar:2015yza}, the vanishing of the Hubble rate at
the bounce leads, via the Friedmann Eq.~\eqref{friedrho}, to $\rho_m=0$.
However, this is not the case here. The classical Friedmann equations are not
valid around the bounce, which is dominated by quantum cosmological effects, and
$\rho_m \propto Y^{-3}$ always. However, this does not guarantee that the model is free from backreaction. Let
us examine this point in more detail in this section.

As the electromagnetic power goes as $Y^{-4}$, and $\rho_m \propto Y^{-3}$, the first obvious critical point to investigate the issue of backreaction is at the bounce itself. As shown in the previous section, we have near the bounce that $\vert A_k \vert  \propto k^{-1/2}$ and $\vert \Pi_k\vert  \propto k^{3/2}$. Furthermore, $\vert A_k\vert $ does not depend on $x_b$ and $C$, and $\vert \Pi_k\vert  \propto C^2/\sqrt{x_b}$. This can be seen by inspecting the integral appearing in the first term of Eq.~\eqref{iterations2}, where after integration, and evaluating at the bounce, we get the constants $C^2 x_b/\alpha = C^2/x_b^{1/2}$. 

After integrating the magnetic and electric energy densities at the bounce, see Eqs.~\eqref{magnetic spectral density} and \eqref{electric spectral density}, and denoting the cut-off scale as $k_f$ (which we will refer to galactic scales, where this simple treatment may cease to be valid due to short range interactions leading to dissipation and other effects), we obtain
\begin{align}
\rho_{B,b} = \frac{3C^2 x_b^4}{32 \pi^2 R_{H_0}^4} k_f^4 \;, \quad \rho_{E,b} = \frac{C^2 x_b^3}{9 \pi^2 R_{H_0}^4} k_f^6 \;.
\end{align}
The ratio of magnetic energy density over electric energy density is then simply
\begin{align}
\frac{\rho_{B,b}}{\rho_{E,b}} \approx \frac{x_b}{k_f^2} \;,
\end{align}
and the magnetic field is dominant when $\rho_{B,b} \gg \rho_{E,b}$, or $\sqrt{x_b}\gg k_f$. As $x_b\gg 1$, this condition is always satisfied.

In units of Hubble radius, the matter energy density reads
\begin{align}
\rho_{m,b} = \frac{7.8 \Omega_m 10^{120}}{R_{H_0}^4 Y^3} \;.
\end{align}
At the bounce, the matter energy density is given by
\begin{align}
\rho_{m,b} = \frac{7.8 \Omega_m 10^{120} x_b^3}{R_{H_0}^4} \;.
\end{align}
Then, comparing the magnetic density to the matter density, and requiring the ratio be small enough gives
\begin{align}
\frac{\rho_{B,b}}{\rho_{m,b}} < 10^{-4}  \quad \implies C^2 x_b k_f^4 < 10^{118} \;. 
\label{magnetic backreaction}
\end{align}
Choosing the galactic scale (tens of kiloparsecs), $k\approx 10^5$, gives $C^2 x_b < 10^{98}$. The values given in Fig.~\ref{parameter space 1Mpc} all respect this constraint. In conclusion, there is no electromagnetic backreaction at the bounce.

As we have seen in Figs.~\ref{spectra evolution cc} and \ref{spectra evolution xb}, and discussed when commenting them, the electric density overcomes the magnetic density after the bounce for some time during the period $1/\alpha < t < C/\alpha$. The coupling behaves as
\begin{align}
f \propto t^{-2} \;, \quad \frac{1}{\alpha} < t < \frac{C}{\alpha} \;,
\end{align}
and the scale factor as $Y \propto t^{\frac{2}{3}}$ in this region. This can be also be seen in Fig.~\ref{ymf}. Then, the electric density goes as $\rho_E \propto t^{-2/3}$. This is to be compared to the matter density $\rho_m \propto t^{-2}$, giving the ratio evolution
\begin{align}
\frac{\rho_E}{\rho_m} \propto t^{\frac{4}{3}} \;.
\end{align}
To get an estimate of the electric backreaction, let us evaluate the initial conditions at the bounce and evolve this ratio in the considered time range. Performing a procedure similar to the one leading to \eqref{magnetic backreaction}, we obtain
\begin{align}
\frac{\rho_{E,b}}{\rho_{m,b}} = 10^{-122}\:C^2 k_f^6  . 
\end{align}
Then, the ratio will evolve as
\begin{align}
\frac{\rho_{E}}{\rho_{m}} = 10^{-122}\:C^2 k_f^6  \left(\frac{t_f}{t_i} \right)^{\frac{4}{3}} \;.
\end{align}
Choosing the initial time $t_i\equiv1/\alpha$ and the final time $t_f\equiv C/\alpha$ and imposing once again that the backreaction be small, we finally obtain
\begin{align}
\frac{\rho_{E}}{\rho_{m}} < 10^{-4} \quad \implies C^{\frac{10}{3}} k_f^6 < 10^{118} \;. 
\end{align}
Once again, $k_f \approx 10^{5}$ is compatible with the maximum value $C\approx10^{26.3}$ allowed in Fig.~\ref{parameter space 1Mpc}. Again, there is no backreaction problem in our model
\footnote{To discuss the (absence of) backreaction in our model, we have shown that the electromagnetic energy density is always smaller than the matter energy density. In other models of bounce,
	such as those based on the Lee-Wick theory \cite{Cai:2008qw}, there are mechanisms preventing \emph{ab initio}
	the uncontrolled growth of the electromagnetic energy density.}.

\subsection{Conclusions on Quantum Cosmology}

We have investigated in this chapter quantum effects described in the de Broglie-Bohm interpretation of quantum mechanics. We showed that a bounce is produced by such quantum effects. More precisely, the background is homogeneous and isotropic, filled with pressureless (dark) matter. Motivated by observations of cosmological magnetic fields, we presented the generation of primordial magnetic fields in the context of a cosmological bounce, through a coupling between curvature and electromagnetism, predicted by 
QED in curved spacetimes \cite{Drummond:1979pp}. Unlike inflationary magnetogenesis scenarios, bouncing magnetogenesis is free of the strong coupling problem, which makes this type of models very appealing. 

The model is characterised by three parameters, namely the presureless (dark) matter density today, $\Omega_m$, the scale factor at which the bounce happens, $x_b$, and the mass scale of the coupling $m_{\star}$. In particular, the first parameter is tightly constrained by observations. Using the Hamiltonian framework, we showed that an adiabatic vacuum can be defined as initial condition for the electromagnetic field in the far past of the contracting phase.This allowed us to explain analytically the behaviour of the electric and magnetic modes, summarised in Eqs. \eqref{historyA} and \eqref{historyP}. We confronted these analytical results with a numerical integration of the modes, given in Figs.~\ref{mode evolution cc} and \ref{mode evolution xb}, and presented in Figs.~\ref{spectra evolution cc}
and \ref{spectra evolution xb} the time evolution of the magnetic and electric power spectra. We illustrated the scale dependence of both spectra  in Fig~\ref{powerspectrum}, finding they behaved as a power-law with the same spectral index $n_E=n_B=6$. This result is reminiscent of
non-helicoidal, causally generated magnetic fields from phase transitions in the
early Universe for which the magnetic spectral index must be even and positive \cite{Durrer:2003ja}. In Figs.~\ref{magfield} and \ref{magfield2}, we showed the amplitude of the magnetic field today was found to be strong enough on a wide range of scales to pass the current limits from observations. Interestingly, we note that the same coupling in the context of power-law inflation does not generate large enough magnetic fields  \cite{Campanelli:2008qp}. At the scale of 1~Mpc, we derived constraints on $x_b$ and $m_{\star}$, summarised in Fig.~\ref{parameter space
	1Mpc}. Finally, we also demonstrated that backreaction is not a problem in our model, therefore the presence of electromagnetism can be safely assumed to leave the background evolution unchanged.

%% file: sections/chapter6.tex
\chapter{Affine Quantisation and Cosmology}
\label{affine quantisation and cosmology}

%In signal analysis, a scale change of time corresponds to an inverse scale change in the frequency domain. Therefore, a signal $s(t)$ with narrow support has its Fourier transform with wide support, and conversely. The spread about an initial time $t_0$ of a signal normalised to unity is
%\begin{align}
%	D_{t_0} = \int_{-\infty}^{\infty} \left(t-t_0\right)^2 |s(t)|^2 \textup{d}t \;,
%\end{align}
%bearing the consequence
%\begin{align}
%	D_{t_0} (s) D_{\omega_0} (\hat{s}) \geq \frac{1}{4} \;.
%\end{align}
%This is a Fourier ``uncertainty principle", holding for Gaussian signals.

In this chapter, we present the integral affine quantisation method, which allows the choice of scale at which the quantisation is performed and also brings powerful features to probe a physical system compared to other quantisation schemes. We show the affine quantum version of general relativity, and we study in detail the affine quantum Brans-Dicke Theory (BDT), tackling the problem of quantum equivalence between frames. We use the material covered in \cite{Gazeau:2009zz} to present this chapter.

\section{The Affine Quantisation}
\label{Aff. Quantisation}

\subsection{Motivation from Signal Analysis}

The affine quantisation is the direct quantum correspondence of the continuous wavelet transform used in signal processing. The strength of  wavelet analysis lies in the combination of complementary wavelets used to recover data of a damaged signal. More specifically, we can extract information from the unknown parts of the signal, that can be viewed as singularities in the signal, by using convolution of wavelets and the known parts, what can be seen as using different mathematical approaches to describe one physical system. Rigorously, wavelet analysis makes use of wavelet series, \textit{i.e.} a representation of square-integrable functions with respect to an overcomplete set of a vector space. The set of wavelets then defines a Hilbert basis. The difference between the wavelet transform compared to the Fourier transform is that wavelets are localised both in time and frequency, whereas the Fourier transform is only localised in frequency, and are therefore more appealing for signal analysis. 

The energy of a signal is conserved iff there exists a vector $\psi$ satisfying the condition
\begin{align}
c_{\psi} := 2\pi \int_{-\infty}^{\infty} |\hat{\psi}(k)|^2 \frac{\textup{d}k}{|k|} < \infty \;.
\label{admissibility condition}
\end{align}
The generating function $\psi$ is called a \textit{mother wavelet}. The admissibility condition \eqref{admissibility condition} implies the signal is zero in average. Indeed, there is a divergence in  \eqref{admissibility condition} for $k=0$, and we must impose 
\begin{align}
\hat{\psi}(0)=0 \Leftrightarrow \int_{-\infty}^{\infty} \psi (t) \textup{d}t = 0 \;.
\end{align}
Thus, a wavelet is necessarily an oscillating function, hence the name. Wavelets derived from the mother wavelet by rescaling the signal or shifting it in time are called \textit{child wavelets}, and are defined by
\begin{align}
\psi_{a,b} = \frac{1}{\sqrt{a}} \psi \left(\frac{t-b}{a}\right) \;,
\label{child wavelets}
\end{align}
where $a$ is positive and defines the scale and $b$ is real and defines the shift. 

We now explain how the admissibility condition is crucial in signal analysis. Let us note $s \in L^2(\mathbb{R})$ a function viewed as a signal. We first introduce the representation of $s$ in a timescale half-plane. This representation is called the continuous wavelet transform (CWT) of $s$ with respect to $\psi$, and is given by the scalar product of a child wavelet and the signal
\begin{align}
S(a,b) = \braket{\psi_{a,b}}{s} &= \int_{-\infty}^{\infty} \textup{d}t \, \frac{1}{\sqrt{a}} \overline{\psi \left(\frac{t-b}{a}\right)} s(t)  \;. \\
&= \sqrt{a} \int_{-\infty}^{\infty} \textup{d}k \,  \overline{\psi \left(ak\right)} \hat{s}(k) e^{ibk} \;.
\end{align}
The norm of a signal
\begin{align}
\norm{s}^2 = \frac{1}{c_{\psi}} \int_{-\infty}^{\infty} \textup{d}b \int_{0}^{\infty} \frac{\textup{d}a}{a^2} |S(a,b)|^2 \;,
\label{signal norm}
\end{align}
and therefore its energy, is preserved iff the admissibility condition \eqref{admissibility condition} is respected.

A common feature between signal analysis and integral quantisation methods is the resolution of the identity. Given a group $G$ and a unitary irreducible representation (UIR) of it, the quantisation map transforms a classical function (or distribution) into an operator using a bounded square-integrable operator $M$ and a measure $d\nu$, such as
\begin{align}
\label{identity resolution}
\int_{G} M(g) \,d\nu(g) = I \,,
\end{align}
where $g \in G$, $M(g) = U(g) M U^{-1}(g)$. This is the resolution of the identity for the operator $M$. With this, from a classical observable $f(g)$, we obtain the corresponding operator
\begin{align}
\label{int. quant. map}
A_{f} = \int_{G} M(g) \, f(g) \, d\nu(g) \,.
\end{align}
Similarly, the reconstruction of a signal $s$ is given by
\begin{align}
\ket{s} = \int_{G} M(g) \, \ket{s} \, d\nu(g) \,.
\end{align}

For the affine group  $G= \text{Aff}_{+}(\mathbb{R})$, representing the group of affine transformations over the half-plane used \textit{e.g.} in phase space with variables $(q,p)$, we have two non-equivalent UIR $U_{\pm}$, plus a trivial one $U_{0}$ \cite{Aslaksen,Isham:1984hq}. The fact that $U_{\pm}$ are both square-integrable is the reason why the resolution of the identity allows for a quantisation procedure based on the continuous wavelet analysis. Actually, energy conservation is directly related to the resolution of the identity. Indeed, the norm \eqref{signal norm} is equivalent to
\begin{align}
I=  \frac{1}{c_{\psi}} \int_{-\infty}^{\infty} \textup{d}b \int_{0}^{\infty} \frac{\textup{d}a}{a^2} \ket{\psi_{q,p}} \bra{\psi_{q,p}} \;.
\end{align} 
Therefore, the action of $U:=U_+ \oplus U_-$ allows the identification from 
\begin{align}
\psi_{q,p} (x) = (U(q,p) \,\psi)(x) \;. 
\end{align} 
To summarise, a mathematical model describing a physical system is scale-dependent, and probing the system on different scales using affine quantisation can allow to find ``hidden" physical features. We will illustrate this by showing how affine quantising the Brans-Dicke theory leads to the quantum equivalence of Jordan and Einstein frames.

\subsection{Mathematical Background}
\label{Mathematical backg.}

First, let us introduce the affine quantisation method for a generic phase space equipped with variables $(q,p)$. The half-plane $\Pi_{+}:=\{(q,p)\,|\, q>0\,,\,p\in \mathbb{R} \}$ with a multiplication operation defined by
\begin{align}
\label{multiplication} 
\left(q,p \right) \left(q_{0},p_{0} \right) := \left( qq_{0}, \frac{p_{0}}{q}+p \right) \,; \quad q \in \mathbb{R}_{+}^{*} \,,\quad p \in \mathbb{R} \, ,
\end{align}
is identified with the affine group Aff$_{+}(\mathbb{R})$ of the  real line. The group acts on $\mathbb{R}$ as
\begin{align}
\label{group action}
(q,p) \cdot x = \frac{x}{q} + p \quad,\quad  \forall \, x \in \mathbb{R} \,. 
\end{align}
On a physical level, one can interpret (\ref{group action}) as a contraction/dilation (depending on if $q>1$ or $q<1$) of space plus a translation.
We shall equip the half-plane with the measure $dq \, dp$, which is invariant under the left action of the affine group on itself \cite{Almeida:2018xvj}. 

Rigorously, the affine quantisation is a covariant integral method, that combines the properties of symmetry from the affine group with all the resources of integral calculus. This method makes use of \textit{coherent states} \cite{Gazeau:2009zz} to construct the quantisation map, whose definition is connected with the symmetry of the phase space, as we will see. For the remainder of this part, we choose $U = U_{+}$, which acts on the Hilbert space 
$L^{2}(R_{+}^{\ast},dx/x^{\alpha +1})$, with $\alpha \in \mathbb{N}$, as 
\begin{align}
\label{U+}
(U(q,p) \,\psi)(x)=\frac{e^{i px}}{\sqrt{q^{-\alpha}}}\psi \left( \frac{x}{q} \right)\,.
\end{align}
We choose the operator $M$ such as
\begin{align}
\label{fiducial vectors}
M = | \psi \rangle \langle \psi | \quad; \quad \psi \in L^{2} \left(R_{+}^{\ast},\frac{dx}{x^{\alpha +1}} \right) \cap L^{2} \left(R_{+}^{*},\frac{dx}{x^{\alpha +2}} \right) \,.
\end{align}
The normalised vectors $\psi$ are arbitrarily chosen providing the square-integrability condition (\ref{fiducial vectors}). They are the mother wavelets mentioned before, but are also known as \textit{fiducial vectors}. For simplicity, we will consider only real fiducial vectors and will choose $\alpha=-1$. The action (\ref{U+}) of the UIR of $U$ over fiducial vectors produces the quantum states
\begin{align}
\label{ACS definition}
|q,p\rangle := U(q,p)|\psi\rangle \,.
\end{align}
These states are called \textit{affine coherent states} (ACS) or \textit{wavelets}. It is easy to show that 
\begin{align}
\label{affine res. identity}
\int_{\Pi_{+}}|q,p\rangle\langle q,p|\,\frac{dq \, dp}{2\pi c_{-1}}=I\,,
\end{align}
where the constant $c_{-1}$ depends on the choice of $\psi$, and is defined as
\begin{align}
\label{c_gamma}
c_{\gamma} = c_{\gamma}(\psi) := \int_{0}^{\infty} |\psi(x)|^{2} \, \frac{dx}{x^{2+\gamma}} \,.
\end{align}
Hence, the quantisation maps (\ref{int. quant. map}) becomes
\begin{align}
\label{aff. quant. map} 
f(q,p)\ \mapsto\ A_{f}=\int_{\Pi_{+}}f(q,p) \,|q,p\rangle\langle
q,p| \, \frac{dq dp}{2\pi c_{-1}}\,.
\end{align}
With this, one can easily verify that the quantisation of the elementary functions position $q^{\beta}$ (for any $\beta$), momentum $p$ and kinetic energy\footnote{Up to some factor.} $p^{2}$ yields
\begin{align}
\label{quantum operators} 
A_{q^{\beta}} = \frac{c_{\beta-1}}{c_{-1}} \, \hat{Q}^{\beta}\quad;\quad A_{p}=-i\frac{\partial}{\partial x}= \hat{P} \quad ; \quad
A_{p^{2}}= \hat{P}^{2} + \frac{c^{(1)}_{-3}}{c_{-1}} \hat{Q}^{-2} \,,
\end{align}
with $\hat{Q}$ being the position operator defined by $\hat{Q}f(x)=xf(x)$ and the constant $c^{(1)}_{-3}$ is defined as
\begin{align}
\label{c^beta_gamma}
c^{(\beta)}_{\gamma} (\psi) := \int_{0}^{\infty} \Big|\psi^{(\beta)}(x) \Big|^{2} \, \frac{dx}{x^{2+\gamma}} \,.
\end{align}
In appendix \ref{example affine quantisation}, we show explicitly how to compute a quantum operator in the affine quantisation. Note that this quantisation procedure does not involve the Planck constant, and has not been put equal to unity. This is exactly the point of the procedure; the fiducial vector allows for a scaling of the system, thus we can quantise at any desired scale. In other words, when writing the commutation relation
\begin{align}
\left[A_q, A_p\right] = \frac{c_0}{c_{-1}} i I\;,
\end{align}
we can set the arbitrary prefactor equal to $\hbar$.

Notice that, in this affine quantisation method, the only dependence on the fiducial vector $\psi$ is in the constant coefficients of the quantum operators. Thus, the arbitrariness of $\psi$ does not play a fundamental role in the quantisation. This is an advantage to be explored. For example, we can adjust the fiducial vectors to regain the self-adjoint character of the operator $p^{2}$ \cite{Almeida:2018xvj} . Choosing $\psi$ such that $ 4c^{(1)}_{-3} \geq 3c_{-1}$, the kinetic operator becomes essentially self-adjoint \cite{Reed:1975uy}, which is a desired characteristic since an Hermitian operator must be self-adjoint. An Hermitian operator can be obtained by imposing boundary conditions. However, there is a continuous infinity of possible boundaries, thus the choice of a representation is arbitrary (this is the \textit{operator ordering} problem of the canonical quantisation). In the affine quantisation, the choice of a fiducial vector can naturally result in an essentially self-adjoint operator, which means there is only one possible extension of it and, therefore, no need to impose boundary conditions. We stress, however, that choosing fiducial vectors is not the same as choosing boundary conditions. Self-adjointness is a well-known problem in the canonical quantisation of this theory, and it has been studied extensively in \cite{Almeida:2017gvx}. However, with the affine quantisation we naturally recover the quantum symmetrisation of the classical product momentum position 
\begin{align}
\label{operator qp} 
qp \quad \mapsto \quad A_{qp} = \frac{c_0}{c_{-1}}\frac{\hat{Q} \hat{P} + \hat{P} \hat{Q}}{2} \,,
\end{align}
up to a constant that once again depends on the choice of the fiducial vector.

\subsection{Quantum Phase Space Portraits}
\label{semclassport}

The construction of the affine quantisation method presented in the previous section using coherent states allows us to define a ``de-quantisation" map, named quantum phase space portrait, in a very obvious way: by calculating the expectation value of an operator with respect to the coherent states. That is, given a quantum operator $A_{f}$, we obtain a classical function $\check{f}$ such that 
\begin{align}
\label{checkf}
\check{f} (q,p) = \langle q,p | \, A_{f} \, | q,p\rangle \,.
\end{align}
If the operator is obtained from a classical function $f$, as suggested in the notation, then $\check{f}$ is a quantum correction or lower symbol of the original $f$ \cite{Lieb}. It corresponds to the average value of $f(q,p)$ with respect to the probability density distribution  
\begin{align}
\label{prob. distr.} 
\rho_{\phi}(q,p)=\frac{1}{2\pi c_{-1}} | \langle q,p|\phi \rangle |^{2} \,,
\end{align}
with $|\phi \rangle = |q^{\prime},p^{\prime} \rangle $. We can also define the time evolution of the distribution (\ref{prob. distr.}) with respect to time through a Hamiltonian operator $\hat{H} = A_{H}$, using the time evolution operator $e^{-i \hat H t}$. Then,
\begin{align}
\label{time prob. distr.} 
\rho_{\phi}(q,p,t):=\frac{1}{2\pi c_{-1}}|\langle q,p|e^{-i \hat H t}|\phi\rangle|^{2} \,.
\end{align}
Thus, if you consider the operator $M=\rho$, the lower symbol of $A_f$ becomes
\begin{align}
\check{f}(z)=\int \textup{tr}\left(\rho(z) \rho(z')\right) f(z')\frac{\textup{d}^2 z^{'}}{\pi} \, ,
\end{align}
with $tr$ the trace. From the resolution of the identity \eqref{identity resolution}, one finds $\textup{tr}\left(\rho(z) \rho(z')\right)$ is a probability distribution of the phase space, and $\check{f}$ is indeed an average measurement of the classical $f$.

From equation (\ref{checkf}), using the quantisation map (\ref{aff. quant. map}), the quantum correction $\check{f}$ of a classical function $f$ is then
\begin{eqnarray}
\nonumber
\check{f}(q,p) = \frac{1}{2\pi c_{-1}} \int_{-\infty}^{\infty} \int_{0}^{\infty} \frac{dq^{\prime} \, dp^{\prime}}{qq^{\prime}} \int_{0}^{\infty} \int_{0}^{\infty} dx \, dx^{\prime} f(q^{\prime}, p^{\prime}) \left[ e^{ip(x^{\prime} - x)}  \right.
\\
\label{checkf formulae}
\left. \times \, e^{-ip^{\prime}(x^{\prime} - x)} \psi \left(\frac{x}{q}\right) \psi \left(\frac{x^{\prime}}{q}\right) \psi \left(\frac{x}{q^{\prime}}\right) \psi \left(\frac{x^{\prime}}{q^{\prime}} \right) \right] \,.
\end{eqnarray}
Thus, it is not necessary to find the operator $A_{f}$ of a classical function $f$ to obtain its lower symbol. One can use the above formula (\ref{checkf formulae}) to do so. For example, the quantum correction of the classical functions $q^{\beta}$, $p$ and $p^{2}$ are given by
\begin{align}
\label{semi-class. correc.} 
\check{q^{\beta}} = \frac{c_{\beta-1}c_{-\beta-2}}{c_{-1}} \, q^{\beta} \quad; \quad  \check{p}= p \quad; \quad \check{p^2}= p^2 + \left(c_{-2}^{(1)} + \frac{c_{0} c^{(1)}_{-3}}{c_{-1}}\right) \frac{1}{q^2}\,,
\end{align}
with the constants $c_{\gamma}$ and $c^{(\beta)}_{\gamma}$ defined in (\ref{c_gamma}) and (\ref{c^beta_gamma}), respectively. Notice that the corrections also depend on the choice of specific fiducial vectors to determine these constants. We present a list of all quantum operators and their semi-classical counterparts used in this chapter in appendix \ref{compendium}.

\section{Affine Quantum General Relativity and Brans-Dicke}
\label{Introduction}

Now that we have introduced the affine quantisation method and the quantum phase space portrait coming from it, we can apply the method to GR, since the scale factor is positively defined. However, the Schutz variable associated to the fluid has the whole real line as its domain and therefore we cannot apply the affine method in it. Nevertheless, we can use another integral quantisation method based on the Weyl-Heisenberg group, which acts on the real line \cite{Gazeau2}. Here we could also use the canonical quantisation for this variable, since it works just fine for parameters in the whole line, a domain that does not have any singularity and, therefore, no problems of self-adjointness.\footnote{Using the Weyl-Heisenberg method can give us the advantage of introducing another constant that depends on the fiducial vector chosen in the quantisation. This can be an asset used to adjust energy levels, for example.} In both cases, we have
\begin{align}
\Pi_{T}  \quad \mapsto \quad \pi_{T} = -i \, \frac{\partial}{\partial T} \quad; \quad \Pi_{T} \quad \mapsto \quad \check{\pi}_{T} = E \,.
\end{align}

Let us name the fiducial vector associated to the scale factor $\psi_{a}$. Then, the coherent states are given by
\begin{eqnarray}
\label{coherent state a}
|a, p_{a} \rangle = U_{a} \, |\psi_{a} \rangle  \quad &\Rightarrow& \quad \langle x \, |a, p_{a} \rangle = \frac{e^{ip_{a} x}}{\sqrt{a}} \, \psi_{a} \left( \frac{x}{a} \right)
\end{eqnarray}

Applyting the quantisation rules \eqref{quantum operators} on the Hamiltonian \eqref{hamiltonian gr}, we find the constraint
\begin{align}
-\frac{c_{-1}(a)}{24 c_{0}(a)}\frac{1}{a} \left(\pi_a^2+\frac{c_{-3}^{(1)}(a)}{c_{-1}(a)a^2}\right) - 6k \frac{c_{0}(a)}{c_{-1}(a)} a + \frac{c_{-1}(a)}{c_{3w-1}(a) a^{3w}} \pi_T \simeq 0 \;.
\end{align}
Without detailing further (we will explain with great care the BDT), we see that there is a new term compared to the canonical quantisation of the form $1/a^2$. This term is purely quantum, and regularises the theory. As shown in \cite{Bergeron:2013ika}, a smooth big bounce is expected as the semi-classical level thanks to this quantum effective potential.

Now, let us use the same procedure on the Brans-Dicke theory, prototype of modified theories of gravity. The following sections present the results we obtained in \cite{Frion:2018oij}. The relevance of such models has been pointed out in section \ref{hint modified gravity}, where we discussed that the inflationary model which best fits Planck data is the Starobinsky model, a $R^2$-modification to the Einstein-Hilbert action. Modified gravity theories are also present in late-time cosmology, as they can account for the observed accelerated expansion. One of the oldest modifications of GR is the Brans-Dicke theory, proposed in the early 1960s by Carl H.~Brans and Robert H.~Dicke \cite{Brans:1961sx}, in which there is a non-minimal time-dependent coupling of the long-range scalar field with geometry, that is, with gravity. The BDT also introduces a dimensionless constant $\omega$ such that, for a constant gravitational coupling, GR is recovered at the limit  $\omega \rightarrow \infty$ if the trace of the energy-momentum tensor is not null \cite{Faraoni:1998yq,Faraoni:1999yp,Chauvineau:2003hv}. Today it is well known that, classically, the BDT is practically indistinguishable from GR, with the constant $\omega$ estimated to be over $40,000$ \cite{Avilez:2013dxa,Will:2014kxa}. Interestingly, the Brans-Dicke scalar field arises naturally in superstring cosmology, associated with the dilaton, which couples directly with the matter field \cite{Lidsey:1999mc}. In spite of the fact that the BDT is classically equivalent to GR, the quantum treatment can reveal new dynamics for the primordial Universe. There are also claims that the BDT can not reproduce GR for a scale-invariant matter content. In fact, in this case, it has been shown that $\omega$ can display various effects depending on its value, such as a symmetry breaking resulting in a binary phase structure. However, for a strong coupling $\omega \rightarrow \infty$, the BDT reproduces GR only in the quantised version \cite{Pal:2016hxt}.

We choose to explore the quantisation of the BDT with the affine quantisation instead of the canonical one because the domain of the variables involved (scale factor and scalar field) is the real half-line. Indeed, the scalar field represents the gravitational force, and can only be positive in the BDT.

\section{The Brans-Dicke Theory with a Perfect Fluid}
\label{BDT}

The Brans-Dicke theory is characterised by the introduction of a scalar field non-minimally coupled to gravity, and it is described by the gravitational Lagrangian
\begin{align}
\label{Lagrangian}
\mathcal{L}_{G} = \sqrt{-g}\left\{ \varphi R-\omega \frac{\varphi _{;\rho }\varphi^{;\rho }}{\varphi }\right\}\, .
\end{align}
The Brans-Dicke coupling parameter $\omega$ is chosen to be a constant. We show in appendix \ref{gravitational lagrangian BDT} that in a homogeneous and isotropic universe, the Lagrangian (\ref{Lagrangian}) becomes
\begin{align}
\label{Lagrangian 2}
\mathcal{L}_{G}=\frac{1}{N}\left\{ 6\left[ \varphi a\dot{a}^{2}+a^{2}\dot{a}%
\dot{\varphi}\right] -\omega a^{3}\frac{\dot{\varphi}^{2}}{\varphi }\right\} \,,
\end{align}
where we have already discarded the surface terms. The Lagrangian of the system is completed with a matter component, which we will consider to be a radiative perfect fluid, defined by the equation of state $p=\rho/3$.

The classical Hamiltonian constraint $\mathcal{H} \approx 0$ still holds for the BDT with a perfect fluid. Using the procedure described in section \ref{minisuperspace approximation}, we have
\begin{align}
\label{Hamiltnian constraint}
\frac{\omega}{12 \varphi} \Pi_{a}^{2} + \frac{1}{2a} \Pi_{a} \Pi_{\varphi} - \frac{\varphi}{2a^{2}} \Pi_{\varphi}^{2}  = (3+ 2\omega) \Pi_{T} \,.
\end{align}

The model requires the scale factor and the scalar field to be positive. Indeed, $\varphi>0$ implies an attractive gravitational force. Thus, the phase space is a four-dimensional space which is the Cartesian product of two half-planes,\footnote{In the case of radiative matter, at least \cite{Almeida:2017gvx}.} 
\begin{align}
\Pi^{2}_{+}:=\left\{(a,p_{a}) \times (\varphi, p_{\varphi}) \,|\, a>0, \varphi >0 \,,\, p_{a}, p_{\varphi} \in \mathbb{R} \right\} \,.
\end{align}
Since it is a Cartesian product, we can analyse each half-plane separately. We now present the affine quantisation of the Brans-Dicke theory.

\section{The Affine Quantisation of the BDT}
\label{BDT quantisation}

\subsection{Quantisation in the Jordan Frame}

With this, the quantisation of equation (\ref{Hamiltnian constraint}) results in the following Wheeler-DeWitt equation:
\begin{eqnarray}
\nonumber
\left\{- \omega \lambda_{1} \frac{1}{\varphi} \partial_{a}^{2} + \left(\omega \lambda_{2} - \lambda_{3} \right) \frac{1}{\varphi a^{2}} - \lambda_{4} \frac{1}{a} \partial_{a} \partial_{\varphi} + \lambda_{5} \frac{\varphi}{a^{2}} \partial_{\varphi}^{2} + \quad \quad \quad \quad \right.
\\
\label{WDW equation}
\left.  + \lambda_{6} \frac{1}{a^{2}} \partial_{\varphi} \right\} \Psi(a, \varphi, T) = -i \left(3 + 2\omega\right) \partial_{T} \Psi (a, \varphi, T) \,,
\end{eqnarray}
where $\Psi (a,\varphi,T)$ is the wave function. The constants $\lambda_{i}$ are given by
\begin{eqnarray}
\nonumber
\lambda_{1} = \frac{1}{12 c_{-1}(\varphi)} \quad&;& \quad \lambda_{2} = \frac{1}{12 c_{-1}(\varphi)} \frac{c_{-3}^{(1)}(a)}{c_{-1}(a)} \,;
\\
\label{lambdas}
\lambda_{3} = \frac{1}{2} \frac{c_{-3}(a)}{c_{-1}(a)} \frac{c_{-2}^{(1)}(\varphi)}{c_{-1}(\varphi)} \quad&;& \quad \lambda_{4} = \frac{1}{2c_{-1}(a)}\,;
\\
\nonumber
\lambda_{5} = \frac{1}{2} \frac{c_{-3}(a)}{c_{-1}(a)} \frac{c_{0}(\varphi)}{c_{-1}(\varphi)} \quad &;& \quad \lambda_{6} = \frac{1}{2} \frac{c_{-3}(a)}{c_{-1}(a)} \frac{c_{0}(\varphi)}{c_{-1}(\varphi)} + \frac{1}{4c_{-1}(a)} \,,
\end{eqnarray}
and we defined
\begin{align}
%\label{c_gamma a and varphi}
c_{\gamma}^{(j)}(a) = \int_{0}^{\infty} [\psi_{a}^{(j)}(x)]^{2} \frac{dx}{x^{2+ \gamma}} \quad; \quad c_{\gamma}^{(j)}(\varphi) = \int_{0}^{\infty} [\psi_{\varphi}^{(j)}(x)]^{2} \frac{dx}{x^{2+ \gamma}} \,.
\end{align}

If we choose $\psi_{a} = \psi_{\varphi}$, then $c_{\gamma}^{(j)}(a) = c_{\gamma}^{(j)}(\varphi) = c_{\gamma}^{(j)}$. So, let us choose a fiducial vector such that
\begin{align}
\psi_{a} = \psi_{\varphi} = \frac{9}{\sqrt{6}} \, x^{\frac{3}{2}} \, e^{-\frac{3x}{2}} \,.
\end{align}
With these vectors, we have $c_{-2} = c_{-1} = 1$, and $c_{-3}^{(1)} = 3/4$, which, as mentioned before, is a necessary condition for the quantised kinetic energy to be an essentially self-adjoint operator \cite{Reed:1975uy}. In turn, this gives the us the Wheeler-DeWitt equation in the Jordan frame
\begin{eqnarray}
\label{WDW equation Jordan}
\left\{- \frac{\omega}{12} \frac{1}{\varphi} \partial_{a}^{2} + \left( \frac{\omega}{16} - \frac{3}{4} \right) \frac{1}{\varphi a^{2}} - \frac{1}{2a} \partial_{a} \partial_{\varphi} + \frac{\varphi}{a^{2}} \partial_{\varphi}^{2} + \frac{5}{4a^{2}} \partial_{\varphi} \right\} \Psi = -i \left(3 + 2\omega\right) \partial_{T} \Psi \,.
\end{eqnarray}
From this equation, absorbing the constant $12\left(3 + 2\omega\right) \omega^{-1}$ into the temporal parameter, that is, accounting it as energy, we find the Hamiltonian for the BDT in the Jordan frame to be
\begin{align}
\label{Hamiltonian Jordan}
H_{J} = \frac{1}{\varphi} \partial_{a}^{2} - \frac{12}{\omega} \left( \frac{\omega}{16} - \frac{3}{4} \right) \frac{1}{\varphi a^{2}} + \frac{6}{\omega a} \partial_{a} \partial_{\varphi} - \frac{12}{\omega} \frac{\varphi}{a^{2}} \partial_{\varphi}^{2} - \frac{15}{\omega a^{2}} \partial_{\varphi} \,.
\end{align}
The Hamiltonian (\ref{Hamiltonian Jordan}) is essentially self-adjoint for the usual measure $da \,d\varphi$ on the Hilbert space, as expected. One can notice that equation (\ref{WDW equation Jordan}) is not separable. We can work around this problem by considering the Einstein frame instead.

\subsection{Conformal Transformation of Affine Operators}
\label{Conf. transf.}

The Jordan and Einstein frames are related to each other by a conformal transformation given by $g_{\mu \nu} = \phi^{-1} \, \tilde{g}_{\mu \nu}$, where $g_{\mu \nu}$ and $\tilde{g}_{\mu \nu}$ represent the metric tensors in each frame, respectively. Thus, before analysing the equivalence between these frames, let us first comment on how affine operators change with a conformal transformation. 

As opposed to what happens in the canonical quantisation (see \cite{Almeida:2017xhp}), the affine operators are uniquely defined by equation (\ref{aff. quant. map}). Also, if $A_{f}$ is the operator obtained from a classical function $f(q,p)$, with $q$ being a positive-defined variable and $p$ its associated momentum, then for a general conformal scaling factor  $\Omega(q)$ on the domain, we have
\begin{align}
\label{fAg neq Afg}
\Omega^{2}(q) A_{f} \neq A_{\Omega^{2}(q) f} \,.
\end{align}
Therefore, we need to be careful when we quantise models related by conformal transformations. Even if the constraint obtained from an Hamiltonian is classical, we cannot cancel overall coefficients (for instance, the factor $1/b$ in equation \ref{Hamilt. const. Einstein frame}).To illustrate this, let us give an example. Consider $\Omega^{2}(q) = q$ and $f(q,p) = p$. The operator $A_{\Omega^{2} f}$ is given by (\ref{operator qp}), and then
\begin{align}
A_{\Omega^{2} f} = A_{qp} = \frac{c_{0}}{c_{-1}} \, \frac{\hat{Q}\hat{P} + \hat{P}\hat{Q}}{2} \neq \hat{Q}\hat{P} = qA_{p} = \Omega^{2}(q) A_{f}\,.
\end{align}

This means that classically, it is always possible to cancel non-null coefficients,
%that is,
%\begin{align}
%\label{property multiplication by 1}
%h(q,p)f(q,p) = h(q,p)g(q,p) \quad \Leftrightarrow \quad f(q,p) = g(q,p) \,,
%\end{align}
%if $h(q,p) \neq0$.
however, quantising the constraint in different frames can result in very different scenarios, because of (\ref{fAg neq Afg}). 
%If $h(q,p)f(q,p) = h(q,p)g(q,p)$ holds, then $A_{hf} = A_{hg}$ and
%\begin{align}
% h(q,p) A_{f} \neq A_{hf} = A_{hg} \neq h(q,p) A_{g} \,.
%\end{align}
%Therefore $A_{f}= A_{g}$ does not hold, \textcolor{red}{in general}. 
In conclusion, we cannot cancel out non-null functions before quantising to compare the quantisation of two different frames connected by a transformation of coordinates.

\subsection{Quantisation in the Einstein Frame}
\label{Einstein frame}

Since the seminal paper of Brans and Dicke \cite{Brans:1961sx}, we know that two formulations of the theory (and in fact, for every scalar-tensor theory) are possible. These formulations, related by a conformal transformation, are the target of a long debate on which of these frames is physically relevant. Some authors claim they are equivalent classically but should be different at the quantum level \cite{Artymowski:2013qua,Banerjee:2016lco}, while others claim that both are equivalent at classical and quantum level \cite{Kamenshchik:2014waa,Pandey:2016unk,Almeida:2017gvx,Ohta:2017trn}. Some also claim that the equivalence is broken by off-shell one-loop quantum corrections, but recovered on-shell\cite{Ruf:2017xon}. Since theoretical predictions depend entirely on the conformal frame we are working on, a natural question arising is if there is a preferred frame or not, and which one is the most suitable to observations. In the Jordan frame, we found the differential equation governing the wave function evolution (\ref{WDW equation Jordan}), however as a crossed term appeared in the partial derivatives, finding a solution can be  difficult. Let us now analyse the problem in the Einstein frame instead.

The Brans-Dicke Lagrangian, with a non-minimally coupled scalar field, is given by (\ref{Lagrangian}), and by using the conformal transformation, $g_{\mu \nu} = \varphi^{-1} \, \tilde{g}_{\mu \nu}$, where $g_{\mu \nu}$ is the metric in the non-minimal coupling frame, the Lagrangian reads as
\begin{align}
\label{GR Lagrangian}
\mathcal{L}_{G} = \sqrt{-\tilde{g}} \left[\tilde{R} - \biggr(\omega + \frac{3}{2}\biggl) \frac{\varphi_{;\rho} \, \varphi^{;\rho}}{\varphi^{2}}\right] \,,
\end{align}
which is the Lagrangian for General Relativity with a minimally coupled scalar field. The Lagrangian (\ref{Lagrangian}) is written in the Jordan frame, and (\ref{GR Lagrangian}) is written in the Einstein frame. The conformal transformation is given by the change of coordinates
\begin{align}
\label{conformal trans.}
N^{\prime} = \varphi^{\frac{1}{2}} N \quad; \quad b = \varphi^{\frac{1}{2}} a \quad ; \quad \varphi^{\prime} = \varphi \,,
\end{align}
and, applying these to (\ref{Lagrangian 2}), we obtain
\begin{align}
\mathcal{L}_{G} = \frac{1}{N^{\prime}} \left[ 6b \dot{b}^{2} - \left( \omega + \frac{3}{2}\right)b^{3} \left(\frac{\dot{\varphi^{\prime}}}{\varphi^{\prime}} \right)^{2} \right] \,.
\end{align}
%We choose to drop the prime on $N$ and $\varphi$ for simplicity. 
The total Hamiltonian is thus
\begin{align}
\label{Hamiltonian Einstein}
H_{T} = N^{\prime} \left( \frac{p_{b}^{2}}{24b} - \frac{\varphi^{\prime \, 2}}{2(3+ 2\omega)b^{3}} p_{\varphi^{\prime}}^{2} - \frac{p_{T}}{b} \right) \,,
\end{align}
and the constraint $H_{T} =0$ gives us\footnote{We keep the 1/b factor in order to avoid inconsistences in the quantisation (see the discussion in Subsection \ref{Conf. transf.}).}
\begin{align}
\label{Hamilt. const. Einstein frame}
\frac{p_{b}^{2}}{24b} - \frac{\varphi^{\prime \, 2}}{2(3+ 2\omega)b^{3}} p_{\varphi^{\prime}}^{2} = \frac{p_{T}}{b} \,.
\end{align}

In order to quantise equation \eqref{Hamilt. const.  Einstein frame}, it is necessary to know the Hilbert space in the Einstein frame. From the change of variables \ref{conformal trans.}, the measure becomes
\begin{align}
da \, d\varphi = \varphi^{\prime \,-\frac{1}{2}} db \, d\varphi^{\prime} \,.
\end{align}
Thus, the Hilbert space for the coordinates $(b,\varphi^{\prime})$ is $L^{2} (\mathbb{R}_{+}^{*} \times \mathbb{R}_{+}^{*},\varphi^{\prime \,-\frac{1}{2}} dbd\varphi^{\prime})$. Then, according to definition (\ref{fiducial vectors}), the fiducial vectors $\psi_{\varphi^{\prime}}$ are defined on another Hilbert space:
\begin{align}
\psi_{\varphi^{\prime}} \in L^{2} \left(R_{+}^{\ast},\frac{dx}{x^{\frac{1}{2}}} \right) \cap L^{2} \left(R_{+}^{*},\frac{dx}{x^{\frac{3}{2}}} \right) \,.
\end{align}
With this measure, the operator associated with the kinetic energy, is given by
\begin{align}
A_{p^{2}} = -\partial_{\varphi^{\prime}}^{2} + \frac{1}{2\varphi^{\prime}} \partial_{\varphi^{\prime}} + \left( \frac{c_{-5/2}^{(1)}(\varphi^{\prime})}{c_{-1/2}(\varphi^{\prime})} - \frac{3}{8} \right) \frac{1}{\varphi^{\prime \,2}} \,,
\end{align}
which is already self-adjoint. 

Now, for the coordinate $b$, using (\ref{aff. quant. map}), we obtain
\begin{eqnarray}
A_{b^{-1}p_{b}^{2}} &=& - \frac{1}{c_{-1}(b)} \frac{1}{b} \partial_{b}^{2} + \frac{1}{c_{-1}(b)} \frac{1}{b^{2}} \partial_{b} - \left(\frac{1-c_{-4}^{(1)}(b)}{c_{-1}(b)} \right) \frac{1}{b^{3}} \,.
\end{eqnarray}
For the coordinate $\varphi^{\prime}$, we get
\begin{eqnarray}
A_{\varphi^{\prime \, 2} p_{\varphi^{\prime}}^{2}}  &=& -\frac{11}{8}\frac{c_{3/2}}{c_{-1/2}} +\frac{c_{-1/2}^{(1)}}{c_{-1/2}}-\frac{3}{2}\frac{c_{3/2}}{c_{-1/2}} \varphi^{\prime} \partial_{\varphi^{\prime}} -\frac{c_{3/2}}{c_{-1/2}}\varphi^{\prime 2} \partial_{\varphi^{\prime}}^2 \,. 
\end{eqnarray}
Then, the quantisation of equation (\ref{Hamilt. const. Einstein frame}) results in
\begin{eqnarray}
\label{WDW equation Einstein gen.}
\left\{- \varpi \partial_{b}^{2} + \frac{\varpi}{b} \partial_{b} + \left( \tilde{\lambda}_{1} \varpi  + \tilde{\lambda}_{2} \right) \frac{1}{b^{2}} + \frac{\tilde{\lambda}_{3}}{b^{2}} \left(\varphi^{\prime \, 2} \partial_{\varphi^{\prime}}^{2} 
+ \frac{3}{2} \varphi^{\prime} \, \partial_{\varphi^{\prime}} \right) \right\}\Psi = -24\varpi i \,\partial_{T}\Psi \,, 
\end{eqnarray}
with $\varpi = \omega + \frac{3}{2}$, and $\tilde{\lambda}_{i}$ are given by
\begin{eqnarray}
\nonumber
\tilde{\lambda}_{1} &=& c_{-4}^{(1)}(b) -1  \,;
\\
\tilde{\lambda}_{2} &=& \frac{3}{4} \frac{ c_{-4}(b)}{ c_{-1/2}(\varphi^{\prime})} \left( \frac{11}{8}c_{3/2}(\varphi^{\prime}) - c_{-1/2}^{(1)}(\varphi^{\prime}) \right) \,;
\\
\nonumber
\tilde{\lambda}_{3} &=& \frac{c_{-4}(b) \, c_{3/2}(\varphi^{\prime})}{c_{-1/2}(\varphi^{\prime})} \,.
\end{eqnarray}

On the other hand, one can change variables as in (\ref{conformal trans.}) directly on (\ref{WDW equation}). This yields
\begin{eqnarray}
\nonumber
\label{WDW equation Einstein}
\left[- \left(\omega \lambda_{1} + \frac{\lambda_{4}}{2} - \frac{\lambda_{5}}{4} \right) \partial_{b}^2 +\frac{\lambda_{5}-\lambda_{4}}{4}  \frac{1}{b} \partial_{b} + \left( \omega \lambda_{2} - \lambda_{3} \right) \frac{1}{b^2} \,+  \right.
\\
\left.  +\, \left(\frac{\lambda_{5}}{2}-\lambda_{4} \right) \frac{\varphi^{\prime}}{b}\partial_{b} \partial_{\varphi^{\prime}} + \lambda_{5} \frac{\varphi^{\prime 2}}{b^2} \partial^{2}_{\varphi^{\prime}} + \lambda_{6} \frac{\varphi^{\prime}}{b^2} \partial_{\varphi^{\prime}}  \right] \Psi = -i(3+2\omega)\partial_{T} \Psi \,,
\end{eqnarray}
with $\lambda_{i}$ given in (\ref{lambdas}). Notice that the coefficients $\lambda_{i}$ are in terms of $c_{\lambda}^{(i)}(a)$ and $c_{\lambda}^{(i)}(\varphi)$, while the coefficients in equation (\ref{WDW equation Einstein gen.}) are in terms of $c_{\lambda}^{(i)}(b)$ and $c_{\lambda}^{(i)}(\varphi^{\prime})$. Considering the freedom in the choice of the fiducial vectors,\footnote{The quantisation is not determined by this choice, although there is an inequality constraint ($4c_{-3}^{(1)} \geq 3c_{-1}$) in order to obtain a Hermintian operator (see discussion at the end of section \ref{Mathematical backg.}).} and comparing equations (\ref{WDW equation Einstein gen.}) and (\ref{WDW equation Einstein}), we conclude that there is equivalence between Einstein and Jordan frames only if
\begin{align}
\frac{\lambda_{5}}{2}-\lambda_{4} = 0 \quad \Rightarrow \quad c_{-3}(a) = 2 \frac{c_{-1}(\varphi)}{c_{0}(\varphi)} \,.
\end{align}
In a way, this result is similar to the one found in \cite{Almeida:2017gvx}, where it is concluded that the equivalence depends on the choice of \textit{ordering factors} for the canonical quantisation, which are related to the coefficients of the Hamiltonian operator. In our case, the unitary equivalence is then obtained if we impose some constraints on the fiducial vectors:
\begin{align}
4\,c_{-3}^{(1)} \geq 3\,c_{-1} \, \quad \text{for} \quad \psi_{a}, \psi_{b}, \psi_{\varphi} \,; \quad \text{and} \quad c_{-3}(a) = \frac{2c_{-1}(\varphi)}{c_{0}(\varphi)} \,.
\end{align}

Let us solve, without loss of generality, equation (\ref{WDW equation Einstein gen.}). We suppose the following separation of variables: $\Psi (b,\varphi ,t) := X(b)\,Y (\varphi)\, P(T)$. We obtain, for the function of time
\begin{align}
P(T)=A \exp \left[{i \frac{ET}{24}} \right] \,,
\end{align}
where $E/24$ is the energy constant. This results in the following system of partial differential equations:
\begin{eqnarray}
\label{Bessel equation}
\left\{-\partial_{b}^{2} + \frac{1}{b} \partial_{b} + \frac{1}{\varpi} \left[ \tilde{\lambda_{1}} \varpi + \tilde{\lambda_{2}} - \tilde{\lambda_{3}} k^{2} \right] \frac{1}{b^{2}}\right\} X(b) = E \, X(b) \,; \nonumber 
\\
\label{Euler equation}
\left\{\varphi^{2} \partial_{\varphi}^{2} + \frac{3}{2} \varphi \, \partial_{\varphi} \right\} Y(\varphi) = -k^{2} \, Y(\varphi) \,, 
\end{eqnarray}
with $k^{2}$ being a separation constant. The general solutions are given by
\begin{eqnarray}
\label{X_b}
X(b) &=& C_{1}\, b\, J_{\nu} \left(\sqrt{E}\,b\right)	+ C_{2} \, b\, Y_{\nu} \left(\sqrt{E} \, b\right) \,,
\\
\label{Y_varphi}
Y(\varphi) &=& D_{1} \, \varphi^{-\frac{1}{4}(\sqrt{1-16k^2}+1)} + D_{2} \, \varphi^{\frac{1}{4}(\sqrt{1-16k^2}-1)} \,,
\end{eqnarray} 
with $J_{\nu}$ and $Y_{\nu}$ the Bessel functions of first and second kind, respectively, $C_{1,2}$, $D_{1,2}$ are integration constants,
\begin{align}
\label{Bessel index}
\nu = \sqrt{\frac{\left(1+\tilde{\lambda_{1}}\right) \varpi + \left(\tilde{\lambda_{2}} - \tilde{\lambda_{3}} k^{2}\right)}{\varpi}} \,,
\end{align}
and $k^{2}<1/16$.
The wave-function of the Universe $\Psi_{T}(b, \varphi) = X(b)Y(\varphi)$ must be square-integrable. This is the reason for the choice of the limit set for the separation constant. Equation (\ref{Euler equation}) is known as Euler equation and the solution (\ref{Y_varphi}) corresponds to said limit of $k^{2}$. The solution for $k^{2} = 1/16$ gives similar results, however $k^{2}>1/16$ results in a non square-integrable wave-function. This is also the reason why we choose a negative sign for the separation constant. Also, since $Y_{n}$ blows up at the origin, we must take $C_{2}=0$. Now, let us consider the following transformation for the variable $\varphi$:
\begin{align}
\sigma =  \ln \varphi \quad \Rightarrow \quad d\sigma = \frac{1}{\varphi} d\varphi \,.
\end{align}
With this, the solution (\ref{Y_varphi}) becomes\footnote{With this, it becomes more evident why it is only square-integrable for $k^{2} < 1/16$.}
\begin{align}
Y(\sigma) = D_{1} \, e^{-\frac{\sigma}{4} \left(\sqrt{1-16k^2}+1\right)} + D_{2} \, e^{\frac{\sigma}{4} \left(\sqrt{1-16k^2}-1\right)} \,.
\end{align}
For the sake of simplicity, let us consider $D_{2} =0$. We construct the wave packet as 
\begin{align}
\Psi = N \int_{-\frac{1}{4}}^{\frac{1}{4}} dk b\, J_{\nu} \left(\sqrt{E}\,b\right) e^{-\frac{\sigma}{4} \left(\sqrt{1-16k^2}+1\right)} \, e^{i\frac{E}{24}T} \,,
\end{align}
where $N$ is a normalisation constant. Therefore, the norm of the wave packets is 
\begin{eqnarray}
\nonumber
\langle \Psi | \Psi \rangle = N^{2} \int\limits_{0}^{b_{0}} \int\limits_{0}^{\infty} \varphi^{-\frac{1}{2}}\, db \, d\varphi   \, \int\limits_{-\frac{1}{4}}^{\frac{1}{4}} \int\limits_{-\frac{1}{4}}^{\frac{1}{4}}  dk \,dk^{\prime} \, b^{2}\, e^{-\frac{\sigma}{2}}  \times
\\
\times\, J_{\nu} \left(\sqrt{E}\,b\right) J_{\nu^{\prime}} \left(\sqrt{E}\,b\right) e^{i \left(\frac{1}{4}\sqrt{|1-16k^{\prime \,2}}|- \frac{1}{4}\sqrt{|1-16k^2|} \right) \sigma} \,,
\end{eqnarray}
or, writing only in terms of $\sigma$,
\begin{eqnarray}
\nonumber
\langle \Psi | \Psi \rangle = N^{2} \int\limits_{0}^{b_{0}} \int\limits_{-\infty}^{\infty} db \,d\sigma \, \int\limits_{-\frac{1}{4}}^{\frac{1}{4}} \int\limits_{-\frac{1}{4}}^{\frac{1}{4}}  dk \,dk^{\prime} \, b^{2} J_{\nu} \left(\sqrt{E}\,b\right)  \times
\\
\times \, J_{\nu^{\prime}} \left(\sqrt{E}\,b\right) e^{i \left(\frac{1}{4}\sqrt{|1-16k^{\prime \,2}}|- \frac{1}{4}\sqrt{|1-16k^2|} \right) \sigma} \,,
\end{eqnarray}
where the prime on the $\nu$ indicates $\nu(k^{\prime})$ and we can take $b_{0}=1$ as the value of the scale factor today. Performing the integrals over $\sigma$ and $k^{\prime}$ gives
\begin{align}
\label{norm}
\langle \Psi | \Psi \rangle = 8\pi N^{2} \int_{0}^{b_{0}} \int_{-\frac{1}{4}}^{\frac{1}{4}} db dk \, b^{2} \, J_{\nu} \left(\sqrt{E}\,b\right) J_{\nu} \left(\sqrt{E}\,b\right) \,.
\end{align}
Now, we shall consider an approximation for the limit $\omega \gg k^2$. This approximation is relevant due to our understanding of today's estimate of the Brans-Dicke constant $\omega$. Notice that, in this limit, the Bessel index (\ref{Bessel index}) becomes $\nu = \sqrt{1 + \tilde{\lambda}_{1}}$ and then (\ref{norm}) becomes
\begin{align}
\langle \Psi | \Psi \rangle = 8\pi N^{2} \int_{0}^{b_{0}} db \, b^{2} J_{\nu = \sqrt{1 + \tilde{\lambda}_{1}}} \left(\sqrt{E}\,b\right) J_{\nu = \sqrt{1 + \tilde{\lambda}_{1}}} \left(\sqrt{E}\,b\right) \,.
\end{align}
The solution is given in terms of the regularised generalised hypergeometric function $ _2\tilde{F}_3$ \cite{NIST:DLMF} as
\begin{align}
\nonumber
\label{wave packet norm}
\langle \Psi | \Psi \rangle =& 4\sqrt{\pi } N^2 b_0^3 \,\, \Gamma \left(\nu+\frac{1}{2}\right) \Gamma \left(\nu+\frac{3}{2}\right) \left(b_0 \sqrt{E}\right)^{2 \nu} \times 
\nonumber 
\\ 
& _2\tilde{F}_3\left(\nu + \frac{1}{2},\nu +\frac{3}{2}; \nu+1, \nu +\frac{5}{2},2 \nu+1;-b_0^2 E\right) \,.
\end{align}
The regularised generalised hypergeometric functions are defined as the power series
\begin{align}
_p\tilde{F}_q\left(a_{1}, ..., a_{p}; b_{1},..., b_{q}; z\right) := \frac{1}{\Gamma{(b_1)}...\Gamma{(b_q)}} \sum_{n=0}^{\infty} \frac{(a_1)_n ...(a_p)_n}{(b_1)_n ...(b_q)_n} \frac{z^n}{n!}\, ,
\end{align}
with the recurrence relations 
\begin{align}
(a_{j})_{0} = 1 \,; \quad \text{and} \quad (a_{j})_{n} = a_{j} \left(a_{j}+1\right) \left(a_{j}+2\right) ... \left(a_{j}+n-1\right) \,, \quad \text{for} \quad n\geq 1 \,.
\end{align}
The norm of the wave packet becomes
\begin{align}
\label{wave packet norm 2}
\langle \Psi | \Psi \rangle =& A \left(b_0 \sqrt{E}\right)^{2 \nu} \sum_{n=0}^{\infty} \frac{\left(\nu +\frac{1}{2}\right)_n }{\left(\nu +1\right)_n \left(2\nu +1\right)_n} \frac{(-b_0^2 E)^n}{n!} \, , 
\end{align}
where
\begin{align}
A = 4\sqrt{\pi} N^{2} \frac{b_0^3}{\left(\nu +1\right)} \frac{\Gamma \left(\nu+\frac{1}{2}\right)}{\Gamma \left(\nu +2\right)\Gamma \left(2\nu+1\right)} \,.
\end{align}
Then, equation \eqref{wave packet norm 2} suggests that the energy spectrum is discrete. This means we can write $\langle \Psi | \Psi \rangle = \sum_{n} \langle \Psi_{n} | \Psi_{n} \rangle$, and the energy levels satisfy the equations
\begin{align}
\langle \Psi_{0} | \Psi_{0} \rangle = A  \left(b_0 \sqrt{E}\right)^{2 \nu}  \,,
\end{align}
and, for a general $n \geq 1$,
\begin{align}
\langle \Psi_{n} | \Psi_{n} \rangle = A  \left(b_0 \sqrt{E}\right)^{2 \nu} \sum_{n=0}^{\infty} \frac{\left(\nu+\frac{1}{2}\right)_n }{\left(\nu +1\right)_n \left(2\nu+1\right)_n} \frac{(-b_0^2 E)^n}{n!}  \,.
\end{align}

\subsection{Quantum Phase Space Portrait of the BDT}
\label{Quantum phase space BDT}

Let us consider the formalism introduced in section \ref{semclassport}. The constraint (\ref{Hamiltnian constraint}), $\mathcal{H}_{T} = 0$, can be rewritten in its semi-classical version using (\ref{checkf formulae}) to calculate each term. For the sake of simplicity, we will keep the same letter for the energy constant, so  $\check{p_{T}} = E$, and hence
\begin{eqnarray}
\label{energy constratint semiclas.}
\frac{\omega}{12} \, \frac{1}{\varphi} p_{a}^{2} + \left( \omega \kappa_{1} - \kappa_{2} \right) \frac{1}{a^{2} \varphi} + \frac{1}{2a} p_{a} p_{\varphi} - \kappa_{3} \frac{\varphi}{a^{2}} p_{\varphi}^{2} = (3+2\omega) E \,,
\end{eqnarray}
with the constants $\kappa_{i}$ being
\begin{eqnarray}
\nonumber
\kappa_{1} &=& \frac{1}{12} \left(\frac{c_{0}(a)c_{-3}^{(1)}(a)}{c_{-1}(a)} + c_{-2}^{(1)}(a) \right) \,;
\\
\nonumber
\kappa_{2} &=& \frac{1}{2} \, \frac{c_{0}(a) c_{-3}(a)}{c_{-1}(a)} \left(\frac{c_{0}(\varphi)c_{-3}^{(1)}(\varphi)}{c_{-1}(\varphi)} + c_{-2}^{(1)} (\varphi) \right) \,;
\\
\nonumber
\kappa_{3} &=& \frac{1}{2} \, \frac{c_{0}(a) c_{-3}(a)}{c_{-1}(a)} \frac{c_{0}(\varphi) c_{-3}(\varphi)}{c_{-1}(\varphi)} \,,
\end{eqnarray}
where $c_{\gamma}^{(j)}(a)$ and $c_{\gamma}^{(j)}(\varphi)$ are 
\begin{align}
\label{c_gamma a and varphi}
c_{\gamma}^{(j)}(a) = \int_{0}^{\infty} [\psi_{a}^{(j)}(x)]^{2} \frac{dx}{x^{2+ \gamma}} \quad; \quad c_{\gamma}^{(j)}(\varphi) = \int_{0}^{\infty} [\psi_{\varphi}^{(j)}(x)]^{2} \frac{dx}{x^{2+ \gamma}} \,.
\end{align}
If we choose $\psi_{a} = \psi_{\varphi}$, then $c_{\gamma}^{(j)}(a) = c_{\gamma}^{(j)}(\varphi) = c_{\gamma}^{(j)}$. With this in mind, let us choose a fiducial vector such that
\begin{align}
\psi_{a} = \psi_{\varphi} = \frac{9}{\sqrt{6}} \, x^{\frac{3}{2}} \, e^{-\frac{3x}{2}} \,.
\end{align}
With these vectors, we have $c_{-2} = c_{-1} = 1$, and $c_{-3}^{(1)} = 3/4$, the latter being a necessary condition for the quantised Hamiltonian to be an essentially self-adjoint operator \cite{Reed:1975uy}. We want this condition to hold even if we are not doing the quantisation explicitly, since the semi-classical trajectories are probabilistic along the path that a quantum state evolves. Then, (\ref{energy constratint semiclas.}) becomes
\begin{eqnarray}
\label{semiclassical constraint}
\frac{\omega}{12} \, \frac{1}{\varphi} p_{a}^{2} + \frac{9}{8} \left( \omega - 2 \right) \frac{1}{a^{2} \varphi} + \frac{1}{2a} p_{a} p_{\varphi} - 2 \frac{\varphi}{a^{2}} p_{\varphi}^{2} = (3+2\omega) E \,.
\end{eqnarray}
The expression (\ref{semiclassical constraint}) allows us to analyse the expected behaviour of the scale factor $a$ for the early univese, for a given initial value of the scalar field $\varphi (t_{0}) = \varphi_{0}$ and its momentum at this instant $p_{\varphi} (t_{0}) = p_{\varphi \, 0}$.  

Notice that equation (\ref{Hamiltnian constraint}) is the classical Hamiltonian constraint in the Jordan frame. To compare the expected behaviour of the scale factor in the Jordan frame with that in the Einstein frame, let us calculate the quantum phase space portrait of equation (\ref{Hamilt. const. Einstein frame}), the Hamiltonian constraint in the Einstein frame. We have
\begin{eqnarray}
\left(b^{-1} p_{b}^{2} \right)\check{} &=& \frac{p_{b}^2}{b}  +\frac{c^{(1)}_{-1}(b)+c_1(b) \, c^{(1)}_{-4}(b)-c_1(b)}{c_{-1}(b)} \frac{1}{b^3} \,;
\\
\left(\varphi^{\prime 2} p_{\varphi^{\prime}}^{2}\right)\check{} &=& \frac{c_{3/2}(\varphi^{\prime}) \, c_{-7/2}(\varphi^{\prime})} {c_{-1/2}(\varphi^{\prime})}\,  \varphi^{\prime \, 2} \, p_{\varphi^{\prime}}^{2}(\varphi^{\prime}) + \left( \frac{c_{3/2}(\varphi^{\prime}) \, c_{-7/2}^{(1)} (\varphi^{\prime})} {c_{-1/2}(\varphi^{\prime})} + \right.
\nonumber 
\\
& &\left. + \frac{c_{-3/2}(\varphi^{\prime}) \, c_{-1/2}^{(1)}(\varphi^{\prime})}{c_{-1/2}(\varphi^{\prime})} - \frac{11}{8}\frac{ c_{3/2}(\varphi^{\prime}) c_{-3/2}(\varphi^{\prime})} {c_{-1/2}(\varphi^{\prime})} \right) \,.
\end{eqnarray}
Then, the quantum correction of (\ref{Hamilt. const. Einstein frame}) becomes
\begin{eqnarray}
\frac{3+2\omega}{24} \left[ p_{b}^2+\kappa_{4} \frac{1}{b^2} \right]  - \frac{1}{b^2}\left[\kappa_{5}\varphi^{\prime 2} p_{\varphi^{\prime}}^2  + \kappa_{6} \right] = (3+2\omega)E^{'},
\end{eqnarray}
with $E^{'}$ the energy, and the constants
\begin{eqnarray}
\nonumber
\kappa_{4} &=& \frac{c^{(1)}_{-1}(b)+c_1(b) \, c^{(1)}_{-4}(b) - c_1(b)}{c_{-1}(b)} \,; 
\\
\nonumber
\kappa_{5} &=& \frac{1}{2}\frac{c_{-4}(b) \, c_{1}(b)}{c_{-1}(b)} \, \frac{c_{3/2}(\varphi^{\prime}) \, c_{-7/2}(\varphi^{\prime})} {c_{-1/2}(\varphi^{\prime})} \,;
\\
\nonumber
\kappa_{6} &=& \frac{1}{2} \frac{c_{-4}(b) \,c_{1}(b)}{c_{-1}(b)} \, \frac{c_{3/2}(\varphi^{\prime})} {c_{-1/2}(\varphi^{\prime})} \, \left( c_{-7/2}^{(1)} (\varphi^{\prime}) + \frac{c_{-3/2}(\varphi^{\prime}) \, c_{-1/2}^{(1)}(\varphi^{\prime})}{c_{3/2}(\varphi^{\prime})}  - \frac{11}{8} c_{-3/2}(\varphi^{\prime}) \right) \,.
\end{eqnarray}
By choosing the fiducial vectors as before, we find
\begin{eqnarray}
\label{semiclassical constraint einstein}
\frac{3+2\omega}{24} p_{b}^2 + \left[\frac{1296-1500\sqrt{3\pi}+864\omega}{64}\right]\frac{1}{b^2}-\frac{525\sqrt{3\pi}}{16b^2}\varphi^{\prime 2} p_{\varphi^{\prime}}^2  = (3+2\omega)E^{'}.
\end{eqnarray}
Equations (\ref{semiclassical constraint}) and (\ref{semiclassical constraint einstein}) are the quantum corrections of the classical Brans-Dicke Theory described in the Jordan and Einstein frames, respectively. To understand the consequences of these corrections, let us build the quantum phase space of the BDT in both these frames.

\section{Phase Space Portraits}
\label{Results}

As mentioned before, in this section we present the quantum phase space portraits coming from equations \eqref{semiclassical constraint} and \eqref{semiclassical constraint einstein}. The aim is to understand the behaviour of the scale factor $a$, which is connected to the volume of the Universe, so the phase spaces shown here are with reference to this variable. Notice, however, that there are still other free parameters: the scalar field $\varphi$, the energy $E$ and the Brans-Dicke constant $\omega$. These parameters will be varied for the sake of understanding their influence on the issue. Without loss of generality, let us consider the initial state of the scalar field to be $\varphi_{0} =1$. 

\subsection{Jordan Frame}

\begin{figure*}[htp!]
	\centering
	\subfloat{\includegraphics[width=0.5\textwidth]{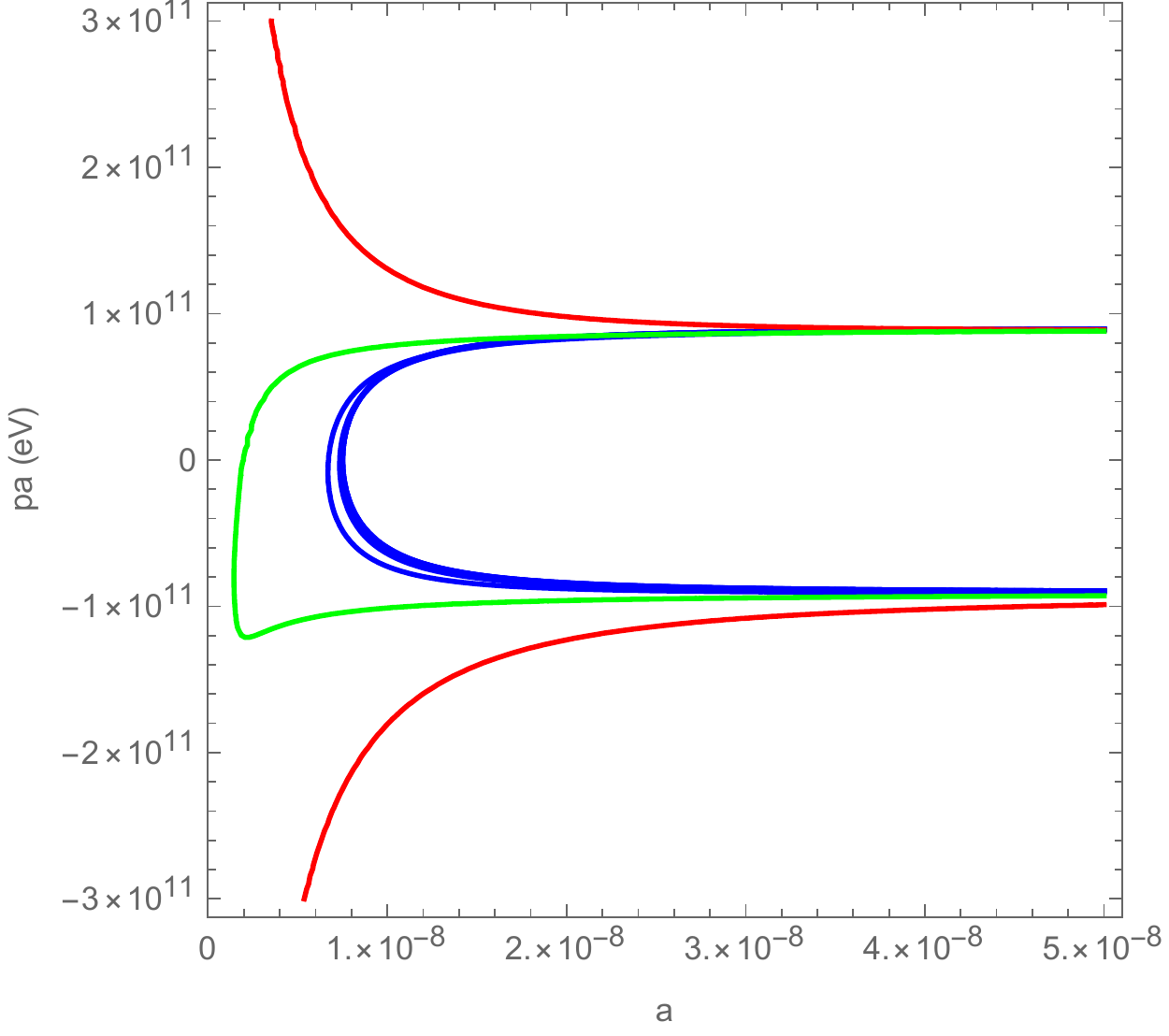}}
	\subfloat{\includegraphics[width=0.5\textwidth]{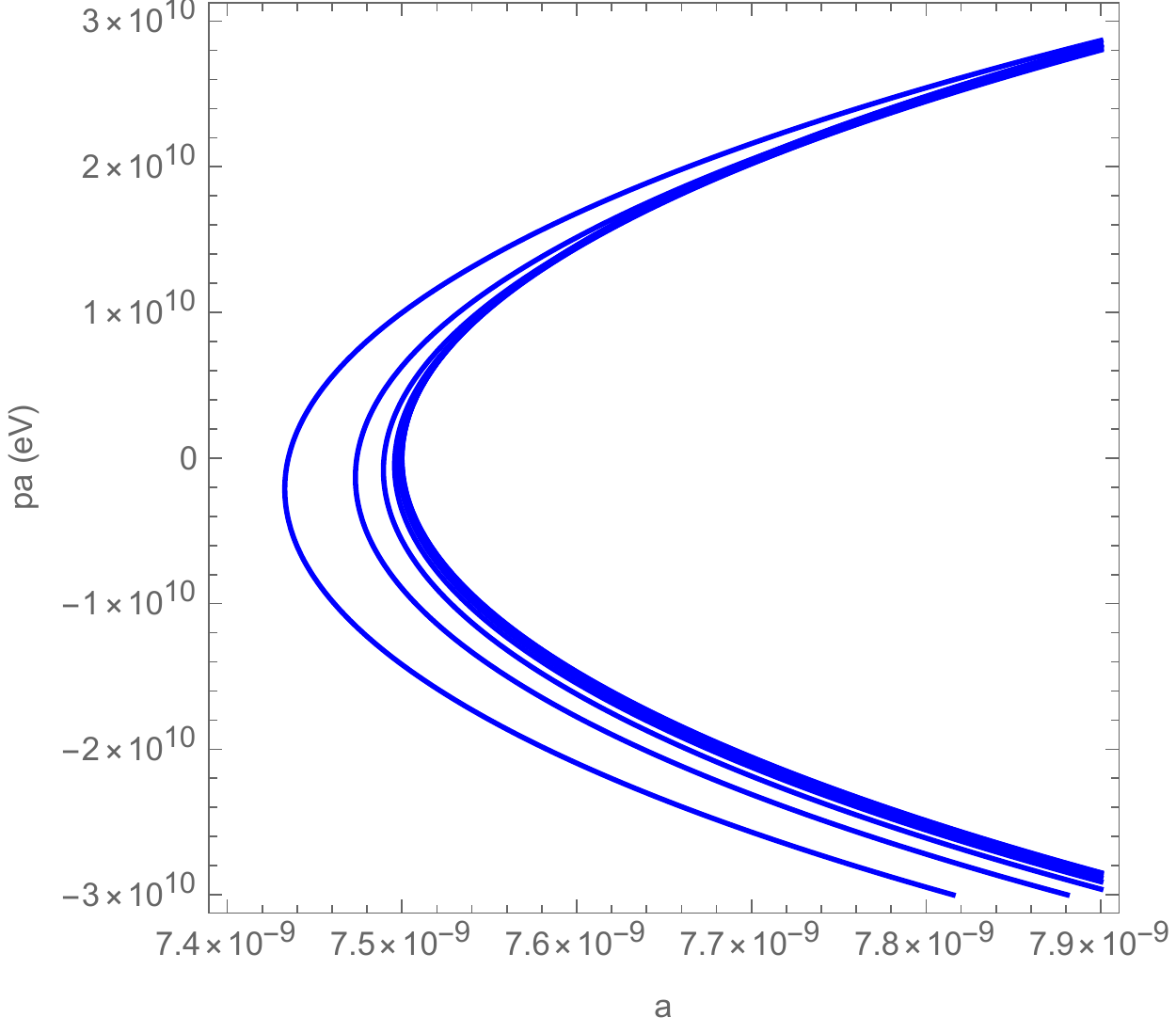}}\hfill
	\caption[Quantum phase space in the Jordan frame with $\omega=410,000$ and $E_{0}=10^{16}$]{Quantum phase space in the Jordan frame, using $\omega=410,000$ and $E_{0}=10^{16}$. The left figure is for a range $1  \leq p_{\varphi} \leq 10^3$, while for the right figure the  range is smaller $1 \leq p_{\varphi} \leq 10^2$.}
	\label{Jordan moment. varying}
\end{figure*}

In the Jordan frame, let us set the energy at $E_{0}$ and construct the phase space for a range of values of $p_{\varphi}$. The results is shown in figure \ref{Jordan moment. varying}. Each curve represents a value for the velocity (momentum) of the scalar field. In each plot, we have a total of ten curves. For each curve, the less the minimum of the scale factor is, the higher $p_{\varphi}$ is. Notice that, up until an upper value for $p_{\varphi}$, the curves are of a smooth bouncing for the Universe, including solutions with a possible inflationary phase. Above a certain value of $p_\varphi$, divergent curves appear. If one assumes that this type of divergence does not describe a physical reality (favoring smoothness), then the scalar field must have a limit in momentum. Otherwise, this model predicts a singularity formed by an accelerated contraction of a prior universe, reaching null volume as the (modulus of the) momentum goes to infinity.\footnote{Notice that, we are reading the graphics in the clockwise direction.}	

\begin{figure*}[htp!]
	\centering
	\subfloat{\includegraphics[width=0.5\textwidth]{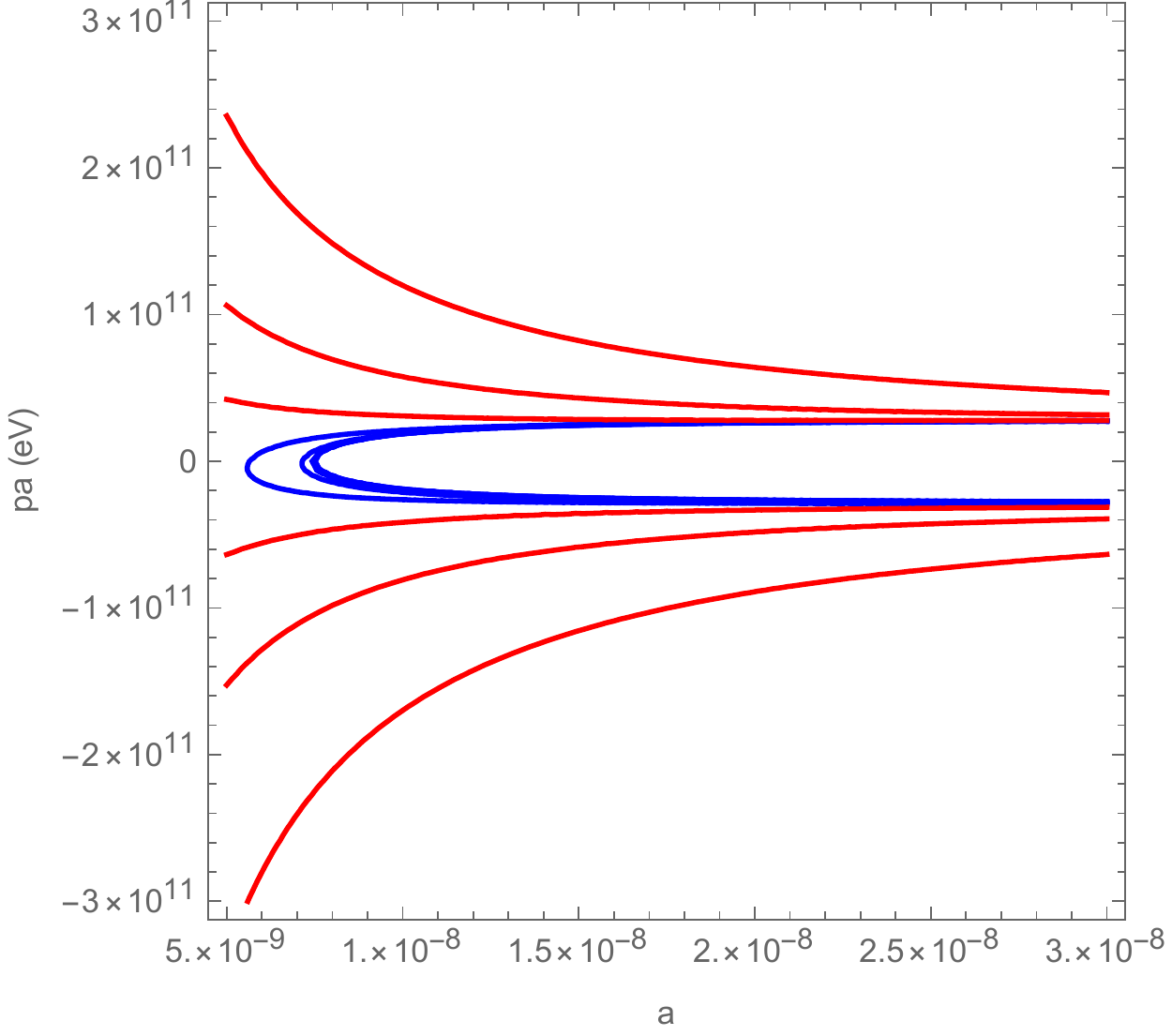}}
	\subfloat{\includegraphics[width=0.5\textwidth]{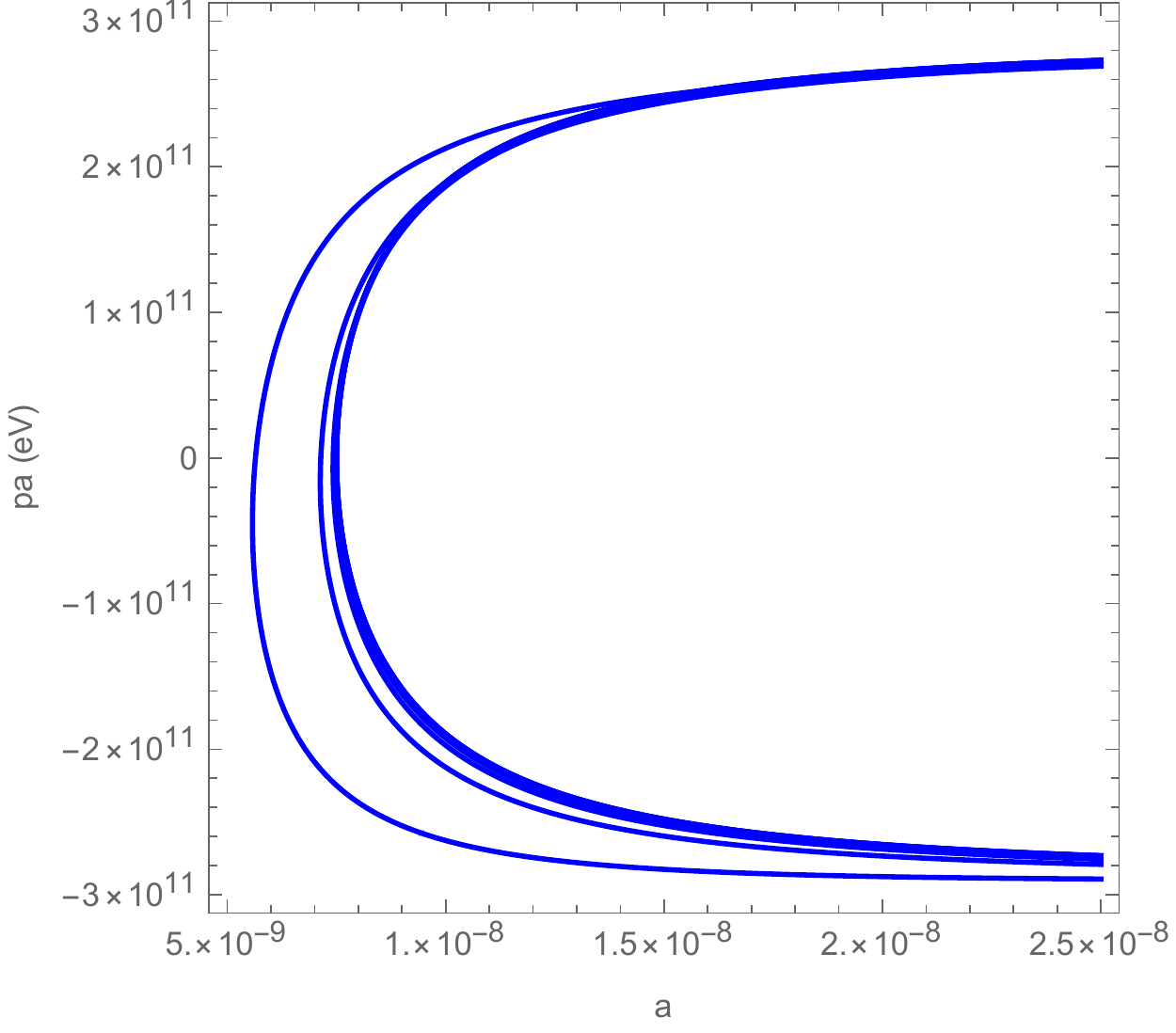}}\hfill
	\caption[Effect of $\omega$ in Jordan frame]{The effect of the Brans-Dicke constant in the scalar field phase space. Once again, we use $E_{0}=10^{16}$ and consider the range $1  \leq p_{\varphi} \leq 10^3$. The left figure is for $\omega=41,000$, and the right figure is for $\omega=4,100,000$.}
	\label{Jordan omega varying}
\end{figure*}

In figure \ref{Jordan omega varying}, we study the effect of the Brans-Dicke parameter $\omega$. In the left figure, we take $\omega=41,000$ and see there are more divergent lines than in the generic case considered in Figure \ref{Jordan moment. varying}. In the right figure, we increased $\omega$ to $4,100,000$. Notice that it requires a much greater initial momentum for the scale factor to obtain divergent solutions. Therefore, a larger $\omega$ seems to lead to a more well-behaved theory. 
This is a result of interest, since the larger $\omega$ is, the greater the coupling between matter and the scalar field, that is, the smaller the effects of the scalar field are. This would correspond to the weak-field limit we observe today. Actually, for a perfect fluid (as in our case), we recover GR in this limit \cite{Paiva:1993qa}. 

\begin{figure*}[htp!]
	\centering
	\subfloat{\includegraphics[width=0.5\textwidth]{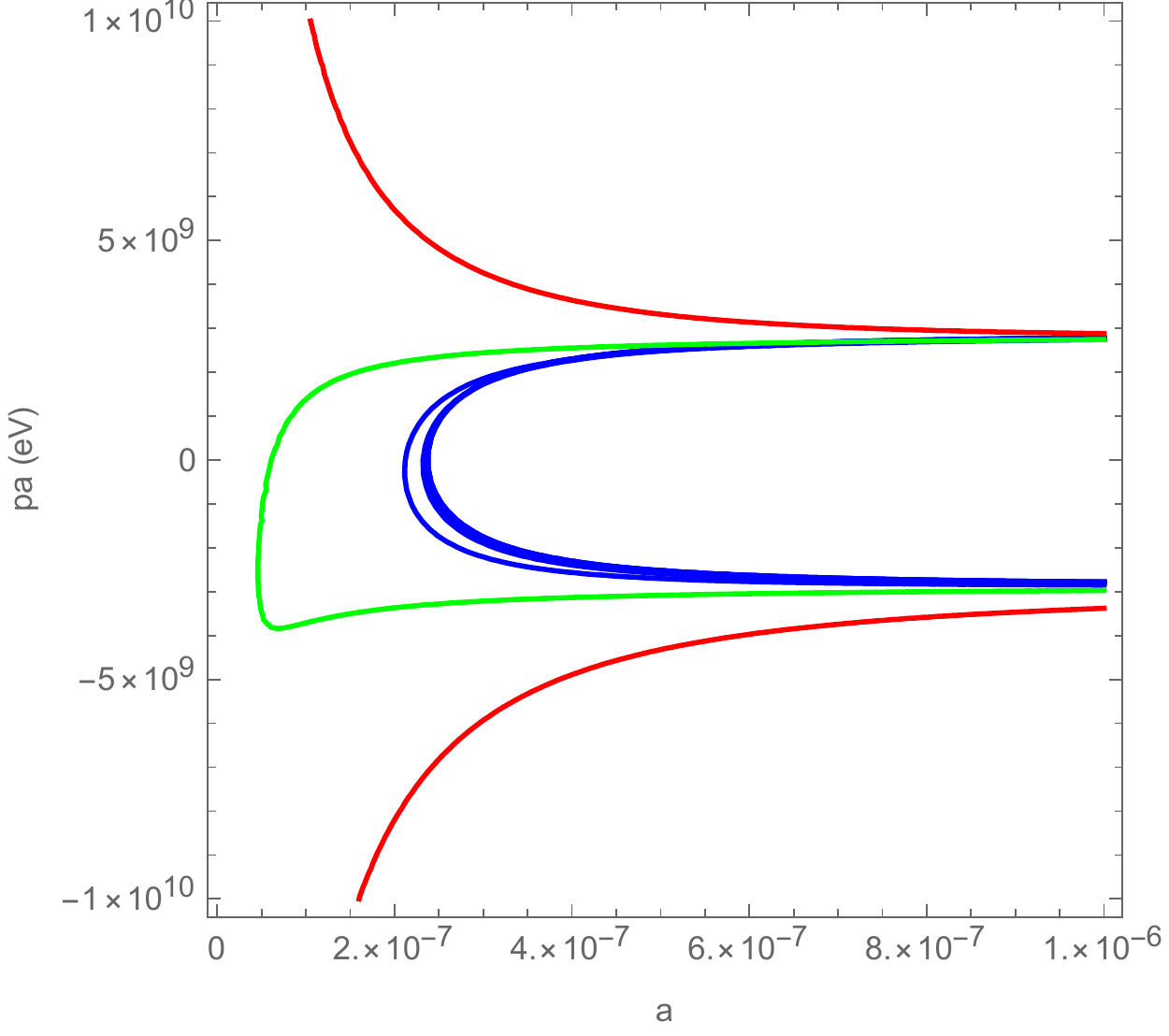}}
	\subfloat{\includegraphics[width=0.5\textwidth]{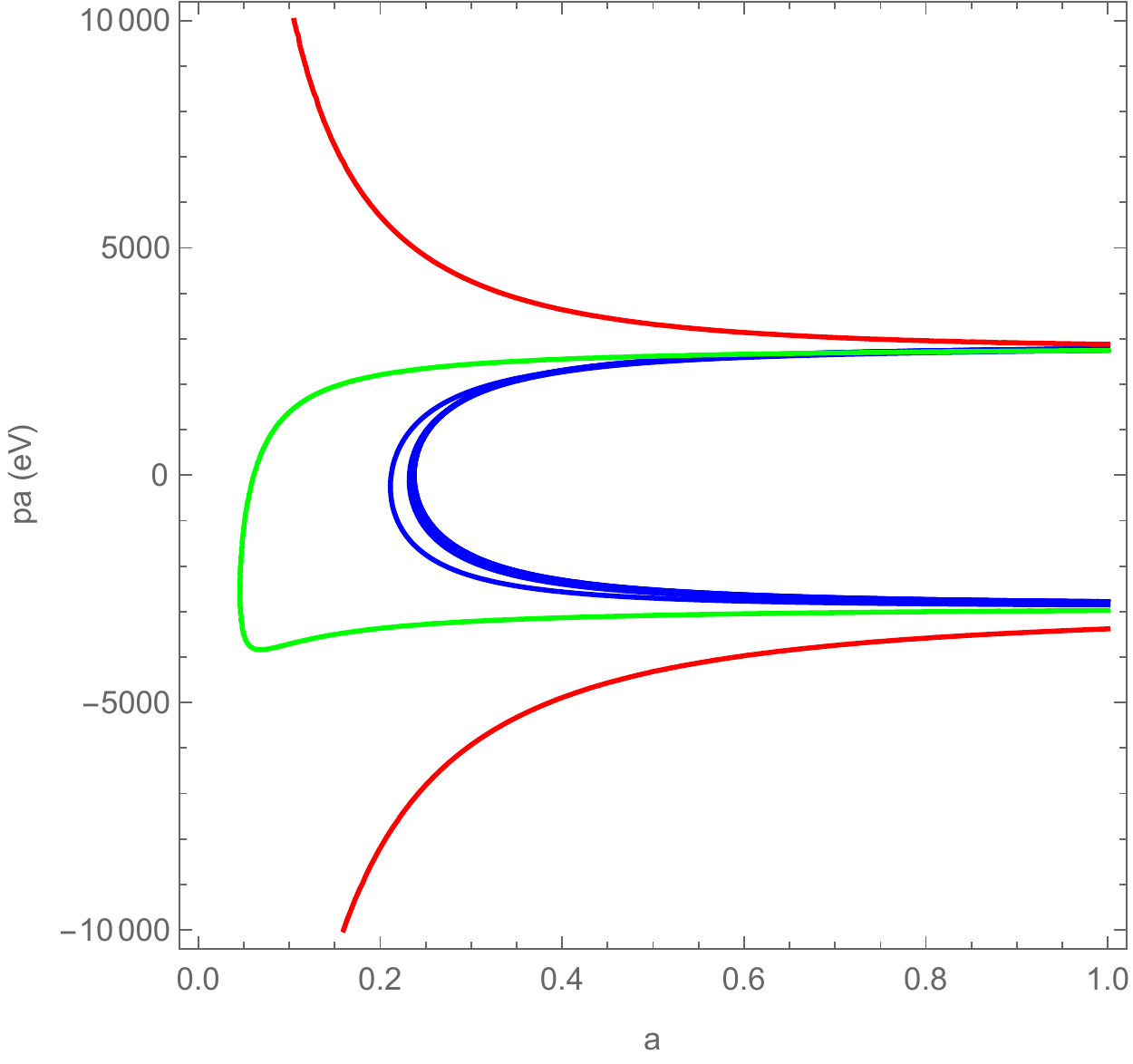}}\hfill
	\caption[Relation between energy and scale in Jordan frame]{The change in the energy of the system results in a change of scale for the solutions. In the left figure we take $E=10^{13}$ and in the right figure we take $E=10$. The same values were used as before for $p_{\varphi}$ and $\omega$: $1  \leq p_{\varphi} \leq 10^3$ and $\omega=410,000$.}
	\label{Jordan E varying}
\end{figure*}

The variation of the energy parameter does not change the behaviour of the solutions, as we can see in figure \ref{Jordan E varying}, but it results in a change of scale in the phase space. So the energy can determine the scale with which inflation happens.

\begin{figure*}[htp!]
	\centering
	\subfloat{\includegraphics[width=0.5\textwidth]{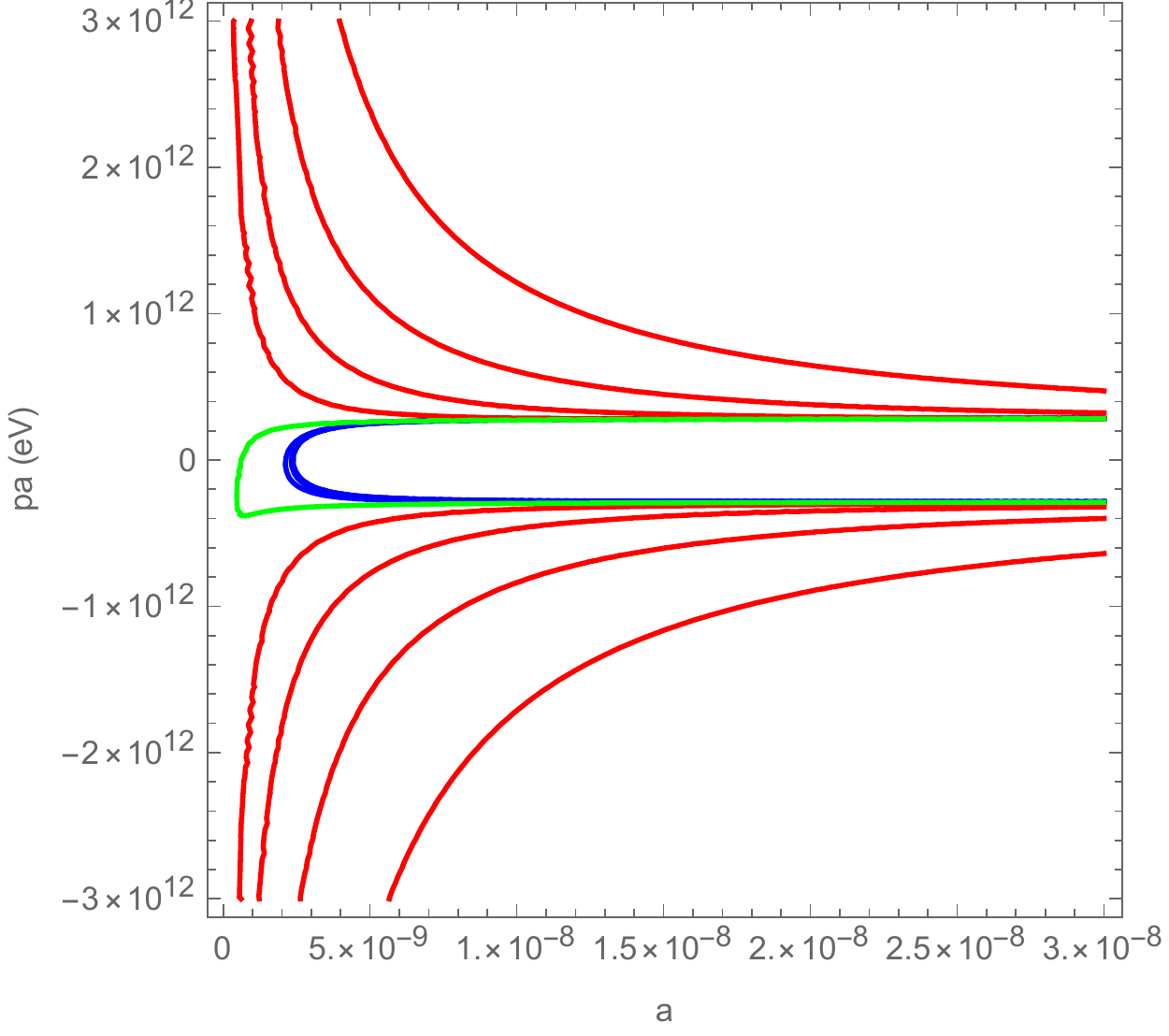}}
	\subfloat{\includegraphics[width=0.5\textwidth]{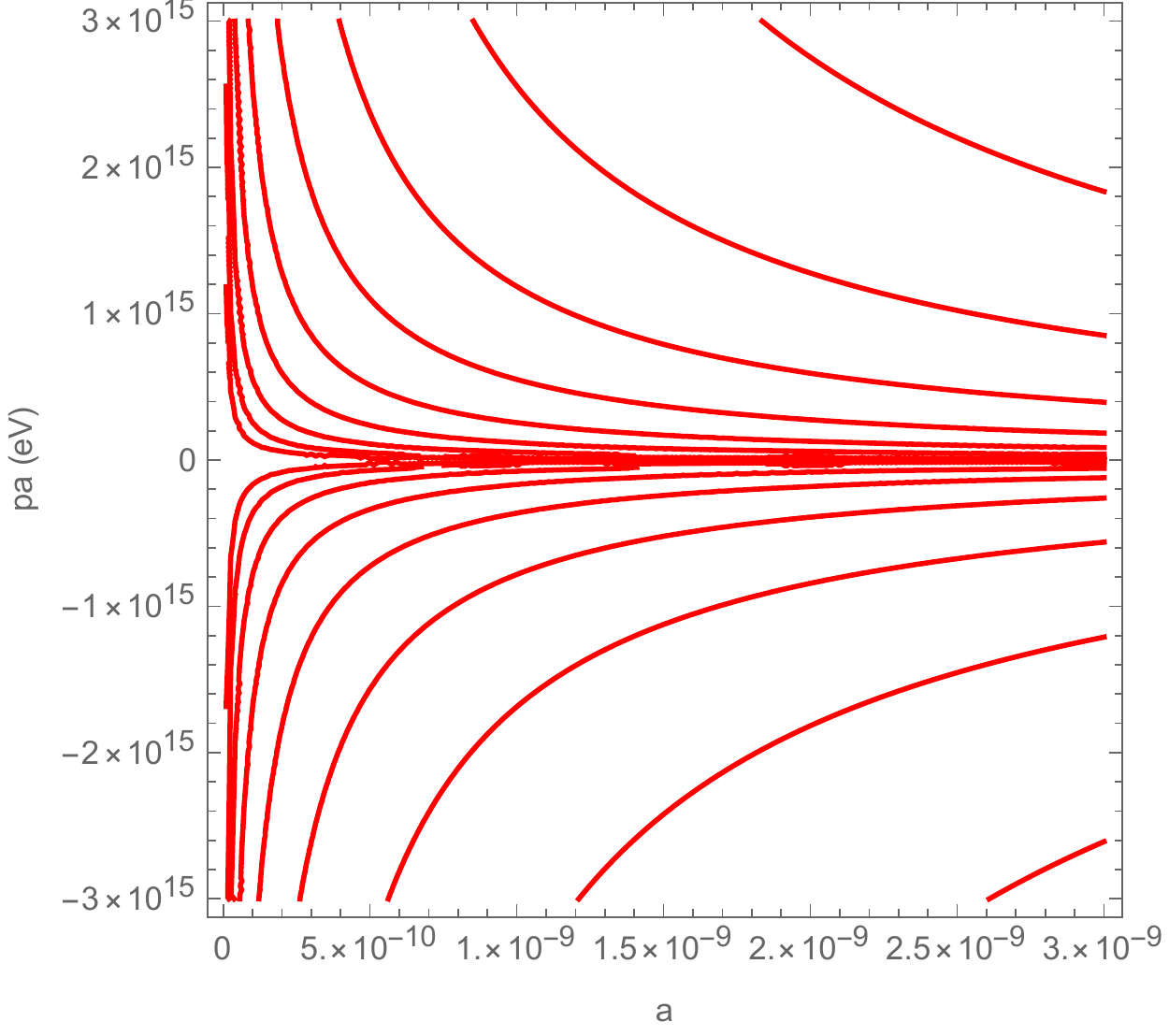}}\hfill
	\subfloat{\includegraphics[width=0.5\textwidth]{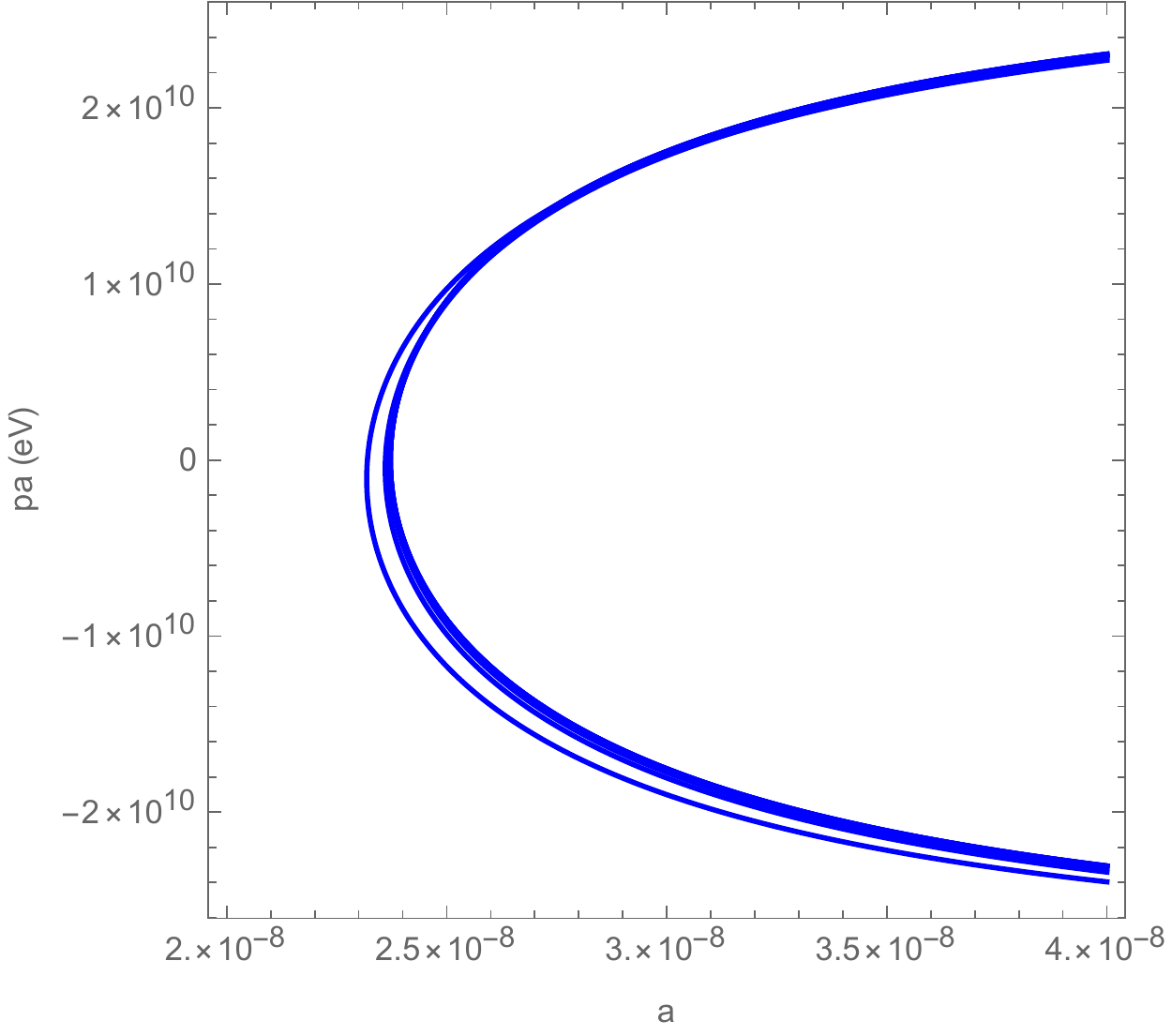}}
	\subfloat{\includegraphics[width=0.5\textwidth]{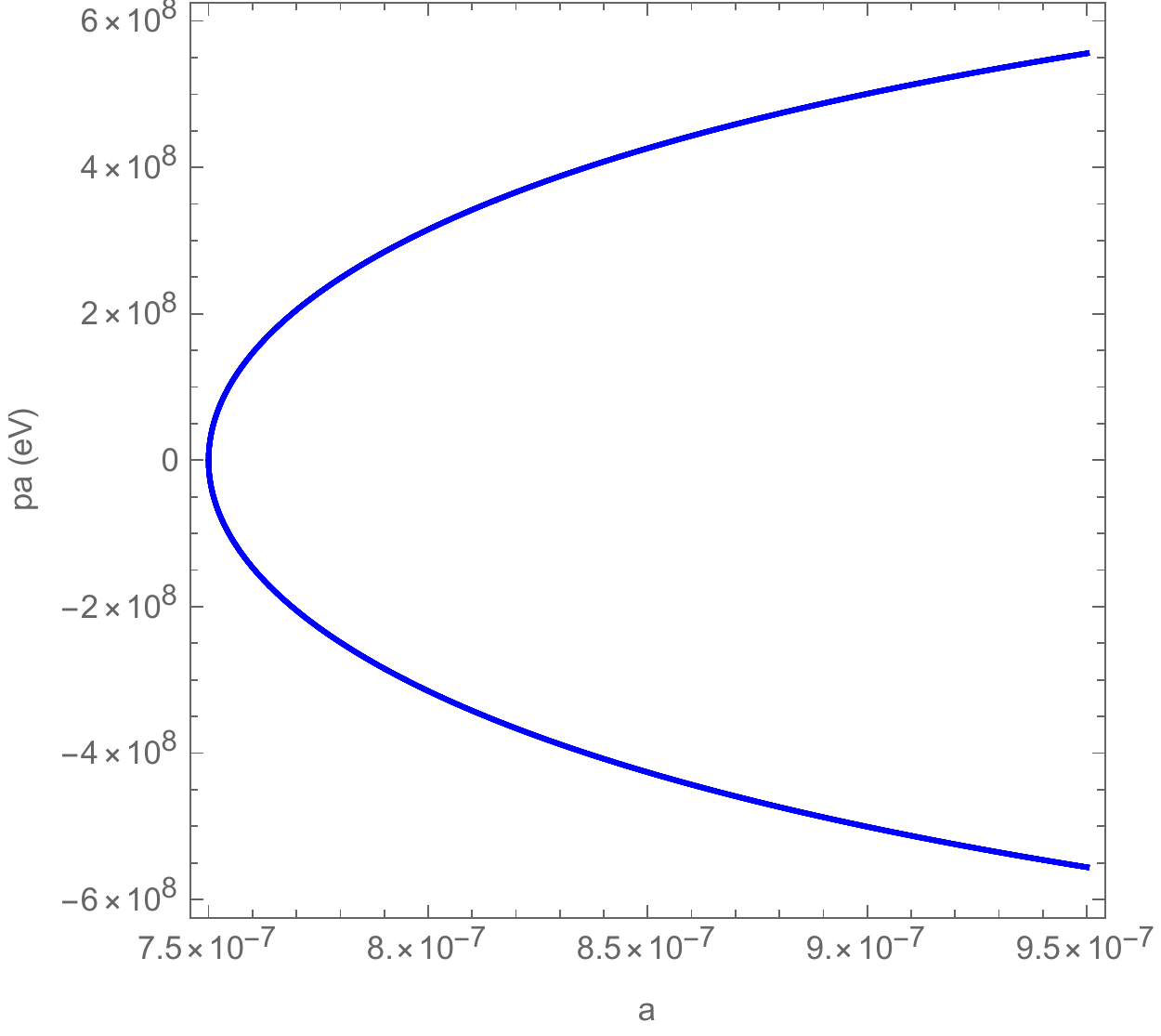}}\hfill
	\caption[Phase space portrait dependence on initial field value in Jordan frame]{The top row shows the solutions for high values of $\varphi_0$: top-left $\varphi_0 = 10$, and top-right $\varphi_0 = 10^4$. The bottom row is for low values of $\varphi_0$: bottom-left $\varphi_0 = 10^{-1}$, and bottom-right $\varphi_0 = 10^{-4}$. For these, we are considering $\omega=410,000$, $E=10^{16}$, and $1  \leq p_{\varphi} \leq 10^3$.}
	\label{plot_pa_phi}
\end{figure*}

Up until now, we have considered the initial value of the scalar field to be $\varphi_{0} =1$, but we also want to understand the effects of the initial condition on the behaviour of the solutions. Thus, in figure \ref{plot_pa_phi}, we show the direct influence of changing the value for the scalar field on the solutions. The top row shows greater values for $\varphi_0 $, from $10$ to $10^4$ (left to right). We notice that the greater $\varphi_0$ is, the more singularities we obtain. Conversely, in the second row, we lower it from $0.1$ to $10^{-4}$. The solutions tend to bounces instead of singularities. As expected, the results are consistent with the study on $\omega$. 
%From the last figure, all solutions are hardly distinguishable, from which we can infer than in the weak field limit $\varphi \rightarrow 0$, the solutions tend to an unique one.

\subsection{Einstein Frame}

\begin{figure*}[htp!]
	\centering
	\subfloat{\includegraphics[width=0.5\textwidth]{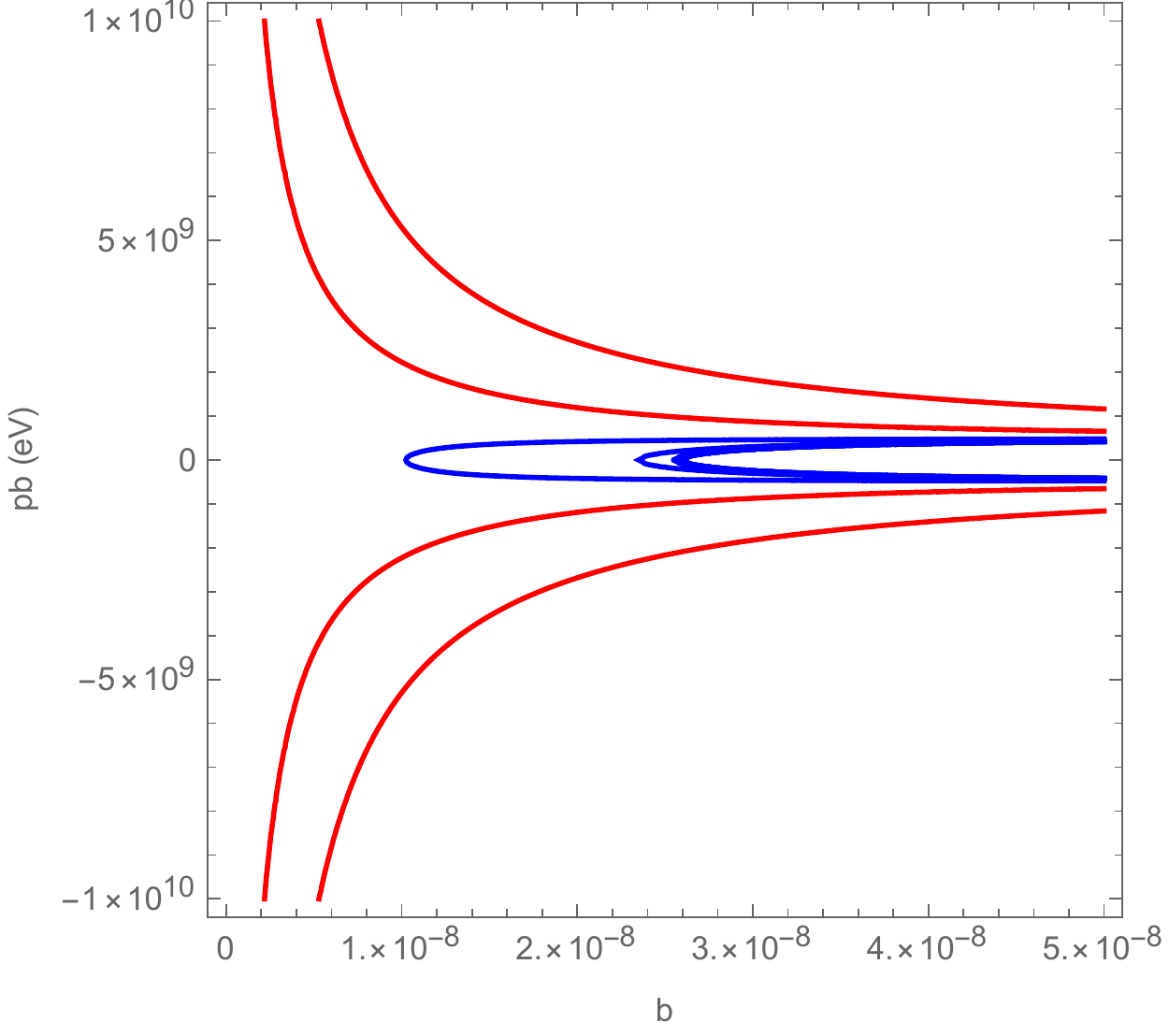}}
	\subfloat{\includegraphics[width=0.5\textwidth]{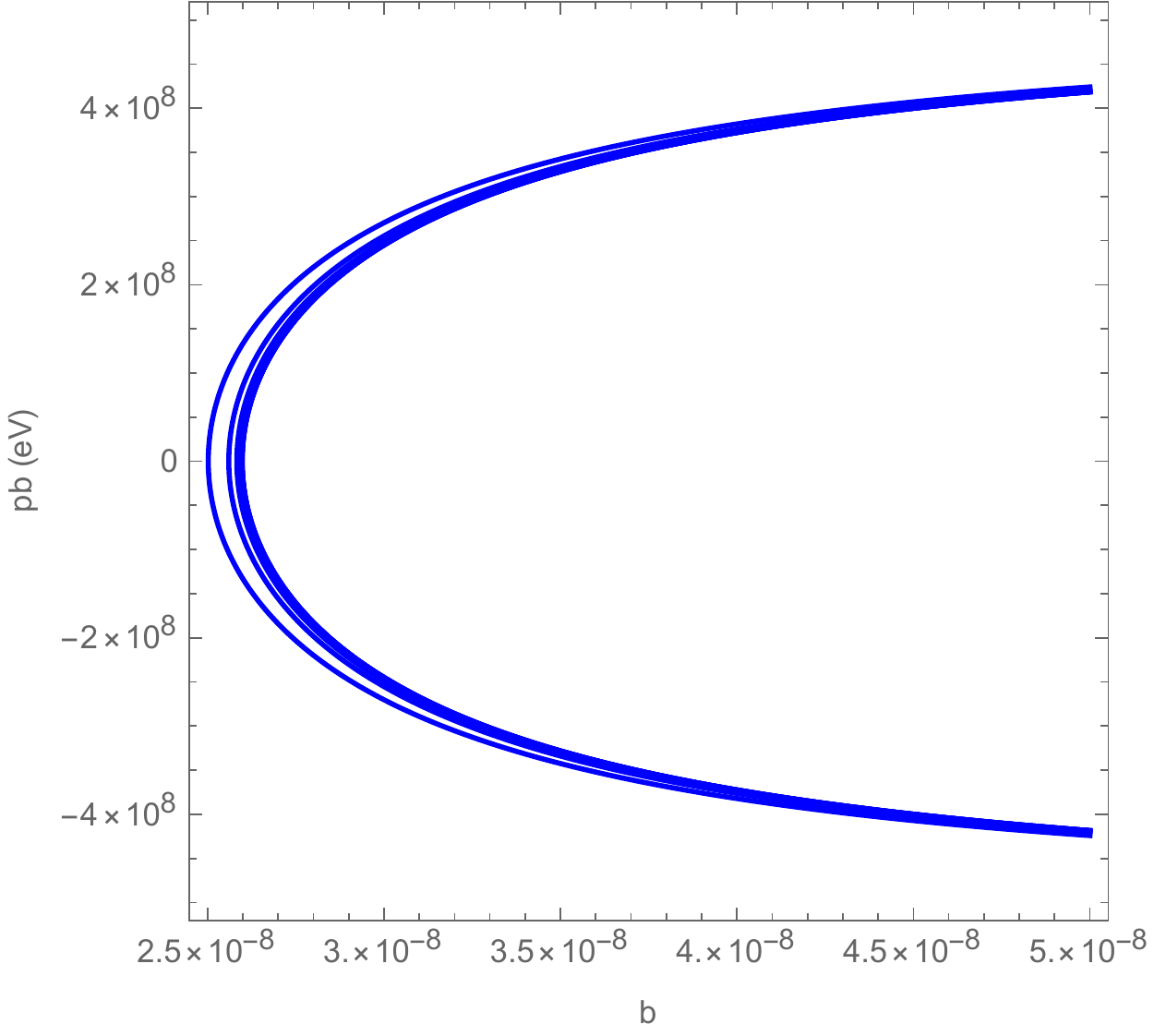}}\hfill
	\caption[Quantum phase space in the Einstein frame with $\omega=410,000$ and $E_{0}=10^{16}$]{Quantum phase space in the Einstein frame, using $\omega=410,000$ and $E_{0}=10^{16}$. The left figure is for a range $1  \leq p_{\varphi} \leq 10^3$, while for the right figure the  range is smaller $1 \leq p_{\varphi} \leq 10^2$.}
	\label{Einstein moment. varying}
\end{figure*}

\begin{figure*}[htp!]
	\centering
	\subfloat{\includegraphics[width=0.5\textwidth]{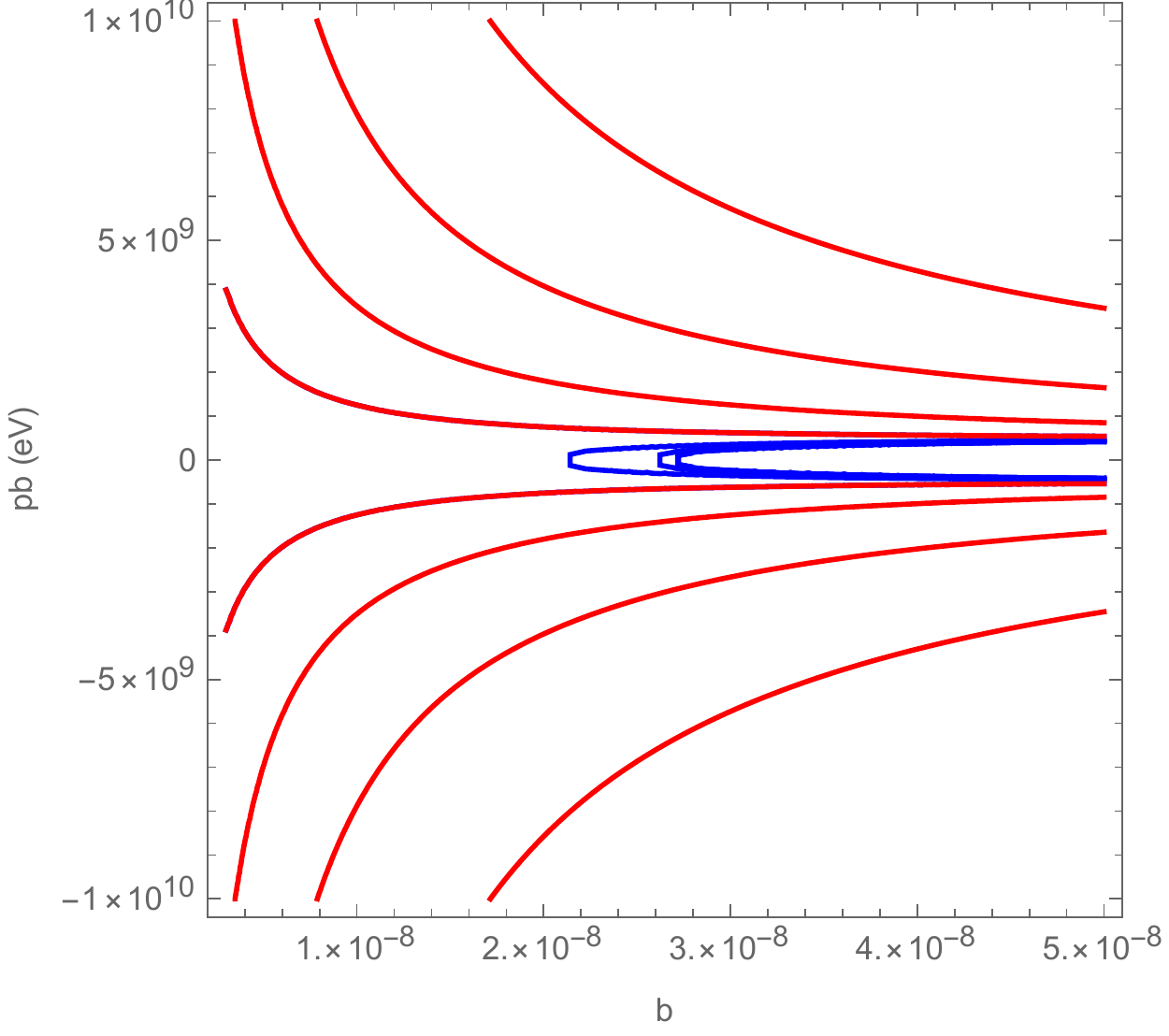}}
	\subfloat{\includegraphics[width=0.5\textwidth]{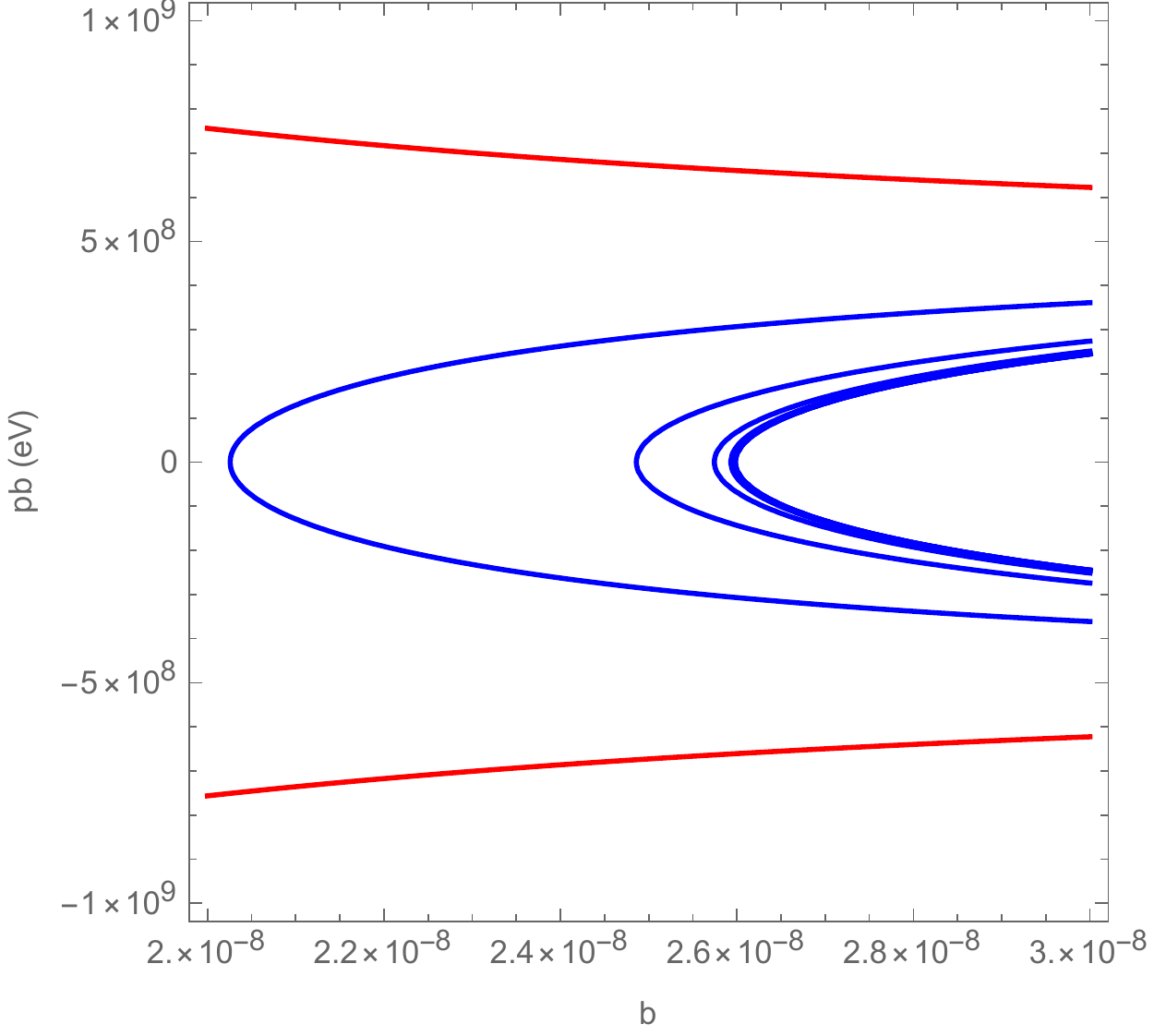}}\hfill
	\caption[Effect of $\omega$ in Einstein frame]{The effect of the Brans-Dicke constant in the scalar field phase space. Once again, we take $E_{0}=10^{16}$ and consider the range $1  \leq p_{\varphi} \leq 10^3$. The left figure is for $\omega=41,000$, and the right one is for $\omega=4,100,000$.}
	\label{Einstein omega varying}
\end{figure*}

\begin{figure*}[htp!]
	\centering
	\subfloat{\includegraphics[width=0.5\textwidth]{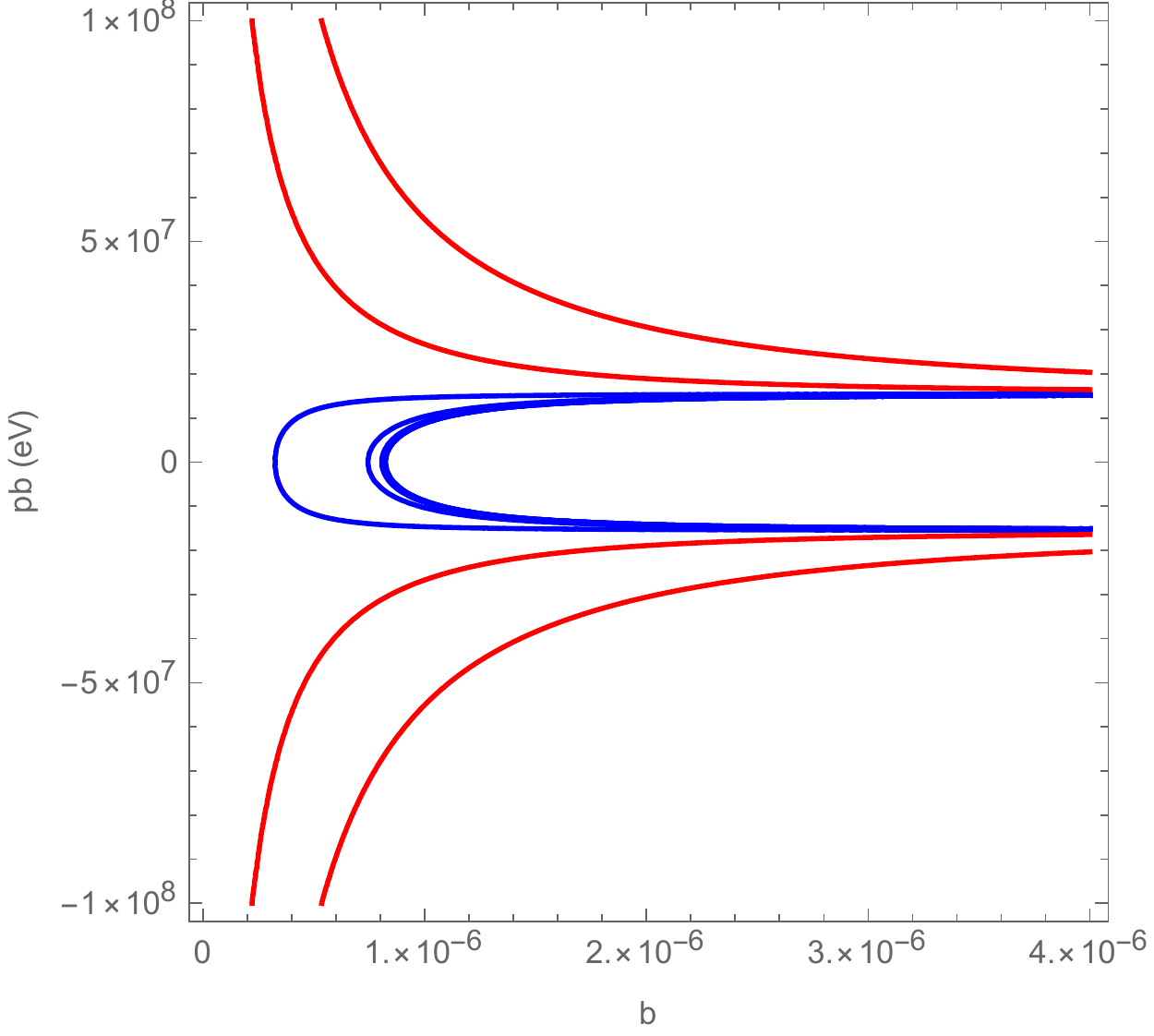}}
	\subfloat{\includegraphics[width=0.47\textwidth]{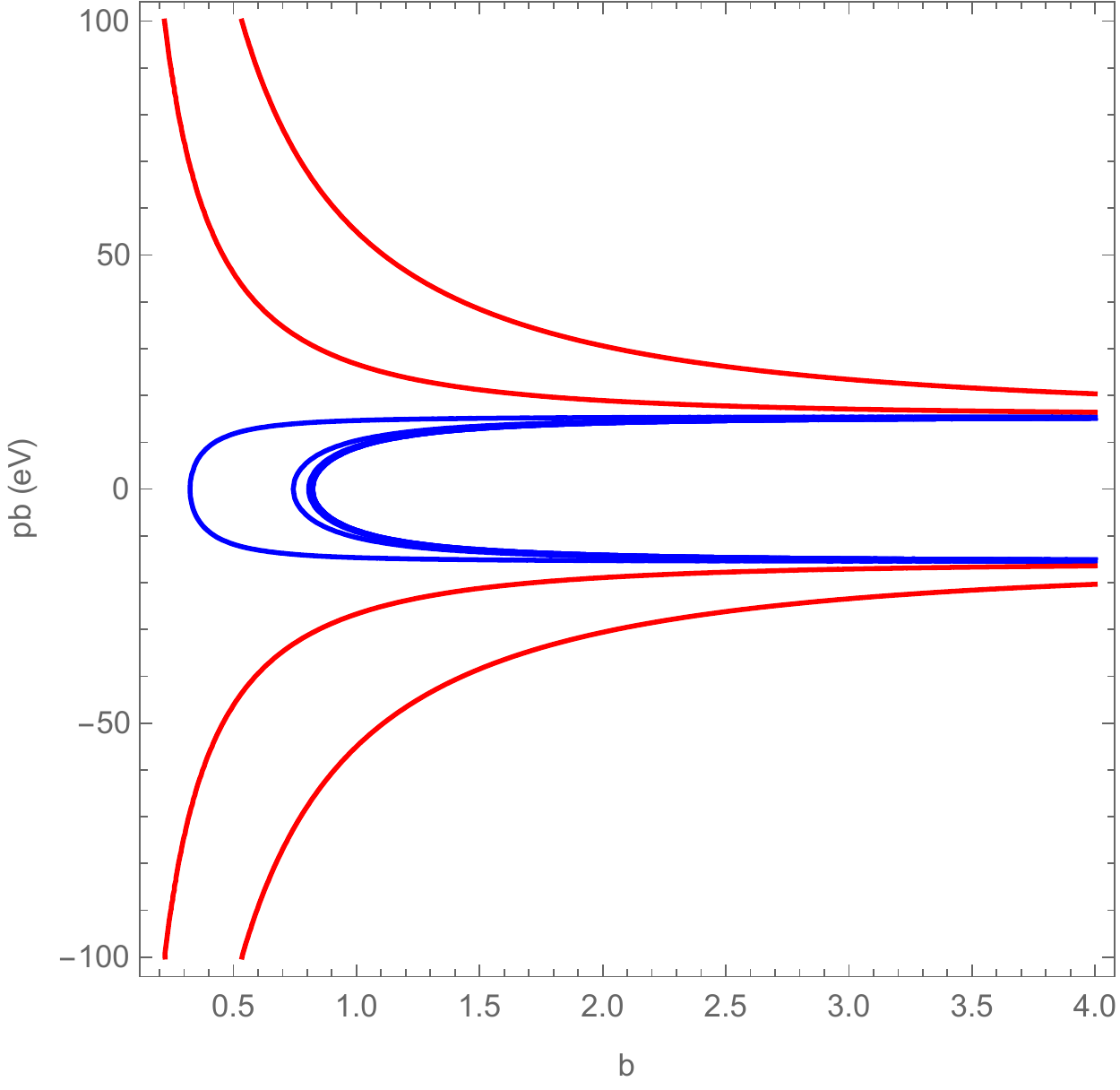}}\hfill
	\caption[Relation between energy and scale in Einstein frame]{The change in the energy of the system results in a change of scale for the solutions. In the left figure we take $E=10^{13}$ and in the right figure we take $E=10$. The same values were used as before for $p_{\varphi}$ and $\omega$: $1  \leq p_{\varphi} \leq 10^3$ and $\omega=410,000$.}
	\label{Einstein E varying}
\end{figure*}

\begin{figure*}[htp!]
	\centering
	\subfloat{\includegraphics[width=0.5\textwidth]{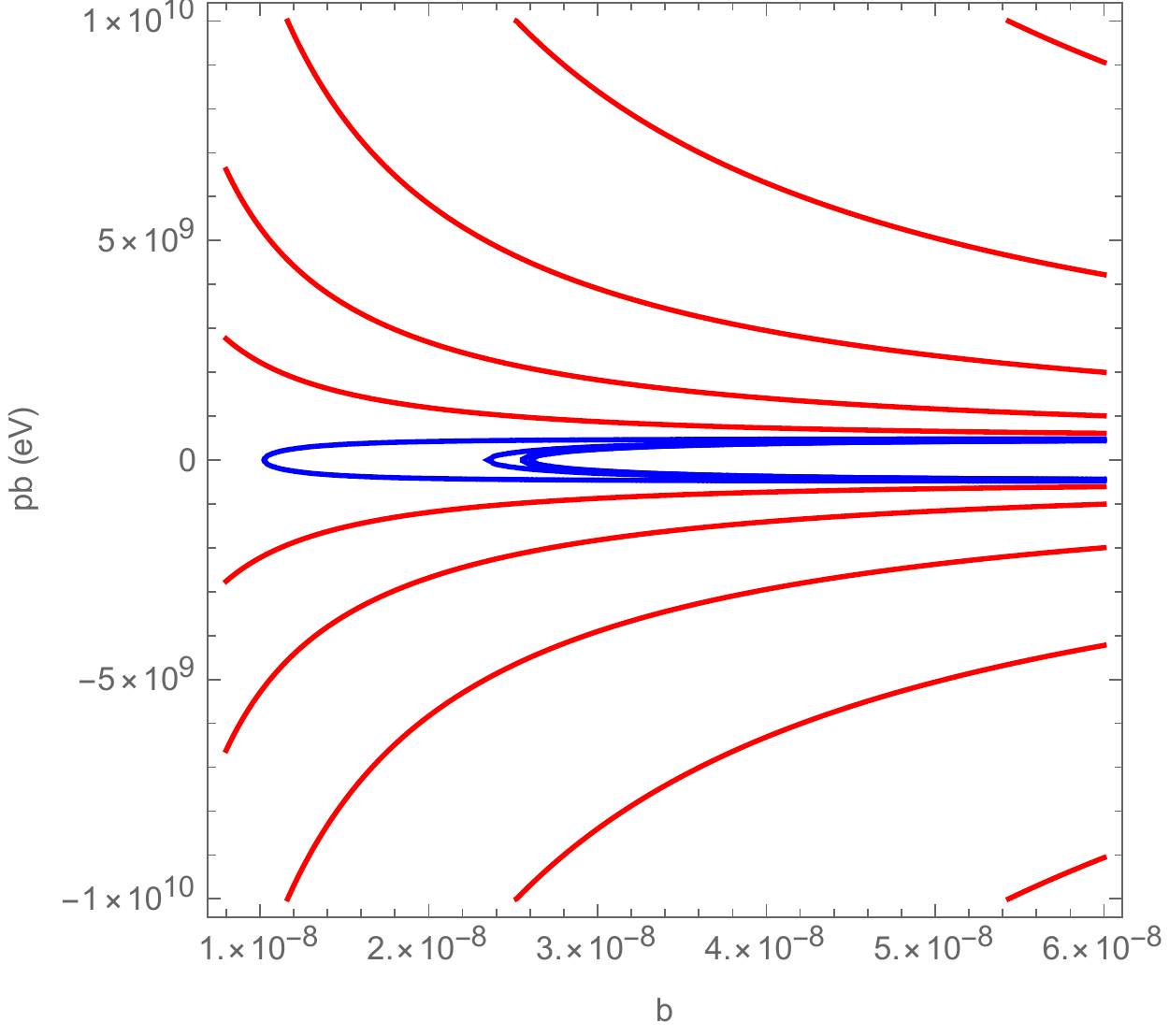}}
	\subfloat{\includegraphics[width=0.5\textwidth]{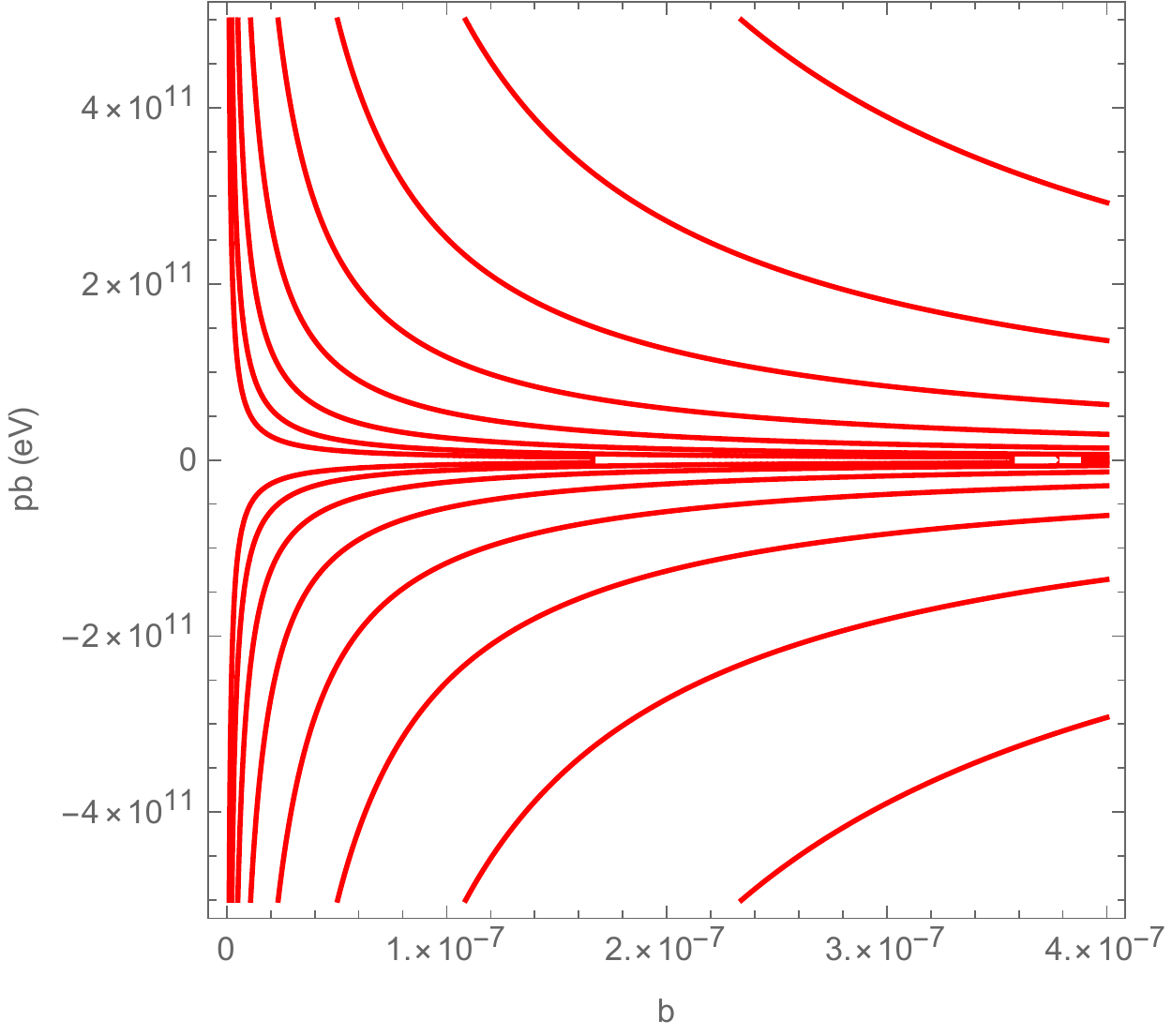}}\hfill
	\subfloat{\includegraphics[width=0.5\textwidth]{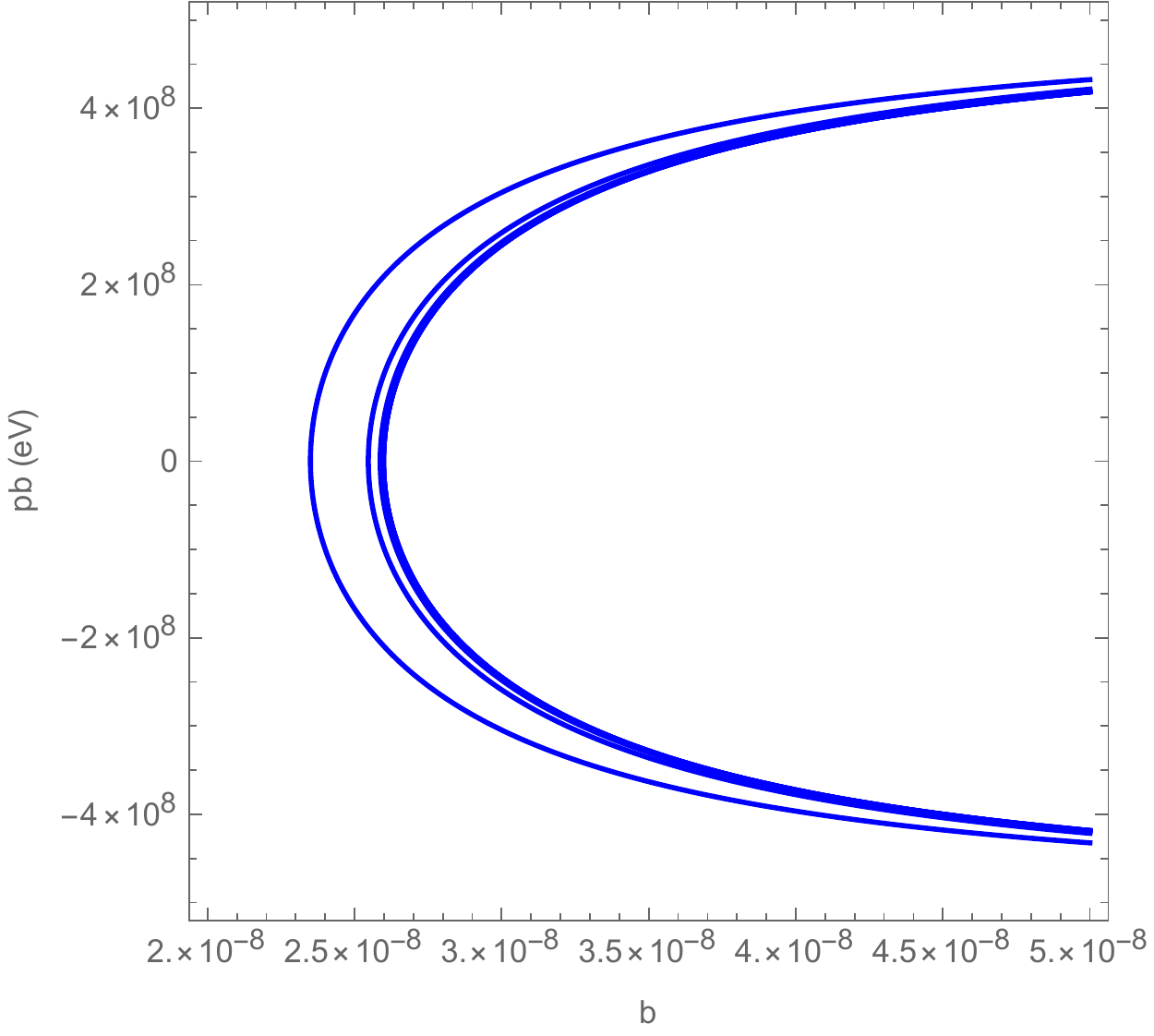}}
	\subfloat{\includegraphics[width=0.5\textwidth]{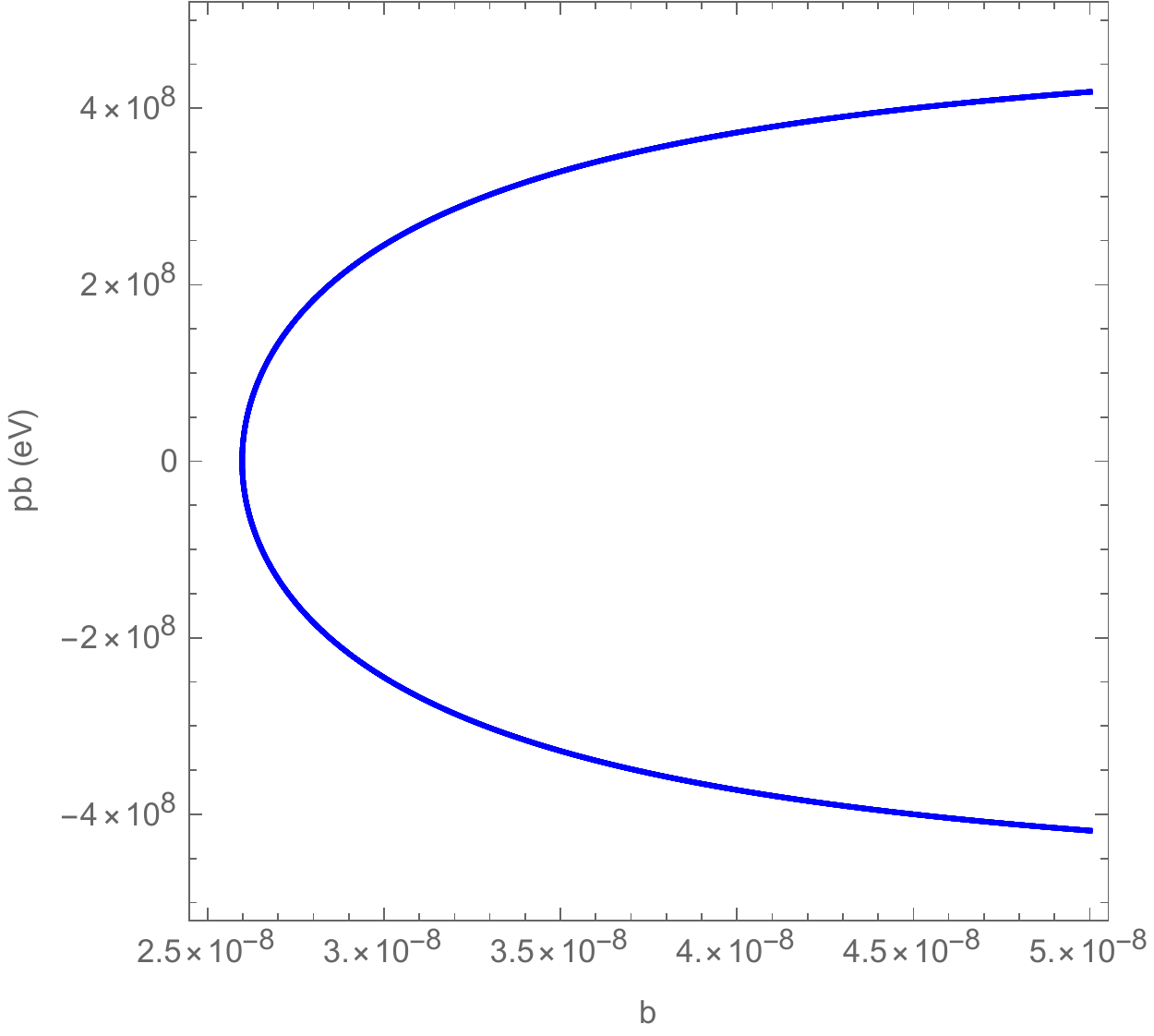}}\hfill
	\caption[Phase space portrait dependence on initial field value in Einstein frame]{The top row shows the solutions for high values of $\varphi_0$: top-left $\varphi_0 = 10$, and top-right $\varphi_0 = 10^4$. The bottom row is for low values of $\varphi$: bottom-left: $\varphi_0 = 10^{-1}$, and bottom-right $\varphi_0 = 10^{-4}$. For these, we are considering $\omega=410,000$, $E=10^{16}$, and $1  \leq p_{\varphi} \leq 10^3$.}
	\label{plot_pb_phi}
\end{figure*}

In the Einstein frame, we have symmetric bounces without any inflationary epoch\footnote{Inflation may be interpreted as a ``stretching" of the solutions induced by the conformal transformation by going from the Einstein frame to the Jordan frame.}, as we see in Figure \ref{Einstein moment. varying}. By varying once again $\omega$ (Figure \ref{Einstein omega varying}) and the energy (Figure \ref{Einstein E varying}), we arrive to the same conclusions as in the Jordan frame, \textit{i.e.} that the larger $\omega$ is, the less divergent the curves we obtain, and varying the energy induces a scaling in the phase space. We also show the effect of the scalar field in Figure \ref{plot_pb_phi}.

Notice that these results are consistent with what was found in the Jordan frame, which provides further evidence that the frames are equivalent. Remember, though, that in spite of choosing specific fiducial vectors, this analysis is still qualitative, since one can always choose different wavelets and also restore the unities (we chose $c=\hbar=1$). For our purpose, this qualitative analysis is enough.

\section{Conclusions on Affine Quantisation and Cosmology} 

In this chapter, we presented the quantisation of the Brans-Dicke Theory using the affine covariant integral method, and the cosmological scenarios arising from it.
We introduced the classical Hamiltonian formalism of the BDT and the mathematical foundations of this quantisation method, in order to familiarise the reader with the concepts used later on. Our model is completed with the addition of a radiative matter component in form of a perfect fluid, introduced via the Schutz formalism, which we adopted as the clock. The affine quantisation is based on the symmetry of the phase space of the system, and we can choose the free parameters, namely the fiducial vectors, in a way to build an essentially self-adjoint Hamiltonian operator. The quantisation of the Hamiltonian constraint results in the Wheeler-DeWitt equation, from which we obtain a Schr\"odinger-like equation \eqref{WDW equation}, with the radiative matter providing the time parameter. One expected setback of this quantisation is that it results in a non-separable partial differential equation. We can work around this problem by changing frames, making a conformal transformation of the coordinates.

The BDT is described in the Jordan frame and a conformal change of coordinates transforms the BDT into GR with a scalar-field, \textit{i.e.} the Einstein frame. The equivalence between these frames is still debatable (see \textit{e.g.} \cite{Artymowski:2013qua,Banerjee:2016lco,Kamenshchik:2014waa,Almeida:2017gvx,Pandey:2016unk}), and our results may contribute to this debate. In the Einstein frame, the Schr\"odinger-like equation is separable, and becomes easier to deal with. We presented the classical GR with a scalar-field model corresponding to the BDT in the Einstein frame and quantised it using the affine method. We also performed a change of coordinates in the already quantised Schr\"odinger-like equation in the Jordan frame. Considering the freedom in the choice of the fiducial vectors, we found an equivalent equation. However, we conclude that the Hamiltonian operator in the Einstein frame is only essentially self-adjoint if we consider different fiducial vectors while quantising the theory in each frame, or if we change the domains (i.e. the measure) of the operators in the respective Hilbert space. In any case, one may argue that, because of this, there is no equivalence between the frames. However, the role of the fiducial vectors during the quantisation is precisely to open up opportunities for adjustment, since it is based on a statistical method ($|\langle q,p| \phi \rangle|^{2}$ is interpreted as the probability density distribution of the function $\phi$, see for example \cite{Gazeau:2013dta}). Thus, considering different fiducial vectors in different frames should not invalidate the equivalence between them. We choose to solve the Wheeler-deWitt equation obtained from the classical BDT in the Einstein frame, in order to do a qualitative analysis, since this equation has a relatively simple solution. From it, we were able to conclude that the energy spectrum of the Hamiltonian operator in Einstein frame is discrete.

The affine quantisation method is completed with a ``de-quantisation", known as the quantum phase space portrait or lower symbol, that transforms the quantised operator into a classical function, by means of their fiducial vectors expectation values. This de-quantisation provides a quantum correction for classical observables, from which we can analyse the behaviour of these observables in semi-classical environments. Even if we cannot find the wave-function of the Universe in the Jordan frame, we can use the  quantum phase space to compare the results with the ones from the quantum phase space in Einstein frame. Thus, we find quantum corrections for the Hamiltonian constraint in both frames in section \ref{Quantum phase space BDT} and compare the results in Section \ref{Results}, drawing the phase space portrait for the scale factor, to better understand the behaviour of the (volume of the) Universe in earlier stages.

We obtained two types of solutions in both frames: bounces and singularities. For both types, we predict a prior universe. For the singular cases in the Jordan frame, there is an accelerated contraction, with a singular point where the volume of the Universe becomes null. However, if we limit the momentum of the scalar field, we obtain only bouncing solutions. Thus, we may argue that the scalar field should have a limited velocity, since this discards the singular solutions. We also analysed the influence of other parameters in the solutions. In the limit $\omega \rightarrow \infty$, in which we expect to reproduce GR (for our model, at least), bounces become more expected. It is interesting to see that an inflationary stage also appears for bounces in this frame. In the Einstein frame, however, we do not have any inflationary era, but similar conclusions can be drawn, with the exception that both singular and bouncing solutions are symmetric. 

While being a fairly recent subject of interest in cosmology, the affine quantisation points to interesting applications, such as the removal of divergences in non-renormalisable theories \cite{Klauder:2010eg,Klauder:2012ij} or the non-singular expanding (and possibly cyclic) universes \cite{Fanuel:2012ij}. Other results on various cosmological scenarii (see \cite{Bergeron:2013ika,Bergeron:2013jva,Bergeron:2015jpa,Bergeron:2015lka,Bergeron:2015ppa,Bergeron:2015sqt,Gozdz:2018aai} also lead to bouncing solutions. 

%% file: sections/chapter7.tex
\chapter{Conclusions}

\section{Summary of Achievements}

The inflationary paradigm and its numerous successes proved the inflationary phenomenology is a viable window to the first instant of the Universe. Hundreds of inflationary models exist in the literature, and summing all alternative or complementary mechanisms to inflation makes finding which particular mechanism led to large-scale structures observed today a difficult task. The increasing precision of observational cosmology will help in discriminating among this plethora of models.

This thesis is devoted to the study of quantum effects in bouncing models. We dedicated the first part of this thesis to the development of the stochastic formalism, describing a non-perturbative semi-classical approach used to calculate the infrared behaviour of light and massless fields, in the case of a contracting universe. %Stochastic inflation shows that infrared divergences can be resummed, rendering the theory infrared finite. 
In a contraction phase, perturbations grow, and the necessity to quantify the backreaction from the noise produced by quantum fluctuations onto the geometry arises. We laid the first stone of stochastic collapse using a simple exponential potential, and showed that quantum effects can modify the evolution of the background. 

The second and last parts are devoted to the quantisation of the whole Universe. The second objective was to argue that the origin of magnetic fields present in galaxies and large-scale structures can be generated contemporaneously with those structures. In particular, we showed that dark matter in the framework of quantum cosmology with a coupling between curvature and electromagnetism can generate magnetic fields within current observational constraints. 

The last work was focused on a minimalist quantisation procedure, the integral affine quantisation, and the description of the quantum Brans-Dicke theory. In particular, we focused on the quantum equivalence between the Jordan and the Einstein frame. We showed that the equivalence lies within the choice of the mathematical model, and we illustrated the semi-classical behaviour of the theory. In both frames, a smooth bounce is expected, with the only difference being the asymmetry in the Jordan frame bounce.

\section{Prospective Directions}

\subsection{Stochastic Bounce}

Given that we have seen that quantum diffusion can play an important role in the semi-classical dynamics, it is interesting to consider whether quantum effects might be a way to avoid the classical singularity as the Hubble rate diverges. However, the stochastic collapse was developed using the number of e-folds $N$ as time variable, since the two-point correlation functions for noise are usually expressed in terms of $N$. However, $N$ is a monotonic time parameter, and cannot be used for both a contraction phase and a subsequent expansion phase. Hence, another time parameter needs to be used in order to model a bounce and to describe the imprints of stochastic effects on the CMB. Since the stochastic formalism is a semi-classical approach, we must consider a classical background evolution. In this case, a natural choice of time would be the scalar field itself, in analogy with many quantum gravity theories \cite{Makinen:2019jam}. Then, including stochastic noise to the field, we would be able to introduce stochastic fluctuations of the geometry, and \textit{a fortiori} fluctuations in the scale factor itself. This will most likely lead to a non-zero scale factor at all times, in turn implying there will be a bounce joining the contracting and expanding phases \cite{vanHolten:2013yca}.

Once the stochastic bounce has been implemented, we can explore what consequences stochastic effects have on the matter spectrum compared to classical bouncing models \cite{Peter:2008qz}. In particular, comparing the spectral index and anisotropies to CMB observations will allow to constrain cosmological models with an exponential potential. An interesting way to achieve this goal is to consider non-gaussian features from gravitational waves observations, which can probe eventual physics beyond inflation \cite{Bartolo:2018qqn}, and forthcoming (KAGRA, DECIGO, LISA, etc...) observatories are enhancing sensitivity for the detection of B-modes polarisations from gravitational waves in the CMB. 

\subsection{Separate Universe and Gauge Choice in a Contracting Phase}

An important point will be to check how results obtained in a collapsing phase depend on the choice of gauge. It is customary to work in the spatially-flat gauge, since the field perturbations coincide with the Mukhanov-Sasaki variable in this case. However, the local time in this gauge is in general perturbed with respect to the global time. The Langevin equations, giving the evolution of the field variables with stochastic noise, are written in the uniform-$N$ gauge. Then, to relate the stochastic distribution of field values, we need to work out the field fluctuations in the uniform-$N$ gauge. The general method showing how to perform the gauge transformation was developed in \cite{Pattison:2019hef}. We could then apply this method to the case of a collapsing universe, and see whether the field perturbations acquire a time dependence in the uniform-$N$ gauge. If this is the case, we will compare how fast this time dependence grows compared to the time dependence of relevant quantities, such as the Hubble rate. If the Hubble rate grow faster, this will show that results obtained in the spatially-flat gauge are trustworthy in the uniform-$N$ gauge as well.

\subsection{Extensions of Magnetogenesis Model}

The primordial magnetogenesis model was studied within the simple assumption of coupling between curvature and Maxwell electromagnetism, but other non-minimal couplings between the electromagnetic and gravitational fields, possibly involving the Ricci and Riemann tensors, could be considered as straightforward extensions of the work. More concretely, QED polarisation in a curved background gives four terms of this kind, and we have only studied the dominant one. The other couplings were considered in the framework of power-law inflation in \cite{Kunze:2009bs}, and could bring some additional contribution to the final magnetic field.

Another potential observational effect arises from non-minimal couplings between electromagnetism and curvature, whose origin comes from the equivalence principle. Indeed, such couplings violate the equivalence principle, and the fine-structure constant acquires a time dependence that could be testable with observations \cite{Wilczynska:2020rxx}. In turn, this could constrain further the mass scale associated to the coupling, and reduce the physically relevant parameter space.

Also of importance is the parity-violating coupling $RF\tilde F$, which may be associated to the generation of helical magnetic fields. This term is usually invoked in inflation to bring some power from small scales to large scales by the use of an inverse cascade process.

A fourth point of interest is the induction of a stochastic background of gravitational waves due to the presence of an electromagnetic energy density. The production of gravitational waves is expected to grow with the magnetic spectral index. Since the model we studied generates magnetic fields  with a very blue spectrum, the inclusion of theoretical limits on gravitational waves production \cite{Caprini:2001nb} is worth being investigated. This will be even more relevant with the upcoming detections from LISA \cite{Caprini:2009yp,Caprini:2018mtu,Saga:2018ont,Pol:2019yex}.

Primordial magnetic fields can also be used to alleviate the Hubble tension, since the presence of such fields induce inhomogeneities in the baryon content at recombination \cite{Jedamzik:2020krr}. Therefore, a natural extension is to consider the possibility of magnetogenesis models solving this tension.

Lastly, we could also take into account other possible backreaction effects. It has been shown recently that the vacuum polarisation in a dielectric medium, the so-called Schwinger effect, increases the medium conductivity and subsequently stops the magnetic field production \cite{Sobol:2018djj,Sobol:2019xls,Sharma:2017eps,Sharma:2018kgs,Shakeri:2019mnt}. This would lead to weaker magnetic fields than expected, and could constrain further our model. 

\subsection{Affine Quantum Field Theory}

The implementation of affine quantisation in quantum field theory is a promising way to bring some insights in quantum physics. As we have shown, there is no singularity, the quantum operator is unique and the quantum dynamical evolution unambiguous. We expect to find some regularisations in QFTs without the need of counterterms. Indeed, this fact has been observed in a non-integral scheme of affine quantisation, but complementary to our approach, in the case of scalar fields \cite{Klauder:2012ij}. To be able to construct such an affine quantum field theory however, we will need to express our current approach for infinite dimensional theories. Then the first step in this construction will be to find the correct measure. %Once this is done, we will make use of the path integral formulation to study the previously mentioned models.

Once this is done, we could return to the question of equivalence between frames in scalar-tensor theories. The quantum equivalence was obtained in the minisuperspace approach, a finite dimensional setting. However, several works on the equivalence using QFT \cite{Kamenshchik:2014waa,Falls:2018olk} display quantum inequivalence. Therefore, extending the affine quantisation to an infinite dimensional framework would allow us to tackle this problem from another point of view.

%% file: sections/appendix1.tex
	\chapter{Basics of General Relativity}

\section{Curvature Components}
\label{section curvature components}

From \eqref{1.15a} and \eqref{1.16} we can read off the explicit components of the four-dimensional extrinsic curvature in terms of the variables defined on the three-dimensional hypersurfaces 
\begin{subequations}
	\begin{align}
	K_{ij}=& -\frac{1}{2} \nabla_{(i} n_{j)} \;,\\
	K_{00}=& N^{i}N^{j}K_{ij} \;,\\
	K_{0i}=& N^{j}K_{ij} \;.
	\end{align}
\end{subequations}
Thus, the relevant components of $K_{\mu \nu}$ are the $K_{ij}$, and we can compute them using \eqref{1.15b}
\begin{align}
K_{ij}=-\frac{1}{2} \nabla_{(i} n_{j)} = -\frac{1}{2} n_{(i,j)} + \tensor{\Gamma}{^{\epsilon}_{i j}} n_{\epsilon} = -N \tensor{\Gamma}{^{0}_{ij}} \;.
\end{align}
Using  the metric \eqref{1.13a}, we get
\begin{align}
\tensor{\Gamma}{^{0}_{ij}} =& \frac{1}{2}g^{0\epsilon}\left(g_{\epsilon i,j} + g_{\epsilon j,i} - g_{ij,\epsilon}\right),    \\
=&\frac{1}{2}g^{00}\left(g_{0i,j} + g_{0j,i} - g_{ij,0}\right) + \frac{1}{2}g^{0k}\left(g_{ki,j} + g_{kj,i} - g_{ij,k}\right),\\
=&\frac{1}{2}\left( -\frac{1}{N^2} \right)\left(N_{i,j}+N_{j,i} - \dot{h_{ij}} \right) + \frac{1}{2}\frac{N^{k}}{N^2}\left(h_{ki,j} +h_{kj,i} - h_{ij,k} \right),\\
=&\frac{1}{2N^2}\left[\dot{h_{ij}} - N_{i,j} -N_{j,i}+ N_{a}h^{ka}\left( h_{ki,j} + h_{kj,i} -h_{ij,k}\right) \right]. 
\end{align}
Defining the intrinsic covariant derivative to the three-dimensional hypersurface with metric $h_{ij}(\vec{x})$ as
\begin{subequations}
	\begin{align}
	N_{i;j}&= N_{i,j} - \tensor{^3\Gamma}{^{a}_{ij}} N_{a}, \\
	\tensor{^3\Gamma}{^{a}_{ij}} &=  \frac{1}{2}h^{ak}\left(h_{ki,j} + h_{kj,i} - h_{ij,k}         \right) \;,  \label{1.20b}  
	\end{align}
\end{subequations}
with $\tensor{^3\Gamma}{^{a}_{ij}}$ the intrinsic connection to the hypersurface satisfying $h_{ij;l}=0$, we can also write the connection as
\begin{align}
\tensor{\Gamma}{^{0}_{ij}} = \frac{1}{2N^2}\left[\dot{h}_{ij} -N_{(i;j)}\right] \;.
\end{align}
Therefore, we obtain
\begin{align}
K_{ij}=-\frac{1}{2N}\left[\dot{h}_{ij} -N_{(i;j)}\right] \;.
\end{align}

The contravariant version of $K_{ij}$ can be computed using the inverse spatial metric $h^{ij}$ and reads
\begin{align}
K^{ab} := h^{ai}h^{bj}K_{ij} = \frac{1}{2N}\left( \dot{h}^{ab} + N^{(a;b)} \right) \;,
\end{align}
where we used $\dot{h}_{ai} h^{ib}=-h_{ai}\dot{h}^{ib}$, that comes from $h_{ai} h^{ib} =\delta_{a}^{b}$. Hence, $K_{ij}$ will be essentially related to the first temporal derivative of the metric. 

\section{ADM Action}
\label{section adm action}

We are now able to compute the Ricci scalar by assuming that it is possible to foliate the whole spacetime with space-like submanifolds. Using \eqref{1.13a} and \eqref{1.22}, the Christoffel symbols read

\begin{subequations}\label{1.24}
	\begin{eqnarray}
	\tensor{\Gamma}{^{0}_{00}} &=& \frac{\dot{N}}{N} + \frac{N^i N_{,i}}{N} - \frac{N^iN^j}{N}K_{ij} \;, \\
	\tensor{\Gamma}{^{0}_{0i}} &=& \frac{N_{,i}}{N} - \frac{N^{j}}{N}K_{ij} \:, \\
	\tensor{\Gamma}{^{0}_{ij}} &=& -\frac{K_{ij}}{N} \;, \\
	\tensor{\Gamma}{^{i}_{00}} &=& N h^{ij}\left( \frac{N_{j}}{N} \right)^{\dot{\,}} +  \frac{1}{2}h^{ij}\left(N^{2} -N_{m}N^{m}\right)_{,j} -\frac{N_{,j}}{N}N^{i}N^{j} + \frac{N^{i}N^{j}N^{k}}{N}K_{jk} \;,\\
	\tensor{\Gamma}{^{i}_{j 0}} &=&N\left[ -\tensor{K}{^{i}_{j}} + \left(\frac{N^i}{N} \right)_{;j} + \frac{N^{i}N^{m}}{N^2}K_{jm}\right] \;, \\
	\tensor{\Gamma}{^{i}_{jk}} &=&  \tensor{\,^3\Gamma}{^{i}_{jk}} + \frac{N^{i}}{N}K_{jk} \;.
	\end{eqnarray}
\end{subequations}
Now we can use equations \eqref{1.24} to write the components of the Ricci tensor
\begin{align}
R_{\mu \nu} := \tensor{R}{^{\lambda}_{\mu \lambda \nu}},   
\end{align}
which read
\begin{subequations}\label{1.26}
	\begin{eqnarray}
	R_{00} &=& Nh^{ij}\dot{K}_{ij} + N \tensor{N}{^{,i}_{;i}} - 2N \tensor{N}{^{i}_{;j}} \tensor{K}{^{j}_{i}} -2NN^{i} \tensor{K}{^{j}_{i;j}} + N N^{i}K_{,i}  +N^{2}K^{ij}K_{ij} \nonumber \\
	&\space & + N^{i}N^{j} \,^3R_{ij} + N^{i}N^{j}K_{ij}K -2N^{i}N^{j} \tensor{K}{_{i}^{k}}K_{kj} -\frac{N^j N^i}{N}\dot{K}_{ij} - \frac{N^{j} N^{i}}{N} N_{,i;j} \nonumber \\
	&\space& + 2\frac{N^j N^i}{N}\tensor{N}{^{k}_{;i}} K_{kj} + \frac{N^{i} N^{j} N^k}{N}K_{ik;j} \;,  \\
	R_{0i} &=& -\frac{N^{j}}{N}\dot{K}_{ij} - \frac{N^j}{N}N_{,i;j} + K_{kj}\frac{N^{j}}{N} \tensor{N}{^{k}_{;i}} 
	+ K_{ki}\frac{N^j}{N}\tensor{N}{^{k}_{;j}}
	-2N^{j} \tensor{K}{_{j}^{k}} K_{ki} -N \tensor{K}{^{j}_{i;j}} \nonumber \\
	&\space & + N K_{,i} + N^{j} \,^3R_{ji}
	+ N^{j}K_{ji}K + \frac{N^{j}N^{k}}{N}K_{ki;j} \;, \\
	R_{ij} &=& \frac{1}{N}\left(-\dot{K}_{ij} -N_{,i;j} + \tensor{N}{^{k}_{;i}}K_{kj} + \tensor{N}{^{k}_{;j}}K_{ik}\right) -2\tensor{K}{_{i}^{k}}K_{kj} + \,^3R_{ij} + \frac{N^{k}}{N}K_{ij;k} + K K_{ij} \;. \nonumber\\
	\end{eqnarray}
\end{subequations}
where $K := h^{ij} K_{ij}$, $h_{ij;k}=0$ and $\,^3R_{ij}$ is the Ricci tensor for three-dimensional hypersurfaces constructed from $\tensor{\,^3\Gamma}{^{a}_{ij}}$ defined in \eqref{1.20b}.

It is important to stress once again that all quantities in the right-hand side of equations \eqref{1.26} are built from tensors defined on three-dimensional hypersurfaces, so they are invariant under a general spatial diffeomorphism $x^{i} \rightarrow {x'}^{i} = {x'}^{i}(x^{j})$ (and $x'^{0}=x^{0}$). Thus, it turns out that the components of the 4-dimensional Ricci tensor, $R_{00},R_{0i},R_{ij}$  transform, respectively, as a scalar, a three-vector and a three-tensor under this change of coordinates.

We can now compute the Ricci scalar using equations \eqref{1.13a} and \eqref{1.26}
\begin{align}\label{1.27}
R = \frac{1}{N}\left( -2\dot{K} -2\tensor{N}{^{,j}_{;j}} + 2N^{i}K_{,i}\right) + K_{ij}K^{ij} + K^2 + \,^3R \;,
\end{align}
where we have defined the three-dimensional Ricci scalar $\,^3R$ using $\tensor{\,^3\Gamma}{^{a}_{ij}}$, and where was also used the identity (that implies \eqref{1.23})
\begin{align}
h^{ij}\dot{K}_{ij} = \left(h^{ij}K_{ij}\right)^{\dot{\,}} - K_{ij}\dot{h}^{ij} = \dot{K} -2NK^{ij}K_{ij} + 2N^{i;j}K_{ij} \;.
\end{align}

We will need as well the expression for the variation of the spatial metric determinant. From \eqref{1.13a}, we have  $\sqrt{-g}=Nh^{1/2}$, and using \eqref{1.22} allows us to obtain
\begin{subequations}
	\begin{align}
	\label{1.29a}
	\delta h^{\frac{1}{2}} = -\frac{1}{2} h^{\frac{1}{2}}h_{ij}\delta h^{ij}=\frac{1}{2} h^{\frac{1}{2}}h^{ij}\delta h_{ij} \;,
	\end{align}
	from which follows that
	\begin{align}
	2\dot{h}^{\frac{1}{2}} = {h}^{\frac{1}{2}}h^{ij}\dot{h}_{ij} = -2N{h}^{\frac{1}{2}}K + 2{h}^{\frac{1}{2}} \tensor{N}{^{i}_{;i}}\;.
	\end{align}
\end{subequations}

However, since $h^{\frac{1}{2}} \tensor{A}{^{i}_{;i}}= \left(h^{\frac{1}{2}}A^{i}\right)_{;i}$, with $A^{i}$ being a general three-vector defined on constant hypersurfaces $t$, we can use the whole technology developed until now to write the Einstein-Hilbert Lagrangian density as
\begin{eqnarray}
\cal{L} &:= & \sqrt{-g} R = N h^{\frac{1}{2}}R = -2h^{\frac{1}{2}}\dot{K} -2h^{\frac{1}{2}} \tensor{N}{^{,i}_{;i}} + 2h^{\frac{1}{2}}N^{i}K_{,i} \nonumber \\    
&\space & + h^{\frac{1}{2}} N\left(K^2 + K_{ij}K^{ij} + \,^3R \right) -2\dot{h}^{\frac{1}{2}}K - 2Nh^{\frac{1}{2}}K^2 + 2h^{\frac{1}{2}} \tensor{N}{^{i}_{;i}} K \nonumber \;, \\
&=& Nh^{\frac{1}{2}}\left( -K^2 + K_{ij}K^{ij} + \,^3R \right) -2 \left(h^{\frac{1}{2}}K\right)^{\dot{\,}} + 2\left( h^{\frac{1}{2}}KN^{i} -h^{\frac{1}{2}}h^{ji}N_{,j} \right)_{;i} \;.
\end{eqnarray}
Finally, the action takes the form
\begin{eqnarray}\label{1.31}
S &=& \int dt d^3x Nh^{\frac{1}{2}}\left(-K^2 + K_{ij}K^{ij} + \,^3R \right) - 2\int dt d^3x \left(h^{\frac{1}{2}}K\right)^{\dot{\,}} \nonumber \\
&\space &  + 2\int dt d^3x \left( h^{\frac{1}{2}}KN^{i} -h^{\frac{1}{2}}h^{ji}N_{,j}        \right)_{;i} \;.
\end{eqnarray}
Let us rename the last two terms as
\begin{subequations}\label{1.32}
	\begin{eqnarray}
	S_1 &:= & -2\int dt d^3x \left(h^{\frac{1}{2}}K\right)^{\dot{\,}} \;, \\
	S_2 &:= & 2\int dt d^3x \left( h^{\frac{1}{2}}KN^{i} -h^{\frac{1}{2}}h^{ji}N_{,j} \right)_{;i} \;. \label{1.32b}
	\end{eqnarray}
\end{subequations}
$S_{1}$ and $S_{2}$ are surface terms and, as a consequence, they will not contribute to the equations of motion; we can then safely neglect them.

\subsection{ADM Equations of Motion}
\label{section equations of motion}

Let us write down the equations of motion. Starting from the lapse function, we have:
\begin{align}\label{1.33}
\frac{\delta S}{\delta N} = 0 \quad \rightarrow \quad \frac{d\cal{L}}{dN} - \left(\frac{d\cal{L}}{d \dot{N}}\right)^{\dot{\,}} - \left( \frac{d\cal{L}}{dN_{,i}}\right)_{;i} = \frac{d\cal{L}}{dN} = \frac{\partial \cal{L}}{\partial N} + \frac{\partial \cal{L}}{\partial K_{ij}}\frac{\partial K_{ij}}{\partial N} = 0 \;,
\end{align}
where we have used the independence of the action from derivatives of the lapse function, and the explicit and implicit (through $K_{ij}$) dependence of $N$. To compute the variation of the Lagrangian density with respect to $K_{ij}$, we can lower the index of the contravariant extrinsic curvature $K^{ij}$ and  express its squared trace $K^2$, using the inverse of the spatial metric $h^{ij}$
\begin{align}
{\cal{L}}_{K} := {h^{\frac{1}{2}}} N \left( {h^{ki}}{h^{jl}}{- h^{ij}}{ h^{kl}} \right) K_{ij} K_{kl} \;,
\end{align}
and using this into \eqref{1.33}, we obtain
\begin{align}
\label{1.34}
\frac{\partial \cal{L}}{\partial K_{ab}} = \frac{\partial \cal{L}_{K}}{\partial K_{ab}} = h^{\frac{1}{2}}N\left(h^{ki}h^{jl} - h^{ij}h^{kl} \right)K_{kl}\frac{\partial K_{ij}}{\partial K_{ab}} + h^{\frac{1}{2}}N\left(h^{ki}h^{jl} - h^{ij}h^{kl} \right)K_{ij}\frac{\partial K_{kl}}{\partial K_{ab}} \;.
\end{align}

With the definition
\begin{align}
\delta^{ab}_{ij} := \frac{1}{2}\left( \delta^{a}_{i}\delta^{b}_{j} + \delta^{a}_{j}\delta^{b}_{i}  \right) = \frac{\partial K_{ij}}{\partial K_{ab}} \;,
\end{align}
together with
\begin{align}
\frac{\partial K_{ab}}{\partial N} = -\frac{1}{N}K_{ab} \;,
\end{align}
we obtain
\begin{align}
\frac{\partial \cal{L}}{\partial K_{ab}}\frac{\partial K_{ab}}{\partial N} = -2h^{\frac{1}{2}}\left(K^{ij}K_{ij} - K^2  \right) \;,
\end{align}
and we can write down the first dynamical equation as
\begin{align}
\frac{\delta S}{\delta N} = 0 \rightarrow \frac{d \cal{L}}{d N}=  \left( -K^2 + K_{ij}K^{ij} - {^{3}R} \right) = 0 \;.
\end{align}

Now, in order to get the second set of equations of motion, we have to apply the least action principle to the shift vector function $N^i$
\begin{align}
\frac{\delta S}{\delta N_i} = 0,  \quad \rightarrow \quad
\frac{d\cal{L}}{dN_i} - \left(\frac{d\cal{L}}{d\dot{N}_{i}}\right)^{\dot{\,}} - \left( \frac{d\cal{L}}{dN_{i;j}}\right)_{;j} = 0, 
\quad  \rightarrow \quad  \left( \frac{d\cal{L}}{dN_{i;j}}\right)_{;j} = 0 \;,
\end{align}
from which we can see that the action does not contain derivatives of the shift function $N_{i}$, that appears only in the definition of $K_{i,j}$
\begin{align}
\frac{d\cal{L}}{dN_{i;j}} = \frac{\partial \cal{L}}{\partial N_{i;j}} + \frac{\partial \cal{L}}{\partial K_{ab}} \frac{\partial K_{ab}}{\partial N_{i;j}} = \frac{\partial \cal{L}}{\partial K_{ab}} \frac{\partial K_{ab}}{\partial N_{i;j}} \;,
\end{align}
and the field equations are reduced to
\begin{align}
\left( \frac{\partial\cal{L}}{\partial K_{ab}}\frac{\partial K_{ab}}{\partial N_{i;j}} \right)_{;j} = 0 \;.
\end{align}
Using $\frac{\partial K_{ab}}{\partial ( N_{i,j})} = \frac{1}{N}\delta^{ij}_{ab}$, we can write down the second dynamical equation
\begin{align}
\frac{\partial S}{\partial N_{i}} = 0 \quad \rightarrow \quad 2h^{\frac{1}{2}} \left(K^{ij} - h^{ij}K  \right)_{;j} = 0 \;.
\end{align}

Finally, in order to obtain the last dynamical equation (the one referred to the spatial metric $h_{ij}$), it is useful to consider the following expressions for the variation of the three-dimensional Christoffel symbols and the spatial Ricci tensor
\begin{eqnarray}
\delta \,^3R_{ij} &=& \delta \tensor{\,^3\Gamma}{^{a}_{ij;a}} - \delta \tensor{\,^3\Gamma}{^{a}_{ia;j}} \;, \label{1.39} \\
\delta \tensor{\,^3\Gamma}{^{a}_{ij}} &=& \frac{1}{2}h^{ak}\left( \delta h_ {ki;j} + \delta h_{kj;i} -\delta h_{ij;k} \right) \;, \label{1.40}
\end{eqnarray}
from which we compute
\begin{align}
\frac{\delta S}{\delta h_{ij}} = 0  \;. 
\end{align}
In the end, we obtain
\begin{align}
\dot{K}_{ij} =  N\left( ^{3}{R_{ij}} + KK_{ij} -2\tensor{K}{_{i}^{m}} K_{mj}\right) -N_{,i;j} + N_{m;i}\tensor{K}{^{m}_{j}} + N_{m;j} \tensor{K}{^{m}_{i}} + N^{m}K_{ij;m} \;.
\end{align}

\section{Hamiltonian for General Relativity}
\label{section hamiltonian gr}

\subsection{Finding the Hamiltonian}

We show in this appendix how to find the total Hamiltonian in GR, and that the constraints are conserved in time. From \eqref{2.2c}, we have
\begin{align}
\Pi := h_{ij}\Pi^{ij} =2h^{1/2}K \qquad\rightarrow\qquad \Pi^{ij}=-h^{1/2}\left( K^{ij}-\frac{h^{ij}}{2h^{1/2}}\Pi \right) \;, \nonumber 
\end{align}
leading to
\begin{subequations}\label{2.33}
	\begin{eqnarray}
	\label{2.33a} K^{ij}&=&-h^{-1/2}\left( \Pi^{ij}-\frac{h^{ij}}{2}\Pi \right) \;,\\
	\label{2.33b}K_{ij}&=&-h^{-1/2}\left( \Pi_{ij}-\frac{h_{ij}}{2}\Pi \right) \;. 		
	\end{eqnarray}
\end{subequations}
Plugging \eqref{1.22} in \eqref{2.33b}, we find
\begin{align}
\label{2.34}
\dot{h}_{ij}=2Nh^{-1/2}\left(\Pi_{ij}-\frac{h_{ij}}{2} \Pi \right) + N_{(i;j)} \;.
\end{align}
Substituting \eqref{2.2}, \eqref{2.33} and \eqref{2.34} in \eqref{2.32}, we get
\begin{eqnarray}
\mathcal{H}_{C}&=&2Nh^{-1/2} \left( \Pi^{ij} \Pi_{ij} -\frac{\Pi^{2}}{2} \right)+2\Pi^{ij}N_{i;j}\nonumber \\ &\space&-Nh^{1/2}h^{-1}\left[\left(\Pi^{ij}-\frac{h^{ij}}{2}\Pi\right)\left(\Pi_{ij}-\frac{h_{ij}}{2}\Pi \right)-\frac{1}{4}\Pi^{2} \right]-Nh^{\frac{1}{2}}\,^3R \nonumber \; \\
&=& N\left( G_{ijkl} \Pi^{ij} \Pi^{kl} -h^{\frac{1}{2}} \,^3R \right) + 2 \Pi^{ij} N_{i;j} \;,
\end{eqnarray}
where
\begin{align}
G_{ijkl} := \frac{1}{2} h^{-\frac{1}{2}} \left( h_{ik}h_{jl}+h_{il}h_{jk}-h_{ij}h_{kl}\right) \;.
\end{align}
The total Hamiltonian then takes the form
\begin{eqnarray}
\label{2.37}
H_{T}&=& \int d^{3}x \mathcal{H}_{C} + \int d^{3}x \left(\lambda P+\lambda^{i}P_{i} \right) \nonumber \\
&=& \int d^{3}x \left( N\mathcal{H}_{0}+N_{i} \mathcal{H}^{i}+\lambda P+\lambda^{i}P_{i} \right) \;,
\end{eqnarray}
where we defined 
\begin{subequations}\label{2.38}
	\begin{eqnarray}
	\mathcal{H}_{0} &:= & G_{ijkl} \Pi^{ij} \Pi^{kl}-h^{\frac{1}{2}} \,^3R \;, \label{2.38a}\\
	\mathcal{H}^{i} &:= & -2 \tensor{\Pi}{^{ij}_{;j}} \;. \label{2.38b}
	\end{eqnarray}
\end{subequations}
Note that, since we are working with closed spaces, we neglected the surface term 
\begin{align}
S_{3} := \left(2\Pi^{ij} N_{i}\right)_{;j}=\left(2\Pi^{ij}N_{i}\right)_{,j} \;.\label{2.39}
\end{align}

Noting that $\Pi_{ij}$ is a three-space tensor density with weight 1 (see \eqref{2.2c}), we get
\begin{align}
\label{2.40}
\tensor{\Pi}{^{ij}_{;a}} := \tensor{\Pi}{^{ij}_{\textcolor{blue}{,}a}}+\tensor{\,^3\Gamma}{^{i}_{ka}}\Pi^{kj}+\tensor{\,^3\Gamma}{^{j}_{ka}}\Pi^{ik}-\tensor{\,^3\Gamma}{^{l}_{al}}\Pi^{ij} \;.
\end{align}

\subsection{Time Invariance of the GR Constraints}

We must verify the two constraints \eqref{2.2a} and \eqref{2.2b} are conserved in time. Using the following Poisson's brackets
\begin{subequations}\label{2.41}
	\begin{eqnarray} 
	\left\lbrace N\left(x^{i},t \right), P\left(y^{i},t\right) \right\rbrace &=& \delta^{(3)} \left(x^i-y^i\right) \;, \\
	\left\lbrace N_{i}\left(x^{l},t \right), P^{j}\left(y^{l},t\right) \right\rbrace &=& \delta^{j}_{i} \delta^{(3)} \left(x^l-y^l\right) \;, \\
	\left\lbrace h_{ij}\left(x^{a},t \right), \Pi^{kl}\left(y^{a},t\right) \right\rbrace &=& \delta^{kl}_{ij} \delta^{(3)} \left(x^a-y^a\right) \;, \label{2.41c}
	\end{eqnarray}
\end{subequations}
where the others are equal to zero, we obtain
\begin{subequations}
	\label{2.42}
	\begin{eqnarray}
	\dot{P}\left(x^{i},t\right) &=& \left\lbrace P\left(x^{i},t \right),H_{T}\right\rbrace =\int d^{3}y \left\lbrace P\left(x^{i},t \right), N\left(y^{i},t\right) \right\rbrace \mathcal{H}_{0}(y^{i})  \nonumber \\
	&=& -\int d^{3}y \delta^{(3)} \left(x^{i}-y^{i} \right) \mathcal{H}_{0}(y^{i}) = -\mathcal{H}_{0}(x^{i}) \simeq 0  \nonumber \\
	&\rightarrow &\mathcal{H}_{0}(x^{i}) \simeq 0  \;, \label{2.42a}\\
	\dot{P}^{i}\left(x^{i},t\right) &=& \left\lbrace P^{i}\left(x^{i},t \right),H_{T}\right\rbrace =\int d^{3}y \left\lbrace P^{i}\left(x^{i},t \right), N_{j}\left(y^{i},t\right) \right\rbrace \mathcal{H}^{j}(y^{i}) \nonumber \\
	&=& -\int d^{3}y \delta^{(3)} \left(x^{i}-y^{i} \right) \delta^{i}_{j} \mathcal{H}^{j}(y^{i}) = -\mathcal{H}^{i}(x^{i}) \simeq 0 \nonumber \\
	&\rightarrow & \mathcal{H}^{i}(x^{i}) \simeq 0 \;.  \label{2.42b}
	\end{eqnarray}
\end{subequations}
Thus, the time conservation of \eqref{2.2a} and \eqref{2.2b} implies new relations between $\Pi^{ij}$ and $h_{ij}$, that is, \eqref{2.42a} and \eqref{2.42b}, where $ \mathcal{H}_{0}(x^{i})$ and $\mathcal{H}^{i}(x^{i})$ are defined by \eqref{2.38a} and \eqref{2.38b} respectively. The constraints \eqref{2.42a} and \eqref{2.42b} are, therefore, \textit{secondary constraints}. Similarly to what we have done before, time conservation of \eqref{2.42a} and \eqref{2.42b} must now be verified. For this computation, the following formulas will be useful
\begin{subequations}
	\label{2.43}
	\begin{eqnarray}
	\frac{\delta h_{ij}(x)}{\delta h_{kl}(x')} &=& \delta^{kl}_{ij} \delta^{(3)} \left( x-x' \right) = \frac{\delta \Pi^{kl}(x)}{\delta \Pi^{ij}(x')} \;, \\
	\frac{\delta \left( h_{ij,m}(x)\right)}{\delta h_{kl}(x')} &=& \delta^{kl}_{ij} \frac{\partial \left[\delta^{(3)} \left( x-x' \right)\right]}{\partial x^{m}} = \frac{\delta \tensor{\Pi}{^{kl}_{,m}}(x)}{\delta \Pi^{ij}(x')} \;, \\
	\delta \left( h^{\frac{1}{2}} \,^3R \right) &=& h^{\frac{1}{2}} h^{ij} h^{kl} \left( \delta h_{ik;j;l} -\delta h_{ij;k;l}\right) -h^{\frac{1}{2}}\left( \,^3R^{ij}-\frac{1}{2}h^{ij}\,^3R\right) \delta h_{ij} \;, \\
	\tensor{\Pi}{^{ia}_{;a}}&=& \tensor{\Pi}{^{ia}_{,a}}+\tensor{\,^3\Gamma}{^{i}_{ka}}\Pi^{ka} \;,
	\end{eqnarray}
\end{subequations}
where we used three-dimensional versions of Eqs.~\eqref{1.39}, \eqref{1.40}, \eqref{1.29a} in the third equation, and Eq.~\eqref{2.40} in the fourth one. The following properties of the Dirac delta will also be useful
\begin{subequations}
	\label{2.44}
	\begin{eqnarray}
	\frac{\partial \delta^{(3)}\left(x-x'\right)}{\partial x^{m}}&=&-\frac{\partial \delta^{(3)}\left(x-x'\right)}{\partial {x'}^{m}} \;, \\
	F(x) \delta^{(3)}\left(x-x'\right)&=& F(x')\delta^{(3)}\left(x-x'\right) \;. \label{2.44b}
	\end{eqnarray}
\end{subequations}
The first Poisson bracket is
\begin{eqnarray}
\left\lbrace \mathcal{H}_{0}(x),\mathcal{H}_{0}(x') \right\rbrace &=& \left\lbrace G_{ijkl}(x)\Pi^{ij}(x)\Pi^{kl}(x)-h^{\frac{1}{2}}(x) \,^3R(x), G_{abcd}(x')\Pi^{ab}(x')\Pi^{cd}(x')-h^{\frac{1}{2}}(x')\,^3R(x')\right\rbrace \nonumber \\
&=& 2 G_{ijkl}(x)\Pi^{ij}(x)\Pi^{ab}(x') \Pi^{cd}(x') \left\lbrace \Pi^{kl}(x), G_{abcd}(x') \right\rbrace \nonumber \\
&\space& -2 G_{ijkl}(x)\Pi^{ij}(x) \left\lbrace \Pi^{kl}(x), h^{\frac{1}{2}}(x') \,^3R(x') \right\rbrace \nonumber \\
&\space& -2 G_{abcd}(x')\Pi^{ab}(x') \left\lbrace \ h^{\frac{1}{2}}(x) \,^3R(x), \Pi^{cd}(x') \right\rbrace \nonumber \\
&\space& +2 G_{abcd}(x') \Pi^{ab}(x') \Pi^{ij}(x) \Pi^{kl}(x) \left\lbrace G_{ijkl}(x), \Pi^{cd}(x') \right\rbrace \;.
\end{eqnarray}
Using equations \eqref{2.41c} and \eqref{2.44b}, it can be shown that the first and last terms cancel each other. Then, we obtain
\begin{align}
\left\lbrace \Pi^{kl}(x), h^{\frac{1}{2}}(x') \,^3R(x') \right\rbrace = -\int d^{3}z \frac{\delta \Pi^{kl}(x)}{\delta \Pi^{ab}(z)} \frac{\delta \left[h^{\frac{1}{2}}(x') \,^3R(x')\right]}{\delta h_{ab}(z)} \;.
\end{align}
Using this result with relations \eqref{2.43} and \eqref{2.44}, we obtain, after a lengthy calculation
\begin{align}
\label{2.47}
\left\lbrace \mathcal{H}_{0}(x),\mathcal{H}_{0}(x') \right\rbrace = \left[h^{ij}(x) \mathcal{H}_{j}(x)+h^{ij}(x') \mathcal{H}_{j}(x')\right] \frac{\partial \delta^{(3)}\left(x-x'\right)}{\partial x^{i}} \simeq 0 \;,
\end{align}
where $\mathcal{H}_{j} := h_{jk} \mathcal{H}^{k}$ is given by \eqref{2.38b}. Using the same methods described above and eq. \eqref{1.40} when terms implying $\left\lbrace \Pi^{ab}(x), \tensor{\,^3\Gamma}{^{i}_{jk}}(x') \right\rbrace$ appear, we find for the other Poisson brackets
\begin{subequations}
	\label{2.48}
	\begin{eqnarray}
	\left\lbrace \mathcal{H}_{i}(x),\mathcal{H}_{0}(x') \right\rbrace &=& \mathcal{H}_{0}(x) \frac{\partial \delta^{(3)}\left(x-x'\right)}{\partial x^{i}} \simeq 0 \;,\\
	\left\lbrace \mathcal{H}_{i}(x),\mathcal{H}_{j}(x') \right\rbrace &=& \left[\mathcal{H}_{j}(x) \frac{\partial}{\partial x^{i}} +\mathcal{H}_{i}(x') \frac{\partial}{\partial x^{j}} \right] \delta^{(3)}\left(x-x'\right) \simeq 0 \;.
	\end{eqnarray}
\end{subequations}

Inserting equations \eqref{2.47} and \eqref{2.48} in the computation of $\dot{\mathcal{H}}_{0} := \left\lbrace \mathcal{H}_{0}, H_{T} \right\rbrace$ and $\dot{\mathcal{H}}_{i} := \left\lbrace \mathcal{H}_{i}, H_{T} \right\rbrace$, we verify that the constraints \eqref{2.42} are identically conserved in time. Thus, the constraints for the theory will be given by \eqref{2.2a}, \eqref{2.2b} and \eqref{2.42}. Furthermore, from \eqref{2.47} and \eqref{2.48} and from the fact that the constraints \eqref{2.2a} and \eqref{2.2b} have null Poisson  brackets when evaluated with constraints \eqref{2.42}, we conclude that they are all first-class constraints, \textit{i.e.} the Poisson brackets with all other constraints vanishes, so they are generators of gauge transformations.

\subsection{Physical Interpretation of GR constraints}

The total GR Hamiltonian is given by \eqref{2.37}. To verify this assertion, we compute Hamilton equations coming from \eqref{2.37} and compare them with the EFE. We have
\begin{subequations} \label{2.49}
	\begin{eqnarray} 
	\dot{N}&=&\left\lbrace N, H_{T} \right\rbrace=\lambda \;, \label{2.49a}\\
	\dot{N}^{i}&=&\left\lbrace N^{i}, H_{T} \right\rbrace=\lambda^{i} \;, \label{2.49b} \\
	\dot{h}_{ij}(x)&=&\left\lbrace h_{ij}(x), H_{T} \right\rbrace=\int d^{3}y \left[\left\lbrace h_{ij}(x), \mathcal{H}_{0}(y) \right\rbrace N(y) + \left\lbrace h_{ij}(x), \mathcal{H}^{k}(y) \right\rbrace N_{k}(y) \right] \nonumber \\
	&=& \int d^{3} y \left[2 G_{abcd}(y) \Pi^{ab}(y)\left\lbrace h_{ij}(x),\Pi^{cd}(y)\right\rbrace N(y) + 2 \delta^{ak}_{ij} N_{k;a} \delta^{(3)}\left(x-y\right)\right] \nonumber \\
	&=& \int d^{3} y \left[2 G_{abcd}(y) \Pi^{ab}(y) \delta^{cd}_{ij} N(y)+N_{i;j}+N_{j;i} \right] \delta^{(3)}\left(x-y\right) \nonumber \\
	&=&  h^{-\frac{1}{2}} \left[2 \Pi_{ij}(x)-h_{ij}(x)\Pi(x)\right]N(x) + N_{i;j}(x)+N_{j;i}(x) \;, \label{2.49c} \\
	\dot{\Pi}^{ij} &=& \left\lbrace \Pi^{ij},H_{T} \right\rbrace = -N h^{\frac{1}{2}} \left(\,^3R^{ij}-\frac{1}{2} \,^3R h^{ij}\right) 
	+\frac{1}{2} N h^{-\frac{1}{2}}h^{ij} \left(\Pi_{kl} \Pi^{kl}- \frac{1}{2}\Pi^{2}\right) \nonumber\\
	&\space&-2N h^{-\frac{1}{2}}\left(\Pi^{ac} \tensor{\Pi}{_{c}^{b}}-\frac{1}{2} \Pi \Pi^{ab}\right)+h^{\frac{1}{2}} \left(N^{;i;j}-h^{ij}\tensor{N}{^{;c}_{;c}}\right) \nonumber \\
	&\space& +h^{\frac{1}{2}} \left(h^{-\frac{1}{2}}N^{k}\Pi^{ij}\right)_{;k}-2 \Pi^{k(i} \tensor{N}{^{j)}}_{;k} \;. \label{2.49d}
	\end{eqnarray}
\end{subequations}
Equations \eqref{2.49a} and \eqref{2.49b}, which come from equation \eqref{2.41}, combined with constraints \eqref{2.2a} and \eqref{2.2b}, show us that $N$ and $N^i$ are, actually, mere Lagrange multipliers for the secondary constraints \eqref{2.42a} and \eqref{2.42b} since their dynamical evolution, given by \eqref{2.49a} and \eqref{2.49b} is completely arbitrary. It is possible to consider, then, $h_{ij}$ and $\Pi^{ij}$ as the real dynamical variables of the theory. A rigorous proof of this statement may be found  on reference \cite{Hanson:1976cn} \footnote{On page 240.}. However, this result was already expected since the only variable that possesses a dynamical equation is $h_{ij}$ (see equation \eqref{1.41}).

Equation \eqref{2.49c} simply corresponds to the velocity definition $\dot{h}_{ij}$ in terms of the momenta $\Pi^{ij}$. It is equivalent to the equation defining $\Pi^{ij}$, given by \eqref{2.2c}, using on it the definition of $K_{ij}$, stated in \eqref{1.22}. Equation \eqref{2.49d} corresponds to equation \eqref{1.41}. It can be checked by multiplying \eqref{2.49d} by $h_{ij}$ (using $h_{ij} \dot{\Pi}^{ij} = \dot{\Pi} - \dot{h}_{ij} \Pi^{ij}$), substituting the result back in \eqref{2.49d} and, using the constraints \eqref{2.42} and equations \eqref{2.49c} and \eqref{1.22}, re-obtain \eqref{1.41}. At last, let us consider constraints \eqref{2.42}. Using \eqref{2.2c} (which, as it has previously been seen, is equivalent to equation \eqref{2.49c}), it may be shown that \eqref{2.42a} corresponds to the Lagrangian constraint \eqref{1.33}, and \eqref{2.42b} to the Lagrangian constraint \eqref{1.38}. Thus, it has been shown that the Hamiltonian \eqref{2.37} provides all Einstein equations. But we still have to determine what is the meaning of constraints \eqref{2.42}. In order to answer this, it must be examined what kind of transformations these constraints induce on the canonical variables $h_{ij}$ and $\Pi^{ij}$.

\begin{itemize}
	\item The constraint $\mathcal{H}_i \approx 0$
\end{itemize}
\begin{eqnarray}
\delta h_{ij}(x) &=& \left\{ h_{ij}(x), \int d^3y \xi^k(y) \mathcal{H}_k(y) \right\} \nonumber \\ 
&=& -2 \left\{ h_{ij}(x), \int d^3y \xi^k h_{kl} \tensor{\Pi}{^{la}_{;a}}(y) \right\} \nonumber \\
&=& -2 \int d^3y \xi_l(y) \left\{ h_{ij}(x), \tensor{\Pi}{^{la}_{;a}}(y) \right\} \nonumber \\
&=& -2 \int d^3y d^3z \xi_l (y) \frac{\delta h_{ij}(x)}{\delta h_{mn}(z)} \frac{\delta \tensor{\Pi}{^{la} _{;a}}(y)}{\delta \Pi^{mn}(z)} \nonumber \\ 
&=& -2 \int d^3y d^3z \xi_l (y) \delta^{mn}_{ij} \delta^3 (x-z) \frac{\delta \tensor{\Pi}{^{la} _{;a}}(y)}{\delta \Pi^{mn}(z)} \nonumber \\
&=& -2 \int d^3y \xi_l (y) \left( \frac{\delta \Pi^{la}(y)}{\delta \Pi^{ij}(x)} \right)_{;a} \nonumber \\
&=& 2 \int d^3y \xi_{l;a}(y) \frac{\delta \tensor{\Pi}{^{la}}(y)}{\delta \Pi^{ij}(z)} \nonumber \\ 
&=& 2 \int d^3y \xi_{l;a}(y) \delta^{la}_{ij} \delta^3(x-y) \nonumber \\
&=& \xi_{i;j} (x) + \xi_{j;i}(x) \;,\label{2.50}
\end{eqnarray}
matches with the previous result. The computation of $\delta \Pi^{ij}$ yields

\begin{eqnarray}
\delta \Pi^{ij}(x) &=& \left\{ \Pi^{ij}(x), \int d^3y \xi^k(y) \mathcal{H}_{k}(y) \right\} \nonumber \\
&=& -2 \left\{ \Pi^{ij}(x), \int d^3y \xi^k h_{kl} \left[ \tensor{\Pi}{^{la}_{,a}} + \tensor{\,^3\Gamma}{^l_{ab}} \Pi^{ab} \right] \right\} \nonumber \\
&=& -2 \int d^3y \xi^k \left( \tensor{\Pi}{^{la}_{,a}} + \tensor{\,^3\Gamma}{^l_{ab}} \Pi^{ab} \right) \left\{ \Pi^{ij}(x), h_{kl}(y) \right\} - 2\int d^3y \xi_l \Pi^{ab} \left\{ \Pi^{ij}(x), \tensor{\,^3\Gamma}{^l_{ab}}(y) \right\} \nonumber \\ 
&=& -2\int d^3y \xi^k \left( \tensor{\Pi}{^{la}_{,a}} + \tensor{\,^3\Gamma}{^l_{ab}} \Pi^{ab} \right)(-)\delta^{ij}_{kl} \delta^3(x-y)  \nonumber  \\
&\space &- 2 \int d^3y \xi_l \Pi^{ab} \frac{1}{2} \left( 2 h_{ma,b} - h_{ab,m} \right) \left\{ \Pi^{ij}(x), h^{lm}(y) \right\}  \nonumber \\
&\space &- \int d^3y \xi^m \Pi^{ab} \left[ 2 \left\{ \Pi^{ij}(x), h_{ma,b}(y) \right\} - \left\{ \Pi^{ij}(x), h_{ab,m}(y) \right\} \right] \nonumber 
\end{eqnarray}
\begin{eqnarray}
&=& \xi^{(i} \tensor{\Pi}{^{j)a}_{,a}} + \xi^{(i}\tensor{\Gamma}{^{j)}_{ab}} \Pi^{ab} -2 \int d^3y \xi_l \Pi^{ab} h_{md} \tensor{\,^3\Gamma}{^d_{ab}} \frac{1}{2} \left( h^{il} h^{jm} + h^{im}h^{jl} \right) \delta^3 (x-y)  \nonumber  \\
&\space &- \int d^3y \xi^m \Pi^{ab} 2 \left[ (-) \delta^{ij}_{am} \frac{\partial \left(\delta^3 (x-y)\right)}{\partial y^b} + \delta^{ij}_{ab} \frac{\partial \left(\delta^3 (x-y)\right)}{\partial y^m} \right] \nonumber \\
&=& \xi^{(i} \tensor{\Pi}{^{j)a}_{,a}} + \xi^{(i} \tensor{\,^3\Gamma}{^{j)}_{ab}} \Pi^{ab} - \xi^{(i} \tensor{\,^3\Gamma}{^{j)}_{ab}} \Pi^{ab} - \frac{\partial}{\partial x^b} \int d^3y \xi^{(i}\Pi^{j)b} \delta^3 (x-y) \nonumber  \\
&\space &+ \frac{\partial}{\partial x^m} \int d^3y \xi^m \Pi^{ij} \delta^3 (x-y) \nonumber \\
&=& \left( - \tensor{\xi}{^{(i}_{,b}} \Pi^{j)b} + \tensor{\Pi}{^{ij}_{,b}} \xi^b + \Pi^{ij} \tensor{\xi}{^b_{,b}} \right) (x) \;. \label{2.51} 
\end{eqnarray}
The same result may be obtained using \eqref{1.40} and from an analogous procedure to the one used on the second way to calculate $\delta h_{ij}(x)$.

The results \eqref{2.50} and \eqref{2.51} show that the constraint $\mathcal{H}_i \approx 0$ is the generator of infinitesimal general transformations of spatial coordinates $(x'^i = x^i + \xi^i (x))$ since they imply, respectively, in $ \delta h_{ij}(x) = \underset{\xi^{l}}{\pounds} (h_{ij})$ and $\delta \Pi^{ij}(x) = \underset{\xi^{l}}{\pounds} (\Pi^{ij})$, where $\underset{\xi^{l}}{\pounds}(A^{ij})$ represents the Lie derivative of $A^{ij}$ three-tensor on the direction $\xi^l (x)$. Because the Hamiltonian \eqref{2.37} and the action \eqref{2.1} are invariant under diffeomorphisms on the three-surface, we conclude that the first class constraints $\mathcal{H}_i \approx 0$ are generators of those gauge transformations.

\begin{itemize}
	\item The constraint $\mathcal{H}_0 \approx 0$ yields
\end{itemize}
\begin{eqnarray} \label{2.52}
\nonumber \delta h_{ij}(x) &=& \left\{ h_{ij}, \int d^3y \epsilon(y) \mathcal{H}_0(y) \right\} \\ \nonumber
&=& \int d^3y \epsilon(y) G_{abkl}(y) \Pi^{ab}(y) \left\{ h_{ij},\Pi^{kl}(y) \right\}\\ \nonumber
&=& \epsilon(x) h^{-1/2}(x) \left( 2 \Pi_{ij} - \Pi h_{ij} \right)(x) \\ \nonumber
&=& -\epsilon(x) 2K_{ij}(x) \\ 
&=& - \epsilon(x) 2 \underset{n^{\mu}}{\pounds} (h_{ij}) \;.
\end{eqnarray}
Thus, this constraint generates a translation in the orthogonal direction to the three-surface $t= \text{constant}$, that is, it generates the system dynamics along the direction of the $0$ component of $n^\mu (x)$. This constraint arises due to invariance under GR time re-parametrisation.%, it is the generator of time translations ($t$ in \eqref{2.31} and $\tau (x)$ associated to $n^\mu (x)$ in \eqref{2.52}\footnote{Whereas the total Hamiltonian is the generator of translations of the time parameter ($\tau$, in \eqref{2.21} and $t$ on \eqref{2.37}).}), and the associated Lagrange multiplier\footnote{The shift function $N$, see \eqref{2.22} and the 00 component in \eqref{1.14} and its subsequent discussion.} corresponds to the rate of change of time with respect to the parameter.

%% file: sections/appendix2.tex
\chapter{Modern Cosmology}

\section{Einstein equations in FLRW spacetime}
\label{AppFLRWEinstein}

We give the derivation of Einstein equations in a homogeneous and isotropic spacetime. We start from the metric \eqref{flrw}
\begin{align}
\textup{d}s^2= -\textup{d}t^2+a^2(t)\left[\frac{\textup{d}r^2}{1-kr^2}+r^2\textup{d}\theta^2+r^2 \sin^2\theta \textup{d}\phi^2\right] \;.
\end{align}
The Christoffel symbols are given by
\begin{align}
\tensor{\Gamma}{^0_1_1} &= \frac{a \dot{a}}{1-kr^2} \;, \quad  \tensor{\Gamma}{^0_2_2} = a\dot{a}r^2 \;, \quad \tensor{\Gamma}{^0_3_3} =  a\dot{a}r^2 \sin^2\theta \;, \quad
\tensor{\Gamma}{^1_1_1} = \frac{kr}{1-kr^2} \;, \nonumber \\ \tensor{\Gamma}{^1_2_2} &= -r\left(1-kr^2\right) \;, \quad \tensor{\Gamma}{^1_3_3} =  -r\left(1-kr^2\right) \sin^2\theta \;, \quad
\tensor{\Gamma}{^2_3_3} = -\sin \theta \cos \theta \;, \nonumber \\  \tensor{\Gamma}{^1_0_1} &= \tensor{\Gamma}{^2_0_2} = \tensor{\Gamma}{^3_0_3} = \frac{\dot{a}}{a} \;, \quad \tensor{\Gamma}{^2_1_2} = \tensor{\Gamma}{^3_1_3} = \frac{1}{r} \;, \quad
\tensor{\Gamma}{^3_2_3} = \cot \theta \;.
\end{align}
The Ricci tensor takes the form
\begin{align}
R_{00} &= -3\frac{\ddot{a}}{a} \;, \quad R_{11} = \frac{a\ddot{a}+2\dot{a}^2+2k}{1-kr^2} \;, \nonumber \\
R_{22} &= r^2 \left(a\ddot{a}+2\dot{a}^2+2k\right) \;, \quad R_{33} = r^2 \left(a\ddot{a}+2\dot{a}^2+2k\right) \sin^2 \theta \;.
\end{align}
The Ricci scalar $R=g^{\mu \nu} R_{\mu \nu}$ is
\begin{align}
R = 6 \left(\frac{\ddot{a}}{a}+\frac{\dot{a}^2}{a^2}\right) \;.
\label{ricci scalar}
\end{align}
Then, the Einstein tensor $G_{\mu \nu} = R_{\mu \nu} -R/2g_{\mu\nu}$ reads
\begin{align}
G_{00} &= 3 \left[ \left(\frac{\dot{a}}{a}\right)^2 +\frac{k}{a^2}\right] \;, \quad  G_{11} = - \left[ \left(\frac{\dot{a}}{a}\right)^2 + 2 \frac{\ddot{a}}{a} +\frac{k}{a^2}\right] \frac{a^2}{1-kr^2} \;, \nonumber \\
G_{22} &= - \left[ \left(\frac{\dot{a}}{a}\right)^2 + 2 \frac{\ddot{a}}{a} +\frac{k}{a^2}\right] a^2 r^2 \; \quad G_{33} = - \left[ \left(\frac{\dot{a}}{a}\right)^2 + 2 \frac{\ddot{a}}{a} +\frac{k}{a^2}\right] a^2 r^2\sin^2 \theta \;.
\end{align}

%% file: sections/appendix3.tex
	\chapter{Stochastic Effects in Cosmology}

\section[Mapping]{Mapping between $p$, $\lambda^{2}$, $w$, $\nu$ and $n_{s}-1$}\label{AppMapping}
Throughout this work, we use the quantities $p$, $\lambda^{2}$, $w$ and $\nu$ because, even if they are connected to the others, each one of them is more appropriate for a specific analysis. In order to facilitate the understanding of the reader, we show the explicit mapping between them in 
Table \ref{table3}.
\begin{table}[h!]
	\centering
	\begin{tabular}{c|c|c|c|c|c}
		&\large $p$ \hspace{0.2cm}& \large $\lambda^{2}$ & \hspace{0.2cm} \large $w$ \hspace{0.2cm} & \large $\nu$ \hspace{0.2cm} & \large $n_{s}-1$ \\\hline\hline
		\large $p$ \hspace{0.2cm} & $p$ \hspace{0.2cm}& $\frac{2}{\lambda^{2}}$ \hspace{0.2cm}& $\frac{2}{3(1+w)}$ \hspace{0.2cm}& $\frac{2\nu -1}{2\nu - 3}$ \hspace{0.2cm}& $\frac{4-\left(n_{s}-1\right)}{6-\left(n_{s}-1\right)}$\\ \hline
		\large $\lambda^{2}$ \hspace{0.2cm} & $\frac{2}{p}$ \hspace{0.2cm}& $\lambda^{2}$ \hspace{0.2cm}& $3(1+w)$ \hspace{0.2cm}& $\frac{4\nu - 6}{2\nu -1}$ \hspace{0.2cm}& $\frac{2\left(n_{s}-1\right)-12}{\left(n_{s}-1\right)-4}$\\ \hline
		\large $w$ \hspace{0.2cm} & $\frac{2-3p}{3p}$ \hspace{0.2cm}& $\frac{\lambda^{2}-3}{3}$ \hspace{0.2cm}& $w$ \hspace{0.2cm}& $\frac{-2\nu - 3}{6\nu - 3}$ \hspace{0.2cm}& $\frac{\left(n_{s}-1\right)}{12-3\left(n_{s}-1\right)}$\\ \hline
		\large $\nu$ \hspace{0.2cm} & $\frac{3}{2}+\frac{1}{p-1}$ \hspace{0.2cm}& $\frac{3}{2}+\frac{\lambda^{2}}{2-\lambda^{2}}$ \hspace{0.2cm}& $\frac{3}{2}-\frac{3(1+w)}{1+3w}$ \hspace{0.2cm} & $\nu$ \hspace{0.2cm}& $\frac{3\left(n_{s}-1\right)}{8}-\frac{3}{2}$ \\ \hline
		\large $n_{s}-1$ \hspace{0.2cm}& $\frac{6p-4}{p-1}$ \hspace{0.2cm}& $\frac{4(3-\lambda^2)}{2-\lambda^2}$ \hspace{0.2cm}& $\frac{12w}{1+3w}$ \hspace{0.2cm}& $\frac{4(2\nu+3)}{3}$ \hspace{0.2cm}& $n_{s}-1$
	\end{tabular}
	\caption[Relation between $p$, $\lambda^{2}$, $w$, $\nu$ and $n_{s}-1$ in terms of each of them]{Table showing how to write $p$, $\lambda^{2}$, $w$, $\nu$ and $n_{s}-1$ in terms of each of them.}\label{table3}
\end{table}

\section{Kinetic-dominated solution}
\label{AppKinetic}

We show in this appendix the solutions for the other critical point, namely the kinetic-dominated regime. This is interesting for two reasons. First, this regime corresponds to the critical value $\nu=0$, which is the interface between purely adiabatic perturbations (at first-order) and non-adiabatic perturbations. Second, perturbations in this regime act as a stiff fluid and go as $a\propto (-\eta)^{-6}$. Such behaviour is usually invoked in the classical resolution of the initial singularity, see for instance the reviews \cite{Peter:2008qz,Battefeld:2014uga,Lilley:2015ksa,Brandenberger:2016vhg}. 

To begin, note we can rewrite (\ref{eq:delta x total}) in terms of the variables $x$ and $y$ as
\begin{align}
\delta x = \frac{\kappa}{6}\left[\left(1 - x^{2}\right)\frac{\dot{\delta\varphi}}{H} + \left(3x^{4}-3x^{2}+\frac{\lambda^{2}}{2}y^{2}\right)\delta\varphi\right]\;.
\end{align}
In the case of the kinetic-dominated solution, the fixed points are $x_a =\pm 1$, $y_a=0$. In this configuration, we have $w=1$, or equivalently $\lambda^2=6$, resulting in the trivial expression
\begin{align}
\delta x = 0 \;,
\end{align}
regardless of the value of the solution $\delta\varphi$. Hence, any first-order linear field perturbation leads to adiabatic perturbations in the kinetic-dominated regime.

\section{A different approach to find the noise}

\label{AppPerturbedMomentum}
\label{appendixc}	

We know that $\dot{\varphi}$ can be related to its momentum $\pi\equiv \partial {\cal L}/\partial \dot\varphi$ using the ADM formalism, as shown in \cite{Grain:2017dqa}, by
\begin{align}
\dot{\varphi} = \frac{1}{a(t)^{3}}\pi_{\varphi}\;,\label{eq:phipi2}
\end{align}
where the lapse function $N$ was chosen as the cosmic time, which means that $N=1$. Also, the evolution for $\pi_{\varphi}$ is given by
\begin{align}
\dot{\pi}_{\varphi} = - a(t)^{3} V_{,\varphi}%+a(t)\delta^{ij}\partial_{i}\partial_{j}\varphi
\;,
\end{align}
where ``,$\varphi$" represents the derivative with respect to $\varphi$.

The scalar field and its momentum can be split into a long-wavelength part and small-wavelength part as
\begin{align}
\varphi = \overline{\varphi} + \varphi_{Q} \quad \pi = \overline{\pi} + \pi_{Q}\;,
\end{align}
where the subscript ``$Q$" describes the small-wavelength part, which will allow us to calculate the quantum noise for $\varphi$ and $\pi$. 

We get the linearly perturbed momentum including scalar field perturbations
\begin{align}
\pi + \delta \pi = \frac{\partial \left( \mathcal{L}+\delta \mathcal{L}\right)}{ \partial \left(\frac{1}{1+A}\dot{\varphi} \right)} = \left(1+A\right) \frac{\partial \left( \mathcal{L}+\delta \mathcal{L}\right)}{ \partial \dot{\varphi}}
\end{align}
where we have also perturbed the lapse function $t\rightarrow(1+A)t$.
The perturbed momentum is then
\begin{align}
\delta \pi &= A \frac{\partial \mathcal{L}}{\partial \dot{\varphi}} + \frac{\partial \delta \mathcal{L}}{\partial \dot{\varphi}} \nonumber \\
%&= A a^3 \dot{\varphi} + a^3 \left(\dot{\delta \varphi}-2A\dot{\varphi}\right) \nonumber \\
&= a^3 \left(\dot{\delta \varphi}-A\dot{\varphi}\right)
\end{align}
We may use the constraint $A = \kappa^{2}\dot{\varphi}\delta\varphi/2H$ to eliminate the perturbed lapse function since we are working in the spatially-flat gauge \cite{Pattison:2019hef}.  

Using the definition of the coarse-graining scale \eqref{cgscale} in the expressions for the field and its conjugate momentum
\begin{align}
\delta\varphi &= \frac{i}{\sqrt{4\pi}}\left(\frac{2}{2\nu -1} \right) \frac{2^{|\nu|}\Gamma(\lvert\nu\rvert)}{k^{|\nu|}} \frac{H}{(-\eta)^{|\nu|-3/2}}\;,\label{eq:deltaphisol}\\
\delta\pi_{\varphi} &= \frac{i}{\sqrt{4\pi}}\frac{2^{|\nu|}\Gamma(\lvert\nu\rvert)}{k^{\lvert\nu\rvert}}\left(\nu - \frac{1}{2}\right)^{4}\left[\left(\frac{2}{2\nu - 1}\right)\left(\lvert\nu\rvert - \nu\right) -  \frac{\kappa^2 \dot{\varphi}^{2}}{2H^{2}}\right]\frac{1}{H(-\eta)^{\lvert\nu\rvert+3/2}} \;,\label{eq:pisol}
\end{align}
we get
\begin{align}
\lvert\delta\varphi\rvert^2 &= \frac{\Gamma^{2}(\lvert\nu\rvert)2^{2\lvert\nu\rvert}H^{2}}{4\pi \sigma^{2\lvert\nu\rvert}} \left(\frac{2}{2\nu - 1}\right)^{2\lvert\nu\rvert + 2}\frac{1}{(-\eta)^{-3}}\;, \\
\lvert\delta\pi\rvert^2 &= \frac{\Gamma^{2}(\lvert\nu\rvert)2^{2\lvert\nu\rvert}}{4\pi \sigma^{2\lvert\nu\rvert} H^{2}}\left(\frac{2}{2\nu - 1}\right)^{2\lvert\nu\rvert -4}\left[\left(\frac{2}{2\nu - 1}\right)\left(\lvert\nu\rvert - \nu\right) - 4\pi G \frac{\dot{\varphi}^{2}}{H^{2}}\right]^{2}\frac{1}{(-\eta)^{3}}\;, \\
\delta\varphi \delta\pi^{*} & = \frac{\Gamma^{2}(\lvert\nu\rvert)2^{2\lvert\nu\rvert}}{4\pi \sigma^{2\lvert\nu\rvert}} \left(\frac{2}{2\nu - 1}\right)^{2\lvert\nu\rvert -1} \left[\left(\frac{2}{2\nu - 1}\right)\left(\lvert\nu\rvert - \nu\right) -4\pi G \frac{\dot{\varphi}^{2}}{H^{2}}\right]\;.
\end{align}

From this we can work out the two-points correlation matrix associated with the quantum noise with respect to conformal time $\xi_{\varphi}$ and $\xi_{\pi}$ \cite{Grain:2017dqa}
.\begin{align}
\Xi_{\varphi, \varphi}^{(\eta)} =& \frac{1}{(2\pi)^{3}}\frac{\Gamma^{2}(\lvert\nu\rvert)2^{2\lvert\nu\rvert}}{\sigma^{2\lvert\nu\rvert - 3}} \left(\frac{2}{2\nu -1}\right)^{2\lvert\nu\rvert -1}\frac{H^{2}(\eta)}{(-\eta)} \;,\\
\Xi_{\pi, \pi}^{(\eta)} =& \frac{1}{(2\pi)^{3}}\frac{\Gamma^{2}(\lvert\nu\rvert)2^{2\lvert\nu\rvert}}{\sigma^{2\lvert\nu\rvert - 3}}\left(\frac{2}{2\nu -1}\right)^{2\lvert\nu\rvert -7}\nonumber\\
&\times \left[\left(\frac{2}{2\nu - 1}\right)\left(\lvert\nu\rvert - \nu\right) -  4\pi G \frac{\dot{\varphi}^{2}}{H^{2}(\eta)}\right]^{2}\frac{1}{H^{2}(\eta)(-\eta)^{7}} \;,\\
\Xi_{\varphi, \pi}^{(\eta)} = \Xi_{\pi, \varphi}^{(\eta)} =& \frac{1}{(2\pi)^{3}}\frac{\Gamma^{2}(\lvert\nu\rvert)2^{2\lvert\nu\rvert}}{\sigma^{2\lvert\nu\rvert - 3}} \left(\frac{2}{2\nu -1}\right)^{2\lvert\nu\rvert -4}\nonumber\\
&\times \left[\left(\frac{2}{2\nu - 1}\right)\left(\lvert\nu\rvert - \nu\right) -  4\pi G \frac{\dot{\varphi}^{2}}{H^{2}(\eta)}\right]\frac{1}{(-\eta)^{4}}\;,
\end{align}
with the dependence in conformal time is left explicitly in the subscript. We can write a stochastic version for $x$ of the form
\begin{align}
\dot{x} = \bar{x}+\xi_x \;,
\end{align}
and relate the noise of $x$ to the noises of $\varphi$ and $\pi_{\varphi}$. By doing so, it can be shown the correlation matrix in $x$ is a combination of those contributions,
\begin{align}
\label{xix}
\left\langle \xi_{x}  \xi_{x} \right\rangle := \Xi_{x,x}^{(\eta)} = \frac{a^{12}}{(\bar{\pi}_{\varphi}^2+2a^6V)^{3}} \left[4V^2 \Xi_{\pi, \pi}^{(\eta)}+(\bar{\pi}_{\varphi}V^{\prime})^2\Xi_{\varphi, \varphi}^{(\eta)} -4VV^{\prime}\bar{\pi}_{\varphi} \Xi_{\varphi, \pi}^{(\eta)}\right] \;,
\end{align}
and using the expression of the correlation matrix \eqref{eq:Xifg} in terms of number of e-folds as
\begin{align}
\Xi_{x,x}^{(N)}&=\frac{d\eta}{dN} \; \Xi_{x,x}^{(\eta)} \nonumber\\
&= (-\eta_{\star}) \frac{1}{\nu-1/2} \exp\left[\frac{1}{\nu-1/2}\left(N_{\star}-N\right)\right]  \Xi_{x,x}^{(\eta)}\;,
\end{align}
it is straightforward to show we can recover our result \eqref{deltaxN}.

%Then, applying the above expressions in (\ref{xix}) we obtain exactly the same result as the (\ref{eq:Xixx}) at the critical point.

% We can also note that the perturbation in $x$ can be expressed in terms of perturbations of $\varphi$ and $\pi$ as
%\begin{align}
%\label{eq:delta x}
%\delta x= \frac{a^6}{(\pi_{\varphi}^2+2a^6V)^{3/2}}\left(2V\delta \pi-V^{\prime} \pi \delta \varphi\right) \;.
%\end{align}
%We note from \eqref{eq:delta x} and \eqref{eq:noise x} that the perturbation in $x$ and its noise are consistent with one another.
%We also note the noise associated to the EOS can be written as
%\begin{align}
%\xi_{w}&=4 \frac{\bar{\pi}_{\varphi}} {\sqrt{\bar{\pi}_{\varphi}^2+2a^6V}} \xi_{x} \nonumber \\
%&=\frac{4\bar{\pi}_{\varphi}a^6}{\left(\bar{\pi}_{\varphi}^2+2a^6V\right)^{2}}\left(2V \xi_{\pi}  - \bar{\pi}_{\varphi}  V^{\prime} \xi_{\varphi}\right) \;.
%\end{align}

\section{Next-to-leading order field contribution}
\label{appnextorder}
We expand the field solution \eqref{eq:delta_phi} and its derivative to third order to get all terms contributing to second order. Then the field is now
\begin{align}
\delta \varphi &= \frac{i}{a} \frac{\Gamma\left(|\nu|\right)2^{|\nu|}}{\sqrt{4\pi}k^{|\nu|}} \left[1+\frac{\left(-k\eta\right)^2}{4\left(|\nu|-1\right)}+\frac{\left(-k\eta\right)^4}{32\left(|\nu|-1\right)\left(|\nu|-2\right)}\right] \frac{1}{\left(-\eta\right)^{|\nu|-1/2}} \nonumber \\
&= \frac{i}{\sqrt{4\pi}}\left(\frac{2}{2\nu -1} \right) \frac{2^{|\nu|}\Gamma(\lvert\nu\rvert)}{k^{|\nu|}} \nonumber \\
&\quad \times \left[ \frac{H}{(-\eta)^{|\nu|-3/2}}+ \frac{k^2H}{4\left(|\nu|-1\right) (-\eta)^{|\nu|-7/2}} + \frac{k^4H}{32\left(|\nu|-1\right)\left(|\nu|-2\right) (-\eta)^{|\nu|-11/2}}\right] \;, \\
\dot{\delta \varphi} &= \frac{i}{\sqrt{4\pi}}\left(\frac{2}{2\nu -1} \right)^2 \frac{2^{|\nu|}\Gamma(\lvert\nu\rvert)}{k^{|\nu|}} H^2 \nonumber \\
&\quad \times \left[\frac{\left(|\nu|-\nu\right)}{ (-\eta)^{|\nu|-3/2}}+ \frac{k^2 \left(|\nu|-\nu-2\right)}{4\left(|\nu|-1\right) (-\eta)^{|\nu|-7/2}} + + \frac{k^4 \left(|\nu|-\nu-4\right)}{32\left(|\nu|-1\right)\left(|\nu|-2\right) (-\eta)^{|\nu|-11/2}}\right] \;.
\end{align}

The contribution to the noise in $x$ becomes
%\begin{align}
%\delta x = \frac{i }{\sqrt{24\pi}}\left(1-\frac{\lambda^{2}}{6}\right)\left(\frac{2}{2\nu - 1}\right)^{2}\frac{\Gamma(\lvert\nu\rvert)2^{\lvert\nu\rvert}}{k^{\lvert\nu\rvert}} \kappa H(\eta) \left[\frac{\left(|\nu|-\nu\right)}{ (-\eta)^{|\nu|-3/2}} + \frac{k^2 \left(|\nu|-\nu-2\right)}{4\left(|\nu|-1\right) (-\eta)^{|\nu|-7/2}}\right] \;,
%\end{align}
\begin{align}
\Xi_{x,x}(N) &= \bar{g}(\nu,\sigma) \kappa^{2} H^{2}_{\star} \exp{\left[-\frac{3-2\nu}{\nu-1/2} \left(N_{\star}-N\right)\right]}
\;,
\end{align}
with
\begin{align}
\bar{g}(\nu,\sigma) :=  &\frac{\Gamma^{2}(\lvert\nu\rvert)2^{2\lvert\nu\rvert +2}}{(12\pi)^{3}}\left(\frac{\nu^{2}}{\sigma^{2\lvert\nu\rvert - 3}}\right)\left(\frac{2}{2\nu - 1}\right)^{2\lvert\nu\rvert+4} \left[\left(\lvert\nu\rvert - \nu\right)^{2} +\frac{\sigma^2}{2} \frac{\left(\lvert\nu\rvert - \nu\right)\left(\lvert\nu\rvert - \nu -2\right)}{\left(\lvert\nu\rvert - 1\right)} \left(\nu - \frac{1}{2}\right)^2 \right. \nonumber \\
&\left. +\frac{\sigma^4}{16} \frac{\left[\left(\lvert\nu\rvert - \nu\right)^2-4\left(\lvert\nu\rvert - \nu\right)\right]\left[2\lvert\nu\rvert-3\right]+4\left(\lvert\nu\rvert-2\right)}{\left(\lvert\nu\rvert - 1\right)^2\left(\lvert\nu\rvert - 2\right)} \left(\nu - \frac{1}{2}\right)^4 \right]\;.
\end{align}
For positive $\nu$, the above equation is simplified to
\begin{align}
\bar{g}(\nu > 0,\sigma) :=  &\frac{\Gamma^{2}(\lvert\nu\rvert)2^{2\lvert\nu\rvert +2}}{(12\pi)^{3}\sigma^{2\lvert\nu\rvert -7}}\left(\frac{\nu}{\lvert\nu\rvert - 1}\right)^{2}\left(\frac{2}{2\nu -1}\right)^{2\lvert\nu\rvert} \;.
\end{align}
The variance in the case of quasi-de Sitter inflation  ($\nu=3/2-\epsilon$) is given by
\begin{align}
\sigma_{x,qu}^2 &= \frac{1}{24 \pi^2}  H_{\star}^2 \kappa^2 \sigma^{4-\epsilon}\left(1+4\epsilon\right) \left\lbrace 1- \exp{\left[-6(N-N_{\star})\right]}\right\rbrace \;.
\end{align}
Since $N$ is growing with time the exponential vanishes quickly and eventually the time at which $\sigma_{x,qu}^2=1$ is 
\begin{align}
H(N)=\sqrt{3\pi}\left(1-2\epsilon\right)\frac{M_{pl}}{\sigma^{2-\epsilon/2}} \;,
\end{align}
and since $\sigma<1$ we see quantum diffusion drives us away only if the Hubble rate is far above the Planck scale. We note this result stays true for $N\approx N_{\star}$ since in this case we have
\begin{align}
H(N)=\sqrt{\frac{\pi}{2}}\left(1-2\epsilon\right)\frac{M_{pl}}{\sigma^{2-\epsilon/2}} \;.
\end{align}

\section{Fourier transform on a finite domain}
\label{App2}
This Appendix is dedicated to show explicitly the solution given by (\ref{variance}), while pointing out that the final result depends on a conventional factor. From (\ref{variancedifeq}), we easily get 
\begin{align}
\label{stochastic variance}
\sigma_x^2 (N) = \sigma_x^2(N_{\star}) e^{2m\left(N-N_{\star}\right)} + 2\int_{N_{\star}}^{N}  dS \, e^{2m(N-S)} \left\langle \hat{\xi}_x(S) \left(\bar{x}(S)-x_c\right) \right\rangle.
\end{align}
We can re-express the expectation value in the second term using \eqref{stochastic solution x} to get
\begin{align}
\left\langle \hat{\xi}_x(S) \left(\bar{x}(S)-x_c\right) \right\rangle = \left\langle\int_{S_{\star}}^{S} e^{m\left(S-U\right)} \hat{\xi}_x(S)\hat{\xi}_{x}(U) dU \right\rangle\;.
\end{align}
Using the Fubini theorem, we get
\begin{align}
\left\langle \hat{\xi}_x(S) \left(\bar{x}(S)-x_c\right) \right\rangle &= \int_{S_{\star}}^{S} e^{m\left(S-U\right)} \left\langle \hat{\xi}_x(S)\hat{\xi}_{x}(U)  \right\rangle dU \;.
\end{align}
The variance \eqref{stochastic variance} is then
\begin{align}
\sigma_x^2 (N) = \sigma_x^2(N_{\star}) e^{2m\left(N-N_{\star}\right)} + 2\int_{N_{\star}}^{N} \int_{S_{\star}}^{S} dS dU \, e^{m(2N-S-U)} \left\langle \hat{\xi}_x(S)\hat{\xi}_{x}(U) \right\rangle \;.
\end{align}
The resolution of \eqref{stochastic variance} leads us to consider the following integral:
\begin{align}
\int_{S_{\star}}^{S} dU e^{m\left(2N-S-U\right)} \delta 
\left(S-U\right) \;.
\end{align}
For a general function, we have
\begin{align}
\int_{a}^{b} f(x) \delta (x-x^{\prime}) dx &= \int_{-\infty}^{\infty} dx \, f(x) \left[\theta(x-a)-\theta(x-b)\right]\delta (x-x^{\prime})  \nonumber \\
&= f(x^{\prime}) \left[\theta(x^{\prime}-a) -\theta(x^{\prime}-b) \right] \;,
\end{align}
what gives in our case
\begin{align}
\int_{S_{\star}}^{S} dU e^{m\left(2N-S-U\right)} \delta 
\left(S-U\right) &= e^{2m\left(N-S\right)} \left[\theta(S-S_{\star}) -\theta(S-S)\right] \nonumber \\
&= e^{2m\left(N-S\right)} \left[\theta(S-S_{\star}) -\theta(0)\right] \;. 
\end{align}
Using the half-maximum convention for the unit step function, we obtain
\begin{align}
\int_{S_{\star}}^{S} dU e^{m\left(2N-S-U\right)} \delta 
\left(S-U\right) &= \frac{1}{2} e^{2m\left(N-S\right)} \;.
\end{align}
Now, we are able to write the full solution for the variance in $x$ as
\begin{align}
\sigma_x^2 (N) = \sigma_x^2(N_{\star}) e^{2m\left(N-N_{\star}\right)} + \int_{N_{\star}}^{N}  dS \, e^{2m(N-S)} \Xi_{x,x}(S)  \;.
\end{align}

%	\section{Solutions for $a(t)$ and $\varphi(t)$ for the scalar field in a flat universe}\label{App1}
%	The Friedmann and Klein-Gordon equations, given by (\ref{eq:friedmann1}) and (\ref{eq:KGback}), can be combined to derive the continuity equation for matter
%	\begin{align}
%	\dot{\rho_{\varphi}} + 3 H (1+w)\rho_{\varphi} = 0\;,
%	\end{align}
%	which provides $\rho_{\varphi}= \rho_{0} a^{-3(1+w)}$. Now we can use this solution in the Friedmann equation in order to obtain the scale factor in terms of time $a(t) = a_{0}t^{2/3(1+w)}$. 
%	
%	Following the addition of first and second Friedmann equations, we get
%	\begin{align}
%	\left(\frac{\dot{a}}{a}\right)^{2} - \frac{\ddot{a}}{a} = \frac{\kappa^{2}}{2}\dot{\varphi}^{2}\;,
%	\end{align}
%	here we can apply the above scale factor solution to give
%	\begin{align}
%	\dot{\varphi}^{2} = \frac{4}{3\kappa^{2}(1+w)}\frac{1}{t^{2}}\;,
%	\end{align}
%	and it has as solution
%	\begin{align}
%	\varphi (t) = \sqrt{\frac{4}{3\kappa^{2}(1+w)}} \ln t + C\;,
%	\end{align}
%	where $C$ is an integration constant.

%% file: sections/appendix4.tex
	\chapter{Quantum Cosmology}

\section{The Schutz Formalism}
\label{schutz}

The theory of relativistic perfect fluids with velocity potential was first formulated by Schutz \cite{Schutz:1970my,Schutz:1971ac}. Later, Lapchinskii and Rubakov used this framework to discuss the physics of a quantised Friedmann universe filled with a perfect fluid \cite{Lapchinsky:1977vb}. The idea is to introduce five scalar fields  as independent variables: $\alpha$ and $\beta$, describing the vortex motion and set to zero in a Friedmann universe because of its symmetry, the entropy $S$, and $\theta$, $\epsilon$ whose physical meaning is unclear.

The fluid's four-velocity is a combination of these potentials
\begin{align}
U_{\nu} &= \mu^{-1} \left(\epsilon_{,\nu}+\alpha\beta_{,\nu}+\theta S_{,\nu}\right)\\
&= \mu^{-1} \left(\epsilon_{,\nu}+\theta S_{,\nu}\right) \;.
\end{align}
where $\mu$ is the specific enthalpy of the fluid
\begin{align}
\mu= \frac{\rho + p}{\rho_{r}} \;,
\end{align}
with $\rho$ the particle number density, $p$ the pressure and the subscript $r$ means we evaluate in the rest frame. Since the four-velocity only has one non-null component in this model $U_{\nu}=(N,0,0,0)$, with $N$ the lapse function, we can relate the enthalpy to the potentials through
\begin{align}
\mu = \frac{\dot{\epsilon}+\theta \dot{S}}{N} \;.
\end{align}

For a perfect fluid with a barotropic equation of state $p=w\rho$ with $w$ constant, the introduction of a deviation $\Pi$ in the rest-mass density leads to the thermodynamic relation
\begin{align}
\rho=\rho_{r}\left(1+\Pi\right) \,,  \quad \mu=1+\Pi+\frac{p}{\rho_{r}}  \,,  \quad \tau dS=\left(1+\Pi\right)d\left[\ln\left(1+\Pi\right)-w\ln \rho_{r}\right] \;,
\end{align}
with $\tau$ the temperature, and by identification
\begin{align}
\tau = 1+\Pi \,, \quad  S=\ln\left(1+\Pi\right)-w\ln \rho_{r} \;.
\end{align}
Finally, it is straightforward to express the pressure in terms of the enthalpy and the entropy as 
\begin{align}
p=w \left( \frac{\mu}{1+w} \right)^{1+\frac{1}{w}} e^{-\frac{S}{w}} \;.
\end{align}

We are now able to compute the Lagrangian for a perfect fluid in the Friedmann model
\begin{align}
\mathcal{L}_{M} &= 16 \pi \sqrt{^{4}g} p(\epsilon,\theta,S) \\
&=16 \pi N^{-\frac{1}{w}} a^3 \frac{w}{(1+w)^{1+\frac{1}{w}}}\left(\dot{\epsilon}+\theta \dot{S}\right)^{1+\frac{1}{w}}e^{-\frac{S}{w}} \;.
\end{align}
with $\sqrt{^{4}g}$ the volume element and $a$ the scale factor.

\section{Initial Conditions for the Universe Wave Function}
\label{section initial conditions wave function}

Since the domain of the scale factor, and equivalently of $\chi$, is $\mathbb{R}_+^{\star}$, boundary conditions on the wave function must be defined in order to retain self-adjointness for the Hamiltonian operator. In the case $k=0$ leading to \eqref{wave function schrodinger}, the operator $\hat{H}$ is self-adjoint in the inner product
\begin{align}
\left(\psi, \phi\right) = \int_{0}^{\infty} \textup{d}\chi \, \psi^{\star}(\chi) \phi(\chi) \;,
\label{inner product}
\end{align}
iff (if and only if) it is symmetric
\begin{align}
\left(\psi, \hat{H} \phi \right) = \int_{0}^{\infty} \textup{d}\chi \, \psi^{\star}(\chi) \frac{1}{24} \frac{\partial^2 \phi(\chi) }{\partial^2 \chi} = \int_{0}^{\infty} \textup{d}\chi \,  \frac{1}{24} \frac{\partial^2  \psi^{\star}(\chi) }{\partial^2 \chi}  \phi(\chi) = \left(\hat{H} \psi, \phi\right) \;.
\end{align}
It is known that the squared-momentum operator is symmetric on the set of square-integrable smooth functions $L^2(0,\infty)$, and the wave functions obey the relation
\begin{align}
\frac{\partial \Psi (\chi,T)}{\partial \chi} \bigg\rvert_{\chi =0} = \alpha \Psi (\chi,T) \bigg\rvert_{\chi =0} \;,
\label{condition symmetry}
\end{align} 
with $\alpha$ an arbitrary number, generally chosen null or infinite. Choosing $\alpha=0$ amounts to considering a boundary condition of the form
\begin{align}
\frac{\partial \Psi (\chi,T)}{\partial \chi} \bigg\rvert_{\chi =0} = 0\;.
\end{align}
This follows from the conception of an extended superspace, where the scale factor could be extended to negative values as well. The second choice, $\alpha=\infty$, implies that
\begin{align}
\Psi (\chi,T) \bigg\rvert_{\chi =0} = 0 \;.
\end{align}
Then, any wave function obeying \eqref{condition symmetry} also obeys the condition
\begin{align}
\left(\Psi^{\star} \frac{\partial \Psi}{\partial \chi} - \Psi \frac{\partial \Psi^{\star}}{\partial \chi} \right) \bigg\rvert_{\chi=0} = 0 \;.
\end{align}

It is instructive to have a look at the stationary solutions of \eqref{wave function schrodinger}, whose solutions are of the form
\begin{align}
\Psi (\chi,T) = A(\chi) e^{iET} \;,
\end{align}
with $E$ a real parameter.  We now show under what conditions this parameter is always positive. This is of importance since we can consider $E$ as the total energy of the system, and a positive value is thus necessary to maintain the system stable. Some authors even consider $E$ as a cosmological constant \cite{hildebrand1962advanced}. Inserting the solution into \eqref{wave function schrodinger} leads to
\begin{align}
\frac{1}{24} \frac{\partial^2 A(\chi)}{\partial \chi^2} = - E A(\chi) \;.
\label{differential equation A}
\end{align}
Multiplying this result by $A^{\star}$ and integrating over the domain of $\chi$, the half-line, we obtain
\begin{align}
\frac{1}{24} \int_{0}^{\infty} \textup{d}\chi \, A^{\star}\frac{\partial^2 A(\chi)}{\partial \chi^2}  = - E \int_{0}^{\infty} \textup{d}\chi \, |A(\chi)|^2 \;.
\end{align}
Then, integrating by part the left-hand size and using condition \eqref{condition symmetry}, we obtain
\begin{align}
E = \frac{1}{24} \frac{\alpha |A(0)|^2 + \int_{0}^{\infty} \textup{d}\chi \, |\partial A(\chi) / \partial \chi|^2}{\int_{0}^{\infty} \textup{d}\chi \, |A(\chi)|^2} \;.
\end{align}
Then, $E\geq 0$ iff we have $\alpha \geq 0$.

The general solution of \eqref{differential equation A} is 
\begin{align}
A(\chi) = C_1 e^{\sqrt{24E}\chi} + C_2 e^{-\sqrt{24E}\chi} \;,
\end{align}
with $C_1$, $C_2$ constants. Choosing the first boundary conditions leads to 
\begin{align}
A(\chi)^{(1)} = \sinh(\sqrt{24E}\chi) \;,
\end{align}
while the second reads
\begin{align}
A(\chi)^{(2)} = \cosh(\sqrt{24E}\chi) \;,
\end{align}
with the superscript referring to which boundaries we use. These two solutions have an infinite norm with respect to the inner product \eqref{inner product}. They can be seen as an analogy with planes waves in quantum mechanics. Then, a solution with finite norm can be obtained by superposing them.

%% file: sections/appendix5.tex
	\chapter{Affine Quantisation and Cosmology}

\section{Example of Affine Quantisation}
\label{example affine quantisation}

We show in this section a step-by-step example of how to compute the quantum operator of the classical function $f(q,p)=q^2 p^2$ in the affine quantisation framework. Starting from the quantisation map \eqref{int. quant. map} and the unitary irreducible representation \eqref{U+} with $\alpha=-1$, we find

\begin{eqnarray}
\nonumber
\langle x| A_{q^2p^{2}} | \phi \rangle &=& \int_{-\infty}^{\infty} \int_{0}^{\infty} \frac{dq}{q} \, \frac{dp}{2\pi \,c_{-1}} \int_{0}^{\infty} dx^{\prime} \, q^2p^{2} e^{-ip(x^{\prime} - x)} \psi \left(\frac{x}{q}\right) \psi \left(\frac{x^{\prime}}{q}\right) \phi(x^{\prime})
\\
\nonumber
&=& - \int_{-\infty}^{\infty} \int_{0}^{\infty}  \frac{dq \, dp}{2\pi \,c_{-1}} q \int_{0}^{\infty} dx^{\prime} \, \partial_{x^{\prime}}^{2} \left[ e^{-ip(x^{\prime} - x)} \right] \psi \left(\frac{x}{q}\right) \psi \left(\frac{x^{\prime}}{q}\right) \phi(x^{\prime})
\\
\nonumber
&=& - \int_{-\infty}^{\infty} \int_{0}^{\infty}  \frac{dq \, dp}{2\pi \,c_{-1}} q \int_{0}^{\infty} dx^{\prime} \, e^{-ip(x^{\prime} - x)} \psi \left(\frac{x}{q}\right) \partial_{x^{\prime}}^{2} \left[\psi \left(\frac{x^{\prime}}{q}\right) \phi(x^{\prime}) \right]
\\
\nonumber
&=& -\int_{0}^{\infty}  \frac{dq}{c_{-1}} q\, \psi \left(\frac{x}{q}\right) \left[\psi^{\prime \prime} \left(\frac{x}{q}\right) \phi(x) + 2 \psi^{\prime} \left(\frac{x}{q}\right) \phi^{\prime}(x) + \psi \left(\frac{x}{q}\right) \phi^{\prime \prime}(x) \right]
\end{eqnarray}
Let us make a change of variables:
\begin{eqnarray}
u &=& \frac{x}{q}  \quad \Rightarrow \quad du = -\frac{x}{q^{2}} dq \quad \Rightarrow \quad dq = -\frac{x}{u^{2}} du
\\
\frac{\partial}{\partial x} &=& \frac{\partial u}{\partial x}\frac{\partial}{\partial u} = \frac{1}{q} \frac{\partial}{\partial u} = \frac{u}{x} \frac{\partial}{\partial u}
\end{eqnarray}
Thus, we obtain
\begin{eqnarray}
\nonumber
\langle x| A_{q^2p^{2}} | \phi \rangle &=& -\frac{1}{c_{-1}} \int_{0}^{\infty} du \frac{x}{u^{2}} \frac{x}{u}\, \psi \left(u\right) \left[\frac{u^{2}}{x^{2}} \ddot{\psi} \left(u\right) \phi(x) + 2 \frac{u}{x}  \dot{\psi} \left(u\right) \phi^{\prime}(x) + \psi \left(u\right) \phi^{\prime \prime}(x) \right]
\\
\nonumber
&=& -\frac{1}{c_{-1}} \int_{0}^{\infty} du \left[ \frac{1}{u} \psi \left(u\right) \ddot{\psi} \left(u\right) \phi(x) + 2 \frac{x}{u^2} \psi \left(u\right) \dot{\psi} \left(u\right) \phi^{\prime}(x) + \frac{x^2}{u^{3}} |\psi \left(u\right)|^{2} \phi^{\prime \prime}(x) \right]
\end{eqnarray}
where dots are the derivative with respect to $u$ and primes are derivatives with respect to $x$. We proceed further by integrating by parts and remembering that $\psi \in L^2\left(\mathbb{R}^{\star}_{+},dx\right)$:
\begin{eqnarray}
\nonumber
\langle x| A_{q^2p^{2}} | \phi \rangle &=& -\frac{1}{c_{-1}} \Bigg[\left(- \int_{0}^{\infty} |\dot{\psi} \left(u\right)|^{2}  \, \frac{du}{u} + \int_{0}^{\infty} \psi (u)\dot{\psi} \left(u\right) \, \frac{du}{u^2}\right) + 2 \int_{0}^{\infty} \frac{du}{u^{2}} \psi (u)\dot{\psi} \left(u\right) \, x \partial_{x} \\
& & + \int_{0}^{\infty} \frac{du}{u^{3}} |\psi \left(u\right)|^{2}  \,  x^2\partial_{x}^{2} \Bigg] \phi(x)
\\
\nonumber
&=&- \frac{1}{c_{-1}} \Bigg[\left(- c_{-1}^{(1)} + c_{1}\right) + 2 c_{1} \, x \partial_{x}  + c_1  \,  x^2\partial_{x}^{2} \Bigg] \phi(x)
\\
\nonumber
&=&  \left[ \frac{c_{-1}^{(1)}}{c_{-1}} - \frac{c_{1}}{c_{-1}} \left(1+2 x \, \partial_{x}+ \, x^2 \partial_{x}^{2} \right) \right] \phi(x)
\end{eqnarray}
where $\partial_{x}$ is a shortcut notation for $\partial/\partial x$

\section{Compendium of Quantum Operators and Their Semi-Classical Counterpart }
\label{compendium}

\subsection{Quantum Operators}

\begin{eqnarray}
\langle x| A_{q^{\beta}} | \phi \rangle &=&  \left[ \frac{c_{\beta-1}}{c_{-1}}x^{\beta} \right] \phi(x) \\
\langle x| A_{p} | \phi \rangle &=&  \left[ -i\frac{\partial}{\partial x} \right] \phi(x) \\
\langle x| A_{p^2} | \phi \rangle &=&  \left[-\frac{\partial^2}{\partial x^2} +\frac{c_{-3}^{(1)}}{c_{-1}}\frac{1}{x^2} \right] \phi(x) \\
\langle x| A_{qp} | \phi \rangle &=&  \left[ \frac{c_{0}}{c_{-1}} \left( -ix\frac{\partial}{\partial x} -\frac{i}{2} \right) \right] \phi(x) \\
\langle x| A_{qp^{2}} | \phi \rangle &=&  \left[ \frac{c_{-2}^{(1)}}{c_{-1}} \, \frac{1}{x} - \frac{c_{0}}{c_{-1}} \, \frac{\partial}{\partial x}  - \frac{c_{0}}{c_{-1}} \, x \frac{\partial^2}{\partial x^2} \right] \phi(x) \\
\langle x| A_{q^2p^{2}} | \phi \rangle &=&  \left[ \frac{c_{-1}^{(1)}}{c_{-1}} - \frac{c_{1}}{c_{-1}} \left(1+2 x \, \frac{\partial}{\partial x}+ \, x^2 \frac{\partial^2}{\partial x^2} \right) \right] \phi(x)\\
\langle x| A_{q^{-1}p} | \phi \rangle &=&  \left[-\frac{i}{c_{-1}}\frac{1}{x}\frac{\partial}{\partial x} +\frac{i}{2c_{-1}}\frac{1}{x^2} \right] \phi(x) \\
\langle x| A_{q^{-1}p^2} | \phi \rangle &=& \left[ \frac{1}{c_{-1}} q^{-1} p^2 - \frac{i}{c_{-1}} q^{-2} p - \frac{\left(1-c_{-4}^{(1)} \right)}{c_{-1}} q^{-3} \right] \phi(x)
\\
\langle x| A_{q^{\beta}p^{\beta}} | \phi \rangle &=&   -\frac{i^{\beta}}{c_{-1}} \int dq \, dx^{'} q^{\alpha-1} \delta(x^{'}-x) \Psi\left(\frac{x}{q}\right) \frac{\partial^{\beta}}{\partial x^{' \beta}} \left[\Psi\left(\frac{x^{'}}{q}\right)\phi(x^{'})\right] \nonumber\\
\end{eqnarray}

\subsection{Semi-Classical Operators}

\begin{eqnarray}
\check{q^{\beta}} &=&  \frac{c_{\beta-1} c_{-\beta-2}}{c_{-1}} q^{\beta} \\
\check{p} &=& p \\
\check{p^2} &=& p^2 + \left(c_{-2}^{(1)}+\frac{c_0 c_{-3}^{(1)}}{c_{-1}}\right) q^{-2}\\
\check{q p} &=& \frac{c_0 c_{-3}}{c_{-1}}q p   \\
\check{q^2 p^2} &=& \frac{c_1 c_{-4}}{c_{-1}}q^2 p^2 +\frac{c^{(1)}_{-1}+c_1 c^{(1)}_{-4}-c_1}{c_{-1}}   \\
\check{q^{-1}p} &=& q^{-1}p -i\frac{c_{0}}{2c_{-1}} q^{-2} \\
\check{q^{-1}p^2} &=& q^{-1}p^2 +\frac{c^{(1)}_{-1}+c_1 c^{(1)}_{-4}-c_1}{c_{-1}} q^{-3} \\
\end{eqnarray}

\section{Gravitational Lagrangian of the BDT}
\label{gravitational lagrangian BDT}

In a homogeneous and isotropic universe, the Ricci scalar and the metric determinant can be written as
\begin{align}
R &= - \frac{6}{a^2 N^3} \left[-aa^{\prime}N^{\prime}+N\left(a^{\prime 2}+aa^{\prime \prime}\right)\right], \nonumber \\
g &= -N^2a^6
\end{align}
hence the gravitational part of the Lagrangian \eqref{Lagrangian} becomes
\begin{align}
\mathcal{L}_{G} &= \sqrt{-g}\left\{ \varphi R-\omega \frac{\varphi _{;\rho }\varphi^{;\rho }}{\varphi }\right\} \nonumber \\
&=	Na^3 \left[-\frac{6\varphi}{a^2 N^3} \left(Naa^{\prime \prime}+Na^{\prime 2}-aa^{\prime}N^{\prime}\right)-\omega \frac{\dot{\varphi}^{2}}{N^2 \varphi }\right] \nonumber \\
&= 	 -6\varphi \left(\frac{a^2a^{\prime \prime}}{N}+\frac{aa^{\prime 2}}{N}-a^2a^{\prime}\frac{N^{\prime}}{N^2}\right)-\omega a^3 \frac{\dot{\varphi}^{2}}{N \varphi } \nonumber \\
&= 	 -6 \left(\varphi\frac{aa^{\prime 2}}{N}+\varphi a^2 \frac{d}{dt} \left[\frac{a^{\prime}}{N}\right]  \right)-\omega a^3 \frac{\dot{\varphi}^{2}}{N \varphi } \nonumber \\
&= 	 -6 \left(\varphi\frac{aa^{\prime 2}}{N}-\frac{a^{\prime}}{N} \frac{d}{dt} \left[a^2\varphi \right]\right)-\omega a^3 \frac{\dot{\varphi}^{2}}{N \varphi } \nonumber \\
&= \frac{1}{N} \left(6\varphi a a^{\prime 2} + 6 a^{\prime} a^2 \varphi^{\prime} - \omega a^3  \frac{\dot{\varphi}^{2}}{\varphi }\right)
\end{align}